%% file: ms.tex
\documentclass[manuscript,longabstract]{aastex}
\usepackage{graphicx}
\usepackage{natbib}
\usepackage{mydefs}
\usepackage{lscape}

\newcommand{\noprint}[1]{}
\newcommand{\figsetstart}{{\bf Fig. Set} }
\newcommand{\figsetend}{}
\newcommand{\figsetgrpstart}{}
\newcommand{\figsetgrpend}{}
\newcommand{\figsetnum}[1]{{\bf #1.}}
\newcommand{\figsettitle}[1]{ {\bf #1} }
\newcommand{\figsetgrpnum}[1]{\noprint{#1}}
\newcommand{\figsetgrptitle}[1]{\noprint{#1}}
\newcommand{\figsetplot}[1]{\noprint{#1}}
\newcommand{\figsetgrpnote}[1]{\noprint{#1}}

\shorttitle{Flare Spectra}
\shortauthors{Kowalski et al.}

\begin{document}


\title{Time-Resolved Properties and Global Trends in \lowercase{d}M\lowercase{e} Flares from
  Simultaneous Photometry and Spectra\footnote{\lowercase{\uppercase{B}ased on observations obtained with the \uppercase{A}pache \uppercase{P}oint \uppercase{O}bservatory 3.5-meter telescope, which is owned and operated by the \uppercase{A}strophysical \uppercase{R}esearch \uppercase{C}onsortium.}}}


\author{Adam F. Kowalski\altaffilmark{2, 8},
  Suzanne L. Hawley\altaffilmark{2},  John P. Wisniewski\altaffilmark{3},
  Rachel A. Osten\altaffilmark{4}, Eric J. Hilton\altaffilmark{5}, Jon
  A. Holtzman\altaffilmark{6}, 
   Sarah J. Schmidt\altaffilmark{7}, James R. A. Davenport\altaffilmark{2}}


\altaffiltext{2}{University of Washington, Astronomy Department, Box 351580, U.W.
    Seattle, WA 98195-1580, USA}
\altaffiltext{3}{HL Dodge Department of Physics \& Astronomy, University of Oklahoma, 440 
W Brooks St, Norman, OK 73019 USA}
\altaffiltext{4}{Space Telescope Science Institute, 3700 San Martin Drive
Baltimore, MD 21218, USA}
\altaffiltext{5}{Universe Sandbox, Seattle, WA, USA}
\altaffiltext{6}{New Mexico State University, Department of Astronomy,
  Box 30001, Las Cruces, NM 88003, USA}
\altaffiltext{7}{Department of Astronomy, Ohio State University, 140 West 18th Avenue, Columbus, OH 43210}
\altaffiltext{8}{current address:  NASA Postdoctoral Program Fellow, NASA Goddard Space Flight Center, Code 671, Greenbelt, MD 20771, USA; email: adam.f.kowalski@nasa.gov}

\begin{abstract}
We present a homogeneous analysis of line and continuum emission from
simultaneous high-cadence spectra and photometry covering 
near-ultraviolet and optical wavelengths for twenty M dwarf flares. These data were
obtained to study the white-light continuum components at bluer and
redder wavelengths than the Balmer jump. Our goals were to break the
degeneracy between emission mechanisms that have been fit to broadband
colors of flares and to provide constraints for radiative-hydrodynamic
(RHD) 
flare models that seek to reproduce the white-light flare
emission. The main results from the analysis are the following: 1) the
detection of Balmer continuum (in emission) that is present during all
flares and with a wide range of relative contributions to the
continuum flux at bluer wavelengths than the Balmer jump; 2) a blue
continuum at flare maximum that is linearly decreasing with wavelength
from $\lambda = 4000-4800$\AA, indicative of hot,
blackbody emission with typical temperatures of
\TBB$\sim9\,000-14\,000$ K; 3) a redder continuum apparent at
wavelengths longer than \Hb\ ($\lambda \gtrsim 4900$\AA) which becomes
relatively more important to the energy budget during the late gradual
phase. The hot blackbody component and redder continuum component
 have been detected in previous
 studies of flares.   However, we have found that although the hot blackbody emission component
 is relatively well-represented by a featureless, single-temperature
 Planck function, this component 
 includes  absorption features and has a continuum shape strikingly similar to the 
  spectrum of an A-type star as directly observed in our flare
  spectra. New model constraints are presented for the time-evolution among the
Hydrogen Balmer lines and between Ca II K and the blackbody
continuum emission.   We calculate Balmer jump flux ratios and compare to the solar-type
 flare heating predictions from RHD models. The model ratios are too large and the blue-optical ($\lambda = 4000-4800$\AA)
slopes are too red in both the impulsive and gradual decay phases of
all twenty flares.  This discrepancy implies that further work is needed to
understand the heating at high column mass during dMe flares.

\end{abstract}


\section{Introduction} \label{sec:intro}
Optical and near-ultraviolet (NUV) continuum radiation during stellar flares is a commonly
observed phenomenon, yet its origin remains unknown despite decades of
investigation.  This emission is termed the 
white-light continuum because it is detected in broadband filters,
such as the TRACE white-light filter during solar flares and Johnson
$UBVR$ bands during stellar (especially M dwarf) flares.
Broadband color investigations suggest that the
white-light energy distribution peaks within the $U$ band
($\lambda\sim 3250-3950$\AA) or just shortward of the $U$-band near
$\lambda \sim 3000$\AA.  
Accurately flux-calibrated, time-resolved spectra in the blue and
NUV are important for understanding the white-light 
emission processes, which 
encode information about the depths, temperatures, and densities where
the emission is formed, and ultimately the heating
mechanism(s).  Understanding white-light
emission therefore also necessitates radiative-hydrodynamic (RHD) flare
model atmospheres that are produced self-consistently with a
realistic flare heating mechanism, with the goal of reproducing the
observed NUV/optical spectrum.

\subsection{dMe Flares}
Magnetically active M dwarfs are those with a persistent chromosphere 
often diagnosed by \Ha\ or Ca \textsc{ii} H and K line emission even outside of flares.
Chromospheric line emission is attributed to strong magnetic fields (\s few thousand Gauss)
covering sometimes \s50\% or more of the stellar surface \citep{Saar1985, JohnsKrull1996}.  These active M dwarfs
regularly produce flare emission across the electromagnetic spectrum, from
soft X-rays (\s$0.4-50$ keV) to the radio (\s10 MHz \,--\,10 GHz).  Due to the low
photospheric background at blue and NUV wavelengths, white-light flares on M dwarfs
produce a large contrast which facilitates flare detection and
reduces the contribution of quiescent (non-flare) emission.  The contrast of the flare
emission against the quiescent background is known as the ``flare
visibility'' \citep{Gershberg1972} and makes the Johnson $U$-band
filter preferred for flare studies \citep{Moffett1974}.  The $U$-band
flare energy comprises \s 1/6 of the white-light energy
\citep{HawleyPettersen1991}, which in turn dominates the energy
observed at shorter wavelengths, such as in the EUV and soft X-ray \citep{Hawley1995, Fuhrmeister2011}.
Active M dwarfs with spectral subtypes
dM3e-dM6e\footnote{``e'' indicating that in quiescence H$\alpha$ is in
  emission.} have been found to flare frequently with good
visibility \citep{Lacy1976} and thus are the main targets for flare
monitoring.

Flare light curves are typically divided into an impulsive phase
and a gradual decay
phase \citep{Moffett1974, MoffettBopp1976}.  The impulsive phase
consists of a fast rise lasting tens of seconds or more, a peak, and a fast decay.  The
quasi-exponential, gradual decay phase begins with a transition from fast to slow
decay \citep{HawleyPettersen1991} and can last from minutes to hours. These two phases comprise the classical
flare light curve morphology, although much more complex
light curves are observed \citep[e.g.,][]{Moffett1974, Kowalski2010}.

Stellar flares produce greatly enhanced emission in chromospheric
lines, such as the Hydrogen Balmer series, Ca \textsc{ii} H and K, and
He \textsc{i}.  These lines are typically associated with
chromospheric temperatures ranging from \s6000\,--\,$20\,000$ K.  The
Hydrogen lines have a fast rise phase but typically peak several minutes after the peak of
the ($U$-band) continuum emission \citep{Kahler1982,
  HawleyPettersen1991, GarciaAlvarez2002}.  The energy in the emission
lines is only a small percentage (\s4\%) of the total FUV to optical flare energy in
the impulsive phase but larger (\s17\%) in the gradual decay phase
\citep{HawleyPettersen1991}, indicating that the major channel of
atmospheric cooling occurs through continuum radiation for the flare duration.   Some flares produce
line emission that contributes a larger percentage, \s30\,--\,50\%, of the total energy \citep{Hawley2007}.
 Broadening of the Balmer line profiles has been observed, with full widths
 at the continuum level that approach \s20\AA\ for large flares \citep{HawleyPettersen1991, Fuhrmeister2008}.  The broadening of
 Hydrogen (and Helium) lines has
 been interpreted as an indication of turbulent or directed 
 mass motions of tens to several hundred \kms\
 \citep{Doyle1988,Eason1992,Fuhrmeister2008} or Stark broadening due to
 the electric fields from increased electron densities on the
 order of $10^{13}-10^{14}$ cm$^{-3}$ \citep[][ see also \cite{Kurochka1970}]{Svestka1972, Worden1984}.  Broadening
 of the Ca \textsc{ii} lines is not observed \citep{Paulson2006}, but the
 total Ca \textsc{ii} K flux exhibits a characteristic late peak after the
 Balmer lines at the beginning of the continuum gradual phase \citep{HawleyPettersen1991}.
 In almost all previous studies, the entire Balmer
series has not been captured simultaneously in order to achieve
wavelength coverage in the blue, usually at the expense of \Ha.
 \cite{Eason1992}, \cite{Crespo2006}, \cite{Fuhrmeister2011} have provided data
covering most of the Balmer series (the \Hb\ line was not included in
the studies of 
\cite{Eason1992} and \cite{Fuhrmeister2011}), but with the red and blue data 
obtained at significantly different cadence.  The relative flux
in each Hydrogen line (the Balmer decrement) over the duration of the flare is an important
diagnostic for the evolution of electron densities \citep{Drake1980}.

\subsection{Emission Mechanisms from Broadband Colors}
Candidates for the emission mechanism that produces the white-light
continuum have been described by \cite{CramWoods1982}, \cite{Giampapa1983}, and
\cite{Nelson1986}, and include blackbody (BB), Hydrogen free-free (ff), Hydrogen
bound-free (bf), and H$^-$ bound-free radiation processes.  Inverse Compton scattering of
quiescent infrared radiation from relativistic electrons has
also been proposed \citep{Gurzadyan1988}, but this ``fast-electron
hypothesis'' has not been confirmed with X-ray observations
\citep{Mullan1990}.  
The emission types have been constrained using multicolor Johnson
 broadband photometry (hereafter, ``colorimetry'') of flares with
 energies ranging from 
 $E_U \sim 10^{31}$ ergs  to $E_U \sim 10^{34}$.
\cite{HawleyPettersen1991} and \cite{Hawley1992} used $UBVR$ photometry and IUE SWP/LWP data to study
the continuum shape evolution during a larger ($10^{34}$ erg) flare on
the dM3e star AD Leo, finding that the peak 
flux occurs in the $U$ band and a significant amount of flux (27\% of
the total) is observed in the NUV.
The broadband distribution was fit very well by a blackbody with
$T=9000-9500$ K in the impulsive (rise, peak) phase and $T = 8400-8800$ K in the
gradual decay phase.  The data from this flare also indicated a
 reddening of the continuum in the gradual decay phase, and this was
suggested to be the result of the presence of two (or more) competing
emission mechanisms, including a contribution from Hydrogen (Paschen continuum)
recombination radiation.  

 A two-component model was first proposed using simultaneous colorimetry 
and spectra of dMe flares by \cite{Kunkel1970} -- see also
\cite{MoffettBopp1976} -- 
who concluded that a single, isothermal ($T_e = 3000-30\,000$ K)
optically thin hydrogen emission (bf$+$ff)
model was too blue to explain the observed flare colors, nor could it account for
the spread of colors among a sample of flares. Instead, they proposed a model
consisting of a dominant component of Hydrogen bf (recombination)
radiation with a secondary
contribution from a heated photosphere, which increases in relative contribution over time during the
flare decay.  Other studies have similarly concluded from colorimetry that flare radiation consists of
a combination of hot blackbody emission (with temperatures as high
as $13\,000-18\,000$ K) and optically 
thin Balmer continuum recombination radiation \citep{DeJager1989, Zhilyaev2007}.  It has been speculated that the blackbody is short-lived 
and the Balmer continuum becomes more dominant in the gradual decay phase
\citep[see also \cite{Abranin1997, Zhilyaev2007}]{DeJager1989,
  Abdul-Aziz1995}, but the direct characterization of both components
(i.e., with spectra)
has thus far not been possible. 
Furthermore, \cite{Allred2006} recently showed that continuum constraints using colorimetry are
fraught with degeneracies; a model spectrum that
has a large Balmer jump (due to Hydrogen bf radiation) may exhibit the shape of a hot, blackbody with \TBB \s 9000 K 
when convolved with broadband filters.

\subsection{Emission Mechanisms from Spectra}
Previously, spectra of the Balmer jump (covering \s1000\AA\ near 3646\AA) have been obtained during several large flares on
dMe stars:  AD
Leo \citep{HawleyPettersen1991}, UV Ceti \citep{Eason1992}, Gl 866
\citep{Jevremovic1998}, AT Mic
\citep{GarciaAlvarez2002}, CN Leo
\citep{Fuhrmeister2008}, and YZ CMi \citep{Doyle1988}.  None of these studies showed conclusive evidence of a 
component that could be attributed to Hydrogen (Balmer) recombination
radiation, in contrast to the findings of \cite{Kunkel1970}.
Interestingly, \cite{Eason1992} noted that
the Balmer continuum appeared in
\emph{absorption}\footnote{Although the authors gave the caveat that the observations
  were obtained at high airmass.}, a property that we study in this paper.

Flux-calibrated spectra at wavelengths redder than the Balmer jump 
has revealed blackbody
temperatures consistent with those subsequently inferred from
colorimetry.  \cite{Mochnacki1980} used
a multichannel spectrophotometer to map the evolution of the hot
blackbody component, which they speculated may be dominant at flare
maximum.  They found that the rise phase could be caused by
increasing area coverage of the hot component while the decay phase is
explained by both rapidly decreasing temperature (from 9500 K at peak to $5500-7000$ K in
the decay) and relatively constant area.  They were unable to accurately observe the Balmer continuum due to
spectral vignetting, but they did note a smaller Balmer jump than predicted by
\cite{Kunkel1970} and found the decay of NUV emission was slower
than in the optical, perhaps indicating two components in action.
Similar temperatures of $8000-11\,000$ K have been directly measured from
blue/optical spectra (at $\lambda > 4000$\AA) \citep{Kahler1982,
  Katsova1991, Paulson2006, Kowalski2010}. \cite{DeJager1989} determined a temperature of $16\,000$ K
from (non-flux calibrated)
spectra of a $\Delta B \sim$ 5 mag flare on UV Ceti.  During a different
$\Delta U \sim$ 5 mag event on UV Ceti, \cite{Eason1992}
concluded that an optically thick thermal-bremsstrahlung (Hydrogen ff)
model with $T\sim13\,000$ K gave the best
fit to their spectrum.  

Direct spectra of the blackbody peak are not yet available due to the
difficulty of observing in the NUV at $\lambda < 3200$\AA, but spectra around the Balmer jump
have provided several important clues.  The
spectra from \cite{HawleyPettersen1991} of AD Leo
showed that the flux distribution 
increases toward the blue with little (if any) observed Balmer jump;
\cite{Hawley1992} suggested that the peak 
lies somewhere between $\lambda=3000-3500$\AA.  Other spectra indicate the possibility that the continuum peaks outside the
atmospheric window, $\lambda < 3250$\AA\ \citep{Fuhrmeister2011}. The largest
observed flux is emitted within the $U$ band \citep{Hawley1992,
  Hawley2003}, and the blackbody fits indicate a peak flux at $\lambda\sim3000$\AA.
 \cite{Fuhrmeister2008} and \cite{Schmitt2008} measured
the NUV shape of the flare continuum down to the atmospheric limit at $\lambda \sim
3250$\AA\ with VLT/UVES and found a temperature of $11\,300$ K; the
temperature fits have large uncertainties (\s4000 K), possibly
due to the long integration times, narrow wavelength range
($\lambda = 3250-3800$\AA), and lack of a robust flux calibration
(without standard stars) for these echelle data.  The same authors also
measured the continuum shape in red, higher cadence 
spectra and found temperatures that were different than from the NUV:  $20\,000 -
27\,000 (\pm 5000$) K at peak and $3200-5600$ K during the decay.  
\cite{Fuhrmeister2011} observed a flare on Proxima Centauri with a
similar VLT/UVES setup and did not find a good blackbody fit to the
flare spectrum.

In \cite{Kowalski2010}, we reported initial results from a
``Megaflare'' on the dM4.5e star YZ CMi.  In particular, we found
evidence for both BaC and hot blackbody emission which varied in 
strength during secondary flare heating events.  See Section
\ref{sec:bluecont} for additional discussion of this flare.

\subsection{Flare Heating Models}
The origin of the $T\sim10\,000$ K blackbody component is
currently unknown, and indeed its existence
has been contested by \cite{vandenoord1996} and \cite{Nelson1986} on 
grounds that it requires very high heating fluxes from a solar-type
flare heating 
beam of \s $5\times10^{11}$ ergs s$^{-1}$ cm$^{-2}$. However, we 
know today that fluxes larger than this are possible even on the Sun \citep{Neidig1993,
  Krucker2011}.  The inferred areal coverages of the blackbody
component are small, $\le 0.5$\%, for
even the largest stellar flares \citep[e.g.,][]{Hawley1992}, supporting the idea that this hot,
blackbody emission component originates from a compact source at
locations
of intense and focused heating, perhaps at the footpoints of
magnetic loops.

The blackbody has been reproduced in static phenomenological models.
In \cite{CramWoods1982}, 
their model atmosphere \#5 features extreme
heating from the chromosphere through the deep photosphere and results
in a \TBB\s$14\,000$ K emission component and an \Ha\ line with a central absorption; they note that this atmosphere most closely matches the continuum observations 
of stellar flares and could represent the stellar-analog of solar
flare kernels where there is deep and concentrated atmospheric
heating.  \cite{Houdebine1992} also produced model spectra that rise
into the NUV, similar to a hot blackbody but with a sizeable
Balmer jump; their models employ very large electron
densities of $10^{16}$ cm$^{-3}$.  They suggest that 
hydrogen recombination radiation and blackbody continua contribute in varying proportions depending 
on the various parameters of the flare atmospheres, but provided 
few details.  Phenomenological models
of an energetic flare observed in the extreme-ultraviolet on a dM4e star gave a
very small Balmer jump and continuum peak in the NUV
\citep[2400\AA;][]{Christian2003}.  \cite{Kowalski2011IAU} used the
static, NLTE RH code \citep{Uitenbroek2001} to model the secondary
heating events during the ``Megaflare'' in \cite{Kowalski2010}.  They
found that a 
Gaussian temperature ``hot spot'' placed below the temperature
minimum in a quiescent M dwarf atmosphere produces an optical spectrum with \TBB\s$11\,000-18\,000$ K and
strong absorption in the Hydrogen Balmer features, as observed during
the secondary events (see Section \ref{sec:astar}).

Self-consistent models that use realistic flare
heating mechanisms typically result in a white-light continuum that is
dominated by a strong Hydrogen recombination component \citep{Hawley1992}.  
The sophisticated one-dimensional RHD models of \cite{Abbett1999} and
\cite{Allred2005, Allred2006}
used the RADYN code \citep{Carlsson1994, Carlsson1995, Carlsson1997} to simulate
flares on an M dwarf and on the Sun
using moderate (10$^{10}$ \ergscm, F10) and 
large (10$^{11}$ \ergscm, F11) fluxes of mildly
relativistic electrons injected at the top of a semi-circular flare loop. The RADYN models employ the thick-target 
formulae of \cite{Emslie1978}, \cite{Ricchiazzi1983}, and \cite{HawleyFisher1994} that
describe how the nonthermal electron beam deposits energy throughout the atmosphere.  The
accelerated electron distribution in the \cite{Allred2006} models
employs a double power-law
energy distribution of beam electrons with a minimum cutoff energy,
$E_c$, which is assumed to be 37 keV (inferred
from solar flare hard X-ray observations with RHESSI
\citep{Holman2003}).   The energy from the electrons is deposited in
the chromosphere, which explosively evaporates into the corona,
illustrating the chromospheric evaporation scenario developed by
\cite{Fisher1985}.   The \cite{Allred2006} F10 and F11 models represent \emph{impulsive} heating,
and they were run with constant beam fluxes for 230 sec and 16 sec, respectively.

The \cite{Allred2006} RHD predictions of Hydrogen Balmer line
emission, such as line broadening and flux decrements, are generally consistent with observations.  
However, an important shortcoming of the model predictions is the lack of hot blackbody emission which is clearly present,
and in fact dominant, in the spectra of stellar flares. Instead, the dominant continuum
components at $\lambda > 2000$\AA\ are a large spectral
discontinuity at the Balmer jump and prominent Balmer (bf) continuum emission.   The optical flare emission is due 
to Paschen (bf) continuum emission and radiation from the moderately heated photosphere.  
The photospheric heating is at most $\sim1000$ K and results from incident NUV backwarming
radiation; direct beam heating contributes relatively little to the
heating of deep layers, and cannot reproduce the heating or densities
implied by phenomenological
models.   The ultimate problem in these physically self-consistent
models is therefore not enough heating at high densities.  
However, as mentioned previously, the broadband colors of the model flare spectrum produce a continuum distribution
with the general shape of a blackbody with $T\sim9000$ K and also appear to match the observed broadband $UV,UBVR$ fluxes
of a moderate sized flare quite well \citep[flare 8 of][]{Hawley2003}.
A recent observation of the flare decay phase for the first time directly
detected a Balmer continuum in emission which matched the shape of
the F11 RHD model Balmer continuum \citep[][see also Section \ref{sec:bluecont}.]{Kowalski2010}.

\subsection{Motivation for the Present Study}
The last study of a large sample of flares with simultaneous spectra and
photometry was that of \cite{Bopp1973} and \cite{MoffettBopp1976}, who revealed several
global trends in the spectroscopic characteristics of flares.  Their sample consisted
of five flares with low-resolution spectral coverage from $\lambda=3700-5700$\AA\ and high cadence $U$-band photometry. 
The exposure times of the spectra were between 30 sec and 3 minutes, with most $> 1$ minute.  Their main result was the demarcation of 
flares into ``spike'' and ``slow'' phases according to the relative
contribution of line and continuum emission;  they speculated that  
a two-component model with Hydrogen recombination could explain these
phases.  Other results were shown for the He \textsc{i}$\lambda4
026$, He \textsc{i} $\lambda4471$, and Mg \textsc{i}b lines, indicating
no apparent relation between flare properties and their
detection/non-detection.  They found a longer time delay between
continuum and emission line maxima for higher luminosity stars, but their use of
equivalent widths for line diagnostics makes interpretation difficult.  
The continuum shapes were not analyzed and the lack of blue wavelength
coverage did not allow for an assessment of possible Balmer continuum radiation.

As a modern day extension of the \cite{MoffettBopp1976} study, 
we have obtained high signal-to-noise spectral observations of flares in the
blue/NUV, including the Balmer jump wavelength,
for a homogenous analysis of the line and continuum properties.  These data are necessary to
break the degeneracy of fitting emission types to broadband
photometry and to determine which continuum processes
contribute (and how much) to the white-light.  
In the past, colorimetry was preferred in order to achieve a high
signal-to-noise at good time resolution, but the $U$-band is difficult to interpret because
it straddles the Balmer jump.  Modern, 4-m class telescopes now
provide good blue/NUV sensitivity and allow time-resolved spectra
to be used to characterize faint levels of flare flux varying on short timescales.  A large, systematic 
study of blue/NUV flare emission will reveal if hydrogen recombination (Balmer continuum)
radiation is present in flares and how it relates to the blackbody component.  Including a range of flare types (e.g.,
fast vs. slow) is useful to assess why this continuum component
does not obviously appear in the most recent, high quality observations
\citep[e.g., ][]{HawleyPettersen1991, GarciaAlvarez2002, Fuhrmeister2008}.  
For example, perhaps the disappearance of the Balmer continuum is
 a phenomenon that only occurs during large impulsive flares, which coincidentally are the only ones that 
have yet been studied in detail with blue/NUV spectra.

We have obtained broadband photometry simultaneously with the spectral
observations to connect with
decades of single-filter and colorimetric flare studies.
Most importantly, simultaneous observations allow us to relate
the spectral continuum characteristics to the diverse types of light curves.
Perhaps coincidentally, most
 of the largest 
flares which have been studied with NUV spectra have time-profiles in the $U$-band that deviate from
the classical model and include a secondary, (usually lower) amplitude continuum enhancement following the fast decay phase of the 
first peak.
Some flares have three or more continuum peaks in the impulsive phase \citep{Kahler1982, Eason1992}, while other flares
have low-amplitudes but gradually emit a large amount of energy over a longer time period \citep{Hawley1995}.

A basic phenomenological question therefore is, ``How do the continuum properties evolve 
through the different phases of the $U$-band evolution, and can we ultimately use these properties to 
create a flare continuum model that can explain the gamut of flare light curves that 
are observed?''  For example, \cite{Osten2005} studied two $U$-band flares on the dM3.5e star, EV Lac:  their durations
were 4.5 and 7 minutes, yet their peak amplitudes were a factor of
twenty different: how is the physics (i.e., flare heating and
subsequent radiation) different between these two flares?
With a large sample of flares that have simultaneous time-resolved spectra and photometry, we can constrain the detailed
continuum parameters over a range of flare characteristics to provide
constraints for future models.

The most general aspect of this question is how the continuum components compare 
between the impulsive and gradual decay phases, and therefore how their
respective ``fast'' and ``slow'' evolution are related to the dominant
flare heating mechanism, atmospheric cooling response, and radiation at those times.
In addition to identifying the differences between the individual phases of flares, we also seek to identify similarities between the same phases of different
flares to clarify important phenomena that must be reproduced by RHD models.

This paper is organized as follows.  In Section \ref{sec:reduction}
we discuss the observations and data reduction.  In Section
\ref{sec:params}, we introduce several 
parameters used to analyze the data.  A general overview of the
flare sample is given in Section \ref{sec:spectra_atlas}, and Section
\ref{sec:emissionlines} contains the detailed emission
line analysis.  We apply the two-component blue continuum analysis of \cite{Kowalski2010} to all flares in the sample in Section
\ref{sec:bluecont}. The gradual decay phase is analyzed in Section
\ref{sec:gradual}, and we introduce a third continuum component in
Section \ref{sec:conundruum}.  
  We compare to the \cite{Allred2006} RHD model predictions
in Section \ref{sec:modelcomparison}.  Finally, we analyze filling
factors and flare speeds in Section \ref{sec:filling_speeds}.  In
Section \ref{sec:wtf}, we summarize our findings and discuss their physical
significance for flare processes.
Finally, in Section \ref{sec:future} we discuss future observations that are
needed.
There are several Appendices where we discuss detailed aspects of the
data and analysis algorithms.  Throughout this paper, we refer
the reader to relevant sections and appendices of the PhD thesis of \cite{KowalskiTh}.

\section{Observations and Data Reduction} \label{sec:reduction}
\subsection{Target Stars}
Over the course of three years, we obtained high-cadence simultaneous photometric and
spectroscopic observations of nearby, dMe
flare stars. 
We monitored the dMe flare stars that had the highest measured 
optical/NUV flare rates ($\sim1 $ / hour, \cite{Lacy1976,
  Pettersen1984}), and our initial sample consisted of four bright,
nearby stars (EV Lac, YZ CMi, AD Leo, EQ Peg) to allow for short exposure times.   The multiwavelength
properties of the flares on these stars have been studied extensively
\citep[e.g.,][]{Kahler1982, Hawley1995, vandenoord1996, Osten2005, Osten2010}.
The targets and basic stellar parameters are given in
Table~\ref{table:starssummary}.

A fifth target star, GJ 1243, is a star that hasn't had its flare rate previously
measured.  It is known to be an active star \citep{Gizis2002} of spectral
type dM4e, and its long
baseline photometric starspot activity was recently measured by
\cite{Irwin2011}.  
It is particularly important to examine GJ 1243 due to its inclusion
in our Kepler GO program
with observations at 1 minute cadence (the Kepler flare properties will be presented in
subsequent papers).

\subsection{Spectral Data} \label{sec:spectra}
Spectra were obtained with the Dual-Imaging Spectrograph (DIS) on the ARC
3.5-m telescope at the Apache Point Observatory (APO).  We employed the
low-resolution B400/R300 gratings, which provided continuous wavelength coverage from
$\lambda \sim3400-9200$\AA, except for a dichroic feature that affected the flux
calibration from $\lambda\sim5200-5900$\AA.   The CCD was binned by 2 
and windowed to \s$130$ pixels along the spatial axis
($\sim$100''), thereby reducing the readout time from 40 sec to
$\sim10$ sec.   Integration times ranged
from 1 second to 45 seconds (most were between 10 and 20 seconds).
Short cadence
($1-8$ second) spectra were occasionally interspersed in the observing
sequence in order to avoid
non-linearity and saturation in the red during the longer
integration times which were necessary to obtain visible counts
on the 2D image at $\lambda\sim 3450$\AA.  This typically
provided an adequate signal-to-noise of $\sim10$ at 3600\AA\ in quiescence.
We obtained data with the 1.5\arcsec\ slit for the first two
years and primarily with the 5\arcsec\ slit for the last year (if the conditions
were clear).  The wide slit facilitated absolute flux calibration
(so that we could check the scaled spectra against the original
flux-calibrated spectra; see below), mitigated the effects from
miscentering the star on the slit and from seeing variations, and
allowed
the exposure times to be reduced.  The slit was automatically rotated
to the parallactic angle in order to
account for atmospheric differential refraction
\citep{Filippenko1982}\footnote{The data from 2008 Oct 01 were not
  obtained at the parallactic angle because both EQ Peg A and EQ Peg B were
  positioned along the slit.  The additional steps taken to flux
  calibrate these data are described in Appendix A.1 of \cite{KowalskiTh}.}.
Care was taken to ensure the star stayed centered on the slit through
the course of the observations, but in some instances small deviations
may have affected the observations.  The observations
were taken under moderate to good weather conditions; the seeing was
estimated with each spectrum but rarely exceeded the narrow (1.5\arcsec) slit width by more
than $\sim$0.5\arcsec.  

The spectra were reduced using standard IRAF\footnote{IRAF is distributed by the National Optical Astronomy Observatories,
which are operated by the Association of Universities for Research
in Astronomy, Inc., under cooperative agreement with the National
Science Foundation.} procedures via a
customized PyRAF\footnote{PyRAF is a product of the Space Telescope
  Science Institute, which is operated by AURA for NASA.} wrapper,
developed from the reduction software of \cite{Covey2008}.
Initial processing included bias, overscan, and flat-field
corrections.  Aperture extraction and background subtraction were
performed and a wavelength solution was applied using HeNeArHg and
HeNeAr lamps.  The resulting dispersions were 1.82 \AA\ per pixel for the blue and 2.3
\AA\ per pixel for the red.  The spectral resolutions were
determined from the He \textsc{i} $\lambda$4471 arc line taken at the beginning
of the night.  For the 1.5\arcsec\ slit, the resolution at this wavelength
was $5.5-7.3$\AA\ (R $\sim 600 - 800$) and
$\sim18$\AA\ (R $\sim 250$) for the 5\arcsec\ slit.   For the very wide (5\arcsec) slit width, the profiles of
arc lines are wider and less gaussian than for point sources, and measuring the quiescent
emission line profiles of the M dwarfs reveals an actual resolution
closer to
$13-15$\AA\ (R $\sim320$).  Although broad, the line profiles for the spectra of M
dwarfs taken with the wide
slit were in general nearly gaussian and allowed line fluxes to
be measured.  

Observations of spectrophotometric standard stars (white dwarf or sdO
stars) were obtained every
night and were used to convert from counts to an energy scale.   An
 airmass correction was applied to the spectra using the atmospheric
extinction curve for APO, published by the Sloan Digital Sky
Survey (SDSS). 
The spectral shape accuracy across the $\lambda=3400-5200$\AA\ range was usually better than
10\%, as calculated from observations of multiple
spectrophotometric standard
stars.  Synthetic $U$, $u$, and $g$ absolute magnitudes obtained from the spectra were
accurate to within $20-25$\% during good conditions; the large
uncertainties likely result from not knowing the
precise blue response of all optical elements.
  The spectrophotometric standard star fluxes were obtained from \cite{Oke1990};
second order corrections to the APO atmospheric extinction curve were
not applied to the data.  The wavelength ranges that we use for
continuum analysis in this
paper are $\lambda=3420-5200$\AA\ and $\lambda = 5900-7500$\AA.  
The He \textsc{i} $\lambda5876$ line may be used in future analysis but
is marginally affected by the far end of the dichroic.  The Ca
\textsc{ii} IR triplet can also be analyzed 
\citep{KowalskiTh}, but spectral fringing affects the red flux at
$\lambda > 8350$\AA\ (with amplitude of $\sim$2\%).  
For the observations between February 2010 and July 2011, the
wavelength region at $\lambda < 3600$\AA\ was affected by a \s2\% ``ripple'' variation in flux, which
is apparent in very high flux levels (such as at flare peak and in the
standard stars).  A similar effect 
has been attributed to inter-pixel variations of quantum efficiency
\citep{Rutten1994}, but we have attributed it as a transient artifact in
instrument performance.  The ripple is visible in flat-fields but is not removed in the
reduction process.  Further details about the data reduction and flux accuracy are
given in Appendix A of \cite{KowalskiTh}.  

The observing log for each target star is given in
Table ~\ref{table:obslogAll}.  The monitoring time each night, number of exposures, exposure times, spectral
resolutions, and available simultaneous photometry are also
provided. For the number of exposures given, the three values indicate
the number used for blue wavelength analysis, the total number of spectra
recorded, and the number of spectra obtained at a shorter exposure
time and lower cadence for red wavelength analysis: ($n$B, $n$T, $n$R).  For a spectrum to be
considered in the blue wavelength analysis
(and be counted in $n$B), the standard deviation of flux divided by
the average flux just blueward of the Balmer jump was required to be less than 15\%.  
This allowed us to select spectra that were (mostly) unaffected by weather
and cosmic rays.  Note, some short-exposure spectra (a fraction of $n$R) are 
included by this requirement and some were excluded explicitly (see caption
for Table \ref{table:obslogAll}); the remaining short
exposure spectra are analyzed independently.

\subsection{Photometric Data} \label{sec:phot}
Photometry was obtained from the NMSU 1-m telescope and the ARCSAT
0.5-m telescope at the Apache Point Observatory.  The 1-m is operated
robotically \citep{Holtzman2010}, and provided continuous Johnson $U$-band
photometry.  The $U$-band data were taken with exposure times of 10
sec for YZ CMi, 4 sec for AD Leo, 4 sec for EV Lac, and 15 sec for EQ
Peg AB (the A and B components were both visible but not completely
resolved). With a readout of $\sim$10 sec\footnote{Every 39 exposures
  an automatic focus check was performed resulting in a slightly
  larger gap in the data.}, the cadence was therefore several
observations per minute. 
  Observations were reduced as part of the
standard 1-m pipeline (and in some cases by hand using standard IRAF procedures).  
We used the Flarecam instrument (Hilton et al. 2011) on the 0.5-m, which
was remotely operated.  Flarecam has enhanced UV sensitivity, a fast
readout ($\sim1$ second), and rapid filter wheel rotation among the available SDSS $ugri$
filters.  The exposure times varied depending on conditions, but $gri$
band exposures were typically $1-2$ sec.  A variety of imaging sequences
were employed during the campaign in order to determine the optimal balance of duty cycle
and wavelength coverage. 
The data were reduced using standard IRAF procedures
with the \emph{ejhphot} reduction wrapper \citep{Hilton2011}.

Differential aperture photometry was performed using a nearby bright star, and a
quiescent window was chosen to normalize the count flux for the night\footnote{Airmass
and color corrections were not applied to the data as differential photometry provides sufficient accuracy.}.
 By comparing the times in the spectra and photometry headers, we
determined that the timing of the Flarecam images was not always
synched with UTC.  This offset ranged between 7\,--\,30 seconds,
and we adjusted the center times of the measurements accordingly.
This timing precision is adequate for our study, since the
spectra have a cadence $\ge11$ seconds.

Table~\ref{table:obslogAll} contains information about the photometry used for each
night. Note that the photometry and spectra for 2009 January 16 are
the same as 
those presented in \cite{Kowalski2010}.

\subsection{Combined Spectral and Photometric Sample}
From the $U(ug)$-band and \Hg\ equivalent width lightcurves of thirty-one nights of observations, we selected eighteen
flares from these five stars to analyze in detail. In order to
facilitate analysis of the NUV continuum flux (which the Earth's atmosphere
efficiently scatters or absorbs), the largest $Uug$ amplitude events
were chosen.  
The flares occurred on fourteen separate nights which comprised 75 hours of
spectral monitoring with 7780 spectra\footnote{There are several more
  nights and additional (small-amplitude) flares and other
  types of variability in the data that are not analyzed
  in this study.}.  In addition,
we consider the data
 obtained during the large flare of 2009 Oct 27 on EV Lac, which was 
 discussed in
 \cite{Schmidt2012}.  The blue spectra were obtained at the Dominion
 Astrophysical Observatory (DAO) and have much lower time resolution (200\,--\,300
 sec) and only cover the wavelength range $\lambda=3550-4700$\AA\ with
 $R\sim750$, but these data
 encompass an unusually fast and large amplitude flare.  We also calculate relevant quantities from
 the Great Flare on AD Leo of 1985 April 12 \citep[][hereafter, HP91]{HawleyPettersen1991}
 for comparison.  The total number of flares in our sample is
 therefore 20.

\subsection{Emission Line Fluxes} \label{sec:calc_lines}
Line fluxes were calculated for hydrogen Balmer $\alpha$, 
$\beta$, $\gamma$, $\delta$, $\epsilon$\ $+ $ Ca \textsc{ii} H, Ca \textsc{ii} K,
several Helium \textsc{i} lines ($\lambda4026$, $\lambda4386$, $\lambda4471$), He \textsc{ii} $\lambda4686$,
and several prominent lines with ambiguous identifications that
possibly represent a combination of Helium \textsc{i}, Fe \textsc{ii}, or Mg \textsc{i}b lines
($\lambda \sim 4924$\AA, $\lambda\sim5018$\AA, $\lambda5171$\AA)\footnote{There is an Fe \textsc{ii} triplet at $\lambda=$
  4924\AA, 5018\AA, and 5169\AA, which has been identified in previous flare
  spectra on M dwarfs \citep{Abdul-Aziz1995} and the Sun \citep{JohnsKrull1997}.}.  A line flux can be calculated 
by measuring the equivalent width of the line and multiplying by the nearby
continuum \citep{Reid1995}. If the continuum is normalized to
absolute photometry or if the observations are spectrophotometric,
this method accounts for the effects of weather and slit-loss.
However, the continuum changes dramatically during flares, especially
in the region of the blue lines.  Alternatively, one could look for a
region of the continuum with an accurate calibration whose value
does not vary during flares and use this to calculate the equivalent
width and flux normalization.   After close inspection, we
found that the entire optical continuum (3400\,--\,9200\AA) experiences
significant flux variations during the largest flares.
The red continuum near $\lambda=$ 8650\AA\ is most nearly constant, but the flux calibration
here is not always reliable due to spectral fringing in the far red.

The calculation of line fluxes during flares is further complicated by
the dramatically changing widths of the line wings during flares
\citep[][HP91]{Doyle1988}.
The integration limits and continuum ranges are given in
Table~\ref{table:linewindows}, and were chosen to be wide enough to
account for the maximum amount of broadening observed in the flare
sample; the same windows were used for all spectra of all flares in
the DIS sample in order to
be consistent.  An example peak flare spectrum with significant broadening is
shown with the integration windows
in Figure \ref{fig:trustme}.

The measurement of line fluxes employed in this paper is as follows.
Starting with the total (flare$+$quiescent)
flux in each spectrum, we define local continuum regions and determine
a linear fit between regions on both sides of each emission line.
The linear fit allows us to estimate a first-order change in the
continuum beneath the line (as is important during flares).
We then compute the flux in the line region (Table \ref{table:linewindows}).  Measurements of the line fluxes 
in H$\gamma$, Ca \textsc{ii} K, and He \textsc{i} $\lambda 4471$ can be reliably
calculated before subtracting a preflare spectrum; flare-only emission
is obtained after calculating the total line fluxes by subtracting the
quiescent or pre-flare line fluxes.
However, for the other lines, it was necessary to subtract the
quiescent spectrum before calculating
the line flux because the lines were either at very low-level and were
not readily visible (e.g., the other He \textsc{i}
lines) or the surrounding continuum is poorly modeled by a linear function due
to ``jagged'' quiescent molecular features (as is the case for H$\alpha$, H$\beta$, H$\delta$). 
For these lines, the line flux was calculated after subtracting a
quiescent spectrum (see Section \ref{sec:scaling}), allowing for a
more precise fit of the line to the local continuum.
For H$\alpha$ (and for flare-emission in faint lines)
this method resulted in  
negative features away from line-center if the quiescent lines were 
not aligned precisely with the flare features (e.g., due to occasional
single
pixel jumps from wavelength instabilities); therefore, we summed the positive
and negative flux values over the H$\alpha$ line.  

The local
continuum near important features such as H$\delta$, H$\gamma$, H$\beta$ and H$\alpha$ also
contains numerous photospheric molecular features which present an
additional complication for defining line and continuum regions.
 Because we integrate over a large wavelength window, we include some
molecular features in the line flux (for H$\gamma$, He \textsc{i}
$\lambda4471$; see above); however, the molecular flux is assumed to
be 
removed by subtracting the quiescent spectrum.  The line calculations
are done with an automatic routine, and they were examined by eye 
to ensure that we accounted for all of the excess flare flux.  The line windows
given in Table~\ref{table:linewindows} were adjusted by a small amount
for every spectrum
depending on the centroid of the line, which was determined initially
for the Balmer H$\alpha$, H$\beta$, H$\gamma$, and H$\delta$ lines;  the
wavelength shifts for the Ca \textsc{ii} K and He \textsc{i} $\lambda$4471 lines were
forced to be the same as for the H$\gamma$ line.  The wavelength centroid stability is
typically $< 1$\AA\, but can vary up to a pixel ($\Delta \lambda =
1.83$\AA\ per pixel) from one spectrum to
the next.

\subsection{The Determination of Absolute Flare-only Fluxes}   \label{sec:scaling}
An additional 
step in flux calibration was necessary to correct for
exposure-to-exposure grey variations in the level of flux due to
variable seeing, variable transparency, and imperfect centering of the
star in the slit.  In \cite{Kowalski2010}, we used simultaneous
$U$-band photometry to apply a single scaling factor to each
spectrum.  Since
then, we have developed an improved method to scale the spectra
which
minimizes the subtraction residuals in the molecular features in the 
red continuum.  Importantly, this new
technique allows us to independently compare the spectra and
photometry, as the integration times for the photometry 
(especially during the fast impulsive phase of a flare) may differ from the spectra.

 For each night, we determined a master quiescent or pre-flare spectrum 
 by identifying non-variable times from the photometry and the
 H$\gamma$ line.
 We scaled the spectra during the quiescent or pre-flare interval to a common flux at
$\lambda=4500$\AA\ in order to account for weather or slit-loss
variations over the course of the spectra within this time window.
Synthetic Johnson $B$ and $V$ fluxes were compared to 
the accepted magnitudes \citep[in Table~\ref{table:starssummary}, obtained
from \cite{Reid2005} who compiled magnitude data from][GJ 1243
measurements were obtained from \cite{Reid2004}]{Bessell1990, Koen2002, Leggett1992}, using the
\cite{Johnson1966} flux zeropoints.  The observed fluxes were then scaled
so that the synthetic fluxes were equal to the accepted fluxes, which is important for placing all nights (for a given star)
on the same baseline flux level.  

A scaling for each flare spectrum relative to the master quiescent spectrum is then performed as follows.  
For each spectrum during the flare, we multiplied by a large 
range of possible scale factors ($0.2-4.0$), subtracted the quiescent spectrum,
and calculated the sum of the 
standard deviation of the subtraction residuals in three spectral
regions (outside of features from the Earth's atmosphere which can
change over time) from
$\lambda = 6600-6800$\AA\ (excluding the region around He I 6678\AA),
$\lambda=7000-7100$\AA, and $\lambda=7350-7550$\AA\footnote{The data on 2009 Oct
  10 had highly non-linear or saturated flux values in the red, and
  the data on 2008 Oct 01 did not have data from the red CCD of DIS.  For these spectra,
  we scaled using molecular features in the blue from $\lambda =
  4745-4770$\AA,  $4947-4960$\AA, and $5159-5172$\AA.   On nights with 
good red data, this gave scalings that were consistent with the red
windows.  For the DAO spectra from 2009 Oct 27 presented in \cite{Schmidt2012}, we used windows:
$\lambda=4572-4589$\AA, $4621-4630$\AA, and $4660-4673$\AA.  We did
not apply scaling corrections to the data from
 1985 April 12 because of limited wavelength range; these data were
 obtained under excellent photometric conditions.}.
   These regions correspond to
strong flux changes in the quiescent spectrum due to the presence of molecular
bandheads; therefore, errors in flux scaling appear as significant over-
or under-subtractions at these wavelengths.  The best scale factor
minimized the sum of the standard deviation of the subtraction
residuals.  We tested the accuracy of this procedure, which we
  describe in Appendix \ref{sec:appendix_scaling}.
Essentially, we generated a model flare spectrum and multiplied by an arbitrary scale
factor to simulate flux loss (from weather or slit-loss).  We found
that our
simple algorithm determines the correct scaling factor for all but
extremely large amplitude flares that increase the $U$-band flux by a
factor of \s100 or more,
which aren't in our sample.  A similar scaling method was employed by
\cite{Abdul-Aziz1995}.  
The principle behind the scaling method is analogous to
PSF subtraction in imagery of protoplanetary disks, where the optimal
subtraction is found by minimizing the subtraction residuals in the
background \citep[e.g.][]{Wisniewski2008}.   

Ultimately, a single scaling factor was
determined for each spectrum.  The final stage in flux calibration was to
multiply the flux density, line fluxes, continuum fluxes, and
synthetic filter fluxes by the scaling factor during the flare times.  The flare-only fluxes were
then obtained by
subtracting the quiescent values.  Figure \ref{fig:scale_power} in Appendix \ref{sec:appendix_scaling}
 demonstrates the recovery of flare variations during times of
 variable cloud cover.

\section{Basic Observational Parameters} \label{sec:params}
In this section, we describe the basic observational parameters that
we use to analyze the data.  

\subsection{Observational Parameters: Photometry} \label{sec:phot_param}
In Figure \ref{fig:thalf_example}, we show a light
curve from 2008 Oct 01 of a large flare on EQ Peg A.   We use this
figure to illustrate several of the following empirical
values that can be directly measured from the photometry.

\begin{itemize}

\item \textbf{$t_{1/2}$}

To describe the time-evolution of a light curve, we define $t_{1/2}$, the
full width of the light curve at half-maximum.  This measures the
``timescale'' of the impulsive phase of the flare, including both fast rise and fast decay times.  The measure of $t_{1/2}$ does not assume a functional
form for the decay, which can be complex, as seen in Figure \ref{fig:thalf_example}.  We
also measure $t_{1/2}$ for the light curves of spectral components
(Section \ref{sec:spectra_param}).   In some cases, the rise
time is fast compared to the photometric or spectral cadence; in these
cases we must interpolate between data points to obtain $t_{1/2}$.
For the example flare in Figure
\ref{fig:thalf_example}, we illustrate the $t_{1/2}$ value.

\item \textbf{$I_f$, $I_{f}+1$}

The measure $I_{f}$ is the familiar quantity in flare studies
\citep{Gershberg1972}. It is the ratio of flare-only count flux (photons cm$^{-2}$ s$^{-1}$) in a given band, to the quiescent
count flux in that band.  $I_f + 1$ is the flux enhancement, or the \emph{total}
 count flux relative to quiescence. 
In solar physics, $I_f$ is used to express the \emph{intensity contrast}\footnote{$I_f$ has traditionally
  been used in stellar flare work also, although we realize that it is not the intensity, but the
  count flux that we are measuring in that case.}. If $C(t)$ is the total count
flux ratio ($\mathrm{counts}_{\mathrm{target}} /
\mathrm{counts}_{\mathrm{comp}}$) in the differential photometry, normalized to 1 during quiescence, then

\begin{equation}
I_f(t) = C(t) - 1;   I_f(t) + 1 = C(t)
\end{equation}

For the example flare in Figure \ref{fig:thalf_example}, $I_{f, U, \mathrm{peak}} = 20.4$. 

\item ED

The equivalent duration (ED) in a given bandpass is the integral of $I_f$ over the
duration of a flare \citep{Gershberg1972}.  The units are \emph{seconds} and
multiplying by the quiescent luminosity in the band gives the energy
of the flare.  

\item \textbf{$\mathcal{I}$}

To characterize the shape of the light curve, we use an
 ``impulsiveness index'', $\mathcal{I}$, which is defined as
\begin{equation}
\mathcal{I} = I_{f,peak} / t_{1/2}
\end{equation}

The quantity $\mathcal{I}$ is a measure of the peak relative flux of a
flare weighted
by how fast it rises to peak and decays.  Both a more luminous-at-peak flare and a
smaller $t_{1/2}$ (faster timescale) can give rise to larger values of $\mathcal{I}$. 
We find that this measure provides a way to quantitatively sort 
the flares by their light curve evolution, while only using
observables measured directly from the light curve.

\item \textbf{$\mathcal{L}$}

The specific luminosity ($\mathcal{L}$, units of ergs s$^{-1}$ \AA$^{-1}$) is useful for
characterizing the luminosity without the ambiguity of using a spectral
window of width, $\Delta \lambda$.  For example, $U$-band luminosities (and energies) assume $\Delta \lambda =
700$\AA\ whereas $B$-band luminosities (and energies) assume $\Delta \lambda = 900$\AA,
making luminosities in the two bands not directly comparable.
In some cases, we present the integrated (over wavelength) quantities
\emph{L, E} in typical bandpasses for comparison to
previous studies.  However, when comparing spectral continuum measurements, we
use $\mathcal{L}$.  All measures (\emph{L, E}, $\mathcal{L}$) assume
an isotropically-emitting source and employ
the distances in Table \ref{table:starssummary}.

\end{itemize}

\subsection{Observational Parameters:  Spectra}  \label{sec:spectra_param}
We now refer to Figure \ref{fig:adleo_E10} to describe the measured
parameters from the spectra.  This spectrum illustrates the flare-only emission 
at the peak of a simple, moderate-amplitude flare on AD Leo from 2010 April
03.

\begin{itemize}

\item Spectral Zones:  

We divide the spectrum into four zones: the
\emph{near-UV (NUV) zone} ($\lambda=3420-3646$\AA), the \emph{intermediate
  zone} ($\lambda=3646-4000$\AA), the \emph{blue-optical zone}
($\lambda=4000-5200$\AA), and the \emph{red-optical zone} ($\lambda=5800-7550$\AA).
These are useful for our analysis of continuum variations during flares.

\item C$3615$ and C$4170$

In each spectrum, we measured the average flux in 
several $\sim$30\AA\ - wide continuum windows across the DIS spectral range.
The continuum windows were chosen to correspond to
spectral windows free of major (and most minor) emission lines that appear
during flares.  The two continuum measures that we use to characterize the blue are
the average flare-only flux in the wavelength region from
$3600-3630$\AA\ (denoted C3615) and the average flux in the wavelength region
from $4155-4185$\AA\ (denoted C4170).  The C3615 region was chosen to be
blue of the Balmer jump at 3646\AA, while also red enough to obtain a reasonable signal-to-noise in
small to moderate-size flares. This measure covers the approximate central wavelength of the $U$-band,
which is much broader.  
 The C$4170$ region was chosen to
emulate the NBF4170 continuum filter, which is a custom continuum
filter that is also employed on the solar camera ROSA \citep{Jess2010} and stellar camera
ULTRACAM \citep{Dhillon2007}.  C4170 provides a measure of the
continuum flux redward of the Balmer jump and unaffected by blending
of high order Balmer lines.  The continuum windows are summarized 
in Table \ref{table:linewindows} (including two other continuum
windows, C4500 and C6010, used in Section \ref{sec:conundruum}).

\item \textbf{$\chi_{\mathrm{flare}}$}

To describe the flare color across the blue and near-UV wavelengths in (mostly) line-free continuum
bands, we use the quantity:

\begin{equation}
\chi_{\mathrm{flare}} = \mathrm{C3615} / \mathrm{C4170}
\end{equation}

The error in this quantity
is obtained by propagating the standard
deviation of the fluxes in C3615 and C4170,

\begin{equation}
\sigma_{\chi, \mathrm{flare}} = \chi_{\mathrm{flare}}  \sqrt{ (\frac{\sigma_{\mathrm{C3615}}}{  \mathrm{C3615}})^2 + (\frac{\sigma_{\mathrm{C4170}}}{\mathrm{C4170}})^2 }
\end{equation}

 Formally, the uncertainties of C3615 and C4170 are
the standard errors of the mean values, but some weak
emission line features (e.g., Fe \textsc{i}, Fe \textsc{ii}) appear in
this spectral region; therefore, a better estimate of the uncertainty in the
continuum level in each window is given by the standard
deviation of the flux.  

\chif\ is similar to the Balmer jump, \textit{J}, that has been measured from
 T Tauri star spectra 
\citep{Valenti1993, Herczeg2008} and also in previous M dwarf flare studies in
the blue to derive electron temperatures assuming an isothermal,
isodensity slab of Hydrogen \citep{Kunkel1970}.

\item BaC3615

The quantity C$3615$ (see \#2 above) is the average flare-only continuum flux from $\lambda =
3600-3630$ \AA\, consisting of Balmer continuum emission from
Hydrogen recombination and other possible components that contribute toward the
continuous emission throughout these wavelengths (such as Paschen continuum and
blackbody continuum).  Our estimate for \emph{only} the flare Balmer continuum emission at 3615\AA,
BaC3615, is obtained by extrapolating and subtracting a continuum
that is fit to the blue-optical zone.  In particular, we fit a
straight line to the wavelength windows listed in Table
\ref{table:bbwindows}  from $\lambda=4000 - 4800$\AA\ (BW1\,--\,BW6), extrapolate to
$\lambda=3600$\AA, and subtract an average of these extrapolated
values at $\lambda=3600-3630$\AA\ from the flare-only flux
average (C3615) to obtain BaC3615. In
Section \ref{sec:bluecont}, we find that a \TBB\s$10\,000$ K blackbody
fits the shape from $\lambda=4000 - 4800$\AA\ well and that a $10\,000$ K
blackbody is approximately linear in this wavelength range. 
This procedure is shown for an example flare spectrum in Figure
\ref{fig:adleo_E10}.  Note that by definition, BaC3615 $\le $ C3615.

\item BaC

The estimate for the wavelength-integrated Balmer continuum energy from
$\lambda=3420-3646$\AA.  The BaC is estimated using the same fitting, 
extrapolation, and subtraction procedure as for BaC3615.  Instead of
averaging the extrapolated line value, the line
extrapolation was subtracted at every wavelength in this region.

\item PseudoC

The intermediate zone (between blue-optical and NUV zones) contains the higher order Balmer lines
(H7 and greater) in addition to the Ca \textsc{ii} H and K lines
($\lambda \sim 3934, 3968$\AA, respectively).  
Although Ca \textsc{ii} H is blended with H$\epsilon$ (H7) in these low resolution
data, Ca \textsc{ii} K is resolved.  Within this zone, we integrate the flare flux
from $\lambda = 3646-3914$\AA\ (from the Balmer jump through H8)
and refer to it as the ``PseudoC'' because many of the Hydrogen lines 
blend together (are partially or completely unresolved) at these
wavelengths to form a pseudo-continuum.  Again, as with the BaC, we use an extrapolation
of the line-fit to the blue-optical to estimate the underlying continuum.

\item S\#

 We refer to the time-sequential spectrum number
  (starting at 0 at the beginning of each night)
  with an S\#.  These numbers refer to the $n$B subcategory of spectra
  used for blue continuum analysis as described in Section
  \ref{sec:spectra} (see also Table
  ~\ref{table:obslogAll}).

\item Times

The times on the light curves indicate the number of hours elapsed on the respective
MJD from Table \ref{table:obslogAll}.  Times always refer to midtimes
of the exposure.
The times for the flare data from 2009 Jan 16 are given in ``elapsed hours from flare start'', as
used in \cite{Kowalski2010}; to obtain the number of hours elapsed on
MJD 54847, add 4.2483 hours to the number of elapsed hours from flare start.
\end{itemize}

\subsection{General Descriptive Terms}  \label{sec:descriptive}

Additional terminology used to describe the photometry and spectra are the following:

\begin{itemize}
\item Impulsive Phase

An impulsive phase consists of a fast rise, peak,
and fast decay of the light curve.  

\item ``Peak'' or ``maximum continuum emission''

``Peak'' or ``maximum continuum emission'' always refer to the 
maximum value of C3615 during a flare.  The peak times are given for
each flare in Table \ref{table:times}.

\item Gradual Decay Phase

Gradual phase emission is observed during times of slowly rising or slowly
decreasing emission.  Following HP91, the
``gradual decay phase'' is 
defined as the turnover from fast to slow
decay. 
 The gradual decay phase spectra
are chosen from the section of the light curve as near to the break
from fast decay to slow decay as possible.  We select three spectra
around a time when C3615 is not changing rapidly so that the spectra
can be coadded to increase the signal-to-noise without largely
affecting the interpretation of atmospheric
parameters. 
The red vertical dashed lines in Figures
\ref{fig:appendix_integ1}\,--\,\ref{fig:appendix_integ17} in Appendix
\ref{sec:appendix_times} 
indicate the times that we selected to represent the gradual
decay phases for the flares in our sample.   The gradual decay phase 
times analyzed for each flare are given in Table \ref{table:times}.

\item ``blackbody continuum component''

This term refers to a continuum \emph{slope} 
 that matches the slope of a Planck
function with temperature \TBB.
 A ``hot blackbody'' is used to designate a blackbody continuum component with \TBB $\gtrsim 8500$ K. 

\end{itemize}

\section{The Flare Atlas} \label{sec:spectra_atlas}
The Flare Atlas refers to the collection of time-resolved
spectra and photometry of the twenty flares analyzed in this paper.  In this
section, we present overview figures and tables, and we briefly
describe the categories of flares based on light curve morphology.  For detailed descriptions of 
each of the flares, we refer the reader to Section 3.4 of \cite{KowalskiTh}.
The spectra (original flux and flare-only flux) and photometry ($I_f+1$)
for all nights are available through the VizieR service.

\subsection{Broadband Light Curves}
The impulsiveness index ($\mathcal{I}$) provides the main light curve morphological classification scheme employed in
the remainder of the paper.  For our flare sample,
$\mathcal{I}$ ranges from $0.02-100$. The \emph{impulsive flares (IF)} are those
that have $\mathcal{I} > 1$, whereas the \emph{gradual flares (GF)} have
$\mathcal{I} < 1$. 
 For flares that are close to this dividing line ($\sim0.6-1.8$), we assign the
classification \emph{hybrid flares (HF)}, as these flares have
a prominent impulsive phase (or several impulsive phases) but also share properties
with the gradual flares.  We also considered the \emph{fast} and
\emph{slow} flare classification scheme from \cite{DalEvren}, but this
grouping employs a total decay time
measurement;  in some cases, poor weather, a standard
star sequence, or secondary flares
interrupted the decay measurements.  Using $t_{\mathrm{1/2}}$ (in the
definition of implusiveness) bypasses the ambiguities with measuring precise start and stop times.

The bluest available photometry (SDSS $u$, Johnson $U$, or SDSS $g$) for the flares in our sample
 are shown in Figure \ref{fig:lcphot_panel_if} (nine impulsive IF
 events), Figure \ref{fig:lcphot_panel_if2} (two impulsive IF events -- IF0
 and IF10 -- with less data), Figure
 \ref{fig:lcphot_panel_hf} (four hybrid HF events), and Figure
 \ref{fig:lcphot_panel_gf} (five gradual GF events). 
In Appendix \ref{sec:appendix_times}, we show figures of each flare with the integration
times of the spectra (Figures \ref{fig:appendix_integ1}\,--\,\ref{fig:appendix_integ17})
and the spectrum numbers (S\#'s) indicated. 
 IF0 on AD Leo from HP91 is known as the ``the Great
 Flare'', and IF1 on YZ CMi \citep{Kowalski2010} is known as ``the Megaflare''.  In
 the decay phase of IF1, we refer to the sub-peak at $t=2.1441$ hours as
 the ``Megaflare decay secondary peak \#2'' or ``MDSF2''\footnote{MDSF2 occurs \s1.7 hours after the
   primary peak of the IF1.  MDSF2 is the fourth large sub-peak in the decay
   phase of IF1, and it is the second large sub-peak within the time of
   the spectral observations.}.

 From the light curves, it is apparent that our sample contains a diverse set of 
 peak amplitudes, total durations, and light curve
 morphologies.  The naming convention 
 (IF, HF, and GF) and the ordering of the flares within this
 classification scheme is based\footnote{IF10 is actually the most
   impulsive flare in the sample, but it is excluded from several areas
   of this study due to slow cadence, long integration time, and
   relatively small spectral coverage.} on the value of
 $\mathcal{I}$ as detailed in Section \ref{sec:phot_param}.
Table \ref{table:flare_summary} summarizes the key properties of the
$U$-band photometry: flare ID (col 1), star name
(col 2), date (col 3), time of peak C3615 (col 4),
$I_{f, U}+1$ at peak photometry
(col 5), equivalent duration in $U$ (col 6), $U$-band energy (col 7), 
$U$-band luminosity at peak photometry (col 8), $t_{1/2,U}$ (col 9), and
$\mathcal{I}$ (col 10).  The time-integrated photometric quantities
are calculated only during the time
period when 
spectra were obtained.  Note that the time of peak photometry and time of
peak C3615 may not precisely coincide due to the different cadences and
integration times.

\subsection{Spectra}
 In Appendix \ref{sec:appendix_atlas}, we present the 
   Flare Atlas with a time-sequence of flare-only spectra from $\lambda=3400-7500$\AA\ for each flare event.
In Figures \ref{fig:peak_panels1} \,--\, \ref{fig:peak_panels3},
the flare-only spectra at maximum continuum emission (i.e., maximum C3615) are presented in the same order as the
photometry in Figures \ref{fig:lcphot_panel_if}\,--\,\ref{fig:lcphot_panel_gf}.  The quiescent levels are shown as
dotted lines for comparison.  Table \ref{table:chitable} gives the
$I_{f,\mathrm{C4170, peak}}$, $t_{1/2, \mathrm{C4170}}$, \chifp, and \chifd\ for each
flare as a reference; these values are important constraints
for flare models (Section \ref{sec:modelcomparison}).  A detailed analysis of the continuum will
be discussed in Section \ref{sec:bluecont}; here, a simple Planck
function (light blue line) has been fit to the windows in the blue optical
zone ($\lambda = 4000-4800$\AA; BW1\,--\,BW6 in Table
\ref{table:bbwindows}) to parameterize the slope of the continuum.  The best-fit temperatures
and \chifp\ values are shown in parentheses.   Except for
GF1 (and possibly GF3 and GF5), the GF events are generally too faint for an accurate
continuum fit in the blue-optical zone.  

There are varying amounts of the excess continuum at $\lambda \lesssim
3640$\AA\ above the extrapolation of the blackbody (blue line),
especially among the IF events.   The HF and GF events have
large amounts of excess continuum at $\lambda \lesssim 3640$\AA.  
The spectral trends in the near-UV zone for
the IF, HF, and GF events are
similar to the underlying blackbody curves.  To illustrate this, we
scale the light blue blackbody curves to
the flux at $\lambda=3600-3630$\AA\ and show these in yellow,
which basically have the same slopes as the light blue fits. 

\subsection{Overview of the Flare Atlas} \label{sec:overview}
The flares in our sample represent relatively large amplitude and high energy events on dMe stars.
For example, the average $U$-band flare energy on YZ CMi is $\sim0.03
\times10^{32}$ ergs \citep{Lacy1976}, which is slightly smaller than the lowest energy
flare (IF8) on this star in Table \ref{table:flare_summary}.  The IF events have a large spread of amplitudes
from low ($I_{f,peak} + 1 \sim 2.5-3$) to very large ($I_{f,peak} +
1 > 10$). The IF events also generally have a classical, simple shape:
a fast-rise, a fast decay, and a more gradual decay beginning at a low level, $\sim20$\% or less,
relative to the peak.  Durations range from several minutes (e.g., IF6, IF8)
to several hours (e.g., IF1, IF3).  
Some IF events have secondary flares but they are usually dominated by a single, large-amplitude peak.  The HF events also have fast rise components,
but they exhibit marked deviations from the classic flare shape, such
as multiple continuum peaks of comparable amplitude during the impulsive phase (e.g., HF1) and an elevated
or prolonged decay phase (e.g., HF2).  These flares are low
to moderate amplitude ($2 < I_{f,peak} + 1 < 5$) and usually longer
lasting ($> 1$ hour) than the IF events of comparable amplitude.  The GF events are low-amplitude
($I_{f,peak} + 1 < 2.2$) except for GF1 which has $I_{f,peak} + 1 \sim 8$.  The rise phases are notably slower,
although they can have distinct periods of faster and slower emission;
and may be accompanied by intermittent peaks (e.g., GF1 and
GF3).  However, these continuum peaks do not significantly contribute to the overall
timescales, which can be several hours for even the low amplitude flares.

There is often another local
maximum (i.e., a secondary flare) 
just after the first peak but before the gradual decay phase.  
Secondary flares are especially evident in IF0, IF1, IF3, IF10, HF2,
and HF4.  Secondary flares usually occur at about half the peak
flux level or less.  Although the four
 largest amplitude events all have secondary flares (with \s15 or more occurring during IF1), lower amplitude events also show them.  IF4
 is a large-amplitude event that shows a stall in the fast decay
 resulting in a nearly constant
 flux level before continuing a fast decay.  This also may be
 interpreted as a relatively low-amplitude secondary flare.  

In addition to the quantitative ``impulsive'', ``hybrid'', and ``gradual'' classification
schemes, we find the following descriptive groupings using the data in Tables
\ref{table:flare_summary}, \ref{table:chitable} in addition to the
light curve data \citep[not shown here; see Chapter 3 of ][]{KowalskiTh} of the spectral components from Section \ref{sec:spectra_param}.

\begin{itemize}
\item \textbf{Simple, classical flares (IF2, IF5, IF7, IF9):}
These flares have moderately large amplitudes
($I_{f, U, \mathrm{peak}}+1\sim5-12$) and 
energies ($E_U \sim 2\times10^{31}-2\times10^{32}$ ergs). A defining
characteristic is that the $U$-band (or bluest
photometry) follows the evolution of C4170, which in fact, holds for
most of the flares.  The different timescales of decay between the spectral components are evident:
although C4170, BaC3615, and H$\gamma$ vary from fastest to slowest, there is a
spread of relative $t_{1/2}$ values, with IF9 having the
smallest ratio of $t_{1/2,\mathrm{C4170}} / t_{1/2,\mathrm{BaC3615}}$.  Except for IF7, these flares don't have obvious
secondary flares in the decay. 

\item \textbf{Low amplitude flares (IF6, IF8, GF2, GF3):}
The low amplitude flares have
$I_{f, U, \mathrm{peak}}+1\sim$2 (GF) and $\sim$3 (IF).  There are both complex
and simple events in this group, and the durations range from minutes to hours.  These flares have lower
energies:  the low-amplitude short-duration flares have $E_U\sim5\times10^{30}$ ergs
and the low-amplitude long-duration flares have $E_U\sim10^{31}$ ergs in
their first main peaks; the complex flares are more than 10
times as energetic as the simple flares, even though the simple flares have larger
peak amplitude in $U$ (or $u$).

\item \textbf{Multiple-peaked, medium amplitude flares (HF1, HF2, HF3,
    HF4):}
These are medium
amplitude flares with multiple peaks in the impulsive phase, having
$I_{f, U\mathrm{peak}}+1 \sim 2.5-5.5$. 
 These flares are all hybrid flares (HF type).   
Generally, in the HF
  flares, the evolution of the $U$-band closely matches the
  evolution of the 
  BaC3615, whereas typically in the IF flares, the evolution in the $U$-band
  matches the evolution in C4170. 
In these flares, \chifp\ is 2.3\,--\,3.

\item \textbf{High energy flares (IF0, IF1, IF3, IF4, IF9, IF10, GF1):}
The high energy flares have $E_U >3 \times
10^{32}$ ergs.  The most impulsive flares (except IF2) tend to be
the most energetic.  C4170 and
$U(ug)$ photometry track each other well in the high energy flares, whereas
the BaC3615 is more similar in evolution to the \Hg\ line.  Secondary flaring to varying
degrees is observed in all cases.

\item \textbf{Low amplitude, simple gradual flares (MDSF2, GF4, GF5):}
These flares have a moderate amount of energy ($E_U \sim
10^{31}-10^{32}$ ergs) for their low peak-amplitudes.
Interestingly, they lack a prominent fast decay phase after the peak emission.  
In GF5, short impulsive events are observed later in the flare
decay.  MDSF2 during the IF1 gradual decay phase is included in this
group; its values of $E_U \sim 10^{32}$ ergs and
$\mathcal{I_U}\sim0.6$ qualify it a high energy, hybrid/gradual flare.

\end{itemize}

\subsubsection{General Relationships Between Spectral and
  Photometric Properties} \label{sec:general}

The general relationships between spectral properties (\chifp, \Hg/C4170) and
photometric $Uug$ light curve properties ($\mathcal{I}$)
of the Flare Atlas are summarized with three figures,
Figures~\ref{fig:if_vs_chi} \,--\,\ref{fig:FEU_HG_inset}.  Figure
\ref{fig:if_vs_chi} (\chifp\ vs. the impulsive index, $\mathcal{I}$) indicates that 
\emph{the
instantaneous continuum shape at maximum amplitude in the broadband
light curve is 
linked to the overall evolution of the flare.}  From this figure (see
also values in Table \ref{table:chitable}), it
is evident that the most impulsive flares have the lowest
$\chi_{\mathrm{flare,peak}}$, with most $\lesssim 1.8$ and all
$\lesssim 2.2$.
The HF events have
intermediate values, \chifp$\sim 2.3-3$, and the GF events have larger but
more uncertain values (besides GF1), \chifp$\sim3+$.  
The errors on \chifp\ are typically 0.01\,--\,0.12 (corresponding to relative errors
of 3\,--\,10\%),
but some flares have significantly larger errors with the GF events
generally having the
largest uncertainties. We do not consider the \chif\ values with 
$\sigma_{\chi_{\mathrm{flare}}}/$\chif $> 0.2$.  This excludes GF4
from \chifp\ analysis and IF5, IF6, IF8, HF4, GF2, GF3, GF4, and GF5 from
\chifd\ analysis.
 Using standard error propagation, we determine the confidence levels
 by which the IF, HF, and GF sequence is ordered according to the \chifp\ parameter.
We find that IF9 and HF1 are separated by 4.5$\sigma$, HF4 and GF1
are separated by $<$1$\sigma$, HF1 and GF1 are separated by $5\sigma$,
and IF5 and IF9 separated by almost $4\sigma$.   The differences in
\chifp\ are generally more significant between the IF and HF events
than for the HF and GF events;  this is not surprising given that the
IF events tend to have the larger signal-to-noise due to their larger amplitudes.
We conclude that the differences in \chifp\ are significant between
IF, HF, and GF events, and even between certain IF events.

In Figures \ref{fig:FEU_HG}\,--\,\ref{fig:FEU_HG_inset}, we show the H$\gamma$
line flux divided by the continuum C4170 flux
(both taken at peak C4170;  essentially this is 
the equivalent width of the C4170 flare continuum in units of \AA).
The flares are color-coded by the IF/HF/GF designation.  
 Figure \ref{fig:FEU_HG_inset} shows a narrower range of values than
Figure \ref{fig:FEU_HG} where IF2, IF3, IF7 , IF8, IF9, and IF10
cluster together at
\chifp\ $\sim 1.6 - 1.8$.  The first and second peaks of
the Great Flare (IF0) are also included in Figure \ref{fig:FEU_HG_inset}.   Note that IF8 and
IF3 are the smallest and largest amplitude impulsive flares (with full
spectral and time 
coverage) respectively on YZ CMi, yet they
show very similar peak characteristics.  IF4 has the lowest
\chifp\ (\s1.3) and also line-to-continuum ratio (\s6)

In Figures \ref{fig:FEU_HG}\,--\,\ref{fig:FEU_HG_inset}, we see 
that the light curve morphology and \chifp\ are related to the ratio
of \Hg/C4170.  The IF events have \Hg/C4170 $\lesssim$50 (most IF
events fall below 30), the HF
events show 38$\lesssim$\Hg/C4170$\lesssim$75, and the GF events
show 85$\lesssim$\Hg/C4170$\lesssim$165.  We 
find a very strong relationship between \chifp\ (recall, \chifp\ $=$
C3615/C4170) and \Hg/C4170:
\begin{equation}
\chi_{\mathrm{flare,peak}} \approx 0.020(\pm0.002)H\gamma/\mathrm{C4170} + 1.28(\pm0.05)
\end{equation}
Among the IF events, IF3, IF5, IF7, IF8, and IF9 fall closest to the
best fit line.  Although IF5 and IF6 are the impulsive flares (see Figure
\ref{fig:lcphot_panel_if}) with the largest values of $\chi_{\mathrm{flare,peak}} \sim
2.2$, they also have relatively large values of \Hg/C4170
\s40\,--\,50.  IF1 is an outlier (with much more relative \Hg\
radiation for the \chifp\ value predicted by the red line), as the peak data
correspond to the peak of a secondary flare (MDSF2) during the gradual
decay phase.  
 As \chif\ is effectively a measure of the
Balmer jump height, this relation implies that flares
with larger Balmer jumps relative to the C4170 flux have larger Balmer line fluxes
relative to C4170 flux. 
This implies a connection between the relative amount of Balmer line radiation
and Balmer continuum radiation.
\chif\ is thus a very important quantity because it can be measured without
spectra \citep{Kowalski2011CS}, yet apparently it can be used as a
diagnostic of the Balmer line and continuum radiation.

\section{Emission Line Analysis} \label{sec:emissionlines}
Emission lines are used to probe the temperatures and densities,
and therefore different heights, of a flaring atmosphere by matching 
models to the observations.  As current radiative-hydrodynamic (RHD) models predict line flux
decrements and profiles that are relatively in 
agreement with the observations \citep{Allred2006}, heights
of formation \citep[via the contribution function, e.g. ][]{Magain1986,
  Carlsson1997, Carlsson1998} can be used to constrain the time-evolution
of heating at different layers in the atmosphere \citep[e.g.,]{Hawley1992}.  Ultimately,
whatever heating mechanism is used to
explain the continuum properties during flares must also be consistent with the observed emission line
properties.  For example, \cite{CramWoods1982} found that the model
atmosphere that best matched the continuum observations did not match
the corresponding properties of \Ha.  As a
result, they suggested a combination of several models to
explain the observations.

An extensive analysis of the Balmer line broadening, line flux and
energy decrements, and 
\Ha\ time evolution for the Flare Atlas can be found in \cite{KowalskiTh}.  Here, we
present the emission line results that 1) are most relevant to understanding
the origin of the continuum and 2) would most greatly
benefit from future observations (e.g., at higher
cadence).  

\subsection{The \Hg\ line and its relation to the continuum} \label{sec:Hgamma}
As was discussed in Section \ref{sec:general} (Figures 
\ref{fig:FEU_HG}\,--\,\ref{fig:FEU_HG_inset}), a larger ratio of \Hg\ flux to C4170 generally results from 
flares with larger \chifp.  Therefore, the height of the Balmer jump is related to
the relative amount of Balmer line radiation produced at peak
emission.
H$\gamma$ is a useful diagnostic because it is a strong, easily measured line, in both high
and low flaring states.  This line is also the highest order Hydrogen
line calculated in most RHD models to date.  Furthermore, its
properties have been studied extensively in the past for dMe flares.
The $U$-band is widely used for flare monitoring
\citep{Moffett1974}, and it is a 
diagnostic of the continuum flux and energy, since colorimetry studies
have shown the peak of the white-light occurs in the $U$-band or at
shorter wavelengths \citep{Hawley1992}.  

The $U$-band contains higher order Balmer lines and
Ca \textsc{ii} H and K, but $\gtrsim$90\% of the $U$-band energy is
due to continuum radiation
\citep[][see also HP91]{Doyle1988}. 
It is a well-established property that the Balmer lines evolve more slowly
than the continuum, staying elevated longer \citep{Kahler1982} and
sometimes peaking as late as the end of the impulsive phase
\citep[HP91,][]{GarciaAlvarez2002, Gurzadyan1984}.  The time-integrated energies scale over approximately 4.5 orders of magnitude with 
$E_U \sim 25 \times E_{H\gamma}$ (HP91).  Using a larger sample in
this study, we find that the BaC3615 component of the $U$-band is even
better correlated with the
properties \,--\, including peak luminosity, total energy, and time-evolution.

Here, we analyze the timing in detail.  Figure~\ref{fig:thalf_UHg} shows the relation between  $t_{1/2, H\gamma}$ and
$t_{1/2, \mathrm{BaC3615}}$.   We find a linear relation among the IF and HF events \emph{without
multiple peaks} (crosses).  The fit to these flares is shown as a light
blue line, given by
\begin{equation}
 t_{1/2, \mathrm{BaC3615}} \approx 0.54(\pm0.01) t_{1/2, H\gamma} + 0.4(\pm0.1)
\end{equation}
In other words, $t_{1/2}$ is twice as fast in the Balmer continuum as
in \Hg.

The GF events
(shown in open circles) follow a different trend with 
nearly equal timescales in H$\gamma$ and BaC3615. GF1 is a multiple-peaked flare, and it produces copious BaC3615 with
a long timescale. The IF and HF events with
2\,--\,3 peaks (IF0, IF4, HF1, HF2, and HF4) spaced relatively close in time\footnote{These have
  C3615 or $U$-band peaks separated by \s6.3, 1.8, 1.5, 1.9, and 3.0 minutes, respectively.} are shown with red squares and
are labeled.  IF4 and HF2
are 
double-peaked flares and are apparently outliers (the red squares with
$t_{1/2, H\gamma} \sim 16$ minutes).  IF4 falls closer to the GF distribution, and
HF2 (also possibly IF0) have much faster BaC3615 timescales given the H$\gamma$
timescale predicted by the light blue line.  HF1 and HF4 fall near 
the single-peak distribution and the GF distribution, respectively.  Note that $t_{1/2}$
depends on which peak dominates and also how it is measured (e.g.,
the time delay between multiple peaks).  A larger sample of
double-peaked IF and HF events would constrain their different
timing behavior in \Hg\ and BaC3615.

The IF0, IF4, and HF2 events exhibit large delays of \s$10,
4$, and 3 minutes, respectively, between the times of
maximum continuum (C3615) and maximum line (\Hg) emission, suggesting a 
difference in the heating and cooling timescales of the continuum and
line radiation during the impulsive phase.   We find that
these relatively common, large lags result when secondary
continuum peaks lag the primary events, as discussed above for Figure \ref{fig:thalf_UHg}.  For most
flares without a relatively large secondary event following closely
after the first peak,
however, there is a difference of less than one minute (and in most
flares, no lag within the time resolution of the spectra) between the peak times of the continuum and H$\gamma$ line.
The short delay in peak times of $<1$ minute suggests that a common 
heating mechanism produces the impulsive phase of line and continuum
radiation during high energy flare events with relatively simple
morphology, such as IF3 and IF9.  The largest (apparent) lags of 18 and 15
minutes result in the GF2 and GF3 events, which have two (or more) primary
peaks in C3615 of similar amplitude separated by large
amounts of time ($\gtrsim 15$ minutes).  However, considering the time only
around the first
impulsive phase in the GF2 and
GF3 events, the lags between the local maxima of \Hg\ and C3615 are 0\,--\,0.5 minutes.
We speculate that lags are the result of superimposed flare events, and
the emission components with different decay timescales 
 (e.g., \Hg\ and C3615) add to produce different timings of
 the peaks.  We plan to investigate this further in a future paper. 

\subsection{The Hydrogen Balmer ``time-decrement''} \label{sec:timing}
We connect the timing properties of the Hydrogen Balmer components
with a simple new relation, the \emph{time-decrement}.
 We use the exceptionally high-quality data covering the time-evolution of the Hydrogen lines in 
flares IF3, IF9, HF2, and GF1 (Figures \ref{fig:lines_IF3_B},
\ref{fig:lines_IF9_B}, \ref{fig:lines_hf2_B}, and
\ref{fig:lines_GF1_B}, respectively)
to show this relationship and how it varies among flare type.  In
addition to \Ha, \Hb, \Hg, and \Hd, we show the continuum evolution of PseudoC,
BaC3615, C4170, Ca \textsc{ii} K, and He \textsc{i} $\lambda4471$.  The fluxes are normalized to their peaks to illustrate
the different decay trends.  Figure \ref{fig:lines_IF3_B} (bottom panel)  also shows
the rise phase in detail for the absolute fluxes of \Ha, \Hb, \Hg, and
\Hd.  

From the data in Figures \ref{fig:lines_IF3_B} \,--\,
\ref{fig:lines_GF1_B}, it is evident that the higher order Balmer lines have a faster decay time
compared to the lower order lines, declining to a lower relative flux by
the end of the impulsive phase.  This effect was noted by
\cite{Doyle1988} and HP91.  According to $t_{1/2}$, the ordering
of the components from fastest to slowest is C4170, He \textsc{i}
$\lambda$4471, BaC3615, PseudoC, \Hd, \Hg, \Hb, \Ha, and Ca
\textsc{ii} K.  \Hg\ and \Hd\ have rather similar decay rates, but \Hg\ is
 apparently slower\footnote{It is possible that the light curve
   evolution, and hence $t_{1/2}$, is affected by the amount
   of \emph{absorption} at peak such as during IF3 (see Section \ref{sec:wing_absorption}).}.
 Ca \textsc{ii} K will be discussed in Section \ref{sec:caiik}.
 
 To connect the timescales across the Balmer series, we plot the
 $t_{1/2}$ value of each transition as a function of the
 wavelength of the transition in Figure \ref{fig:nature_baby} for IF3
 (red asterisks), IF9 (black diamonds), GF1 (black circles), and HF2 (black
 squares).   An estimate of $t_{1/2}$ for the  H10 line (using an extrapolation from the blue-optical
as the underlying continuum) from the PseudoC component is included as
well as the value of $t_{1/2}$ for 
BaC3615.  Remarkably, this ``time-decrement'' relationship among 
the Balmer spectral components appears nearly linear in
wavelength space for the two classical, impulsive (IF) events in the figure.  The
time-decrement for IF3 is fit with a linear
relation (red dashes) to show the trend
\begin{equation}
t_{1/2}=0.014\lambda - 36.96.
\end{equation}
 IF9 has a similar
time-decrement as IF3 but with a factor of $\sim$1.7 shorter
timescales.  The linear relation for IF9 (purple dashes) is  
\begin{equation}
t_{1/2}=0.0077\lambda - 21.83  
\end{equation}

The scaling between the time decrement relationships of these two flares
indicates a fundamental similarity between the heating/cooling
processes of Balmer emission produced in medium and large classical
flares.  A simple discussion of the physical parameters that produce
the linear time-decrement of the IF
events is given in Section \ref{sec:flare_morphology}, but this
phenomenon should be investigated with detailed radiative-hydrodynamic models.

The gradual flare GF1 has the same $t_{1/2}$ for H$\gamma$ and H$\delta$ as IF3, but the lower order 
line evolution is faster and the higher order line evolution is slower: in other words, the time decrement for GF1 is
flatter compared to the impulsive flares.

The time-decrement of HF2 is flat for the lower order
lines and steepens for the higher order lines and BaC3615.
HF2 has two continuum peaks
with a highly elevated gradual decay phase, and its overall time
evolution is a result of the combined (i.e., spatially unresolved)
heating during 
the two emission peaks.  
Recall that in Figure \ref{fig:thalf_UHg} (Section \ref{sec:Hgamma}) we compared the $t_{1/2}$ of \Hg\
 and the BaC3615 between all flares, and found a general relation for the simple events
 while complex events, such as HF2, behaved differently.
Figure \ref{fig:nature_baby} elucidates this difference, 
consistent with its \emph{hybrid} classification.
The time-evolution of HF2 is studied further in Appendix D of
\cite{KowalskiTh}.  To understand the time-decrement of HF2,
 it will be necessary to superpose two simple events (each with
a linear time decrement) with the spacing of the two peaks in HF2.
A study that explores the results of superposing emission of simple events will be presented in a future paper.

The $t_{1/2, \mathrm{C4170}}$ values (Table \ref{table:chitable}) for
these four events are also shown in Figure \ref{fig:nature_baby} as light blue
 symbols.  The very fast evolution ($t_{1/2} = $1\,--\,15 minutes) of C4170 does not follow the
 time-decrement relationship among the Balmer emission components.
 The ratio of the $t_{1/2, \mathrm{C4170}}$ values (not used in the linear
 fits) between IF3 and IF9 is $\sim$3.8 which is larger than 
 the scaling factor of $\sim1.7$ between the time-decrements of the
 respective Balmer components.  This indicates that several heating/cooling
 processes are simultaneously present during the flare -- a dominant
 process for the Balmer emission component, which approximately scales
 (e.g., similar heating over larger area) between classical flares, and a dominant process for the
 C4170 emission component, which does not scale in the same way between
 classical flares.  Therefore, it is possible that these different heating
 processes are present for different lengths of time (e.g., the C4170
 heating process only in the impulsive phase).  If the C4170 originates from
 the same (or similar) heating process that produces the Balmer lines
 \,--\, as was concluded from the similar timing of the peaks of \Hg\ and
 C3615 for these two flares (Section \ref{sec:Hgamma} \,--\,
 the fast timescale of C4170 either implies formation in a denser region of
 the atmosphere where the cooling is more efficient or a threshold in
 the strength of the heating process that can produce C4170.

Additionally, the time-evolution of C4170
 gives important insight into how the heating processes (and
 hence time-decrement) vary between the types
 of flares (IF, HF, GF).  The $t_{1/2,\mathrm{C4170}}$ is
 relatively large for GF1, and this may be related to the general
 flatness of the Balmer time-decrement relation.  In other words, the
 GF1 event has similar timescales for its spectral components,
 implying that they are more closely related in their formation and
 persistence over time.  The $t_{1/2,\mathrm{C4170}}$ for HF2 is
 slightly large compared to the Balmer time-decrement, and this flare
 consists of two temporally resolved, yet spatially unresolved,
 peaks superposed.  The different heating properties in the two peaks
 of HF2 may generate the mixed time-decrement behavior.  The IF
 events show a simple, linear relationship among the Balmer emission
 components, yet they also exhibit the largest difference between
 Balmer emission and C4170 timescales.  The timing of the C4170 and Balmer series are important
 constraints for consistently modeling these spectral components together.

\subsection{The Ca \textsc{II} K Neupert-like Effect} \label{sec:caiik}
The formation and time evolution of the Ca \textsc{ii} K line has long been a
mystery, including why it responds slowly in flares and peaks after
the Balmer lines.
It has been associated with formation in the lower chromosphere in
time-dependent, nonthermal electron heating 
models of the impulsive phase \citep{Abbett1999, Allred2006}, and phenomenological modeling has shown that Ca
\textsc{ii} K emission can also be associated with a hotter, higher
region in the flaring chromosphere \citep{Schmidt2012}.  
Models of coronal X-ray backwarming have shown that relatively large amounts of Ca
\textsc{ii} emission originate from a range of heights in the flare
chromosphere for $T \sim 5000-7600$ K \citep{Hawley1992}. The timing
properties have been interpreted with a scenario in which hot flare
loops cool down to the
temperature of Ca \textsc{ii} K formation
\citep{Gurzadyan1984, Houdebine2003, Crespo2006}, but this
remains to be tested with radiative hydrodynamic models of the gradual
decay phase. 

The Neupert effect \citep{Neupert1968} is an observed relation between the signatures of
impulsive phase nonthermal particles and gradual phase coronal
heating. The Neupert effect is usually reported as the
proportionality between the integral of nonthermal 
microwave, hard X-ray, or white-light emission and the luminosity of thermal
soft X-rays.
The Neupert effect has been observed in solar \citep{Dennis1993} and stellar
\citep{Hawley1995, Gudel1996, Osten2004, Fuhrmeister2011} flares, and is a
fundamental aspect of the standard flare model.  It is usually interpreted in terms of the chromospheric evaporation process
\citep{Fisher1985}, whereby the nonthermal particles impact
chromospheric material, which then ablates into the corona and emits thermal
radiation at millions
of degrees well into the gradual phase of the nonthermal particle
emission.  The models of \cite{Hawley1992} showed that the Ca
\textsc{ii} K line flux evolution during the Great Flare could 
be produced from X-ray backwarming from a flare corona
at $T \sim 10$ MK.
Here we investigate whether Ca \textsc{ii} K follows a Neupert
effect representing the gradual phase emission (like soft X-rays) using
C4170 to represent the impulsive phase emission.

Ca \textsc{ii} K is the most gradual line in the intermediate and
blue-optical spectral zones (Figures \ref{fig:lines_IF3_B},
\ref{fig:lines_IF9_B}, \ref{fig:lines_hf2_B}, and \ref{fig:lines_GF1_B}); well-known characteristics are a longer
rise time than the Balmer lines, a late peak
in the beginning of the gradual decay phase, and a slow return to
quiescence (e.g., HP91).
In Figures ~\ref{fig:CaIIK1} \,--\, \ref{fig:CaIIK2}, we show the flux of Ca
\textsc{ii} K (blue diamonds), the flux of C4170 (crosses) and the cumulative
integral of C4170 (red lines) for flares with ARC 3.5-m/DIS data.   We
calculate the cumulative integral of C4170 until the time of Ca \textsc{ii} K
maximum, which in some cases occurs between the end of the impulsive
phase and beginning of the gradual decay phase
(IF0, HF1, HF2, GF1) and in some cases occurs deep into the gradual
decay phase (IF3, IF9).  This variation in the timing of the Ca
\textsc{ii} K peak has been 
seen in the literature \citep{Houdebine2003}.
There is a varying degree of similarity in the evolution between the Ca
\textsc{ii} K line and the cumulative integral of the continuum.  The closest
similarities are found for the flares IF6, IF7, IF9, HF1, HF2, GF1, and GF2; the largest
differences for IF2, IF3, IF4, and GF3.  For the fastest flares, it would
be useful to test this relation with higher cadence data.   For the flare IF1, we show the cumulative integral of the $U$-band
light curve scaled to match the maximum of the Ca \textsc{ii} K light
curve.  Despite not knowing if the maximum of Ca \textsc{ii} K
within this time window (at $t\sim1.75$ hours) is the absolute maximum
for the flare, the cumulative integral of $U$ follows the Ca
\textsc{ii} K from $t=1.3$ hours to $t=1.75$ hours. 
In some flares, the Ca \textsc{ii} K line evolution does not
obviously respond during the first part
of the impulsive phase, showing evidence of a delay compared to
C4170; this is seen in IF3, IF4, HF1, and HF3.  Ca \textsc{ii} K
shows a 
\emph{decrease} (although by a small amount) relative to the previous
spectrum in the impulsive phase of 
the flares IF2, IF4, IF5, IF9, HF1, and HF3.  In the flare IF2,
a decrease in line flux is most noticeable, disappearing completely from the decay phase value
of HF1 before increasing again in the beginning of the gradual decay
phase of IF2.
We speculate that the lack of response in Ca \textsc{ii} K is a result
of the formation of strong hot, blackbody emission, resulting in 
\textsc{ii} K \emph{absorption} (Section \ref{sec:astar}) effectively canceling the 
amount of emission from a different spatial location.  
Of course, this explanation requires confirmation from detailed models.  

If the Neupert effect underlies the relation between the cumulative
integral of C4170 and the flux of Ca \textsc{ii} K, then
the gradual evolution of Ca \textsc{ii} K could be related to the chromospheric evaporation
process.  Under this interpretation, C4170 represents impulsive phase
heating of the lower atmosphere and material is evaporated into
coronal loops;  
regions of the chromosphere
(perhaps even in distant regions away from the main flare loops) are then heated to \s6000 K
from, e.g.,
incident X-ray and EUV (XEUV) backwarming radiation originating from 
coronal plasma.  
 A connection between the fastest component, C4170, and the slowest component,
Ca \textsc{ii} K, in our stellar flare observations provides important constraints for RHD models that seek to
produce a consistent picture of the flare process whereby the C4170
can be incorporated into the standard solar flare model of, e.g.,
\cite{Martens1989}.
Because we observe this relation in flares relatively independent of 
morphological type, the degree to which the Ca
\textsc{ii} K Neupert relation holds gives additional constraints for how the heating process varies between
flares of a given type.  For example, of the four most impulsive flares in
 the DIS sample, the Ca \textsc{ii} K Neupert relation clearly does
 not hold for IF2 and IF4.  Note that \cite{Osten2005} found a 
 violation of the Neupert effect from soft X-ray observations during a
 (rather impulsive)
 $U$-band flare, suggesting that the relation may break
 down in some heating scenarios.  

\subsection{Hydrogen Balmer Flux Budgets} \label{sec:hydrogen}

The amount of flux that we can attribute to Balmer emission compared
to the total flare emission constrains the amount of unexplained  
energy (e.g., white-light continuum) for model predictions.  
In this section, we characterize the 
amount of non-Hydrogen Balmer radiation as a function of morphological flare type.  We also investigate the 
relationship between the Hydrogen Balmer flux and \chif. 
Current models fail to reproduce \chifp\ (see Section \ref{sec:modelcomparison});  therefore,
 the relative amount of Hydrogen Balmer emission present at times of
 peak flux 
can guide modeling efforts.  

The Hydrogen Balmer (HB) component
is the sum of the fluxes in H$\delta$,
H$\gamma$, H$\beta$, the PseudoC, and the BaC (see Section \ref{sec:spectra_param})\footnote{For a comparative
energy budget among these components, please refer to
\cite{KowalskiTh}, Chapter 4.}.   The HB component does not include H$\epsilon$ (H7) in the
blue-optical because it is blended with Ca \textsc{ii} H\footnote{Including
H$\epsilon$ in the HB flux budget -- by subtracting Ca \textsc{ii} K as an
estimate for Ca \textsc{ii} H -- increases the peak percentages by $\sim$1\% and the
gradual phase percentages by $\sim$2\%.}, and it does not include \Ha\
because it is not available for some flares due to lack of wavelength
coverage or saturated flux values (IF0, IF1, IF10, IF4, IF6, HF4,
and GF3; see \cite{KowalskiTh} for an analysis of the \Ha\ line
evolution in the Flare Atlas).
We investigate the coarse time-evolution of the percentage of HB flux
(``\%HB'' or ``HB flux ratio'') to
total 
($\lambda = 3420-5200$\AA) flare flux at the time of
maximum continuum emission and near the
beginning of the gradual decay phase (these times and spectrum
numbers are given in Table \ref{table:times} and are indicated
 by red
vertical lines in Appendix \ref{sec:appendix_times}, Figures
\ref{fig:appendix_integ1}\,--\,\ref{fig:appendix_integ17}).  The
gradual decay phase
measurements are averaged over three spectra, and the peak
measurements are averaged in several of the GF events to increase
signal-to-noise. The results are shown in Figure \ref{fig:hb_phases} (left panel)
and given in Table \ref{table:hb_phasez} for the impulsive, hybrid,
and gradual flares.  The spectra S\#23\,--\,25 (gradual decay) and \#113 (just before peak of MDSF2)
measurements are shown for IF1 as light blue star symbols and the IF0 peak and
decay spectra are shown as dark purple star symbols.

Nearly every flare shows an increasing percentage of HB flux from the
peak to the gradual decay phase\footnote{Except
   for GF5 which shows equal HB percentages in peak and decay;
  for this flare, the peaks and decay phases produce comparable
  broadband fluxes - see Figure \ref{fig:appendix_integ17} in Appendix
\ref{sec:appendix_times}.}, indicating that the gradual decay phase emission is
marked by an increased relative importance of Hydrogen Balmer emission.   The percentages
increase by
\s20\% for most flares; larger changes by \s30\% occur during IF0, IF3, IF4,
IF5.  Except for IF3, these flares also
have very sudden breaks from impulsive to gradual phases.  Significantly smaller changes, $<$20\%,
occur during HF3, GF1, and GF5 because the gradual decay phase of these flares
begins at a relatively high amplitude compared to the peak, and the 
break from impulsive to gradual phases is much less defined.  

There is also an increasing percentage of Hydrogen Balmer emission during both the peak
and decay phases according to the IF/HF/GF sequence, in agreement with
the flux of \Hg\ divided by C4170 at peak emission in Figure
\ref{fig:FEU_HG}. The impulsive flares show a range of 3\,--\,24\% at
peak, with most between 11\,--\,17\%, but changing to $\sim30-52$\% 
 during the
gradual decay phase.  The HF events are scattered around $25-35$\% at
peak and $40-50$\% in the gradual decay phase, while the GF events are
scattered around $35-50$\% at peak and 40\,--\,65\% in the gradual
decay phase.  The GF events show more scatter and less uniformity in
their behavior compared to the IF and HF events.
Note the light curve morphology of GF1 (Figure
\ref{fig:appendix_integ2} of Appendix \ref{sec:appendix_times}); there
  are impulsive phases at $t\sim$1.92 hours, $t\sim1.94$ hours, and
  $t\sim2.02$ hours prior to the major, broad
  emission peak at $t\sim2.13$ hours.  
The values in these peaks are $\sim$28\%, 25\%, and 36\%, compared to
38\% in the main peak.

The individual flux contributions from \Hd, \Hg, \Hb\, and \Ha\ are also
given in Table \ref{table:hb_phasez} for the peak phases (maximum
C3615) of each
flare.  The values for \Hd, \Hg, and \Hb\ also trace the morphological classification of a
flare:  IF events produce $<$1\,--\,2\%, HF events produce \s2\,--\,3\%, and GF events
produce 3\% or more of the $\lambda=3420-5200$\AA\ flux in
each of these Hydrogen Balmer lines.  When available, the flux contribution from
\Ha\ is generally similar to the other Hydrogen lines at maximum
continuum emission.  The individual contributions are
important constraints for RHD models that calculate the detailed
radiative transfer using a relatively small Hydrogen atom (6 levels is
typically used) and because blending from neighboring Hydrogen lines does not
affect the flux measurements of the lower order transitions.

The evolution of \chif\ (C3615/C4170) from the peak to the gradual phase times is 
shown in the right panel of Figure \ref{fig:hb_phases} and the values
are given in Table \ref{table:chitable}.  Indeed, both \chif\ and the percentage of HB emission
increase during the flare gradual decay phases, implying a connection between
these measures.   The IF events show
rather similar trends: they have a change in the percentage of HB
emission of $\sim20-30$\%
and $\Delta$\chif \s 1 from the peak to the gradual decay
phase times\footnote{It is striking that the largest amplitude flares
have $\sim$40\% HB at the beginning of the gradual flares.
The data for IF1 is far into the gradual decay phase; extrapolating back
to $t=0.8351$ hours at the beginning of gradual decay phase (when
there were no spectra) and using a fit of \%HB(t) $=0.429 - 0.06
\times t$, we predict 38\% HB contribution for
this flare.  IF3 has an HB contribution of 37\% at the beginning of
the gradual decay phase.  The large amplitude
flares IF0 and IF10 have \s44\% of HB emission in the gradual decay phase; however these
flares did not have spectra as red as $\lambda =5200$\AA\ to allow a 
 directly comparable energy budget to be calculated.}. The IF5 and IF6
events stand out as IF events that have a large amount
of HB ($>20\%$) and large \chif\ (\s2.2) at peak (see also Section \ref{sec:general}).  These two flares
have percentages of HB emission that rise to large
values ($42-52$\%) and values of \chif\ that increase to \s4 in the gradual
decay phase, further strengthening the connection between these two
measures (\chif, HB flux / total flux) even for the seemingly peculiar IF events that are not as similar as the other flares
in the IF type.

The peak phase value of the percentage of HB emission 
for IF1/MDSF2 coincidentally falls within the same range as the other
 IF events; however, the percentage of the newly formed 
Balmer \emph{emission} is uncertain due to Hydrogen Balmer
\emph{absorption} during this secondary flare (Section
\ref{sec:astar}).  The \chif\ of IF1 for peak of MDSF2 and decay at the
start of the spectral observations (light
blue star symbols) are also consistent with the other IF events, even though a large amount of
decay emission is present at the start of MDSF2. 

A time-resolved analysis of IF1 reveals the strong connection between
the detailed
evolution of \chif\
and the percentage of HB emission during successive impulsive (e.g.,
MDSF2) and gradual decay phases.  In Figure \ref{fig:megaflare_All1}
(top), the \chif\ evolution is compared to the $U$-band evolution over
1.3 hours of the decay phase of IF1.
The trends are anti-correlated during the secondary flares with
$U$-band peaks
at $t\approx1.6$ hours and $t\approx2.16$ hours (MDSF2), 
as shown in \cite{Kowalski2011CS}.
In Figure \ref{fig:megaflare_All1}  (bottom panel), the percentage of HB 
emission is shown, and it varies similarly with \chif. 
The anti-correlated behavior between the percentage of HB emission and the $U$-band (blue) is very similar to that between 
the $U$-band and \chif.  
The maximum percentage of HB emission is 35\%, but then drops to a minimum of 15\%  at
$t\approx2.17$ hours just after the $U$-band peak of 
MDSF2\footnote{Including H$\epsilon$, Ca \textsc{ii} K and Ca \textsc{ii} H
changes these values to 37\% and 17\%, respectively, but does not
change the trends.}.   
The vertical line in both panels denotes the time of minimum
percentage of HB emission (at S\#116), which occurs just before the minimum \chif\ and just after the maximum $U$-band enhancement.  The time-differences between
  maxima and minima in \chif, percentage of HB emission, and the $U$-band are intriguing
  and should be confirmed 
   with higher cadence data covering the peak and initial fast decay
   phase. 

The percentages of HB flux in flares (Table \ref{table:hb_phasez}, columns 2 and 3)
should be directly compared
to RHD models.
The non-HB flux [1\,--\,HB flux/total flux] is largely due to flux from the
underlying white-light continuum that extends from the NUV through the
optical.  This accounts for the majority ($\gtrsim50$\%) of the total flux in all
flares at peak, and as much as
83\,--\,89\% of the total flux at 
peak in most IF events.  For the first time, we have included the Balmer continuum
at $\lambda > 3420$\AA\ in the energy budget with the other Balmer features, and we
have shown that the Balmer component is not the dominant source of flux in flare
emission in the blue.  At most, the Balmer contribution accounts for $\sim$50\% of the blue-optical
flux, and only in gradual flares and the decay phase of other flares.  The continuum
component that accounts for the non-HB flux is represented by C4170.
In the next section (Section \ref{sec:bluecont}), we will
model this non-HB emission as hot, blackbody emission.

\section{The Impulsive Phase Blue Continuum:  Two-Component Analysis} \label{sec:bluecont}

The relative amount of each emission component as a function of time 
is important for understanding the distribution of flare heating in the stellar atmosphere.  
We extend the two-component continuum analysis from
\cite{Kowalski2010} to the Flare Atlas and find 
that the blackbody and Balmer continuum components account for most of the NUV/blue continuum emission
in both gradual and impulsive phases.  In this section, we present the
blue continuum properties of the flare peak emission.  Note that ``flare peak'' and ``maximum continuum'' always
refer to the times of maximum C3615 (see Section \ref{sec:spectra_param}).

\subsection{Flare Peak Blackbody Emission} \label{sec:bbpeak}
In Section \ref{sec:spectra_atlas} (Figures ~\ref{fig:peak_panels1}
\,--\, \ref{fig:peak_panels3}), we showed the flare spectra at
peak times for all flares.  The flare peak spectra exhibit a steeply
rising continuum towards NUV wavelengths, which is nearly ubiquitous and continues into the NUV beyond
the spectral range of DIS.  In Section \ref{sec:hydrogen}, we found
that typically only 11\,--\,17\% of the flux from
$\lambda=3400-5200$\AA\ in the flare peak spectra of the IF events
 is accounted for by Hydrogen Balmer (line and continuum) emission (Figure
 \ref{fig:hb_phases}).  We find that the spectral shape of non-Hydrogen Balmer emission
 matches that of hot
 blackbody emission which pervades the entire NUV, blue-optical, and
 red-optical zones; C4170 is dominated by the emission from the
 blackbody continuum component.  

The blackbody fitting is performed as follows.  
We calculate a color temperature, $T_{BB}$, of the blue-optical continuum
by fitting a Planck function to the flux in the continuum windows from
$\lambda=4000-4800$\AA\ (BW1\,--\,BW6, similar to those used in
\cite{Kowalski2010}) given in Table~\ref{table:bbwindows}.   We used
the peak flare spectrum from 2011 Feb 24 (IF3) and
a decay spectrum from 2009 Jan 16 (IF1) to guide our selection of line-free
regions.  The same windows were used to fit a straight line to the
continuum in order to calculate BaC3615 (Section \ref{sec:spectra_param}). We do
not include the flux at shorter wavelengths due to the difficulty in
determining bona-fide continuum from blended Hydrogen lines (PseudoC) and due to the contribution from 
Balmer continuum emission.   We do not include flux from
redder wavelengths, because we find evidence for complicated behavior
in the red-optical (Section \ref{sec:conundruum}).   We
also fit a filling factor ($X_{\mathrm{BB}} =
R_{\mathrm{fl}}^2 / R_{\mathrm{star}}^2$; as in \cite{Hawley2003} and
\cite{Kowalski2010}) using the IDL routine \emph{mpcurvefit}, where $X$ is the fraction of the visible hemisphere covered by the
projected area\footnote{The projected flare area is $cos$
  $\theta \times A_{\mathrm{flare}}$, where $\theta$ is the
  angle between the line-of-sight to stellar disk center and the normal vector
at the position of the flare \citep{Mochnacki1980}. } of the flare (discussed in Section \ref{sec:fillingfactors}).  The two parameters are fit
simultaneously, which is an important distinction to the
analysis from \cite{Kowalski2010} where \TBB\ was fixed to $9\,000$ K,
$10\,000$ K, and $11\,000$ K.  The
fits for IF1 are redone here using the new flux scaling (Section
\ref{sec:scaling}), and some quantities change compared to those in
\cite{Kowalski2010, Kowalski2012}.
Good quality fits are obtained for the IF events (except for IF6\footnote{See note on
  calibration in Chapter 3 of \cite{KowalskiTh}.}), the HF events, GF1, GF3\footnote{See note on
  calibration in Chapter 3 of \cite{KowalskiTh}.  For this flare, we
  were able to co-add three spectra around peak to increase the
  signal-to-noise of the blue-optical zone emission.} and GF5.

 The blackbody fits are indicated by light blue
lines in the flare peak spectra panels (Figures ~\ref{fig:peak_panels1}
\,--\, \ref{fig:peak_panels3}) and are given in parentheses.  
In Figure \ref{fig:TXdist}, we show the distribution of $T_{BB}$ at
maximum continuum emission for the seventeen flares with well-determined
temperatures (values are given in Table \ref{table:chitable}).  The
flares are colored by the morphological type, but there is no
discernible trend with type or peak $\mathcal{L}_U$.
 We find that $T_{\mathrm{BB}}=10\,000-14\,000$ K for the
IF events (Figure ~\ref{fig:peak_panels1}).   By comparing to the dotted
(pre-flare) spectra we see that this hot, blackbody
component is present for a variety of flare amplitudes, appearing
in flares with peak $U$-band enhancements greater than $I_{f,U} + 1\sim$3, and
as large as $I_{f,U} + 1\sim$80.  The
high signal-to-noise ratio of our spectra makes the characterization of this
component possible even with relatively small contrast in the
blue-optical at $\lambda > 4500$\AA.
The blackbody component contributes $\sim76-97$\%, and on average 84\%, of the $\lambda=3420-5200$\AA\ flux during the peak times for
the IF events (Section \ref{sec:hydrogen}).

 The flare peak spectra of the HF events
(Figure~\ref{fig:peak_panels2}) also show the hot, blackbody
component, with temperatures between \TBB$\sim9500-12\,000$ K.  The lower
amplitudes of these flares are noticeable, e.g., by comparing to the
pre-flare spectra.  The HF events also have a
conspicuous amount of excess emission above the extrapolated
light blue curve at $\lambda < 3800$\AA.  Of the GF events in
Figure ~\ref{fig:peak_panels3}, the large-amplitude, high
energy gradual flare GF1 shows the most convincing evidence of a
moderately hot blackbody, with
\TBB$\sim9000$ K.  A significant NUV excess is also present during this
flare.  GF1 has a small-amplitude impulsive phase with a strong
response in C4170 that occurs before the
main flare peak (see Figure \ref{fig:lines_GF1_B});  the newly formed flare emission
in this pre-cursor flare peak has \TBB \s $14\,500$ and is included in
Figure \ref{fig:TXdist} as GF1$^\prime$.  GF3 shows evidence of a high
temperature, but these data have a more  
uncertain calibration\footnote{See note on
  calibration in Chapter 3 of \cite{KowalskiTh}.}; however, we averaged three
spectra near the peak and derived a similar value of \TBB\ for this
flare.  We have averaged four spectra
at the flare peak of GF5,
which results in a relatively good fit, but with a lower temperature of
\TBB \s6700 K.  In GF2 and GF4, we cannot accurately ascertain the characteristics of the
slope of the continuum; nonetheless, the very
low-amplitude flares ($I_f + 1 \sim 1.3 - 2.2$, $|\Delta U| < 1$ mag;
GF2 \,--\, GF5) do show excess C4170 emission.

The \s5\% blue-continuum shape uncertainty coupled with the uncertainty in
the scale factor, $R$ (see Section \ref{sec:reduction} here and Appendix
A of \cite{KowalskiTh}), and also the 
contamination of continuum windows from emission lines (Hydrogen
Balmer line wings and low-level Fe and Ti lines) lead to systematic uncertainties in the blackbody
temperatures of about $500-1200$ K.  The statistical uncertainties are
only \s200 K for most events, and are indicated by the error bars in Figure
\ref{fig:TXdist}.  \cite{Kowalski2010} found a $\pm 1000$ K
range in temperatures that well-represented the blue continuum shape.
In Appendix F of \cite{KowalskiTh}, the detailed systematic errors
involved with temperature fitting are examined.  The important result
is that for the 
IF2 event, the extreme amount of line broadening causes the BW1 and BW2
fitting windows to have possible contamination from Balmer line wing emission;  the acceptable range of \TBB\ for this
flare is therefore $11\,700-14\,100$ K, however, the temperature of 
$14\,100$ K better accounts for the shape of the total emission in the
BaC at $\lambda \lesssim 3646$\AA.
 
In summary, we observe a continuum at flare peak times of the impulsive phase with a spectral
slope matching that of a hot, blackbody with \TBB\s $9\,000-14\,000$ K 
(most within $10\,000 - 12\,000$ K) which is relatively independent of
flare amplitude over a large range of peak $U$-band specific luminosities, total energies, and light
curve morphologies.  For this range of temperatures, the
 spectrum would not be expected to strictly follow 
 a $\lambda^{-4}$ slope according to the Rayleigh-Jeans law.  
  Instead, a linearly decreasing fit to the
  blue optical zone
  gives a reasonably good fit and looks very similar to a Planck
  function for these
  temperatures: in the formula for the Planck function, the steeply decreasing $\lambda^{-5}$ part multiplies 
 with the increasing exponential part, giving a linear decrease
 towards the red for $\lambda=4000-5000$\AA.  

 The interflare variation in \TBB\ at flare peak is likely significant
 despite the large systematic uncertainties of $\sim$1000 K.  We
 show convincing evidence that interflare variations are significant
 between two flares, IF2 and HF1, in Appendix 
\ref{sec:stack}, and we discuss the implications for flare heating
differences in Section \ref{sec:wtf}.  Additional high signal-to-noise data would be
useful to determine if the color temperature decreases to $\sim$7000 K
at flare peak during very small amplitude events ($I_f + 1 \lesssim 1.5$), as
implied by the observations of GF5.  NUV data covering the flare peak of the
continuum (likely near $\lambda = 2500-3000$\AA) would help to more precisely determine
 the blackbody temperature at flare peak and its evolution
during the impulsive phase.  

 This hot, blackbody continuum component has been detected
previously using colorimetry and spectra (see references in the Introduction).
Self-consistent flare heating models of
the impulsive phase have thus far not reproduced this (dominant)
emission component; we compare to current models in Section
\ref{sec:modelcomparison}.  In summary, we have found that a \TBB$\sim10^4$ K blackbody
function and linear function are useful mathematical
representations for this continuum component; however, we will shortly
show (Section \ref{sec:astar}) that a single blackbody or line does
not account for important absorption features in the spectrum.

\subsection{Flare Peak Balmer Continuum Emission}  \label{sec:peak_bac}
We find that there is always excess flux in the
NUV zone, above an extrapolation of the Planck function at these
wavelengths.  The slope of the total NUV spectral zone emission at
flare peak generally follows
the same slope as the underlying Planck function.  In Figures \ref{fig:peak_panels1} \,--\,
\ref{fig:peak_panels3}, we scaled the Planck function to the flux at
C3615, and showed in yellow that it matched the flux at $\lambda < 3646$\AA, illustrating that the overall shape
 of the continuum at $\lambda < 3646$\AA\ for the IF, HF, and some GF events can be represented by about the same temperature fit
 as for the blue-optical.  We measure the slope of the NUV flare emission
 at $3420 \le \lambda \le 3630$\AA\ by fitting a line to three
 wavelength bins in this region.  These values are given in Table
 \ref{table:chitable} (columns 9 and 10 are the values for the flare peak
 and gradual decay phase; columns 11 and 12 are the values for the flare
 peak and
 gradual decay phase with the underlying blackbody emission subtracted) are consistent with the
 blue spectral shape indicated by the yellow curves, and will be
 important for comparing to detailed Balmer continuum model predictions in Section
 \ref{sec:modelcomparison}.  Note, the flux calibration errors in the
 NUV are larger (\s5\,--\,10\%), and the statistical uncertainties in
 the fits can also become comparatively large; higher signal-to-noise measurements should be obtained in the NUV.  

Although the \emph{total} flux at $\lambda < 3646$\AA\ follows the same general slope
as the underlying blackbody, we attribute
the \emph{excess} flux in the NUV to chromospheric Balmer continuum (BaC) radiation from
recombination to the $n=2$ level of Hydrogen, as in
\cite{Kowalski2010}.  We find that Balmer continuum emission 
is ubiquitous at flare peak for all flares in our sample, consistent
with our conclusion that Balmer continuum emission affects the
value of \chifp\ and is correlated with the relative amount of \Hg\
emission (Section \ref{sec:general}, Figure \ref{fig:FEU_HG}).  As explained in Section \ref{sec:spectra_param}, we use
both the BaC3615 and BaC quantities as measures of the Balmer continuum
emission.  

In Figure \ref{fig:master0}, we show BaC3615/C3615 (the fraction of
the NUV flux emitted in Balmer continuum emission) and find a general ordering between the IF, HF, and GF events.
There is a decreasing trend with peak specific $U$-band
luminosity with a range of values for each morphology class,
especially for the IF events, which have a standard deviation that is twice the
standard deviation of the percentage of HB emission (Section \ref{sec:hydrogen}). The IF events are
least dominated by BaC3615 with most having values of 0.2 \,--\,
0.35.  
These flares are dominated by the hot blackbody 
emission component which also contributes to the flux of C3615. While the IF5
and IF6 events have larger values, \s0.4, IF4 has a
very low value \s0.05 (but also required additional calibration -- see
Appendix A of \cite{KowalskiTh}). The HF events have about
equal contributions (\s0.5 \,--\, 0.6) and the
GF events have the majority of their peak NUV flux (0.55 \,--\, 0.8) in Balmer
continuum emission.  In column 8 of Table
\ref{table:hb_phasez}, we present these values. We conclude
that the general relationship is that the overall light curve evolution
(flare morphology) is physically connected to the relative importance
of the Balmer continuum flux at the time of maximum continuum emission.

\subsection{An ``A star'' on an ``M star'':  Absorption Features in
  Flare Spectra} \label{sec:astar}  
Several flares in our sample have impulsive phase durations that are
significantly longer than the spectral cadence, thereby allowing
detailed time-resolved analyses of the two continuum components.
In particular, IF1, IF3, and GF1 have the greatest number of spectra
during their respective impulsive phases (see Figures in 
Appendix \ref{sec:appendix_times}).   Note that the impulsive phase of IF1 refers to the
subpeak MDSF2, not the main flare peak which did not have spectral coverage.

In this section, we analyze the rise phase and peak spectra of MDSF2, to show
that the hot, blackbody continuum component is not a featureless
Planck function across our wavelength range.  Here, we establish the presence of
\emph{absorption} features in impulsive phase flare emission, and
thereby redefine how we understand the ``blackbody'' continuum
properties (e.g., Figure \ref{fig:TXdist}).

An indication for the presence of absorption in our spatially unresolved
flare data is an \emph{anti-correlation} between the time-evolution of BaC3615 and the time-evolution of the total amount of the
blue/NUV continuum, as diagnosed by the $U$-band flux or C$4170$ flux. 
This effect has been observed during the secondary flares in the
decay phase of IF1 \citep[Figure 1d of][]{Kowalski2010} and presented graphically
through a sequence of rise-phase spectra \citep[Figure 1
of][]{Kowalski2012}.  Those figures illustrate that as the $U$-band increases during
the secondary flare MDSF2, the entire blue/NUV continuum becomes dominated by the
blackbody component whereas the absolute (and relative) amount of
BaC3615 decreases.

In previous papers on IF1/MDSF2 \citep{Kowalski2011IAU, Kowalski2012}, we have denoted the sum of all (spatially unresolved)
decaying and impulsive flare emission as $F_{\lambda}^\prime$ (obtained
by subtracting the preflare quiescent spectrum), and
newly-formed flare emission isolated from an impulsive event as
$F_{\lambda}^{\prime\prime}$ (obtained from subtracting the preflare
decaying emission)\footnote{In principle, even a ``newly-formed'' flare
emission at any give time could be a sum of ``just-heated'' regions
and ``previously-heated'' regions that are decaying at that time.}.
During MDSF2, we find direct evidence that the newly-formed flare emission ($F_{\lambda}^{\prime\prime}$) during the rise phase 
resembles the spectrum of a hot star.  This was first shown in
\cite{Kowalski2011IAU}; in this section, we present an analysis of
the time-evolution.  

Figure \ref{fig:magnumopus}
shows the newly-formed flare spectra ($F_{\lambda}^{\prime\prime}$)
 during the rise (black; same spectrum as presented in \cite{Kowalski2011IAU}) and just
prior to the sub-peak (grey) at $\lambda < 6500$\AA. 
At both times, the blue-optical shapes are well-matched by the
spectrum of 
Vega \citep[from][]{Bohlin2007} scaled and plotted in red to show the
striking similarities in the spectral slope and absorption features.  
 The difference between the flux at
$\lambda < 3600$\AA\ in the grey spectrum and the flux in Vega could
be due to the spatially unresolved chromospheric Balmer continuum emission forming during the event or due to the
difference in gravity between the hot star and an M dwarf (log $g = 4$
and log $g = 5$, respectively).  Models of deep heating \citep[e.g.,][]{Kowalski2011IAU} in M dwarfs are
required to understand these detailed spectral differences.

\cite{Mihalas} discusses the detailed formation of hot-star spectra:
The lower emission at $\lambda \sim 3615$\AA\ relative to the emission at 
$\lambda \sim 4170$\AA\ (in other words, the BaC is in absorption) is a result of the
non-uniform temperature structure of the atmosphere coupled with
wavelength dependent Hydrogen Balmer and Paschen continua opacities (and flux redistribution) in an optically
thick medium.  In other words, the optically thick emitting material
producing the $\lambda \sim 3615$\AA\ emission is at a lower temperature (e.g., from lower column
mass or higher height for a radiative equilibrium stratification) than the optically thick
emitting material 
at $\lambda \sim 4170$\AA.  We have therefore revealed the multithermal
contributions that comprise the hot blackbody emission
component (see Section \ref{sec:bbpeak} and Figure \ref{fig:TXdist}).
The use of the blackbody function (and \TBB) is 
a useful paramaterization of the $\lambda=4000-4800$\AA\ slope, but it is important to realize that it is
an approximation of the detailed radiation field originating from a flare
atmosphere with an opacity dependence as a function of wavelength and a temperature dependence as a function of height.

Temperature evolution during 
the rise of MDSF2 is qualitatively evident in Figure~\ref{fig:magnumopus} by comparing
the blue-optical spectral shape to Vega from flare rise (black) to flare peak (grey).  The
spectrum becomes steeper (corresponding to a hotter color
temperature), and 
there are three ways to measure this for any given time during
 MDSF2.  For example, we consider the spectrum near
maximum continuum emission (S\#113, the grey spectrum in Figure \ref{fig:magnumopus}):  

\begin{enumerate}
\item \emph{Method 1: Color temperature of total flare-only emission:}
  The color temperature of the total emission with quiescent level
  subtracted ($F_{\lambda}^\prime$)
is \s$11\,000$ K (Figure \ref{fig:TXdist}).  
This represents the color temperature averaged over the entire flaring
area over the duration of the exposure \citep{Kowalski2012}.

\item \emph{Method 2: Color temperature of newly-formed flare-only emission:} The temperature of the newly-formed
flare emission ($F_{\lambda}^{\prime\prime}$) is \s$18\,000$ K just prior
to the $U$-band peak.  This
temperature is obtained by subtracting the previously quiescent$+$decay flare
emission. This technique provides a \emph{better} estimate for the shape of the continuum that is
produced during the times when there is a secondary impulsive flare
event.  Relating the continuum shape to a color temperature with this
technique assumes that the 1) the previously decaying emission is small or
that it does not decay significantly and 2) the newly-formed flare emission
originates from a spatially distinct flare area that was not
previously heated. Otherwise, the non-linear dependence between 
$T$ and intensity (e.g., for the Planck function) results in an incorrect association between \TBB\
and $F_{\lambda}\prime\prime$.

\item \emph{Method 3: Adjusted color temperature of newly-formed flare-only emission:}
In Appendix \ref{sec:appendix_astar}, we estimate the amount by which the flux of the previously decaying emission has
decreases during the secondary flare.  We apply these estimates to the
times of the spectra during MDSF2 which 
 allows the \emph{best} derivation of the properties of
the newly-formed flare emission.  We find that the
color temperatures are \s1800\,--\,2700 K lower than method \#2
(above).  The value at flare peak of MDSF2 (\TBB \s $15\,000$ K) falls just beyond the high end of the
sample's distribution of \TBB\ in Figure \ref{fig:TXdist}.  
\end{enumerate}

We summarize the derived color temperature values at three times in
Table \ref{table:color_astar} using these three methods.  The times
correspond to the time just before MDSF2, halfway
through the rise phase of MDSF2, and just prior to flare peak of MDSF2.
The observed, apparent anti-correlation between the time evolution of
Balmer continuum emission (BaC3615) and blackbody emission \citep{Kowalski2010, Kowalski2011CS,
  Kowalski2012}
 is now explained by the presence of an emission component with
 ``absorption'' features forming during the flare.  If there are significant absorption components in other flares, then
the value of the derived BaC3615 value for these flares (obtained from
a linear extrapolation -- see Section \ref{sec:spectra_param}) is an
underestimate of the intrinsic amount of Balmer continuum emission
originating from the chromosphere.  

We briefly examine whether Balmer continuum absorption has
an obvious effect on the distribution of the peak properties
of the Flare Atlas.  We calculate the peak specific luminosities of
C4170 and BaC3615 (see Sections \ref{sec:phot_param}, \ref{sec:spectra_param}).  We find that in general,
 the IF events (with lower \chifp\ values; Section \ref{sec:general})
 are more luminous in C4170 
whether the
 BaC3615 is emission or absorption (on average,
 $\mathcal{L}_{\mathrm{C}4170}/\mathcal{L}_{\mathrm{BaC}3615} = 2.4$), the GF events are more luminous
 in BaC3615 (on average,
 $\mathcal{L}_{\mathrm{BaC}3615}/\mathcal{L}_{\mathrm{C}4170} = 2.4$) and
 the HF events are slightly more luminous in BaC3615 (on average, $\mathcal{L}_{\mathrm{BaC}3615}/\mathcal{L}_{\mathrm{C}4170} = 1.5$) .  The trend
 between morphological type and the relative amounts of BaC3615 and
 C4170 could be related to the amount of BaC absorption produced by
 the blackbody (C4170) component.  Since the amount of BaC3615 that we
 measure is probably less than the intrinsic BaC3615 emission (due to
 this absorption component) the emission
 and absorption may add together to produce smaller values of BaC3615
 -- and 
\chifp -- for the IF events.   The HF and GF events may produce
intrinsically more BaC3615 than the IF events, they may produce less
BaC3615 absorption than the IF events, or they may have both effects.
However, we
suggest the possibility that absorption can also lead to both low values of \chifp\ and large values of C4170/BaC3615.  The
  Balmer continuum \emph{emission} produced in large amounts in the early
  flare phases, and persisting over a relatively long time compared to
  the blackbody component, effectively veils
  the absorption signatures except in the flares with the most
  extreme heating.  Detailed
modeling of the blackbody component over a large parameter space is needed to understand the role
of absorption in flares of all morphological types.

\subsection{Balmer Wing Absorption} \label{sec:wing_absorption}
We now present cases of line broadening in larger flares
where there is evidence of absorption in the wings. 
In Figure \ref{fig:line_width_5}, we show H$\delta$, H$\gamma$, 
H$\beta$, and He \textsc{i} $\lambda$4471 profiles ($F_{\lambda}^{\prime}$) for IF4, at maximum continuum
emission (black, S\#665)
and at maximum Balmer line emission (S\#672, turquoise).  The profiles are
plotted with the local continuum subtracted with a straight line and then
are normalized to the peak of the line.  The striking effect here is
a ``depression'' (deficit) in the line wings in the black profile (maximum
continuum) between $\pm10$\AA\ to $\pm30$\AA\ from line center.  Furthermore, the amount
of wing depression increases from H$\beta$ to H$\delta$.  
The differences in the widths (measured at the 10\% line flux level)
between the times of maximum continuum flux and maximum line flux are
\s6\AA\ for \Hd\ and \Hg\ and \s2\AA\ for \Hb.  Although the change in the line
widths for \Hd\ and \Hg\ are just \s2$\sigma$ (from 3\AA\ uncertainty in
widths), we observe a flux depression (deficit) over at least 10 contiguous pixels in the
wings on both sides of the line center, which increases the
significance of the flux depression.
During the rise phase spectrum (S\#664, corresponding to a flux of \s40\% of the peak continuum
emission), the Balmer lines are narrower than at both maximum line
  and continuum emission, but there is less
evidence for flux depressions in the line wings.  In the spectrum following flare peak, the
Balmer lines have begun to broaden significantly and no wing
depressions are seen.

A similar wing depression effect is seen in the IF1 spectra during 
 the peak continuum emission of the subpeak MDSF2.  In Figure
\ref{fig:line_width_6}, we show the profiles for
the IF1 flare near the time of maximum Balmer line emission within the
spectral observation window (turquoise) and at
the peak continuum emission (black) during MDSF2.   The wing
depression is present, but relatively less 
than in IF4 (Figure \ref{fig:line_width_5}), and doesn't appear as far
from line center as in IF4. The wing depression increases for the higher order Balmer
lines as in IF4.  The line widths (measured at the 10\% line flux
level) at the time of maximum continuum emission 
  are 2\,--\,3\AA\ narrower than at the time of maximum line emission.  

We find similar evidence for wing depressions during flare peak of IF3
(see Appendix \ref{sec:IF3_wings}), given
that the rise phase emission is subtracted from the flare peak.  

 For IF1, the newly formed flare emission
during MDSF2 was found to have a spectrum like a hot star, with
the Balmer lines in absorption (see Section
\ref{sec:astar}). In particular, note the similarities in the widths of the Balmer absorption
lines between the Vega spectrum and the flare spectra in Figure
\ref{fig:magnumopus}.   We therefore interpret the wing depressions in IF1 and
IF4 as direct signatures of Balmer
line absorption\footnote{Alternatively, this feature could be
    a broader emission component at a lower flux level than the nearby
    continuum and line center.} forming during these flares.  The absorption wings in model hot star spectra with
log $g = 4-5$ are very broad \citep{Kurucz2004} due to Stark broadening
\citep{Peterson1969}. The superposition of an emission
line component and an absorption line component leads to the net
decrease that we observe in the wings in the flare spectra.  As a
result of the superposition of emission and absorption, the measured line
widths (at the 10\% flux level) also decrease.  The Balmer absorption
line flux decrement for Vega is
\Hd\ : \Hg\ : \Hb\ $\approx 1.09 : 1.00 : 0.75$
\citep[using the spectrum from][]{Bohlin2007}; therefore, one would expect the line widths and
wing fluxes of the
higher order Balmer lines in
flare spectra to be
more strongly affected in the presence of
hot star-like absorption.  The Balmer emission line component
originates from the flaring chromosphere \citep{Hawley1992, Allred2006} and
veils the Balmer absorption component in the total, spatially unresolved flare
spectrum.  See \cite{JohnsKrull1997} for a phenomenological
illustration using quiet-Sun \Ha\ absorption and superimposed
chromospheric \Ha\ emission. 

Wing absorption signifies heating in deep layers of the atmosphere, which
is an important constraint for flare heating models.  The heating differences between flares - e.g., between the primary flare of
IF4 and the secondary flare MDSF2 - should be studied with
radiative-hydrodynamic models that successfully produce the range
of blackbody continuum properties.  These flares have two of the three lowest
\chifp\ values in the sample (\chifp \s 1.3\,--\,1.5; Table
\ref{table:chitable}) and also show evidence of Balmer
line wing absorption, implying a common heating scenario.  Further, the data are suggestive of 
differences in \chifp\ and the amount of Balmer line wing absorption (IF4
having a smaller \chifp\ and also more wing absorption, as is most evident for
\Hd).  Also, the peak $U$-band amplitude is much larger in
IF4 ($I_{f, U}$\s20) than in MDSF2 ($I_{f, U}$\s7).  Assuming that IF4
produced the same type of ``blackbody'' emission as MDSF2 (Section
\ref{sec:astar}) with \TBB$=$15\,000 K (Table
\ref{table:color_astar}), the inferred area of this emission would be over 3
times as large in IF4 and formed over a timescale nine times as
fast ($t_{1/2, \mathrm{C4170}}\sim1$ min for IF4,
$t_{1/2, \mathrm{C4170}}\sim9$ min for MDSF2; Table \ref{table:chitable}).  As a result, \emph{more
  ``blackbody'' emission formed over less time leads to more obvious
  absorption wings at peak.}  On the other hand, it is possible that
the ``blackbody'' emission is intrinsically different between the two
flares, related to the speed at which the light curve develops and
ultimately to the depth of primary heating (IF4) compared to the depth at
which primary heating (IF1) may cause secondary flare heating (MDSF2;
see Section \ref{sec:speeds}).
It may also be important to consider that $I_{f, U}$ is much higher \,--\,
therefore with more blackbody emission present \,--\, in
the maximum line emission spectrum for IF1 ($I_{f, U}$ \s 35 at S\#24) than for
IF4 ($I_{f, U}$ \s 4 at S\#671).

Searching for wing depressions is 
a method for directly detecting hot star signatures in $F_{\lambda}^{\prime}$, without needing
to isolate newly formed emission, a technique that relies on several assumptions
(Section \ref{sec:astar}).  This analysis can be employed with
other flare data that cannot be accurately flux-calibrated, or for
spectral data with limited
wavelength range.  A larger sample of primary and secondary peaks during
high energy flares with high time resolution would
be useful for constraining the timing of the absorption wings during
the flare and also the relationship between a very small value of \chifp\
\s1.5 and the presence of absorption wings.

\section{The Gradual Decay Phase Continuum} \label{sec:gradual}
The gradual decay phase begins when the flare light curve
transitions from 
a slow decay after the initial fast decay
phase, marked by either a distinct 
break (e.g., IF4 or the $g$-band light curve of IF5) or a smooth
transition into slowly decaying emission (e.g., IF2 and IF7).  
The $U$-band has typically reached \s50\% of the peak flux in the GF events and
$10-20$\% of the peak flux in the IF events at the start of the
gradual decay phase. 

In dMe flares, the gradual decay phase is energetically important due to its
long duration.  In IF0, 29\% of the total $U$-band energy is emitted
during the gradual decay phase (HP91).  Over 70\% of the $U$-band energy in IF1
and 60\% of the $u$-band energy in IF3 were emitted in the gradual
phases, giving a range of values even for flares of the same type.  In Figure \ref{fig:hb_phases}, we
found that only $30-65$\% of the wavelength-integrated flux at the
beginning of the gradual decay phase can be attributed to Hydrogen
Balmer\footnote{Recall that the HB component as defined in this paper includes the Balmer
  continuum (BaC), the
  PseudoC, \Hd, \Hg, and \Hb.}
radiation.  Given that Ca \textsc{ii}, the Helium \textsc{i} lines,
and various metal lines including Fe \textsc{ii} lines only
account for $\sim5$\% of the emission, this leaves $30-65$\%
of the gradual phase emission due to non-Hydrogen Balmer radiation.  According to Table
\ref{table:hb_phasez} (column 9), $\sim60$\% of the specific
flux in C3615 is due to the
BaC3615 contribution, leaving $\sim$40\% of the specific flux in the NUV
unaccounted for by Balmer continuum emission.  

 Studies of the gradual decay phase continuum during IF1 were presented in
 \cite{Kowalski2010, Kowalski2012}.  A two-component model 
 consisting of a \TBB $=10\,000$ K blackbody and a RHD model Balmer
 continuum was used to derive a filling factor ratio
 ($X_{\mathrm{BaCF11}}/X_{\mathrm{BB}}$) of \s$3-16$.   The IF1
 gradual decay phase showed
 direct evidence for a Balmer continuum in emission, which matched the
 RHD prediction from \cite{Allred2006}.
Thus far in the analysis of the Flare Atlas, we have seen that the
Balmer jump ratio (\chif) and the percentage of flux that
can be attributed to the Hydrogen Balmer (HB) emission increases in the
gradual decay phase (Figure \ref{fig:hb_phases}), indicating a change in the underlying
radiative energy release processes.
 An important step in understanding the gradual decay phase flare heating
 includes determining if the gradual decaying continuum is due to the long cooling timescale
\citep{Cully1994} of material that was heated impulsively in the beginning of the
flare, or if it is due to prolonged, low-level heating leading to 
cooler temperatures than in the beginning of the flare.  If there is prolonged heating, then we must
determine the source (e.g., particle heating, backwarming).

\subsection{Gradual Decay Phase Spectra} \label{sec:gradspec}
To study the gradual phase\footnote{Hereafter, we use ``gradual phase'' and
  ``gradual decay phase'' interchangeably.}, we average three spectra at the times given in Table
\ref{table:times} in order to increase the signal-to-noise in the
blue-optical zone for determining a blackbody temperature fit.
Figures \ref{fig:grad1}\,--\,\ref{fig:grad1b} show the averaged spectra from the
beginning of the gradual phase (Table \ref{table:times} and indicated
by vertical red lines in Appendix \ref{sec:appendix_times}, Figures
\ref{fig:appendix_integ1}\,--\,\ref{fig:appendix_integ17}) for flares
with sufficient signal-to-noise for analysis.  Note that IF4, IF5, and
HF4 have only a single spectrum shown because the data quality was
variable during the low levels of gradual phase emission.  Two
component fits, allowing \TBB, \XBB, and $X_{\mathrm{BaCF11}}$ to
vary, are shown in color with the quiescent spectrum as a dotted line. The gradual phase spectrum
for IF10 is given as S\#32 even though it contains contributions from
(relatively low level)
secondary flares.  In parentheses, we
 have listed \chifd\ and \TBB;  the values are excluded when they are not
well-determined (i.e., very little or no continuum emission in the blue-optical zone).
We remind the reader that \TBB\ is a parametrization of the continuum
 slope; detailed modeling is required to deduce the multithermal
 structure of the atmosphere (see discussion in Section \ref{sec:astar}). 

 A striking similarity among the gradual phase spectra is the value of
 \TBB,  \s7300\,--\,8500 K (typically around 8000 K) in the flares IF1, IF2, IF3, IF7, IF9, IF5,
 GF1, and HF2.  However, some flares have unusual
 characteristics.  The gradual phase of IF4 was
 taken near the peak of the line emission (due to the uncertainty in
 flux calibration later in the flare at very high airmass), and this flare has a larger color temperature
 of 9500 K.  The gradual phases of HF1 and HF3 are also
 hotter with temperatures exceeding $10\,000$ K.  
We have averaged the spectra just following the peak (synthetic $U$-band) emission of
 GF5 to show the remarkable detection of white-light continuum (and
 all the typical flare emission lines) in an
 event with only $I_f + 1 < 1.5$ and $I_{C4170} + 1 \sim $ 1.03.  This
 flare has a gradual phase \TBB \s 4900 K, the lowest detected
 temperature in our sample, and is about 2000 K cooler than at peak.
 
Note that the gradual
 phase spectrum of IF1 (Figure \ref{fig:grad1}) is that shown in
 Figure 1b in \cite{Kowalski2010}, where we modeled this
 spectrum with a $T=9\,000-11\,000$ K blackbody.  \TBB\s8250 K results with our current, better determined fit, with a larger inferred areal coverage, \XBB \s 0.4\% of the
 visible hemisphere \citep[\s2 times the value given in ][]{Kowalski2010}.
 Among the three spectra averaged to form the gradual phase spectrum of IF1 in
 Figure \ref{fig:grad1}, the inferred temperature range is
 between 8000\,--\,8500 K.  The difference between 8250 K and $10\,000$ K 
 is significant, as our estimate of the systematic
 uncertainty in the temperature determinations is \s1000 K 
 \citep[][Appendix F]{KowalskiTh}.

We calculate the specific luminosities of BaC3615 and C4170
  (see Sections \ref{sec:phot_param}, \ref{sec:spectra_param})
and find a trend among the IF
events showing \TBB\s8000 K emission in the gradual decay phase
\emph{and} \chifp $<$ 2:  IF1, IF2, IF3, IF7,
and IF9.  These flares are described by the relation, 
\begin{equation}
\mathrm{log_{10}} \mathcal{L}_{\mathrm{BaC3615}} = -0.63(\pm0.86) + 1.03(\pm0.03) \times
\mathrm{log_{10}} \mathcal{L}_{\mathrm{C4170}}
\end{equation}

A slope of \s1 indicates that the BaC3615 and C4170 specific luminosities increase in equal
percentages during the gradual decay phase according to the size of the
flare.  The flares with \chifp\ $>2$ and which form a \TBB \s 8000 K
component at the beginning of the gradual decay phase -- IF5, HF2, and
GF1 -- may comprise a second distribution which
is offset from 
those with \chifp\ $< 2$ and \TBB \s 8000 K in the gradual decay phase.  However, it is not clear whether the
distribution is offset horizontally (signifying that they are ``deficient'' in C4170) or
vertically (signifying that they have excess BaC3615).  In any case, the data suggest that the persistence of C4170 with \TBB\s 8000
K formation is related to the formation of BaC3615 in the gradual
decay phase.  

The \chif\ value is approximately constant 
in the gradual decay phases of IF3 and GF1 (\chifd$\sim$2.7 and 3.5, respectively).  What
can we learn from this?  This may indicate that the Hydrogen BaC and the
C4170 -- and therefore the \TBB\s8000 K blackbody component -- vary together 
during the gradual decay phase.   However, the interpretation of
\chif\ is complicated because variations in both the blackbody and BaC
components can lead to variations in C3615, and therefore the value of \chif.  
For IF3, we investigate the ratio of
BaC3615 to C4170.  This ratio is similar to the evolution of \chif\ for the
duration of the flare, with both measures being relatively
constant over \s0.6 hours of the gradual phase.   Because BaC3615/C4170 is
nearly constant with values of 
$\sim$1.5\,--\,1.6 over this time period, we conclude
that the local flux in the BaC at $\lambda = 3615$\AA\ does follow the local flux in the blackbody
at $\lambda=4170$\AA\ during the gradual phase.

In summary, we find evidence for a \TBB\s 8000 K component at the
beginning of the gradual decay phase during
eight flares, most of which are IF events.  It is not known if the emission
is formed in the same areal region as the impulsive phase emission
(with \TBB\s $9000-14\,000$). 
We find a strong correlation among different flares between the C4170 
(\TBB\s8000 K) specific luminosity and BaC3615 specific luminosity during
the gradual decay phase, suggesting that the late-phase persistence of 
C4170 and BaC3615 are driven by the same
process. Modeling results indicate that the gradual phase C4170 may be a combination of Paschen
continuum from Hydrogen recombination and increased photospheric
emission (e.g., H$^{-}$ recombination) induced by Balmer continuum \citep{Allred2006} or
X-ray backwarming \citep{Hawley1992}.   However, 
color temperatures as high as 8000 K have not yet been produced
in these models.  We explore the gradual decay phase continuum further
in the analysis of the red 
continuum (Section \ref{sec:conundruum}).

\subsection{A Comparison of Gradual Decay Phase Spectra over Three YZ CMi Flares}
Three flares -- IF1, IF7, and GF5 -- are particularly interesting to compare because
they occurred on the same star and the data were obtained at
the same spectral resolution (1.83\AA\ 
pixel$^{-1}$, using the 1.5\arcsec\ slit).  The gradual decay phase spectra of these three flares on YZ CMi are presented in Figure
\ref{fig:magnumopus2} (from the times given in Table \ref{table:times}).
These comprised the smallest amplitude flare (GF5, blue), a
medium-amplitude flare (IF7, red) and the largest
amplitude flare (IF1, black) in the sample.  At the times of these spectra, 
the $U$-band was emitting at 1.4, 2 and 25 times the quiescent level,
respectively. The best-fit \TBB\ values are 7600 K for IF7 and 8300 K for IF1.  GF5 has very low amounts
of emission, but is fit nonetheless with \TBB \s 4900 K.  

Despite these differences, there are striking similarities across the NUV and blue-optical
zones, including 1) small-scale features at the base of the Hydrogen
lines; 2) the appearance of the prominent Helium \textsc{i} lines at
$\lambda$4471 and $\lambda$4026, He \textsc{ii} at \lam4686, Ca \textsc{i}
at \lam4227$+$Fe \textsc{ii} at \lam4233, and Fe \textsc{ii} (or Mg
\textsc{i}b) at
$\lambda4924+$\lam5018$+$\lam5169;  3) two plateaus in the underlying level of the
pseudo-continuum (one from \lam$=3780-3950$\AA\ and one from
\lam$=3650-3760$\AA); 4) Hydrogen lines resolved through H14 at
\lam3722\AA\ (H15 at \lam3712\AA,
H16 at \lam3704\AA, and He \textsc{i} at \lam3704\AA\ are blended); 5)
 similar small-scale features, which may be blended Fe and Ti
lines in emission in addition to the Balmer continuum emission at
these wavelengths (e.g., $\lambda \sim 3580$\AA, 3680\AA).  These qualitative similarities and the scaling
relationship given in Section \ref{sec:gradspec} illustrate
that the gradual emission is nearly identical, but ``scaled up'' in
flares according to their amplitudes and energies.  Clearly there are
important similarities in the gradual decay phases of flares covering a large
range of impulsive phase peak
amplitudes and energies.

\section{The Third Continuum Emission Component: A ``Conundruum''} \label{sec:conundruum}
Finally, we bring attention to the continuum rise at \lam$> 4900$\AA\
in Figure \ref{fig:magnumopus2}, representing
  excess ``conundruum''\footnote{The ``conundruum'' is a word play
    on ``continuum'' and ``conundrum''.} flux above the best-fit blackbody that is present during the
  gradual decay phase of some flares (IF1, IF2, IF3, IF7, HF1, HF2, GF3, and GF5).  The feature is
also apparent in some peak spectra (MDSF2 of IF1, IF2, IF3, IF4, IF7, HF1, GF2;
Figures \ref{fig:peak_panels1}\,--\,\ref{fig:peak_panels3}).   In the time resolved rise
phase of IF3, this feature is detected starting at S\#25, which is the first spectrum
showing a hot (\TBB \s $10\,000$ K) blackbody component in this flare.
We first present the observational signature and quantitative
properties of the ``conundruum'' flux
(hereafter, Conundruum) during IF3.  Then, we present several possible
explanations of this new continuum component.  

\subsection{The Conundruum Flux Budget and Relation to the Red
  Continuum} \label{sec:conundruum_obs}
The time-evolution of the Conundruum flux compared to other spectral
features provide constraints on its origin. 
For IF3, we show the evolution of the
ratio of HB to total $\lambda=3420-5200$\AA\ flux in the top panel of Figure
\ref{fig:Conundruum} (see Section \ref{sec:hydrogen}). 
The evolution is nearly flat in the gradual decay phase.   If we
consider not just the ratio of HB to total flux, but instead the ratio of
HB to blackbody flux (not shown), the ratio increases steeply in the gradual
decay phase, similar to the trend in \Hg\ flux/C4170, which is shown in black
asterisks in Figure \ref{fig:Conundruum} (bottom).  
Therefore, there must be an extra amount of flux in addition
to the blackbody flux that contributes to the gradual phase emission to
make the HB/total flux ratio in Figure \ref{fig:Conundruum} (top panel) approximately constant.  From the 
excess flux (non-HB, non-blackbody) we also subtract Ca \textsc{ii} K, Ca \textsc{ii} H,
the Helium \textsc{i} lines, and the Fe
\textsc{ii} ($\lambda4924+$\lam5018$+$\lam5169) triplet.  The excess emission, which represents
the Conundruum flux, is shown in Figure \ref{fig:Conundruum} (top panel)
as green squares.  The relative contribution steadily rises into the decay phase
contributing a maximum of \s9\% of the total $\lambda=3420-5200$\AA\
flare flux\footnote{The estimate of 9\% assumes that no Conundruum
  flux 
  contributes to the wavelengths used to fit the gradual phase
  blackbody component;  as such, 9\% is a lower limit.}.  We note that there is also a decreasing value of \TBB\ from
\s8000 K at the beginning of the gradual phase to less than 7000 K after
$t\approx2.7$ hours during this flare (see Figure \ref{fig:IF3_evol},
Section \ref{sec:speeds}), which is probably due to the increasing contribution from the
Conundruum at the redder end of the blackbody fit near $\lambda \sim
4800$\AA, thereby affecting the blackbody fit.  The Ca \textsc{ii} K
line flux reaches its maximum at this time also (Figure \ref{fig:lines_IF3_B}).

In addition to becoming more apparent in the gradual decay phase, the
Conundruum flux also has a larger relative contribution at redder
wavelengths ($\lambda \sim 6000$\AA).  Here, we use a new continuum
measure (C6010) in the red-optical zone (Table \ref{table:linewindows}).
In Figure \ref{fig:Conundruum} (bottom panel), several continuum fluxes are measured
relative to C4170 for IF3 throughout its entire evolution:  C6010,
C4500, and C3615 (Table \ref{table:linewindows}).  In addition, we show the
\Hg/C4170 as a function of time (peak values for the sample were analyzed in Figure
\ref{fig:FEU_HG}).  Figure \ref{fig:Conundruum} (bottom panel) illustrates that, strikingly, C6010/C4170
has a similar time-evolution to \chif\ and the \Hg/C4170 ratio.  The C4500/C4170 ratio, on the
other hand, is relatively flat showing small variations that reflect
the small changes in \TBB\ through the impulsive phase and into the gradual
phase (see Section \ref{sec:speeds}).  The similar behavior seen between C6010/C4170 and \Hg/C4170 suggests a time evolution of red-optical
emission that more closely resembles the blue Hydrogen Balmer
emission.  In the decay phase, the C6010/C4170
ratio is more similar to the \Hg/C4170 evolution than to the
\chif\ evolution.  We conclude that the 
Conundruum has a relatively larger contribution at red wavelengths (C6010) and
that its time-evolution is closer to the lower order Balmer
lines (\Hg, \Hb, and \Ha) and Ca \textsc{ii} K than to BaC3615 (which contributes to the
value of \chif).

For several flares, we can directly measure the color temperature
(spectral shape) of the Conundruum.  
In Figure \ref{fig:megaflare_fullSED} (left), we show for the first time a
complete (gradual phase) flare SED from $\lambda=3400-9200$\AA. This
is an average of the spectra S\#23\,--\,25 from IF1\footnote{Independent confirmation of the very red color at
  late times comes from The MEarth survey \citep{Irwin2011}.  They observed this
flare in their i$+$z filter and the light curve
shows about \s0.2 magnitude enhancement near the time of the S\#24 spectrum (Irwin, private communication).}.
The \TBB\ fit to the blue-optical zone is shown in blue and clearly
does not account for all of the red continuum emission.  
The color temperature of the red optical zone (using windows
RW1\,--\,RW3, Table \ref{table:bbwindows}) for this flare ranges
between \TBB\ $\sim4500-5500$ K (fit shown in red).   In Figure \ref{fig:megaflare_fullSED} (right), we show the complete SED
during the decay phase of IF3 (time listed in Table
\ref{table:times}).  This flare also exhibits a red color, with a best
fit blackbody temperature of \s3700 K. 

At the peak of IF3, the Conundruum is also apparent but with a smaller
relative contribution to the total flux of C6010 compared to the
gradual decay phase.  In Figure \ref{fig:XXX}, we show the extrapolation of a blackbody
curve fit with \TBB$=12\,100$ K to the blue-optical at the peak emission during IF3
(S\#31) in order to illustrate the excess flux of C6010 unaccounted for by a
simple isothermal Planck function that matches the blue-optical spectral
zone.  We tried various combinations of fitting
Planck functions to the entire spectrum to account for the Conundruum.  First, a single blackbody fit to
both the blue and red optical zones reveals a temperature of \TBB \s
$10\,200$ K;  it fits the overall shape well, but misses the
detailed shape of the blue-optical zone.  
If we fit a blackbody to the red
optical zone only (using the continuum windows RW1\,--\,RW3 in Table \ref{table:bbwindows}), we find a temperature of \TBB \s$7700$ K, but this fails
to account for much of the blue-optical flux.  Clearly, an isothermal
fit cannot explain the full flare SED at peak.

Using a \emph{two}-component blackbody to model the peak of IF3, we find
that a combination of a $T$\s$6000-6400$ K (\XBB$\sim1.5$\%) blackbody 
and a superhot blackbody  
$T$\s$10^5$ K\,--\,$3 \times 10^6$ K (\XBB$\sim0.0042$\%, \s 0.00012\%,
respectively) fit
the $\lambda = 4000-6800$\AA\ flux distribution the best using the
continuum windows BW1\,--\,BW6 and RW1\,--\,RW3 (Table \ref{table:bbwindows}).
The optical signature of a very hot temperature ($T\gtrsim10^5$ K)
component can only be confirmed with bluer observations.  
\cite{Hawley1992} discuss that free-free emission from a superhot 1 MK source reproduces
the optical ($UBVR$) colors of the Great Flare well but that it cannot account for the observed turnover
of the broadband flux in the FUV and NUV.  

It is possible that the large number of free parameters for two blackbodies allow an
excellent, yet unphysical, fit to the IF3 peak spectrum.  In particular, the
two bluemost fitting windows (BW1, BW2; Table
\ref{table:linewindows}) may be contaminated from Hydrogen wing
emission in such a large flare, therefore giving unrealistically large
temperatures to fit the rise of the spectrum from $\lambda=4000-4150$\AA.
If we exclude the bluemost two continuum windows for the fit, we find
that a combination of temperatures of $4700\pm500$ K (\XBB$=2.3$\%) and
$13\,700\pm1000$ K (\XBB$=0.17$\%) and scaled by 0.98 fit
the underlying $\lambda = 4150-6800$\AA\ continuum flux distribution
the best (overplotted in red dashes in Figure \ref{fig:XXX}).  This
fitting gives a hot blackbody consistent with a single temperature fit to BW1\,--\,BW6
(\TBB \s $12\,100$ K; Table \ref{table:chitable}) and a single
temperature fit using only BW3\,--\,BW6 \citep[\TBB \s $11\,000$ K; for further
discussion on systematic temperature fitting errors, see Appendix F of][]{KowalskiTh}.

\subsection{Possible Conundruum Interpretations}
Possible explanations of the Conundruum are the following:
\begin{enumerate}
\item \textbf{Calibration Artifact:} We first considered whether the Conundruum was a calibration issue, as the
  DIS dichroic affects the flux calibration in the Johnson $V$-band
  range.  The largest effect is near 5500\AA\ (e.g., in flat-field
  lamp exposures). We took a conservative range and did not consider the flux between
  \lam$=5200-5800$\AA\ due to this issue.

\item \textbf{Minor Species Emission Lines:} It is well-known that spectra of QSO's and Seyfert 1
  galaxies have a forest of Fe \textsc{ii}
  lines between \lam$=5000-5500$\AA\ \citep{Osterbrock1977, Peutter1981}.  There have been detections of
  strong Fe \textsc{ii} lines in stellar flares \citep{Abdul-Aziz1995}
  and Fe \textsc{i} and Fe \textsc{ii} lines from solar
  flares \citep{Severny1960, JohnsKrull1997} at these wavelengths.
  Other authors have aslo seen the presence of Fe lines
  in this range \citep{Eason1992};  many small scale features are
  visible in the high signal-to-noise spectra of IF1, and the Conundruum may be a blend 
  of these lines.

\item \textbf{Flux Redistribution from Minor Species Emission Lines:} Due to the appearance of the iron forest at these wavelengths,
  line blanketing  \citep{Rutten2003}
 may cause the nearby continuum to increase, similar to flux
 redistribution 
generating a higher-than-expected
  effective temperature for the Sun \citep{Bohm1989}.
Higher spectral resolution data of flares from 4500\,--\,6000\AA\ would
help characterize bona-fide continuum regions. Line
blanketing-induced continuum enhancements also need to be understood throughout the
entire blue optical and NUV zones.  

\item \textbf{Non-Hydrogen Continua:} There is a Helium \textsc{ii} continuum edge
  ($n=\infty$ to $n=5$) at $\lambda=$5694 \AA.  Helium \textsc{ii} continuum has been
  detected in solar flares in the EUV \citep{Linsky1976,Milligan2012}.

\item \textbf{Continuum Fitting Artifact:} By including the bluemost windows in the blackbody fitting (BW1,
  BW2), it could be a concern that we misfit the continua at longer wavelengths.
  These continuum windows have low-lying features
  within their narrow range.  In extreme
  cases, the wings of \Hd\ and H$\epsilon$ could contaminate the continuum
  windows.
  Appendix F of \cite{KowalskiTh} tested whether removing BW1 and
  BW2 before fitting the blackbody function can
  explain the apparent Conundruum flux.  In three of the four flares tested, 
  various amounts of Conundruum
  remained after excluding these bluemost continuum
  windows\footnote{In Appendix F of \cite{KowalskiTh}, a plot error
  for IF3 resulted in not showing the blackbody fit after excluding
  BW1 and BW2;  this figure has been corrected, and the resulting
  blackbody fit does not account for all red continuum emission during this flare.}.

\item \textbf{Multithermal Flare Emission:}
 The sum of all flare emission at any given time from a flare is very likely
 not isothermal;  therefore, emission from regions (along all
spatial directions in the atmosphere) with different temperatures
 would superpose to produce a spectrum that is inconsistent with a
 fit of a single blackbody over a large wavelength range.  In Section
 \ref{sec:conundruum_obs}, we were
 able to explain the blue-optical hot blackbody $+$
  Conundruum contribution at the \emph{peak} of IF3 using a double-blackbody
 fit (after minimizing the continuum fitting artifacts described in
  possibility \#5).

\item \textbf{Higher Order Hydrogen Continua:}  In Figure 2 of \cite{Kowalski2012}, they  
  accounted for the Conundruum through a superposition of
  model spectra using hot spot models and the RHDF11
  model. The RHDF11 model ascribes the flux at 
  redder wavelengths as due to a combination of Hydrogen Paschen
  continuum from the chromosphere 
  and increased photospheric continuum that results from chromospheric
  backwarming.  If the Conundruum has a large contribution from Paschen and possibly Brackett continua
(Hydrogen recombination to $n=3$ and $n=4$, respectively); a 
subsequent transition to $n=2$ from either of these two levels would
result in emission of an \Ha\ or \Hb\ photon, respectively.  This scenario could possibly
explain the relation between the Conundruum and the lower order Balmer
lines, which we discussed for \Hg\ in Figure \ref{fig:Conundruum}
(Section \ref{sec:conundruum_obs}). 
Higher order Hydrogen continua have been suggested to be a 
possible contribution to white-light radiation in the infrared
\citep{Tofflemire2012}.

In Section \ref{sec:modelcomparison}, we will discuss the
interpretation of the Conundruum further and show that items 
\#6 and \#7 are the leading possibilities.
\end{enumerate}

\section{Comparison to Radiative-Hydrodynamic Stellar Flare Models} \label{sec:modelcomparison}

The contribution function from RHD models allows one to 
deduce the plasma conditions and detailed opacities that give
rise to the predicted emission.  For example, the BaC originates from a
range of column masses where the electron density
and chromospheric temperature structure are strongly coupled
\citep{Hawley1992, Allred2006}.  Thus a
simple optically thin, isothermal slab \citep[e.g.,][]{Kunkel1970} does not provide an
accurate estimate of the atmospheric parameters or the Balmer
continuum originating from a real atmosphere.  RHD
models are required to give physically self-consistent values of $T$
and $n_e$ as a function of column mass in response to flare heating.

\subsection{Comparison to the BaC and Conundruum Components}
To model the flare Balmer continuum, we used the
impulsive-phase RHD model spectrum at $t=15.9$ sec from the F11 simulation
of \cite{Allred2006}, hereafter A06, following \cite{Kowalski2010}.   We 
subtracted a linear extrapolation of the blue-optical zone in the model to estimate the
amount of flux originating from the Balmer continuum.  We refer to
this model spectrum of the 
Balmer continuum as BaCF11 to distinguish from the published RHDF11
model\footnote{That is, RHDF11 is the total RHD model prediction including
 BaC, PaC, and heated photosphere, while BaCF11 is only the Balmer
 continuum prediction.}. 
We repeated this calculation for the F10 model
at $t=230$ sec from A06, giving an F10 Balmer continuum model
 spectrum BaCF10.  These times (15.9 sec and 230 sec) correspond to
 the maximum amount of emission produced at $\lambda=4170$\AA\ in the models.
We calculated the spectral slopes, $m_{\mathrm{NUV}}$, of the BaCF11 and
BaCF10 model spectra and give these values in 
Table \ref{table:model_table} (columns 4\,--\,5) to compare to the
observations in Table \ref{table:chitable}.  The value
of $m_{\mathrm{NUV}} \sim 5.1-5.5$ indicates that the model flux at 3420\AA\ is
$\sim$90\% of the model flux at 3615\AA.
At the peak of the flares IF2, IF3, and IF9, the slope of the excess
flux is \emph{blue} ($m_{\mathrm{NUV}}<0$; column 11 of Table
\ref{table:chitable}, Section \ref{sec:peak_bac})
and is not consistent with the red slope of BaCF11 or BaCF10 ($m_{\mathrm{NUV}}
> 5$).  Whereas the peak phases of the IF and HF events often show negative
values of $m_{\mathrm{NUV}}$ (see columns 9 and 11 of Table
\ref{table:chitable}), the GF events are more similar to the
RHD predictions.  This finding (determined from above) implies
that our understanding of the impulsive phase Balmer continuum (and
thus the flare chromosphere) is
\emph{not} complete with the current suite of RHD models.  In the
flare gradual decay phases, the IF and HF events tend to be more positive so
the models may be more accurately portraying gradual decay phase (and
GF) flare conditions.

To investigate the Balmer continuum in the gradual decay phase spectra in
more detail, we added
the BaCF11 model spectrum to the blackbody fit at $\lambda <
3646$\AA.  The results are shown in red in Figures
\ref{fig:grad1}\,--\,\ref{fig:grad5}.
We quantitatively compare the slopes of the observed Balmer continuum
emission (after subtraction of the blackbody) to the
slope of the BaCF11 ($\sim5.1$) and BaCF10 ($\sim5.5$) in column 5 of Table
\ref{table:model_table}.  
In all of the gradual decay spectra, there is a relatively good match to the
positive slopes (flux increasing towards redder wavelengths)
of the model spectra, as we found for IF1 \citep[][see also
Figure~\ref{fig:grad1}]{Kowalski2010}.  
The statistical fitting errors on the slopes
(given in parentheses) are quite large, likely due to the difficulty
in modeling the ``true-continuum'' underlying the forest of
low-excitation metallic lines that are likely present in the NUV during the decay
phase (Figure \ref{fig:magnumopus2}).  There are also systematic errors due to flux calibration
of \s5\,--\,10\% at these wavelengths (Appendix A of \cite{KowalskiTh}); this uncertainty
corresponds to about 2\,--\,4 in the units given in the table.  
The best matches to the BaCF11 model are obtained during IF9 and
GF1, which have BaC slopes of 4.9$\pm2.4$ and 5.9$\pm$1.3,
respectively. We conclude that the observed BaC are reflecting flare
chromospheric conditions from the models of A06, which
apparently predict 
the gradual decay phase emission using (impulsive phase) RHD beam models having a
non-thermal electron energy flux of $10^{10} - 10^{11}$ \ergscm, a low
energy cutoff of $E_c = 37$ keV, and a power law distribution of
nonthermal electrons with $\delta=3-4$.  More
accurately flux-calibrated spectra with higher spectral resolution (to
resolve the faint metallic lines) extending further into the NUV will provide better constraints for the models. 
  
The persistence of the BaC during  
late stages and the relatively good match to the shape of \emph{impulsive
phase} RHD model Balmer continuum, implies continued heating
from accelerated
particles during the gradual decay phase of these flares.  Gyrosynchrotron
  microwave radiation has been detected far into the gradual phase of
  white-light flares on M dwarfs, suggesting that nonthermal particles
  are accelerated after the impulsive phase \citep{vandenoord1996,
    Osten2005}.  Simultaneous, high cadence optical spectra,
  photometry, and radio data (e.g., from the EVLA) would better
constrain the timing of gyrosynchrotron emitting particles and 
white-light radiation.  

The impulsive phase RHDF11 model also accounts for the shape
of the red optical Conundruum emission component rather well, with a best-fit blackbody
temperature in the red of $T_{\mathrm{red}} \sim 4600$ K (RHDF10 shows
about the same color in the red).  This color temperature is similar
to the color temperature of the \emph{total emission} in the red-optical zone
in the gradual decay phase
of IF1 and to the color temperature of the \emph{cooler blackbody} using a
double blackbody fit in the peak phase of IF3.
  However, none of the RHD models (F10 or F11)
show evidence for a color temperature of
$\lesssim 4000$ K in the red part of the spectrum, as is needed to explain the
red emission in the IF3 gradual decay phase.  The RHDF11 spectrum predicts increased photospheric radiation
from chromospheric UV backwarming and Paschen continuum from Hydrogen
recombination in the flare chromosphere.  The opacity description in
the RADYN model does not include molecular
features that are important in M dwarfs (e.g., TiO) and which would
likely have a strong influence on the energy balance in the
photosphere, and hence the backwarming physics.
Modeling the backwarming is further complicated due to the
large area over which it may originate \citep{Fisher2012, Isobe2007}.   
Obtaining separate models for the Paschen continuum and backwarming
radiation is not currently possible with RADYN.

A reddening continuum shape during the gradual decay phase of IF0 in the
$V$ and $R$-bands (covering $\lambda \sim 5000-7000$\AA) was reported in
HP91, who speculated about the presence of a Paschen
continuum at these wavelengths.  The best-fit blackbody temperature in the gradual
decay phase to the $UV,UBVR$ photometry was $8400-8800$ K,
similar to the lower temperatures we find in the gradual decay phase of the
DIS sample using
only the blue-optical continuum fitting.  The mismatch to the
blackbody continuum implied by these temperatures in Figure 11 of \cite{Hawley1992} (bottom
two panels) is due to the addition of this red component.  Excess flux in
the $R$-band is also apparent in the moderate size flare on AD Leo described in 
\cite{Hawley2003}.
We speculate that the Conundruum rises into the red and accounts for the excess
$R$-band flux observed in these studies.  

Paschen continuum emission originating from typical temperatures ($6000$\,--\,$8000$ K) of the flare
chromosphere show a characteristic red color 
\citep{Hawley1992}, and we consider the possibility that the red
continuum emission (and hence Conundruum) is primarily due to Paschen
continuum.  From the RHDF11 spectrum, we estimated the height of the Paschen 
jump ($\lambda\sim8204$\AA) as
$F_{\lambda=8200}-F_{\lambda=8300}$, which is
1/16.5 times the Balmer jump height ($F_{\lambda=3640}-F_{\lambda=3655}$).
The predicted flux of the Paschen jump is shown in Figure
\ref{fig:megaflare_fullSED} as red error bars
indicating the height of the Paschen jump (using 1/16.5 times the flux of BaC3615).
We do not find convincing evidence of a Paschen jump to indicate the presence of
Paschen continuum emission, but the
spectral region around the Paschen jump is contaminated by 
features from the Earth's atmosphere and instrumental fringing which make the quiescent subtraction
uncertain and the characterization of flare-only continuum
difficult at these wavelengths.  The Paschen jump has been
(tentatively) detected in a solar flare \citep{Neidig1984Pa}.
 
Using simple, isothermal,
isodensity, optically thin, Hydrogen bf$+$ff slabs from \cite{Kunkel1970} as a guide, we can
explore the shapes of Paschen ($+$ Brackett) continua from $\lambda=3646-8205$\AA.
If we fit only to the red continuum, $\lambda = 6000-6800$\AA, $T_e=$ 
$12\,500-25\,000$ K (where $T_e$
is the electron temperature) is required to achieve a blackbody temperature of \TBB \s4500\,--\,5500
K (which corresponds to the color of the RHDF11 impulsive phase
spectrum and the IF1
gradual phase red spectrum). 
If we
consider the red gradual phase spectrum of IF3, which has a redder \TBB\
compared to IF1 (\TBB \s 3500\,--\,4000 K), its shape is reproduced by an isothermal model with $T_e =
7000-9500$ K.  Clearly, a large range of electron temperatures
($7\,000$\,--\,$25\,000$ K) can
give rise to a relatively small range of red continuum colors (\TBB \s
3500\,--\,5500 K).  More advanced, multithermal, self-consistent, atmospheric 
modeling including photospheric opacity calculations is needed to make further progress in understanding this enigmatic continuum
 component.  

\subsection{Comparison to Peak Phase Blackbody Emission Component}
The largest discrepancies between the current RHD models and the
observations are the shape of the continuum emission in the
blue-optical zone and the height of
the Balmer jump.  We measured the same observational parameters, \chif\
and \TBB, for the F10 and F11 continuum predictions from
A06.  The times analyzed are $t=230$ sec (F10)
and $t=15.9$ sec (F11), respectively.
We find that for the flare-only flux, \chifp$\approx$4.8 and
\TBB$\approx$5000 K for F10 and \chifp$\approx$8.2 and
\TBB$\approx$5300 K for F11 (given in Table \ref{table:model_table}).  The values of \chif\ are much higher than the observations (\chifp\ between 1.5\,--\,4.5; Figure \ref{fig:FEU_HG}), indicating
that the Balmer jump is too large.  The color temperatures are far too
low compared to the observations, which show a range at 
their peak times of \TBB$\sim9\,000-14\,000$ K (Section
\ref{sec:bluecont}, \ref{sec:astar}).  Using state-of-the-art
observations and new diagnostics, we have revisited the
long-outstanding problem that too much flare energy is deposited
in the mid to upper chromosphere, where a large Balmer jump in
emission is
produced.  In the RHD models, there is not enough energy deposited at high column mass
 (in the low chromosphere and photosphere) where a Balmer jump in absorption and/or blue-optical zone continuum with
a shape similar to a  blackbody having temperatures \TBB\s $9\,000-14\,000$ K can
be generated \citep[based on the phenomenological models
of][]{CramWoods1982, Houdebine1992, Kowalski2011IAU}.  Only when the
impulsive phase ``hot, blackbody'' (white-light) emission is successfully
reproduced by the models, can we begin to understand the detailed multithermal origin
of the Balmer continuum, the Paschen continuum, the Conundruum, and the
gradual phase 8000 K blackbody component;  furthermore, we can then
begin to 
understand how physical processes, such as flare heating, differ between GF and IF events and even between individual IF events.
Thus, RHD models incorporating additional physics \,--\, such as
higher energy electron beams and proton beams \,--\, are urgently needed.

\section{Flare Filling Factors and Speeds} \label{sec:filling_speeds} 
\subsection{Continuum Emission Filling Factors} \label{sec:fillingfactors}
A physically significant quantity that can be derived from the spectra is
the filling factor, $X$, which is the fraction of the area of the projected
visible stellar disk that is emitting flare
continuum emission.  Values of $X$ for the white-light emission are
often calculated in the literature
\citep{vandenoord1996, Hawley2003, Kowalski2010, Osten2010, Walkowicz2011,
  Davenport2012}. 
Flare areas allow us to determine the degree to which
individual stellar flares are ``scaled-up'' versions of one another (i.e., same spectral properties with larger
area).   Filling factors allow us to understand what type of heating distribution is
responsible for the observed phenomena -- i.e., using solar flare
terminology, whether the flare
heating is diffuse over a large area such as in extended
\emph{ribbons} or \emph{prominences}, or whether the 
heating is focused into a small area such as in compact
\emph{kernels}. Finally, we find that filling factors are
important for deriving the speeds of stellar flare development.

The area of a stellar flare is strongly dependent on the emission type
used to model the spectrum.  
An important concern is that real stellar atmospheres do not produce
featureless blackbody emission spectra.  In \cite{KowalskiTh}, we
investigated the corrections obtained using more realistic
\cite{Kurucz2004} models of hot star
atmospheres (summarized here in Appendix \ref{sec:appendix_CK}), and we found that the corrected, inferred areal
coverages (compared to those inferred from a Planck function) increase by a factor of $1.4-2.3$ for the peak
temperatures of \TBB \s $9000-14\,000$ K in our sample. 
Moreover, the \emph{effective} temperatures of the hot ``blackbody''
emission\footnote{We still use \emph{blackbody}, but now we mean a hot
  star spectrum with absorption.} 
component can be estimated this way, giving \Teff \s $7700-9400$ K for this range of \TBB\ (for log $g = 5$ atmospheres).  
The bolometric heating flux necessary to sustain the hot ``blackbody'' emission component is therefore
$\approx 2-4 \times 10^{11}$ \ergscm.  Starting in Section
\ref{sec:speeds}, we correct \XBB\ using the algorithm in
Appendix \ref{sec:appendix_CK}.

We now discuss important systematic uncertainties when calculating 
filling factors of flare emission.  First, there may be more than one source of continuum emission contributing
to the total blue-optical (e.g., C4170) flare continuum emission at $\lambda >
3646$\AA\ at peak times during flares.  For example, Conundruum flux
likely extends into the blue (see Figure \ref{fig:megaflare_fullSED}).  The RHDF11 model
predicts continuum flux at $\lambda > 3646$\AA, and this was used in
the modeling of the continuum in \cite{Kowalski2012}.  Therefore, a
single-component hot ``blackbody'' representation places only an upper limit on the areal coverage.
We suspect that if the hot ``blackbody'' was the only source of blue-optical continuum, the
absorption signatures seen in the \cite{Kurucz2004} models would be more prominent in flare spectra; even
in giant flares, such as IF3 and IF4, the absorption phenomena only
become measurable at peak emission when the hot ``blackbody'' has reached an
areal coverage of \s 0.2\% of the visible stellar hemisphere, corresponding to a physical area of \s $3 \times 10^{18}$ cm$^{2}$.

To calculate a filling factor for the (chromospheric) Balmer-continuum emitting 
region, we use the BaCF11 model from A06, as was presented previously 
for IF1 \citep{Kowalski2010}.  Despite the discrepancy between 
the observed slopes and model slopes in the NUV spectral zone (Table
\ref{table:chitable} and \ref{table:model_table}), we use the RHD Balmer continuum flux to
model the observed Balmer continuum as it is the best model we have
currently available.
A major source of uncertainty in the area of the Balmer continuum-emitting region is 
 absorption predicted in the 
blackbody component \citep{Kurucz2004} at $\lambda < 3646$\AA:  the Balmer continuum
emission and absorption components are spatially unresolved in our
stellar flare spectra.  Interpreting the peak phase emission of IF2
using these hot star spectra, the measured amount of Balmer continuum flux (BaC3615) would be 
 three times smaller than the intrinsic amount of Balmer continuum
emission originating from the flare chromosphere if we also account
for the absorption.  In other words, a
simple linear 
extrapolation from the blue-optical (as is used to calculate BaC3615
and therefore the filling factor of Balmer continuum emission) may
vastly underpredict the true amount of chromospheric Balmer
continuum emission.  However, we note that preliminary modeling of ``hot
spots'' in M dwarf atmospheres \citep{Kowalski2011IAU} indicates that
the amount of ``missing'' Balmer continuum flux is only \s 1.5
times the measured BaC3615.

Another important systematic uncertainty in the area of the
Balmer continuum emitting region is our selection of the F11 model to
represent the observed Balmer continuum flux.
For example, using the Balmer
continuum from the F10 RHD model
of A06, we find that the inferred areal coverages of
the BaC-emitting region would then be 8 times larger than inferred
from the F11
model \citep{Kowalski2010}.  The model BaC shape is \emph{not} constrained from the
observations due to the very
small differences in the shape of the BaCF11 and
BaCF10
between $\lambda = 3420-3630$\AA\ (Table
\ref{table:model_table}). Given the discrepancy between NUV slopes
observed at
flare peak (for IF and HF events) and for these
RHD models discussed earlier (Section \ref{sec:modelcomparison}), a
method for discriminating between models would only be applicable
during the gradual phase and peak phase of GF events.  We are seeking
to determine a method for constraining the appropriate model of the
BaC.

Third, a more accurate modeling of the energy deposition by nonthermal
electrons would take into account the pitch-angle distribution of electrons via the
 Fokker-Planck beam formulae \citep[e.g.,][]{McTiernan1990}.  As the energy deposition is not as
localized in the chromosphere, employing these formulae for a given beam flux would likely 
require the areas of the RHD Balmer continuum to be larger to account
for the observed emission \citep[][A06]{Mauas1997}.

Given these considerations, we calculate \XBB, \XRHD, and \XRHD\ /
\XBB\ for all of the flares in the Flare Atlas.  The values of \XBB\
(from fitting a blackbody, before applying the corrections in Appendix
\ref{sec:appendix_CK}) are given in Table \ref{table:chitable} (column 7).
At peak, the blackbody component
exhibits an areal coverage on
the order of \s0.3\% for the large amplitude flares, \s0.03\% for the medium
amplitude 
flares, and \s0.005\% for the small amplitude flares, a range of \s60x in area coverage in the
DIS sample.  We find that larger $U$-band emission at peak is due to a combination of both increasing
blackbody and BaC areal coverage.  The
Balmer continuum emitting region is also compact, though larger than
the blackbody component, originating from
\s0.05\% to 1.7\% of the visible stellar hemisphere.  
The ratio of the inferred surface area coverages (\XRHD/\XBB) is
$1-26$ at peak, with the apparent
amount of BaC originating from a larger area than the blackbody. However, the range of 
$3\le $\XRHD/\XBB$\le15$ is found to represent most flares at peak emission; this range was 
also found during the decay phase of IF1 in \cite{Kowalski2010}, as
the flare went through successive impulsive and gradual phases of
secondary flares.  The
largest luminosity flares have areal ratios of \s$1-10$. 
The ratio of areal coverages, \XRHD/\XBB, after applying the
corrections in Appendix \ref{sec:appendix_CK} to the blackbody emission component, ranges between $2-8$ for twelve out of the seventeen flares with
well-determined blue-optical temperatures.  
We predict that accounting for \emph{all} of the aforementioned
uncertainties in our area calculation would ultimately
\emph{increase} the Balmer continuum emitting area relative to blackbody
emitting area.  It would be instructive to recalculate the ratios of
filling factors using more accurate RHD models of the Balmer continuum
and blackbody emission components.

\subsection{The Speed of an Expanding Flare Area}\label{sec:speeds} 
The IF3 event produced a 
rise phase lasting 2.7 minutes and a fast decay phase lasting 5 minutes, allowing a unique
time-resolved study of the impulsive phase of a large-amplitude
flare.  Following the continuum analysis of Section
\ref{sec:bbpeak}, we fit a blackbody function to each spectrum from
$\lambda=4000-4800$\AA, and the filling factor, \XBB, is corrected
using the algorithm in Appendix \ref{sec:appendix_CK}.  The evolution
of \TBB\ and \XBB\ is shown in Figure \ref{fig:IF3_evol}.  The temperature evolution is
as follows: almost immediately (S\#25\,--\,26) in the fast rise phase, we
measure a color temperature of \s$10\,000$ K, which increases to \s$11\,500$ K by
S\#27 when C3615 is 1/3 of the peak value.  \TBB\ reaches a maximum
value of \s$12\,100$ K at the peak of C3615 at S\#31.  Whereas \TBB\ increases by \s2000 K in the
rise phase, \XBB\ experiences a large increase by a factor of 20
and stays elevated near its maximum value (possibly increasing very
slightly) for three spectra after the peak.  In subsequent
spectra after the peak, the color temperature drops and then
decreases monotonically at an average rate of \s800 K min$^{-1}$ 
during the fast decay phase.  
At the end of the fast decay phase, the temperature reaches the
characteristic gradual decay phase value of
\s8000 K.  For equal C3615 flux levels during
the fast rise and fast decay phases, the fast decay phase is more than
1000 K cooler, which implies that the area
(\XBB) must be greater during the fast decay phase.  In other words, the fast rise is hotter and smaller
than the fast decay.
\cite{Mochnacki1980} observed similar 
impulsive phase trends for \XBB\ and \TBB\ using continuum data during several dMe flares from
$\lambda=4200-6900$\AA\ (e.g., see their Figure 3).  The larger temperatures
inferred from our data likely result from fitting a narrower spectral window
($\lambda=4000-4800$\AA)
to avoid contamination from the Conundruum component.

We now use the rate of areal increase (\XBB) during the rise phase of
IF3 to estimate the \emph{speed} at which the flare area is changing, which
can be used to constrain different scenarios of flare heating.  

We use a simple flare areal model, whereby the white-light emitting
region is circular with radius $r$, and the inferred filling
factor, $X$, is
changing at a rate $dX/dt$.  The speed of the leading
edge (perimeter) of the expanding circular flare area is given by 

\begin{equation}
v(t)_{\mathrm{flare}} = \frac{dr(t)_{\mathrm{flare}}}{dt} = 0.5 \frac{dX(t)}{dt} \frac{1}{r(t)_{\mathrm{flare}}} R_{\mathrm{Star}}^2
\end{equation}
where $r(t_i)_{\mathrm{flare}} = (R_{\mathrm{Star}}^2  X(t_i))^{1/2}$ and
$\frac{dX_i}{dt} = (X_i - X_{i-1}) / (t_i - t_{i-1})$ where $i$ is the
spectrum number with midtime $t$. If instead we assume a more complex
geometry having two expanding flare areas to represent the two footpoint structures observed 
during solar white-light flares \citep{Hudson2006, Maurya2009}, the
inferred speeds decrease by a factor of two.  Note, we use the 
filling factors \XBB\ adjusted using model 
hot star atmospheres of \cite{Kurucz2004} described in Appendix
\ref{sec:appendix_CK}.  
The uncertainties in the speed calculations are estimated to be
\s40\%, primarily due to the systematic uncertainties in the 
temperatures (see Appendix F of \cite{KowalskiTh}).  
An additional uncertainty results from possible projection effects
(see footnote 22 in Section \ref{sec:bbpeak})
such that a flare near the limb would show a smaller expansion
velocity (by an amount proportional to ($cos$ $\theta$)$^{1/2}$, where
$\theta$ is the foreshortening angle)
compared to a flare at disk center.

We show the results for the speeds ($v_{\mathrm{flare}}$) during IF3
in Figure \ref{fig:speed}. According to this
simple, circular flare
model, the leading edge of the flare is moving at a high speed 
($\sim100$ \kms) 
during the rise phase 
and decreases to a small speed ($< 10$ \kms)
 after the peak.  The average speed
during the flare rise is $\sim50$ \kms, and the maximum speed ($\sim$110 \kms)
is attained in the mid rise phase.
The rise-phase speeds are supersonic for the photosphere and chromosphere of an M dwarf
($c_s \sim 5-10$ \kms) and we suggest several mechanisms below that
could be responsible for the expansion rates observed. 
\cite{vandenoord1996} also considered the expansion
speed\footnote{\cite{vandenoord1996} used the area derived from
  $U$-band data at peak emission and the larger area implied from
  radio observations in the gradual phase.} of an erupting filament
during a white-light flare
on YZ CMi, and found \s$100-500$ \kms, but they did not have spectra
which are needed to determine the
speed of the blackbody continuum component.  

IF9 is another (IF) event in our
sample where the blackbody has a rise \emph{and} peak phase detection.
Applying the same analysis, we find velocities of $\sim$130 \kms, 
$\sim$90 \kms, and $\sim$0 \kms\ in the rise, peak, and fast decay
phases, respectively.  Although the uncertainties are again \s40\%, IF9 and
IF3 show a 
similar pattern:  fast speed in the mid rise, decreasing speed at
peak, and very slow speed in the immediate post-peak phase. 

We repeat these calculations for the rise phase of MDSF2 because it
is a flare (albeit a secondary flare during IF1) with a simple, time-resolved rise phase. 
Only the average speed can be calculated for this flare.  We find that during the
rise phase, the area is expanding with an average speed of
\s8($\pm3$) \kms, much smaller than during IF3.  
Although the flux calibration is less certain, IF4 is apparently a
very quickly evolving flare exhibiting speeds of \s65\,--\,150 \kms\
through the rise and peak phases .  We note in passing that both the ``slowest'' and
``fastest'' flares (MDSF2 and IF4, respectively) show signatures of Balmer wing depressions (Section \ref{sec:wing_absorption})
indicative of extreme heating scenarios, discussed in the next section.
In Appendix \ref{sec:stack}, we also consider the speeds during the 
HF1 and IF2 events.

\subsection{The Meaning of Flare Speeds} \label{sec:speeds_meaning}

There are several possible scenarios which could drive the expansion
of a flare area, and we discuss the two most extreme here.  

\begin{itemize}
\item \textbf{Scenario 1 (\emph{Steady bombardment of a single region}): $v_{\mathrm{flare}}\sim c_{s,\mathrm{chrom}}$,
    $dA_{\mathrm{beam}}/dt = 0$ }

For this scenario, we assume that the beam\footnote{``beam'' refers to
  the flare heating source; usually this refers to a 
  flare-accelerated, nonthermal
  electron flux, but we generally mean any focused energy source.  }
bombardment occurs at a fixed location and over a constant area ($dA_{\mathrm{beam}}/dt = 0$).
The sound speed is an important parameter for the expansion of a flare area occuring
on the dynamical timescale in response to persistent beam heating from above (e.g., from the particles driven into
the lower atmosphere
during reconnection).  As pressure equilibrium is achieved with the ambient atmosphere at the
speed of sound,
radiative and conductive heating lead to a temperature
increase in the surrounding region.  The rate at which the flare area expands is therefore related to the
sound speed, and the rate at which the temperature increases is a 
combination of the rates from these heating processes which depend on the detailed physics and chemistry
(e.g., specific heat) of the heated atmosphere.  As the flare area expands, the
average temperature may eventually drop because the beam heating
(over constant area) can no
longer sustain the temperatures against cooling from expansion.  It is known from 1D RHD models that
constant heating of a given region increases the column mass in the
corona so as to inhibit further beam penetration
(A06); future studies of this effect with (3D) RHD models
would illustrate how relevant Scenario 1 may be during flares.  

\item \textbf{Scenario 1b (\emph{Expanding bombardment of a single
      region}): $v_{\mathrm{flare}} \sim v_{\mathrm{reconnection}} \sim c_{\mathrm{s, chrom}}$,
    $dA_{\mathrm{beam}}/dt > 0$ }

Scenario 1b is a variation on Scenario 1.  The flare speed would also
equal the sound speed if a 
sound wave can trigger magnetic reconnection and beam heating in nearby regions. 
Given that the sound wave disturbance persists 
for an extended duration, $dA_{\mathrm{beam}}/dt > 0$.

\item \textbf{Scenario 2 (\emph{Sequential bombardment of multiple
      regions}): $v_{\mathrm{flare}}\sim v_{\mathrm{reconnection}} >
    c_{\mathrm{s, chrom}}$, $dA_{\mathrm{beam}}/dt > 0$}

A second scenario which may be
important during the rise phase of impulsive flares, is
transient, fast heating (\emph{sequential bombardment}) of many small
neighboring regions.  In this scenario, the total area of 
all beams increase.  This scenario allows for speeds larger than $c_s$ in
the lower atmosphere
without having to invoke shock heating.  In fact, this scenario could occur at a speed which is driven
by the speed of reconnection at heights (i.e., in the corona) above
the site of flare heating. 
The emission we see at any given time is 
a superposition of all flare areas, and the increase of flare area during
the rise phase is then related to the increasing \emph{number} of
these individually heated regions --  not to the expansion of a
single region.  However, persistent heating and expansion could
also be taking place simultaneously as each area that is
heated by a beam expands according to the
physics described in Scenario 1.  An alternative explanation for
Scenario 2 would be supersonic expansion of a flare area, 
resulting in heating contributions in neighboring regions from an expanding shock
wave.  

Of the many important physical parameters that would inform this
picture, we consider the depth of continuum formation
and the decay timescale of the continuum to be essential information. Given the depth of continuum
formation, we would know the relevant sound speeds and could test the
relative roles of shock heating, conduction, radiative heating, 
and dynamical expansion.  The
timescale of continuum decay sets how long the 
beam needs to persist in individual flare areas in order to
explain the observed amounts of emission. 
\end{itemize}

We now discuss these scenarios given the derived speeds during IF3 and MDSF2.
 The average speed during MDSF2 is perhaps coincidentally, 
consistent with the speed (5\,--\,10 \kms) of the expanding disturbance from the
initial $|\Delta U|\sim$6 mag at peak event (IF1) that was assumed to trigger 
MDSF2 in \cite{Kowalski2012}.  The slow speeds during the rise phase of MDSF2 are
therefore consistent with Scenario 1b, giving
a connection between the speed of a wave disturbance in the lower M dwarf atmosphere
and the rate at which flare beam heating occurs at a new site.  We
speculate that the slower flare speed in MDSF2 led to more prolonged heating in a
given area.  In order to be consistent with the lack of Balmer line emission
produced in MDSF2, it is also possible that the heating source was
focused deep in the atmosphere \citep[similar to the 
heights at which reconnection is thought to occur in order to produce
Ellerman bombs, mentioned by][]{Kowalski2012}.  These effects 
produced the higher color temperatures (\TBB \s$13\,000-15\,000$
K; Section \ref{sec:astar} and Table \ref{table:color_astar}) and more easily detectable absorption
signatures.

Comparing the speeds for IF3 and MDSF2 signifies a physical connection
between the flare speed
and the type of blackbody spectrum (i.e., type of flare
heating) observed.  The average rise phase flare speed during IF3
is more than 5 times faster than the speed during MDSF2.  
IF3 produced large amounts of chromospheric (e.g., \Hg, BaC3615)
radiation, especially during the first half of the rise phase when the
velocities were higher.  If IF3 is described by Scenario 1, then the flare
heating would have to be higher than the chromosphere ($c_{\mathrm{s, chrom}} \sim
5-10$ \kms) to be consistent with flare expansion occuring at large sound
speeds.  The rise and peak phases of IF3 are therefore most likely described by
Scenario 2, 
implying that the areal increase would be related to the rate of
individual areas being heated.
Understanding the depth of formation of the IF3 continuum would allow
us to place constraints on the heating of individual
flare bursts so that the emission superposes correctly to form the observed light curve.
At the beginning of the fast decay phase of IF3, the speeds decrease
to $<$10 \kms, implying that Scenario 1 (or Scenario 1b) takes over after the peak. 
Unlike in the rise phase of MDSF2 (Scenario 1b) when the color
temperatures are relatively high ($13\,000-15\,000$ K), the color
temperatures in the fast decay phase of IF3 are decreasing and
relatively lower ($\lesssim 11\,000$ K). Slow expansion
is observed for a range of color temperatures, which may be related to
the difference in heating mechanisms during the rise of MDSF2 and initial
fast decay phase of IF3.

Additional, higher time resolution, data during the rise and
fast decay phases of other
types of flares would help us understand the relative importance of
these scenarios and
provide invaluable constraints on future 3D models that attempt
to reproduce the formation of the hot
blackbody continuum.  In Section \ref{sec:solarstellar}, we discuss
the implications of flare speeds for the solar-stellar connection.  

\section{Discussion and Summary} \label{sec:wtf}
In this study, we completed a homogeneous survey of twenty flares with simultaneous optical/NUV photometry
and spectroscopy.  We calculated and analyzed the following
observational parameters:  $t_{1/2}$ (the FWHM of the light curve;
Section \ref{sec:phot_param}), the impulsiveness index
$\mathcal{I}$ (peak $I_{f, U}$ divided by $t_{1/2}$; Section
\ref{sec:phot_param}), \chifp\ (C3615/C4170; Section \ref{sec:spectra_param}),
\chifd, BaC3615 (the Balmer continuum specific flux averaged between
$\lambda=3600-3630$\AA; Section \ref{sec:spectra_param}), \Hg/C4170
(Section \ref{sec:general}), the fraction of
$3420-5200$\AA\ emission in the Hydrogen Balmer (HB) component
(Section \ref{sec:hydrogen}), 
\TBB\ (the color temperature of the continuum between
$\lambda=4000-4800$\AA; Section \ref{sec:spectra_param}), $m_{\mathrm{NUV}}$ (the slope of the
continuum between $\lambda=3420-3640$\AA; Section \ref{sec:modelcomparison}), \XBB\ (the filling factor
of the ``blackbody'' emission; Section \ref{sec:filling_speeds}), and $v_{\mathrm{flare}}$ (the speed of the
expanding ``blackbody''-emitting area; Section \ref{sec:filling_speeds}).  Many of these
parameters require an accurate flux calibration of the spectra, and
we devised a new method for isolating the flare-only emission from
background quiescent emission (Section \ref{sec:scaling}, Appendix \ref{sec:appendix_scaling}).  

We present a summary of the key observational
results from this work in the first
column of Table \ref{table:final_table}.  In the second column, we suggest a possible physical
interpretation of the observation.  In the third column, we list the
most critical
parameter range for reproducing the impulsive
phase hot ($\sim10^4$ K), blackbody emission in future
time-dependent modeling; this parameter range typically corresponds to
the peak phase of the larger amplitude, IF events.

In the following subsections (Section \ref{sec:flare_morphology} \,--\,
Section \ref{sec:astrophysical}), we combine the main observational
parameters to elaborate upon some of the physical interpretations in
column 2 of Table \ref{table:final_table}.  In particular, we discuss
the origin of flare
morphology, the interpretation of the hot, blackbody emission, and
other astrophysical implications including the solar-stellar
connection as revealed by the measured flare speeds.

\subsection{Flare Morphology} \label{sec:flare_morphology}

How are the flare spectral properties related to the $U$-band light curve
``morphology''?  We began our study by separating the flares according
to morphological characteristics from the broadband light curves,
and we showed that the peak properties of the blue/NUV continuum are connected to the overall
time-evolution of the light curve.  We differentiate between
\emph{impulsive} and \emph{gradual} flares according to a new value,
the impulsiveness index, $\mathcal{I}$.  Our designation is not to be
confused with the impulsive and gradual phases of an individual flare:
for example, a flare may be gradual if it has a low peak amplitude and/or
large $t_{1/2}$ even though it may have distinct impulsive and gradual
phases.  The impulsiveness designation instead identifies the emission
type (i.e., fast or slowly evolving emission)
that contributes most to the overall evolution.
The flares that had the largest value of $\mathcal{I}$
are generally the classical, impulsive flares (IF); the gradual flares
(GF) are typically more complex featuring a slowly changing light
curve near peak but also intermittent faster continuum
variations. Some flares exhibited properties of both categories 
and are classified as ``hybrid'' flares (HF); these typically had 
several closely spaced (in time), fast yet resolved, continuum variations.
\cite{Houdebine2003} devised a similar classification scheme between
``impulsive'', ``gradual'', and ``combined'' flares. However, the 
\cite{Houdebine2003} classification does not include white-light continuum
emitting flares in the gradual flare category and our respective
definitions for combined and hybrid flares differ:  our definition
includes the timescales of primarily the impulsive phase.  Also, some of their
combined flares would likely be included as impulsive flares
according to our criteria.

Balmer continuum (BaC) in emission is ubiquitous during stellar
flares, although it is present in varying amounts, contributing as
little as \s20\% of the NUV flux (C3615) during IF events at peak and as much as
\s80\% at peak of the GF events (Section \ref{sec:peak_bac}, Figure \ref{fig:master0}), with the
relative amount also decreasing as a function of flare peak amplitude.  Though it
dominates the relative flux in the GF events, the BaC is seen in large
absolute amounts in some IF
events (e.g., IF1, IF2, IF5, IF9).
For the first time, we provide detailed light curves estimating 
the BaC emission
from flare start to flare finish, and we find that it is generally
similar to the Balmer lines, and evolves quickly, like the
higher order Balmer transitions \citep[as suggested in][]{Doyle1988}.
The $t_{1/2}$ values of the Balmer continuum light curves for the
well-measured flares were $\sim5-20$
minutes, and future models should seek to reproduce this property.

We established a temporal relationship \,--\, the ``time-decrement''
(Section \ref{sec:timing}, Figure \ref{fig:nature_baby})
\,--\, between six of the Balmer features for several flares.  The
time-decrement is a linear relationship between the wavelength
of the Balmer transition and the
$t_{1/2}$ value of the light curve for that transition. The time-decrement is similar for the classical flares IF3 and
IF9, indicating a possible scaling relationship -- similar heating
parameters but over a larger area -- between moderately
large energy flares ($E_U \sim 10^{32}$ ergs) and high energy ($E_U \sim 10^{33}$ ergs) flares for the Hydrogen Balmer emitting region.  
The time-decrement for representative HF and
GF events behaved differently than the
IF events.  For the HF event, the time-decrement
behavior is probably affected by the superposition of two distinct flare
impulsive phases. For the GF event, the flatness of the time-decrement
is related to the larger ratio of $t_{1/2, \mathrm{C4170}}$ to the
$t_{1/2}$ of the Balmer series; for the IF events, this ratio is
normally very small.

 To begin to understand the time-decrement, we look to 
  \cite{Drake1980}, who related the larger energy difference to
  smaller collisional frequency between the upper level and any 
given lower level in the higher order lines compared to lower order lines.
The time-decrement is therefore a result of each transition's individual
sensitivity to the decreasing densities in the flare chromosphere during the
flare decay (see, e.g., Figure 7b and 10c of \cite{Drake1980}) in
addition to the different oscillator strengths in the formula for
$C_{\mathrm{lu}}$ \citep{Rutten2003}.  Due
to the larger optical thickness, the lower order lines are less sensitive to the electron density of the flare
chromosphere.  The higher order lines and Balmer continuum emission are
less optically thick and therefore have contributions from higher column mass (higher densities) where
there are, generally, shorter cooling timescales, resulting in stronger
sensitivity to $n_e$ and thus shorter $t_{1/2}$.
In addition to density, the time decrement is likely dependent on
several other important flare parameters in the decay phase:  
the change in chromospheric temperature structure, the 
sustained level of beam heating in the gradual
phase, and the incident XEUV radiation field and overpressure from the superheated
corona.  Given all of these effects, which are changing as a function
of atmospheric column mass and time during the flare, it is critically
important to investigate the causes of the time-decrement with RHD
models.  Higher spectral resolution data of Balmer lines with high time cadence would help
constrain electron densities (via the widths of the Balmer wings) in order to relate to the
 time-decrement behavior.

The continuum measured redward of the Balmer jump (C4170) is the
fastest component in the blue/NUV and does not follow the
time-decrement of the Hydrogen Balmer series.  
This continuum component contributes a large fraction of the
emission throughout the NUV, blue-optical, and red-optical spectral
zones, and we relate this to the ``blackbody'' emission continuum that has
been detected in previous $\lambda > 4000$\AA\ spectral
\citep{Mochnacki1980, Kahler1982} and $UBVR$ broadband color \citep{Hawley1992,
  Hawley2003, Zhilyaev2007} studies of flares.  The $U$-band amplitude
and time evolution are largely determined by the blackbody emission component
although the $U$ band is also affected by the Balmer continuum,
which adds to the flare emission at $\lambda < 3646$\AA.
The BaC contributes relatively more 
flux in the HF and GF events and is the reason for their smaller
amplitudes and more gradual time evolution as observed in the
$U$ band.  At peak, the percentages of C3615 (total NUV flux) due
to BaC3615 are \s20\,--\,44\% (IF), 50\,--\,60\% (HF), and
55\,--\,80\% (GF; Section \ref{sec:peak_bac}, Figure \ref{fig:master0}).   The four flares with the smallest amount of Balmer
continuum at peak relative to the total flux are the largest amplitude
events:  IF0, IF1, IF3, and IF4.  We therefore conclude that the slower evolution of the GF events is due to a larger
influence from the BaC, which is slower than the blackbody component (as characterized
by $t_{1/2}$); conversely, the faster evolution of the IF events is due to
the dominance of the blackbody emission component (most dominant in
the largest amplitude flares), which evolves the quickest as diagnosed by C4170.
The HF events are intermediate because of the approximately equal
contributions of BaC and blackbody to
the $U$ band at peak. Due to the largely different timescales among the spectral components,
the relative amounts present at peak are therefore related to the overall light curve
morphology.   Clearly, a large sample was necessary for adequately
 understanding that not all stellar flares are identical, ``scaled-up'' versions
 of one another and that there are important spectral differences
 (presumably related to underlying flare heating mechanisms) to
 consider when modeling the gamut of flare behavior.

The IF events with the largest
relative amount of Balmer continuum emission at peak are IF5 and IF6.  These
flares are also unusual for their IF designation with large \chifp
($\sim$2.2$\pm(0.1-0.2)$), large \Hg\ to continuum ratios (\s40\,--\,50), 
and a large fraction of Hydrogen Balmer (HB) flux 
(\s40\%) from $\lambda=3420-5200$\AA.  Coincidentally, these two flares both occurred on EV Lac.  
As illustrated by these two flares, the relative amount of one
continuum component compared to the other is indeed variable even
during the IF events.  Nonetheless, the relative amount of Balmer continuum at peak generally
determines the evolution of the flare as either impulsive (dominated
by blackbody emission), gradual (dominated 
by Balmer continuum emission), or hybrid (both present, but generally more
Balmer continuum emission).  Also, the Balmer continuum does not always have a slow decay but is sometimes quite 
impulsive like the blackbody. This occurs typically with the hybrid
flares (see the time-decrement relation for HF2; Figure
\ref{fig:nature_baby}), or for the impulsive flares with large \chifp.

Successful modeling of the exceptions within each classification scheme (e.g., IF5 and
IF6 for the IF events;  MDSF2 for the HF/GF events) 
undoubtedly will provide invaluable insight into the flare heating
mechanism; this study awaits self-consistent modeling of the blackbody
component which has not yet been attained.

\subsection{The Balmer Jump}
The main parameter we use to describe the blue/NUV continuum is 
\chif, the ratio of flux values to the blue (C3615) and red (C4170) of the
Balmer jump.  The IF events have small \chif\ (and hence small Balmer
jumps) at peak continuum emission whereas the gradual flares have
larger Balmer jumps.  We also measure the ratio of \Hg\ emission to
the 
nearby continuum:  \Hg/C4170.
We connected the size of the Balmer jump to the \Hg/C4170, and find a 
suggestive linear trend such that flares with larger \Hg/C4170 ratios
have larger Balmer jumps (Section \ref{sec:general}, Figure~\ref{fig:FEU_HG_inset}).  Most of the IF events have strikingly
similar \chifp\ values of 1.6\,--\,1.8 over several orders of magnitude
in flare peak amplitudes and energies.   They generally also have
\Hg/C4170 ratios of 15\,--\,25 and only 11\,--\,17\% of the emission
in the Hydrogen Balmer (HB) component at peak.
A narrow range of peak flare properties suggests a common impulsive
heating mechanism among IF events.
The \chif\ and \Hg/C4170 measures indicate the relative importance of each
emission component -- Hydrogen Balmer vs. blackbody emission -- while
also giving information about the flare morphology.  \chif\ is a
measurement that can made \emph{without} spectra, for example at very
high time resolution and high signal-to-noise with the custom
continuum filters used on the stellar instrument ULTRACAM \citep{Dhillon2007, Kowalski2011CS}
and solar instrument ROSA \citep{Jess2010}.

For the IF events, the Balmer jumps increase by the beginning of the
gradual decay phase ($\Delta$\chif$\sim$1) and also show 20\,--\,30\%
more blue and NUV emission in the
Hydrogen Balmer (continuum and line) component compared to peak
(Section \ref{sec:hydrogen}, Figure \ref{fig:hb_phases}).  
These trends indicate that the white-light heating/cooling mechanism in the impulsive and gradual \emph{phases} of flares changes, as
flares become more dominated by Hydrogen Balmer emission in the (fast
and gradual) decay
phases, which has been a phenomenon suggested by several works 
such as \cite{MoffettBopp1976} and \cite{Abdul-Aziz1995}. 

The Balmer jump has not been detected in our low-resolution spectra (although we note that several instances in different flares show suggestive
features near $\lambda \sim 3646$\AA, especially in the gradual
phase).  It remains an open question what causes the large amount of
apparent blending (which we have referred to as the
``pseudo-continuum'' or PseudoC) between $\lambda=3646-3914$\AA, as
the Balmer jump is not detected even in high spectral resolution, $R\sim40\,000$ spectra \citep{Schmitt2008, Fuhrmeister2008}.  We raise the question, \emph{what is the ``Balmer continuum'' that we observe in our low-resolution
spectra?}  We suggest that the amount that we measure may be affected
by the Stark broadening of higher order Balmer line wings and the
continuum edge \citep{Donati1985}.  High spectral resolution, high
time resolution data near $\lambda = 3646$\AA\ (while also
covering some continuum regions at $\lambda > 4000$\AA\ in order to
characterize the blackbody component) together with better modeling of
the Stark broadening in the higher order Balmer lines (following the
work of \cite{Donati1985} for solar flares) are needed to understand the detailed pseudo-continuum properties from $\lambda=3646-3920$\AA.  

\subsection{The Temperature of the Hot, Blackbody Component} \label{sec:temp_meaning}
The large percentage (\s40\,--\,90\%) of the wavelength-integrated blue$+$NUV
(3420\,--\,5200\AA) flux in the peak and gradual decay phases
is emitted in continuum emission other than Hydrogen Balmer radiation
for all flares (Table \ref{table:hb_phasez}, Section \ref{sec:hydrogen}).  We attribute this continuum emission 
to a ``blackbody'' (or blackbody-like) source.
We measured the slope of the blue continuum from $\lambda=4000-4800$\AA\ and found it to be
linearly decreasing with wavelength, thus allowing us to match it well
to a blackbody with a moderately hot color temperature, \TBB\s$10\,000$ K (Section \ref{sec:bbpeak}).  
The range of peak temperatures for the seventeen IF, HF, and GF events
with well-measured slopes is
between \TBB \s $9000-14\,000$ K (with systematic errors of about 1000
K; Figure \ref{fig:TXdist}).  The secondary flare (MDSF2) during IF1 has the highest color
temperature of $\ge 15\,000$ K at peak.  The lowest temperature was
measured as \TBB \s 6700 K for the low-amplitude flare GF5.  The
flares for which we measured blackbody temperatures have a range of
\s2.5 orders of magnitude in flare peak specific luminosity in the $U$-band.  
Why such a narrow range of temperatures?  We speculate that the origin and
transient nature of the blackbody radiation may be due to a thermal
regulation process, when a temperature threshold 
is reached in the deep atmosphere prior to runaway heating whereby 
material heats to higher temperatures and lower
densities (i.e., ``explosive evaporation''; see \cite{Fisher1985}).  The observed Neupert relationship between Ca \textsc{ii} K and C4170
 (the blackbody) may be indicative of an evaporative process
 occurring at high column mass (where C4170 is produced) if there
 is a connection between the formation of Ca
 \textsc{ii} K emission and the evaporated plasma through, e.g.,
 X-ray (from $T > 20$MK) radiative backwarming \citep{Hawley1992},
 condensations \citep{Abbett1999}, or heat
 conduction channels \citep{Canfield1984} from the corona. 

We also explored interflare and intraflare \TBB\ variations.
In Appendix \ref{sec:stack}, we combined spectra from multiple peaks in a 
hybrid flare (HF1) to obtain a spectrum with the same total flux level of a very fast flare
(IF2) and found bona-fide differences in the value of \TBB, the
fraction of HB to total flux, and the evolution of HB to total flux. 
 Of the many parameters that could give rise to interflare \TBB\ variations, we suggested that the speed of the flare
areal increase is an important parameter, which is summarized below.
However, it is imperative that RHD models be used to model temperature
differences between flares. 
With these state-of-the art spectral observations and new analysis
techniques, we also attempted to constrain the
intraflare variation of the blackbody continuum component.  For two flares with
time-resolved rise phases (MDSF2 and IF3; Table
\ref{table:color_astar} and Figure \ref{fig:IF3_evol}), we found an apparent
increase of \s 2000 K in the color temperature of the blackbody component during the rise phases.

The impulsive phase ``blackbody'' continuum component is not a
featureless emission spectrum from $\lambda = 3400-6500$\AA. 
 We detected an A-type star spectrum
during the rise and peak phases of a secondary flare (MDSF2) that
occurred during the Megaflare (IF1; K10).  This finding gives important
insight into the nature of the blackbody component:  the absorption
features and continuum shapes (i.e., \TBB) encode information about the multithermal
stratification of the flaring atmosphere at high densities.  We
hypothesize that the multithermal structure of the atmosphere is a byproduct of 
the thermal regulatory process, described above, such that the
atmosphere responds in such a way that energy is most efficiently (and quickly)
removed from the atmosphere through the continuum.  We will explore
this aspect of flare atmospheres in detail with future RHD models
(Kowalski et al. 2013).

Absorption signatures were also detected in the wings of the Balmer lines
during MDSF2 and IF4.  Note that hot star-like absorption during
flares also affects the Balmer flux and time decrements \citep{KowalskiTh}.  Our analysis techniques
can be applied to other flare observations to understand absorption phenomena
in spectral data that lack broad wavelength coverage or robust flux
calibration. 

We extended the simple blackbody modeling to include the effects of flux redistribution
and wavelength dependent opacities, as in a real stellar atmosphere, resulting in 
``Castelli-Kurucz'' corrections which are described in detail in
\cite{KowalskiTh} and Appendix \ref{sec:appendix_CK}.  In summary, more realistic
filling factors (\XBB) of the ``blackbody'' emission component can be obtained, which are about a
factor of 1.5\,--\,2.3 times larger, than those found using the Planck
intensity to estimate areas.  Also, effective temperatures can be
estimated from the measured color temperatures of \TBB\
($9000-14\,000$ K).
The corresponding range of \Teff\ is \s 7700\,--\,9400 K (with the largest being \s$10\,500$ K
during the MDSF2 peak) indicating that at least $2 \times 10^{11}$ \ergscm\ is required
for the heating flux to sustain the ``blackbody'' component.  Detailed
modeling of these ``hot spots'' in the M dwarf atmosphere would
 constrain the density and
temperature stratification of the lower flaring atmosphere.

Most flares do not show obvious absorption features in the Balmer continuum or Balmer lines, likely due to 
the large amount of (chromospheric) Hydrogen Balmer flux in
\emph{emission} that generates an effective veiling of the
 photospheric absorption.  The veiling leads to an anti-correlation in
 the apparent amount of Balmer continuum emission and the amount 
of blackbody-like continuum emission.  As the absorption becomes
greater \,--\, either through a larger filling factor (areal coverage) or
because of higher blackbody temperature \,--\, the measured Balmer
continuum emission estimated using a
linear extrapolation from the blue-optical zone decreases. Again,
accurate models (e.g., phenomenological ``hot spot'' models) would elucidate how much Balmer
continuum emission is being missed.

An additional benefit of our large sample of flares and 
homogeneous analysis is evident.  Not all flares have the same proportion of
each emission component and therefore some flares reveal more about a
given emission property than others.  For example, MDSF2 exhibited a
spectrum that had an unusually strong contribution from the ``blackbody''
component alone, thereby allowing unambiguous detection of the absorption features.
The IF0 event from HP91 also had such a strong contribution from the
``blackbody'' component that the relative amount of Balmer continuum emission was among the
lowest in the sample.  
In contrast, the IF1 decay phase showed an excessive amount of Balmer continuum
emission, thereby allowing a thorough characterization of its
flux and spectral shape.  We predict that more observations
will reveal additional, important variations in the flare
continuum components that will contribute new constraints on flare models.

\subsection{Flare Speeds and the Solar-Stellar Connection} \label{sec:solarstellar}
We measured the speed at which the inferred ``blackbody'' flare area grows during
the rise phase of IF3.  We discussed ways in which
the inferred flare areas are rather uncertain, and a more realistic determination
was obtained by modeling the ``blackbody'' emission as a hot-star spectrum.
During IF3, the speed is large in the early and mid-rise phase,
with $v_{\mathrm{flare}}$ \s100 \kms, decreasing to \s50 \kms\  at
the peak and  $< 10$ \kms\ in the post-maximum phase.  

These speeds are strikingly similar to the speeds of the development of two-ribbon flares 
parallel and perpendicular to the magnetic neutral line in solar
active regions.  In particular, motions
of white-light, hard X-ray, \Ha\, and C\textsc{iv} kernels are observed on the Sun
\citep{Kosovichev2001, Wang2009, Nishizuka2009, Krucker2011,
  Inglis2012}, and are observed to propagate
at similar speeds of between $50-130$ \kms\ during the development of
two-ribbon structures parallel to
the magnetic polarity inversion line (PIL) \citep{Fletcher2001, Schrijver2006}.
There is also slower motion
perpendicular to the flare ribbons at the locations of the kernels, outward from the PIL with 
velocities of \s15 \kms\ \citep{Fletcher2004, Keys2011} or as small as
a few \kms\ \citep{Qiu2010}, presumably the result 
of reconnection at progressively higher heights.  Perhaps,
the higher speeds in the rise phase are analogous to the formation of
flare ribbons, and the smaller speed after flare maximum represents the
perpendicular motions \citep[e.g., ``spreading moss'',][]{Berger1999}.  This scenario has
been deduced by UV/HXR observations to explain solar flare
light curve morphology \citep{Qiu2010}.

  The heating of new flare
footpoints, resulting in increasing flare area, was a conclusion
from the models of \cite{Hawley1992, HawleyFisher1994} to explain
solar and stellar flare evolution.  
During solar flares, it is an open question what triggers particle
acceleration (a possible source of flare heating) in neighboring flare
regions \citep{Inglis2012}.  Possible scenarios are an unstable flux
rope that erupts sequentially along a flare arcade \citep[the
``tether-cutting scenario''][]{Moore2001} and
magnetoacoustic slow waves that propagate away from the initial flare
site \citep{Nakariakov2011}.  Three-dimensional
models of reconnection predict motions of the reconnection site, which undergo 
changes in velocity to slower speeds due to mass loading \citep{Linton2006}.

The evolution of $v_{\mathrm{flare}}$ during IF3 (Figure
  \ref{fig:speed}) suggests 
that a similar (two-ribbon) flare development process
involving sequentially heated footpoints occurs during solar and
stellar flares, despite our likely over-simplification of the
geometrical morphology of stellar flare regions (i.e., as an expanding
circle).  The quantity $v_{\mathrm{flare}}$ is therefore an exciting new measure that
can be used to further address aspects of the solar-stellar connection.  Broad
wavelength coverage solar flare spectra \citep[e.g.,][]{Neidig1983},
which are not feasible with current standard instrumentation, would be invaluable for providing
detailed insight into the spatially resolved continuum properties during ribbon
and footpoint formation.  

We also found a possible relation between the average speed during the
rise phase and the temperature of the blackbody
component at peak, with a very slow speed (\s8 \kms)
observed for 
hotter temperatures (\TBB \s $15\,000$ K) and prominent Balmer
absorption features, e.g., in MDSF2.  Perhaps the heating occurs by a  
process similar to solar Ellerman bombs on the Sun, where reconnection
is thought to occur in the low chromosphere or photosphere.  The connection to Ellerman bombs and their (possible) formation at high densities and low heights has implications for the heating properties of M dwarf flares that show Balmer absorption signatures.
However, we do not exclude the possibility that the heating during the secondary flare
MDSF2 (which is classified as a hybrid flare event; Figure
\ref{fig:if_vs_chi}) is connected to the Megaflare decay and may exhibit different heating properties than
classical flares.  

Determining the fundamental connection between speed and the type of
white-light continuum emission will require further modeling work on
the spatial development of flare regions.
However, we began by comparing the speeds between flares 
and relating to two possible heating scenarios.  We
present two extreme scenarios: the expansion of a single hot spot in
the lower atmosphere
 (Scenario 1, $v_{\mathrm{flare}} \sim c_{s,\mathrm{chrom}}$) and the formation of
 several hot spots (Scenario 2, $v_{\mathrm{flare}} > c_{s,\mathrm{chrom}}$).  
 Insight into the detailed physics of these scenarios requires 3D
 modeling, but we can gain initial insight by comparing 
  the speed of the flare to the speed of sound in the level of the
  atmosphere where the (blackbody) emission
  is formed.  Higher speeds (50\,--\,130 \kms) are observed
  during the formation of a hot blackbody, \TBB \s
  $10\,000-14\,000$ K (Figure \ref{fig:TXdist}) and values of BaC3615/C3615 of 0.25\,--\,0.35
  (Figure \ref{fig:master0}) for fast rise and peak phases of the classical
  flares IF9, IF3, and IF2 with moderate to large peak $U$-band
  amplitudes and energies.
  In the impulsive phase of these events, we find evidence of Scenario 2-type speeds during the rise and
  peak phases followed by Scenario 1-type speeds during the initial
  fast decay phase.  A
  low speed ($\lesssim 10$ \kms) and high temperature were derived
  from the rise phase of MDSF2, whereas low speeds and relatively
  lower temperatures were derived in the initial fast decay phases of
  IF2, IF3, IF9, and HF1.  Therefore, there is a
   range of heating that is observed at times of low flare speeds.  
  The rise phase of MDSF2 is consistent
  with a variation on Scenario 1 (Scenario 1b), whereby the
  beam heating (and reconnection?) is 
  triggered by a sound wave generated in the lower atmosphere by the
  initial flare peak of IF1.

\subsection{The Conundruum Continuum Component}
The white-light continuum persists well into the gradual decay phase of IF,
HF, and GF events as the
flare spectrum becomes increasingly dominated by Hydrogen Balmer (line and
continuum) radiation.  At the beginning of the gradual decay phase, the blackbody component exhibits a cooler
temperature of \TBB \s 8000 K.  The change in color temperature from
peak to gradual decay phase was first discovered by \cite{Mochnacki1980}.  We 
find that the evolution of C4170 (blackbody flux) begins to follow the Balmer continuum flux in
the gradual decay phase.
  Another continuum component is detected
redward of \Hb.  This is 
observed as an increasing contribution of red flux during the flare gradual
decay phase,
and may be responsible for the redder colors seen in broadband
photometry during the gradual decay phases of flares in the past
\citep[e.g.,][]{Kunkel1970, HawleyPettersen1991, Hawley2003}.
This ``Conundruum'' flux rises into the red spectral zone and has
a color temperature of $\lesssim$5500 K during the peak and gradual decay phases.  
The colors in the red-optical are similar to those (albeit with
large errors) reported during
the gradual phase of a large flare on CN Leo
\citep{Fuhrmeister2008}.  This study found 
large temperatures (\s20\,000\,--\,30\,000 K) in the red during
the impulsive phase, whereas we find a peak phase color
temperature of \TBB \s 7700 K (using a single blackbody fit; Section \ref{sec:conundruum}).
The Conundruum is present in some flares at peak also, but the hotter (blackbody) component dominates 
the white-light continuum shape at peak emission in the blue and, to a
lesser extent, in the red.  According to our analysis in Sections \ref{sec:bluecont}, Section
\ref{sec:gradual}, and Section \ref{sec:conundruum}, the red continuum
($\lambda \gtrsim 6000$\AA) during dMe flares is 
a superposition of the hot, blackbody emission component and Conundruum emission component.
Impulsive phase red emission (e.g., in the $R$ band) is largely due to
the action of the hot blackbody
component, and late phase gradual emission is largely due to the Conundruum.

The red spectral shapes are similar to RHD model predictions
\citep{Allred2006}, from which we infer the origin of the
 Conundruum to be Paschen and Brackett recombination
radiation from the flare chromosphere, and there may also be a
contribution from H$^-$ recombination radiation originating from a
larger backwarmed area of the star \citep[similar to the geometry in][]{Fisher2012}.  
We explored the
Conundruum flux in detail for IF3, and
suggest that it has a physical connection to the lower order
(\Hg, \Hb, and \Ha) Balmer lines.  

In summary, we find three continuum components that are important
at different stages of the flare evolution:  1) Balmer continuum
emission (most important in the mid rise phase and during the gradual
gradual decay phase); 2) Hot ``blackbody'' emission (most important at the peak phase, but
present with cooler temperatures in the fast decay and gradual decay
phases); and 3) Conundruum emission (can be present in all stages, but
most important in the late gradual decay phase).

\subsection{One Flare Model to Rule Them All?}
The emission at flare peak cannot be explained by simply scaling up
small amplitude (e.g., in $U$ band) flares to get large amplitude flares, thereby implying important differences in
heating mechanisms between IF, HF, and GF events. The emission
components originate from different 
atmospheric parameters among the different types of flares, and even
within a given flare type. 
 We have shown
 several ways that IF3 and IF9 have scaled-up Balmer components (via the
 time-decrement; Section \ref{sec:timing}).  However, the C4170 components appear to differ and 
do not follow a scaled up relation in their time evolution, suggesting
that C4170 may be formed differently even among the IF events.
We examined the scaling relationship between two flares in detail in
Appendix \ref{sec:stack} -- HF1 and IF2 on YZ CMi --
 and concluded that one large peak phase is not a simple superposition of
 several smaller peak phases.  Nonetheless, the
 $\mathrm{log}_{10} \mathcal{L}_{\mathrm{BaC3615}}$ vs. $\mathrm{log}_{10} \mathcal{L}_{\mathrm{C4170}}$
 relation among the IF events with \chifp $<$ 2 and \TBB \s 8000 K (IF2, IF3,
 IF7, IF8, and IF9) at the beginning of the gradual
 decay phase follow a linear relation, possibly
 indicating that these flares are scaled
versions of one another after the fast decay of the C4170 (Section \ref{sec:gradspec}).  Further evidence that gradual phase emission follows a
scaling relation were deduced from a qualitative analysis (Figure \ref{fig:magnumopus2}) of the
fine spectral features and continuum shapes spanning the gradual decay phases
of IF1 (a large flare), IF7 (a medium amplitude flare), and GF5 (a
small amplitude flare).  One must independently consider
the degree to which the Hydrogen Balmer (line and continuum) emitting
region and blackbody continuum emitting
regions are scaled between two flares and between individual flare
phases.  Furthermore, the production of the emission
components is not mutually exclusive, as indicated by the short ($<1$
min) delay
between the peaks of Hydrogen line emission and C3615 such as during IF3 and
IF9 (Section \ref{sec:Hgamma}).  In fact, the sources of the
emission components are intricately tied to
one another via the flare heating mechanism and must be
self-consistently accounted for in flare models that attempt to
reproduce the evolution of these components and their variation during
IF, HF, and GF events and during the impulsive and gradual phases of a
given flare.

\subsection{Connection to Other Astrophysical Phenomena} \label{sec:astrophysical}
Transient astrophysical phenomena across the universe may be related
to the underlying physics in the dMe flares, and we 
plan to explore these connections in a future paper.  Here, we briefly note several interesting similarities to gamma ray bursts and
accretion phenomena.  Gamma-ray burst (GRB) light curves have been divided into two classes based on the duration of the hard X-ray light curves:  Class I GRB's with $2$ to several hundred second durations
and Class II GRB's with $<$ 2 second durations \citep{Kouv1993}.   \cite{Kouv1993} further found that the Class II GRB's exhibited harder spectral slopes 
in X-ray emission.  Similarly, we found that \emph{impulsive} flares
have different optical timescales and spectral slopes (given by \chif) compared to \emph{gradual} 
flares.  \cite{Hurley2005} proposed that a Class II GRB originated
from a magnetar flare (MF).  The magnetar flares may be a subclass 
called soft gamma ray repeaters (SGRs) that occur relatively close to the Milky Way and account for only a few percent of the short/hard GRB's \citep{Palmer2005}; the current leading model is the merging of two compact objects, one of which is a neutron star \citep{Blinnikov1984, Nakar2007}.
Interestingly though, the impulsive phase of a MF has been fit well with a blackbody.  In contrast
to flare impulsive phase emission, the blackbody was found to have a
much higher temperature  \s$2\times10^5$ K.  An impulsive phase in the light curve
during times of prominent optically thick (blackbody or blackbody-like) emission is a similarity between dMe flares and magnetar flares and implies a common cooling (and heating?) process
during the most luminous events in the solar neighborhood and the most luminous events in the universe.  

A second astrophysical setting in which we find similarities to dMe
flares is accretion, such as occurs in T Tauri stars and dwarf novae.
The Balmer jump is a common diagnostic in T Tauri spectra \citep[e.g.,][]{Valenti1993, Herczeg2008}. In fact, some T Tauri spectra show small Balmer jump ratios that 
are rather similar to the \chif\ values for flare spectra.   Optical veiling from continua is often seen at $\lambda > 4000$\AA, which can inform our understanding 
of the ``blackbody'' continuum in flare spectra.
Therefore, the physics of accretion models for T Tauri spectra can be used to inform our understanding
of the white-light continuum.  We also note that dwarf novae accretion events produce superposed Balmer line emission and Balmer line absorption features in their spectra, with most conspicuous absorption at times of maximum continuum emission that resembles an A or B type star \citep{Hessman1984}. 

\section{Future Work} \label{sec:future}
Future RHD models should aim to reproduce the basic flare emission
properties shown here:  a hot blue-optical zone color temperature (\TBB) and a
small Balmer jump (\chif).  The time-decrement will be important to
constrain the time-evolution of the various flare types.  Once the
general $\lambda = 3400-5200$\AA\ properties are reproduced
adequately, we should then aim to explain the PseudoC and lack of a
Balmer discontinuity using a Hydrogen atom with many principal levels
and a better treatment of Stark broadening \citep[as in ][]{JohnsKrull1997}.  
Understanding the red-optical continuum radiation will require an 
advanced treatment of photospheric molecular band chemistry and the
 effects of UV
backwarming over large areas \citep{Fisher2012}.  

We propose the following additional observations as future work:

\begin{itemize}
\item Our flare peak spectra (Figures \ref{fig:peak_panels1}\,--\,\ref{fig:peak_panels3}) clearly indicate a rise
  toward bluer wavelengths into the NUV.  
Ultimately, models of blackbody emission must be tested in the
NUV with spectral observations from $\lambda = 2000-3400$\AA.  Unfortunately, there is a current lack of
observational continuum measurements in this wavelength region during the impulsive phase of
flares (of any morphological type) when the unexplained blackbody
component is the brightest.   NUV data at $\lambda < 3400$\AA\ are
critical to constrain the detailed interflare and
  intraflare temperature variations in addition to the peak of the
  white-light (blackbody) emission.

\item Observations of high-energy events with $|\Delta U| \sim3$ magnitude (or greater) 
  are important for establishing the occurrence and timing of small \chifp, Balmer
  continuum absorption, and line wing absorption during primary and
  secondary peaks.

\item Moderate spectral resolution observations of $\lambda \sim
  5000-5500$\AA\ are needed to separate the contributions of Fe \textsc{i} and
  Fe \textsc{ii} emission lines from the Conundruum.

\item High spectral resolution, high temporal resolution observations of $\lambda \sim
  3500-4500$\AA\ are needed to understand the blending of the higher order Balmer
  lines and the blending of the Balmer edge.  Observations (at
  high time cadence) of \Ha\ would also
  be important for determining (directed and turbulent) mass motions during flares.

\item High cadence observations of Ca \textsc{ii} K of very impulsive
  classical flares would help determine the degree to which the
  Neupert effect holds during flares.  These observations would
  also constrain the evolution of flare area (via
  $v_{\mathrm{flare}}$) during this common, yet difficult to observe,
  flare type.

\item High quality blue/optical/NUV spectral observations of the fast 
  decay phase just after flare peak would provide insight into the transition from
  impulsive to decay phase emission.  These observations are also 
  important for understanding the behavior of 
  $v_{\mathrm{flare}}$ (which is small) just after flare peak.  

\item More observations of flares with double-peaked impulsive
  phase broadband light curve morphology (like IF0, IF4, HF2, HF4)
  would constrain the complicated temporal relationship
  between the (Balmer and blackbody) continua and line emission. In
  these large, energetic, and relatively common events, the secondary flare may be
  the result of a triggering mechanism from the primary flare.

\end{itemize}

\input{figures}

\input{tables_final}

\acknowledgments{We thank an anonymous referee for helpful suggestions
  that improved this paper.  This work resulted from the Ph.D. Dissertation of
  A.F.K. at the University of Washington Department of Astronomy.  
  We thank M. Giampapa for suggesting to investigate
  flare speeds and H. Lamers for insight that contributed to
 the ideas developed in this section. 
 We acknowledge many helpful discussions at L. Fletcher's International Space Sciences Institute (ISSI) ``Solar
 Chromospheric Flares'' team meetings, in particular with L. Fletcher,
 H. Hudson, P. Heinzel, G. Cauzzi, and M. Carlsson.  We thank
 M. Mathioudakis for useful discussions.  We thank K. Covey for
 the use of his PyRAF spectral reduction software.  We thank N. Ule for
 obtaining data.  We
 thank J. Allred for the use of his flare model results.  We would like to thank the staff at
 the Apache Point Observatory and especially the Observing Specialists
 (R. McMillan, W. Ketzeback, J. Huehnerhoff, G. Sarage, and J. Dembicky)
 at the 3.5-m for assistance with data acquisition and 
 data calibration feedback.  We also
 thank the SDSS 2.5-m observers at the Apache Point Observatory for on-site assistance with the ARCSAT
 0.5-m telescope.
Finally, we thank E. Agol for useful discussions and feedback on this
work.  A.F.K. acknowledges support from NSF grant AST08-07205, NASA Kepler
GO NNX11AB71G, the ORAU/NASA Postdoctoral Program, and the K-Unit.}






\appendix

\section{Scaling Spectra Using Molecular Features} \label{sec:appendix_scaling}
In this Appendix, we describe how we test the flux scaling algorithm's accuracy (see Section
\ref{sec:scaling}) for recovering a pre-determined flare spectrum. To simulate a flare spectrum, we add a Planck function
(for a given \TBB\ and \XBB) to a quiescent spectrum which has been scaled to
surface flux values using distance and $R_{\mathrm{star}}$ (Table \ref{table:starssummary}).  
Three blackbody temperatures were used -- \TBB$=5000,
10\,000,$ and $50\,000$ K -- to sample a large range of possible flare spectral
shapes.  The flare surface flux was multiplied by a filling factor
(\XBB) to simulate
the surface areal coverage.  We then multiplied the
resulting flare$+$quiescent spectrum by a constant factor, ranging
from $0.7-1.3$
to simulate grey slit-loss or weather-induced flux variations.  
The resulting ``observed'' spectrum was then used as input to the
scaling algorithm (see text) to determine if the artificial factor (0.7\,--\,1.3) applied
initially was recovered.

The largest flares on dM stars have factors of $\sim$100 enhancements
in the $U$-band flux ($I_{f, U} + 1 \sim 100$; we denote the flux
enhancement by $I_f+1$; see Section \ref{sec:phot_param}).  The filling factors necessary to produce this
enhancement of $T_{BB}=5000, 10\,000, $ and $50\,000$ K are 0.76,
0.016, and 0.0004, respectively.  The smallest flares in our sample
increase the $U$-band by a factor of \s 2.  The filling factors
necessary for this enhancement are 0.015, 0.0003, and
8$\times10^{-6}$ for these temperatures, respectively.   In addition to small and extremely
large flares, we also model an intermediate to
large amplitude flare, with $I_{f, U}+1 \sim18$.

The results of testing large and small flare areas with the algorithm
are given in Table \ref{table:scaling} (first and third rows for
each of the three temperatures).  The results indicate that for
small flares, the algorithm determines precisely the correct
predetermined multiple (0.7\,--\,1.3) that was applied to the total
flare$+$quiescent spectrum.  At the other extreme, for the largest possible flares,
there are possibly large errors (overestimations) for
the 5000 K flare, 10\% errors for the $10\,000$ K flare, and 3\%
errors for the $50\,000$ K flare.  However, none of the flares in our sample
are this large ($I_{f, U, max} + 1\sim$80 during IF3; and further it is unlikely that
76\% of the surface is flaring with a temperature of 5000 K).  This
experiment supports the use of a large slit width (with hopefully
good photometric conditions) when obtaining data on large flares.  We note that
the corrections for the data of the large flare on 24 Feb 2011 ($I_{f,
  u} + 1 $\s80 at peak) are very minimal ($< 2$\%), consistent with
robust flux measurements via the use of a wide slit and clear
conditions.   When the algorithm errs, it almost always overestimates the
scale factor.

Finally, we determine the flare amplitude at which the algorithm
begins to show large errors (rows 2, 5, and 8 of Table \ref{table:scaling}).  Only for the 5000 K
flare with an enhancement of $I_{f, u} \sim 40$ is there a significant
($>$5\% error);  for this amplitude flare, the
hotter temperatures (rows 5 and 8 of Table \ref{table:scaling}) give scalings $\lesssim$3\% different from the
predetermined values.

\begin{figure}
\begin{center}
\includegraphics[scale=0.4]{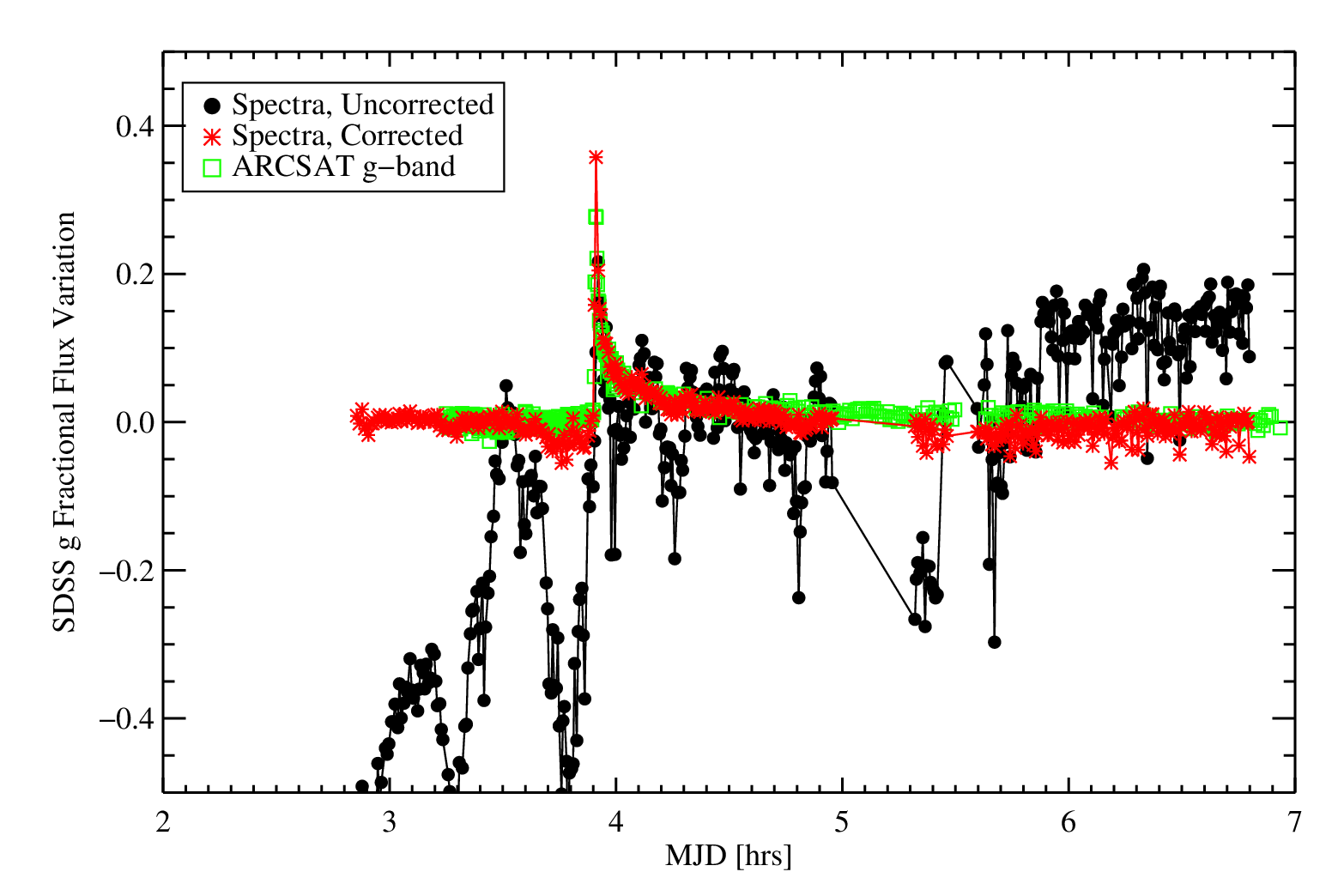}
\caption{Variations due to occasional cloud cover are apparent in the
  raw (synthetic $g$-band) fluxes from the spectra obtained on 2010
  April 03 (IF9) and shown as black circles.  Our simple algorithm predicts corrections
  that adjust for these
  variations, allowing the flare to be characterized at wavelengths
  redder than the $U$-band where the fractional variations are
  smaller. The red asterisks are the adjusted $g$-band fluxes, which
  are a good match to the $g$-band photometry from ARCSAT (green, binned to the
  approximate exposure times of the spectra). }
\label{fig:scale_power}
\end{center}
\end{figure}

In Figure \ref{fig:scale_power}, we show the result of scaling the
spectral fluxes for a night (2010 April 03) with occasional cloud cover
(and variable seeing).  This figure shows the
corrected and uncorrected fractional flux variations for synthetic $g$-band fluxes
obtained from the spectra, compared to simultaneous $g$-band
measurements using the ARCSAT 0.5-m SDSS $g$-band photometry.

\begin{deluxetable}{lccc}
\tabletypesize{\scriptsize}
\tablewidth{0pt}
\tablecaption{Appendix Table: Scaling Spectra}
\tablehead{
\colhead{\TBB\ [K]} &
\colhead{\XBB*} &
\colhead{$I_{f, U} + 1$} &
\colhead{Algorithm Prediction**}
}
\startdata
5000 & 0.015 & 2 & (0.70, 0.95, 1.05, 1.30) \\
5000 & 0.3 & 40 & (0.73, 0.99, 1.1, 1.35) \\
5000 & 0.76 & 100 & (0.82, 1.11, 1.23, 1.52) \\
$10\,000$ & 0.0003 & 2 & (0.70, 0.95, 1.05, 1.30) \\
$10\,000$& 0.0062 & 40 & (0.71, 0.97, 1.08, 1.33) \\
$10\,000$ & 0.016 & 100 & (0.75, 1.02, 1.14, 1.43) \\
$50\,000$ & 8$\times10^{-6}$ & 2 & (0.70, 0.95, 1.05, 1.30)\\
$50\,000$ & 0.00015 & 40 & (0.70, 0.96, 1.06, 1.32) \\
$50\,000$ & 0.0004 & 100 & (0.72, 0.93, 1.08, 1.33) \\
\enddata
\label{table:scaling}
\tablecomments{*The units of \XBB\ are
  fraction of the visible stellar hemisphere.  **Scaling algorithm's prediction for a set of scale
  factors corresponding to the following input, pre-determined scale
  factors: (0.70, 0.95, 1.05, 1.30). }
\end{deluxetable}

\clearpage

\section{The Photometric Flare Atlas and Integration Times of the Spectra} \label{sec:appendix_times}
\begin{figure}
\centering
\includegraphics[scale=0.4]{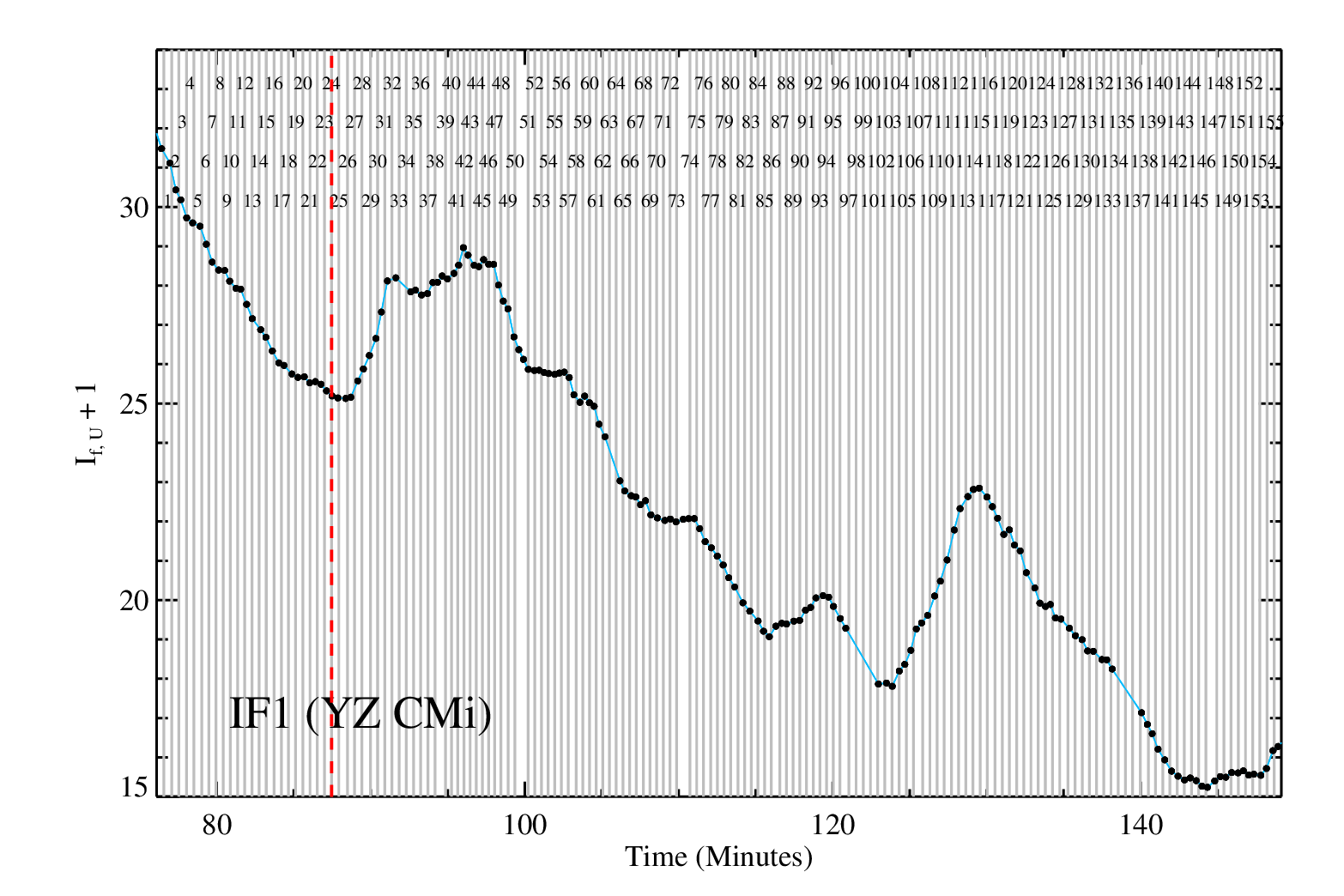}
\caption{The $U$-band photometry (blue line, black
  circles) for IF1 and the spectral integration times
  given as shaded bars;  the S\#'s are indicated on each grey bar. The
  gradual phase spectra shown in Figures
  \ref{fig:grad1} \,--\, \ref{fig:grad5} are the result of averaging
  over the three
  spectra nearest to the vertical red line. }
\label{fig:appendix_integ1}
\end{figure}

In this Appendix, we show the photometry light curves and the
integration windows of the simultaneous spectra (Figures
\ref{fig:appendix_integ1}\,--\,\ref{fig:appendix_integ17}).  
The grey shaded vertical bars show
the integration time windows, and the S\# is given on each 
grey bar.   Note, these spectra correspond to the $n$B subcategory of
spectra (see Section \ref{sec:spectra}).  The vertical red dashed line indicates the time at which
the gradual phase emission is analyzed (Table \ref{table:times}).  The
three spectra nearest to this time are averaged for each flare to
obtain the gradual phase spectra shown in Figures \ref{fig:grad1}
\,--\, \ref{fig:grad5} for the flare sample.  In all figures, times
are given as elapsed number of minutes on the respective MJD from
Table \ref{table:obslogAll};  for IF1 and MDSF2, the times are minutes
elapsed from flare start (see note on times in Section \ref{sec:spectra_param}).

\clearpage

\begin{figure}
\centering
\includegraphics[scale=0.4]{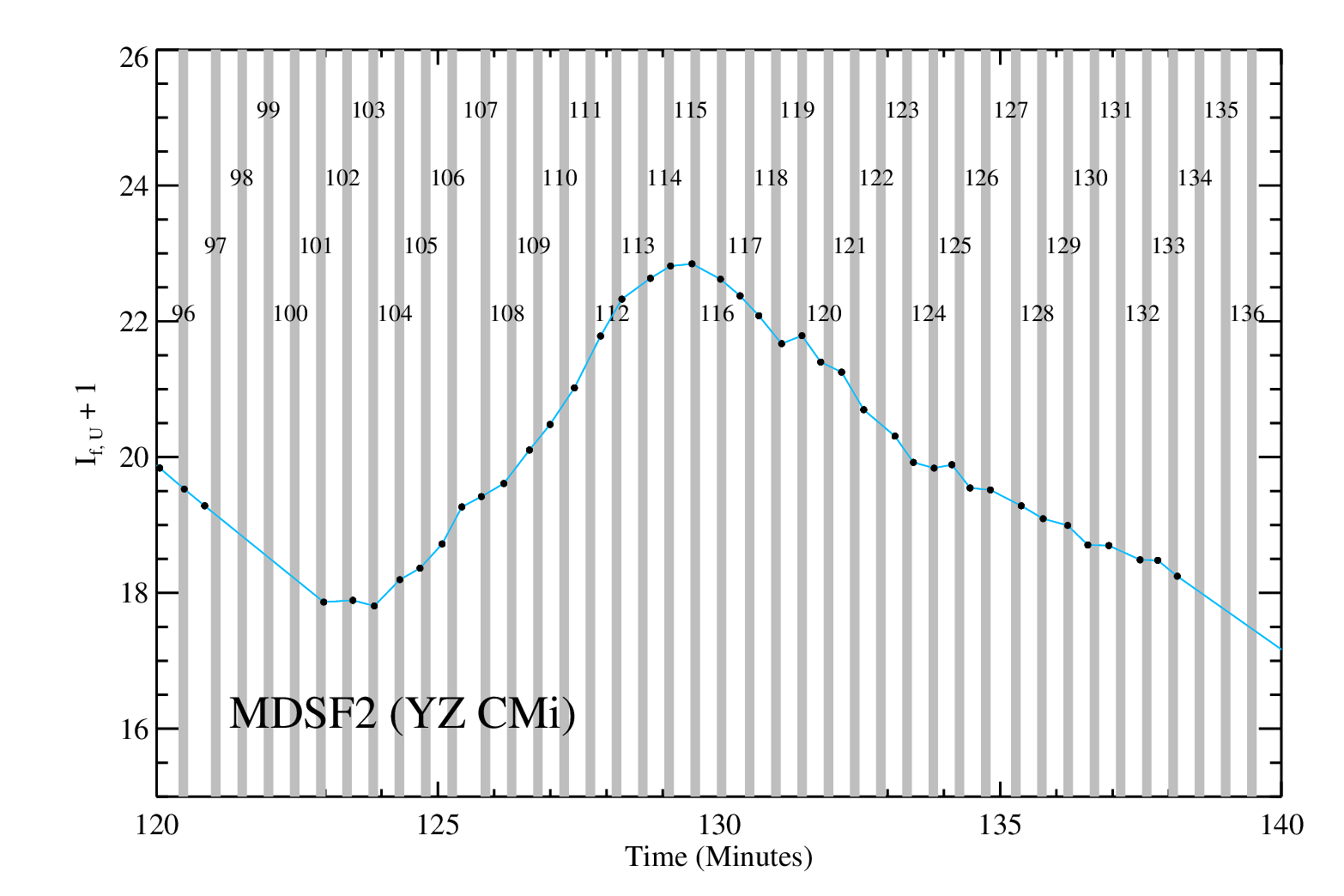}
\caption{Same as Figure \ref{fig:appendix_integ1} but for MDSF2.   }
\label{fig:appendix_integ9b}
\end{figure}

\begin{figure}
\centering
\includegraphics[scale=0.4]{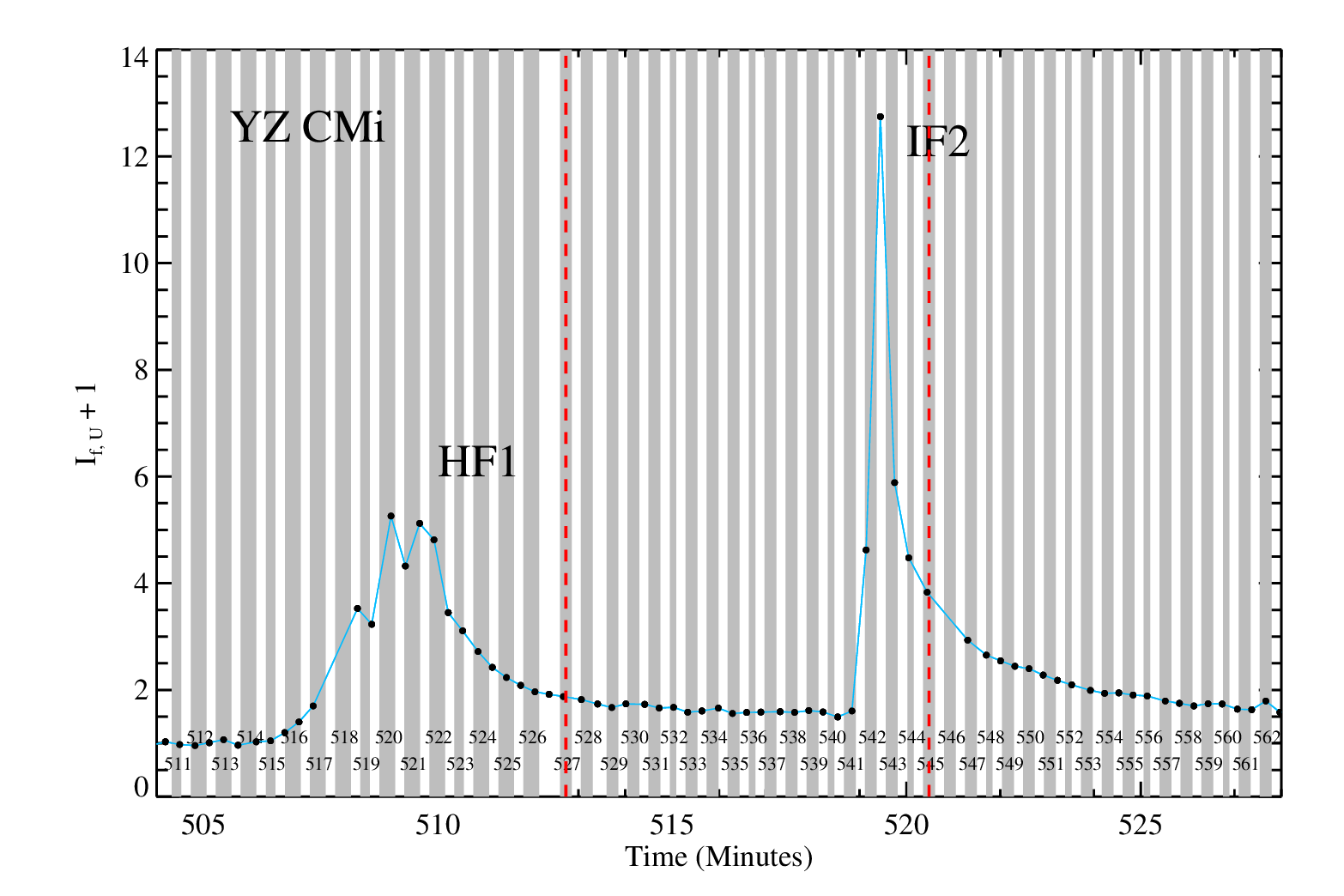}
\caption{Same as Figure \ref{fig:appendix_integ1} but for IF2 and HF4.   }
\label{fig:appendix_integ34}
\end{figure}

\begin{figure}
\centering
\includegraphics[scale=0.4]{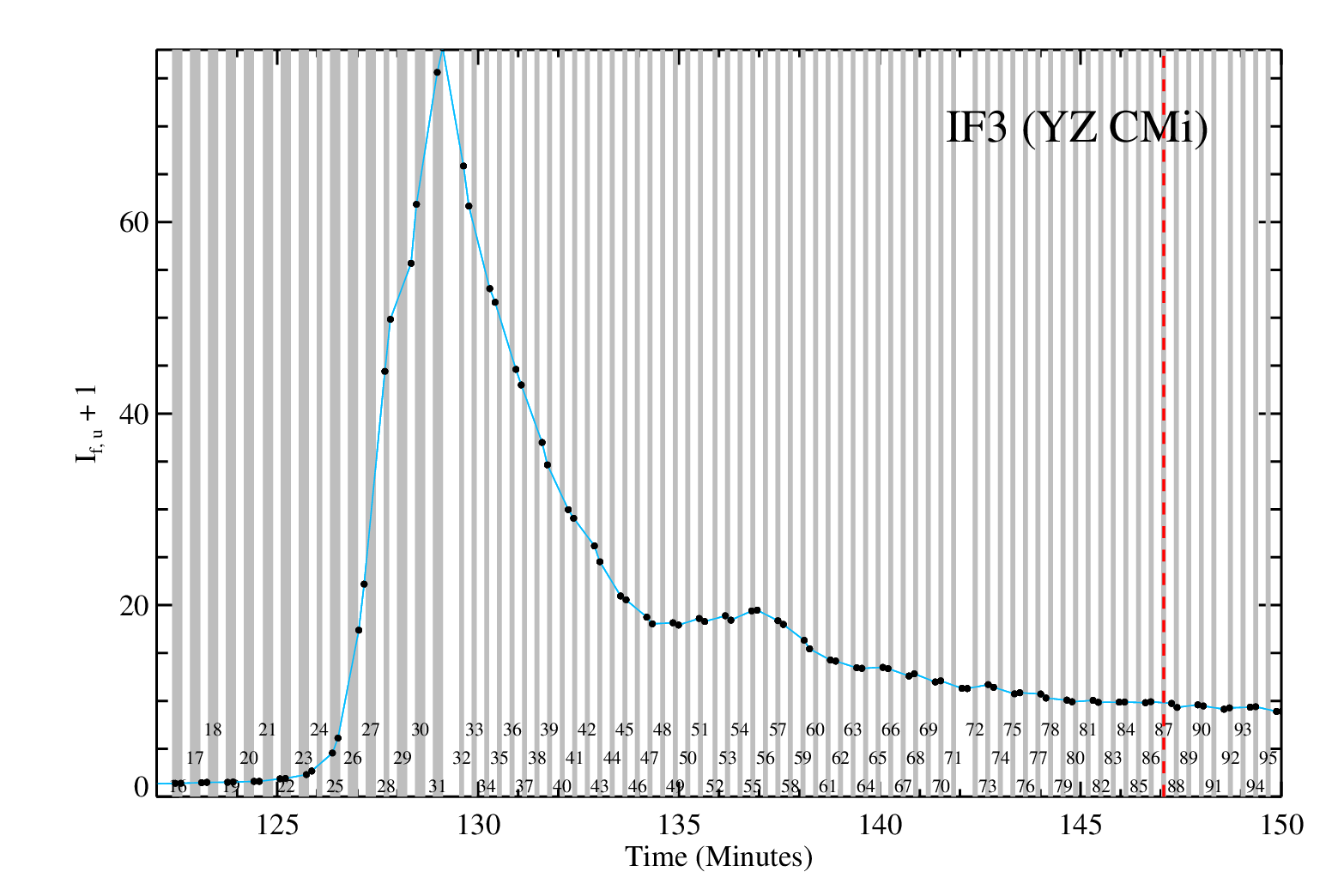}
\caption{Same as Figure \ref{fig:appendix_integ1} but for IF3. }
\label{fig:appendix_integ9}
\end{figure}

\begin{figure}
\centering
\includegraphics[scale=0.4]{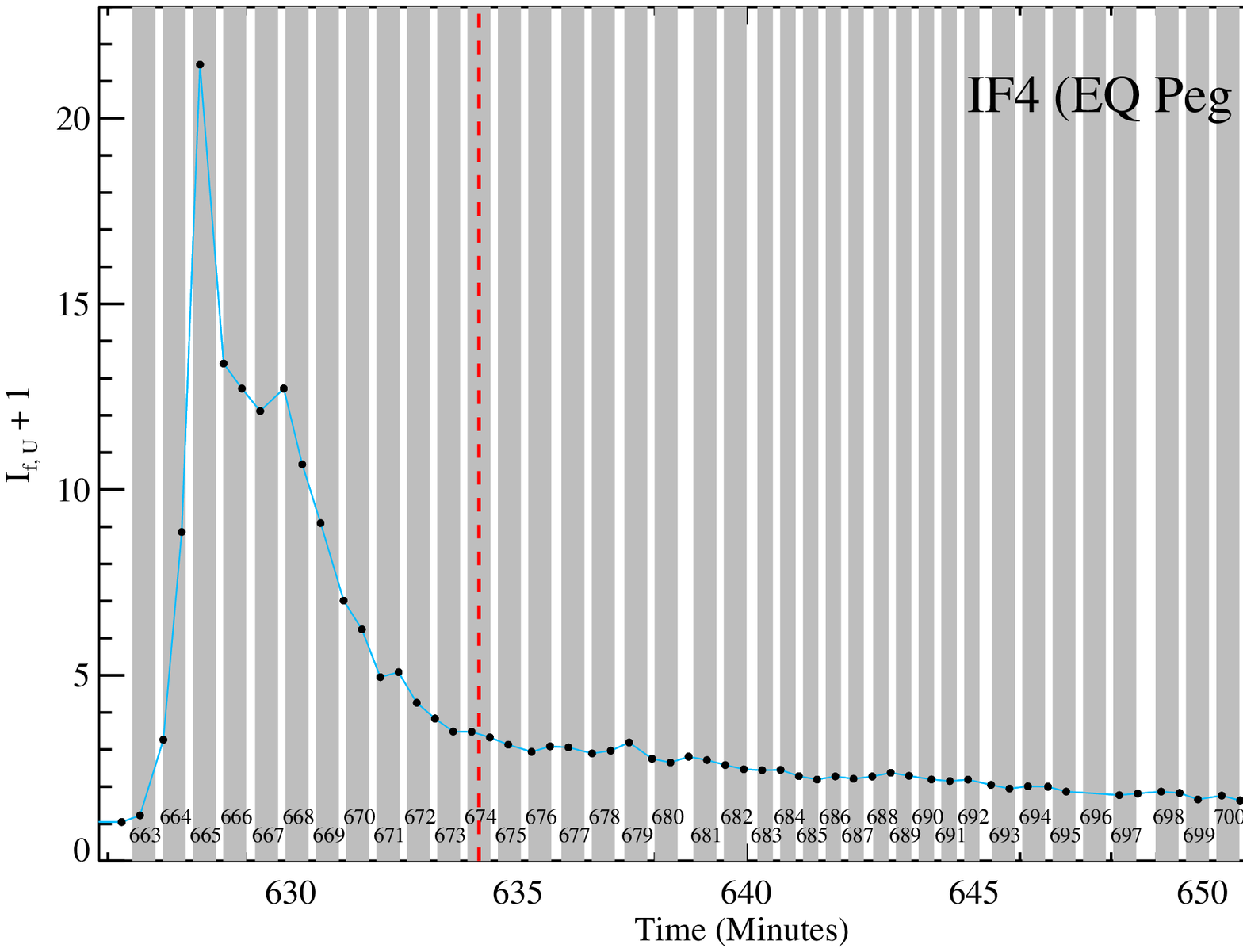}
\caption{Same as Figure \ref{fig:appendix_integ1} but for IF4.   }
\label{fig:appendix_integ12}
\end{figure}

\begin{figure}
\centering
\includegraphics[scale=0.4]{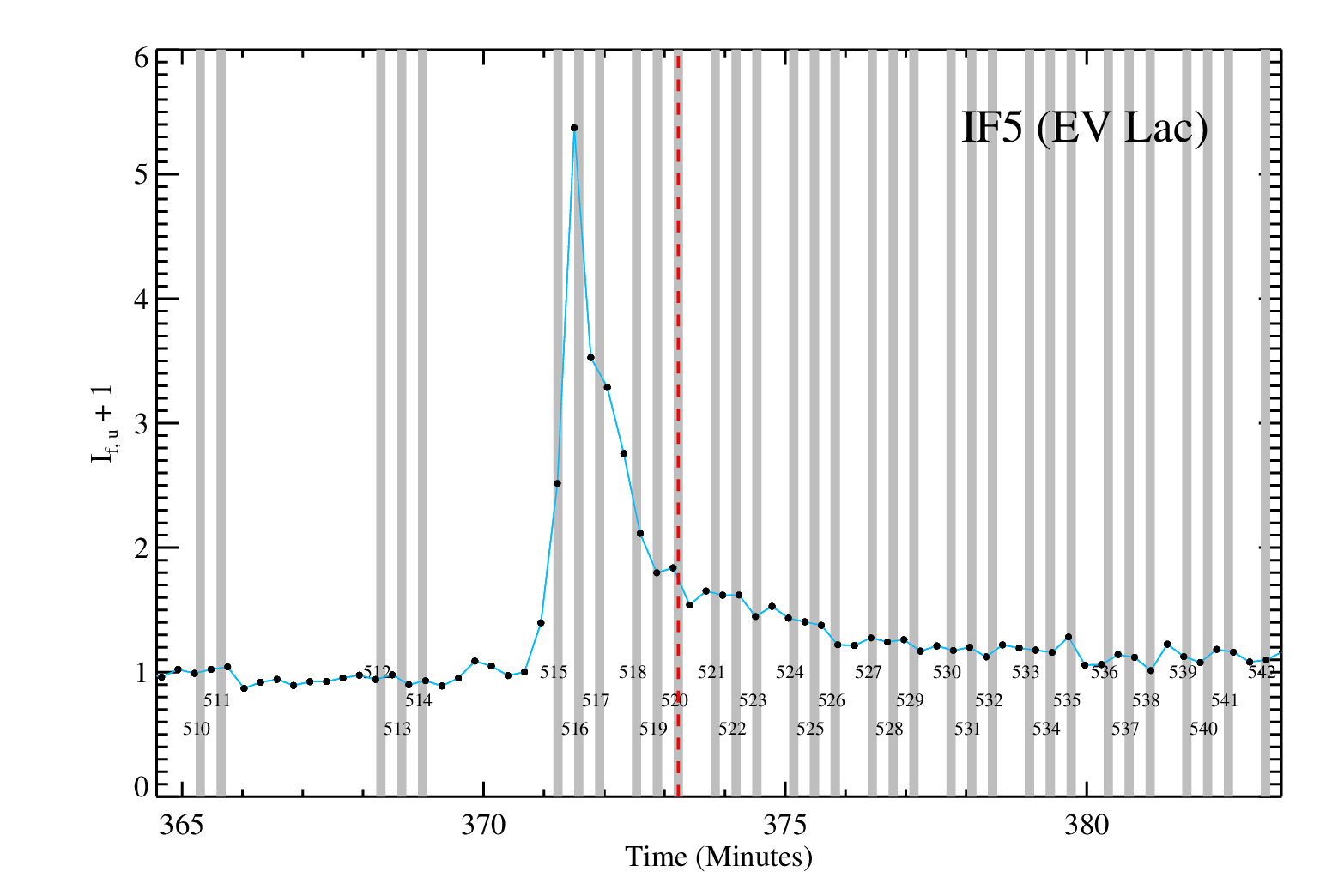}
\caption{Same as Figure \ref{fig:appendix_integ1} but for IF5.   }
\label{fig:appendix_integ20}
\end{figure}

\begin{figure}
\centering
\includegraphics[scale=0.4]{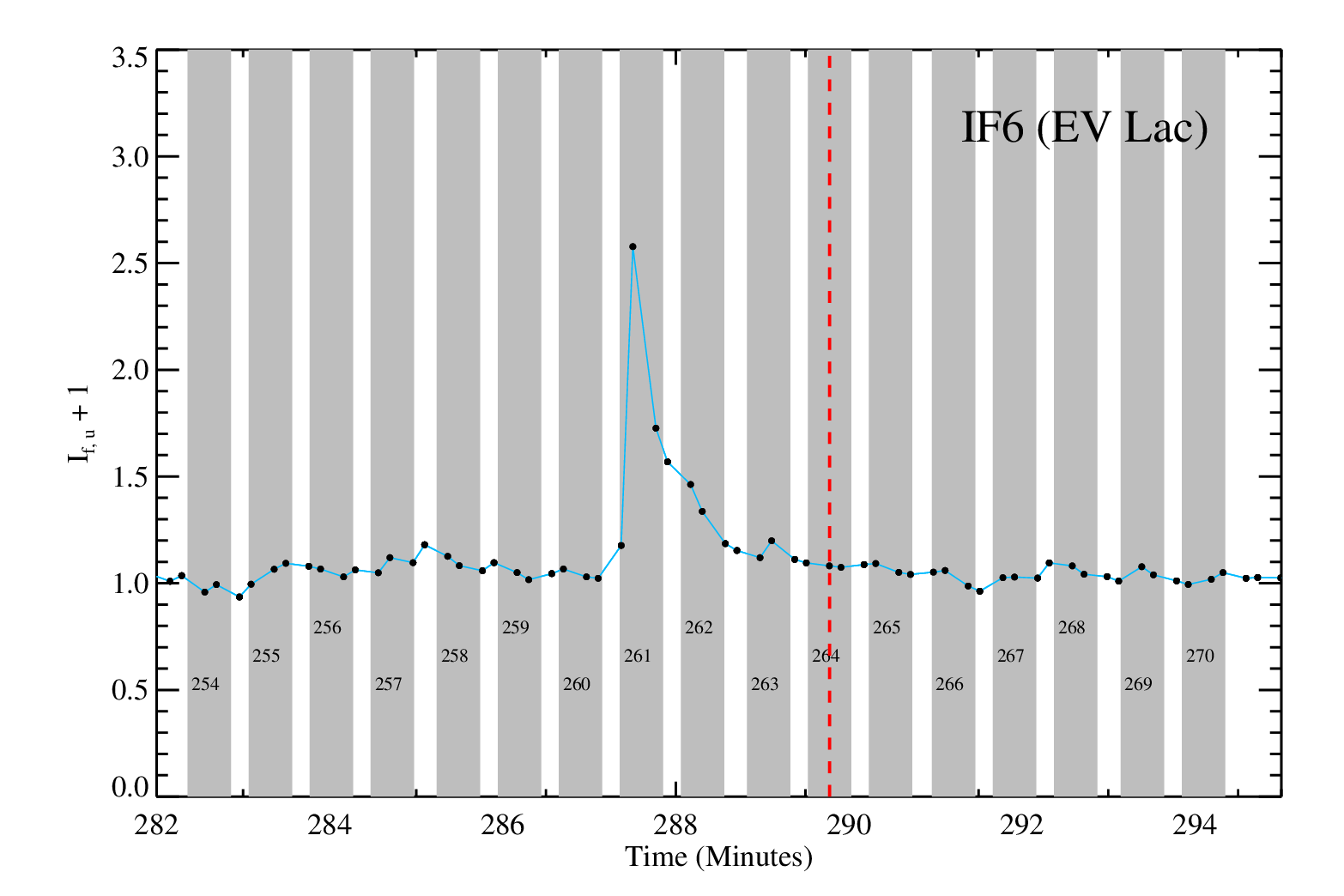}
\caption{Same as Figure \ref{fig:appendix_integ1} but for IF6.   }
\label{fig:appendix_integ6}
\end{figure}

\begin{figure}
\centering
\includegraphics[scale=0.4]{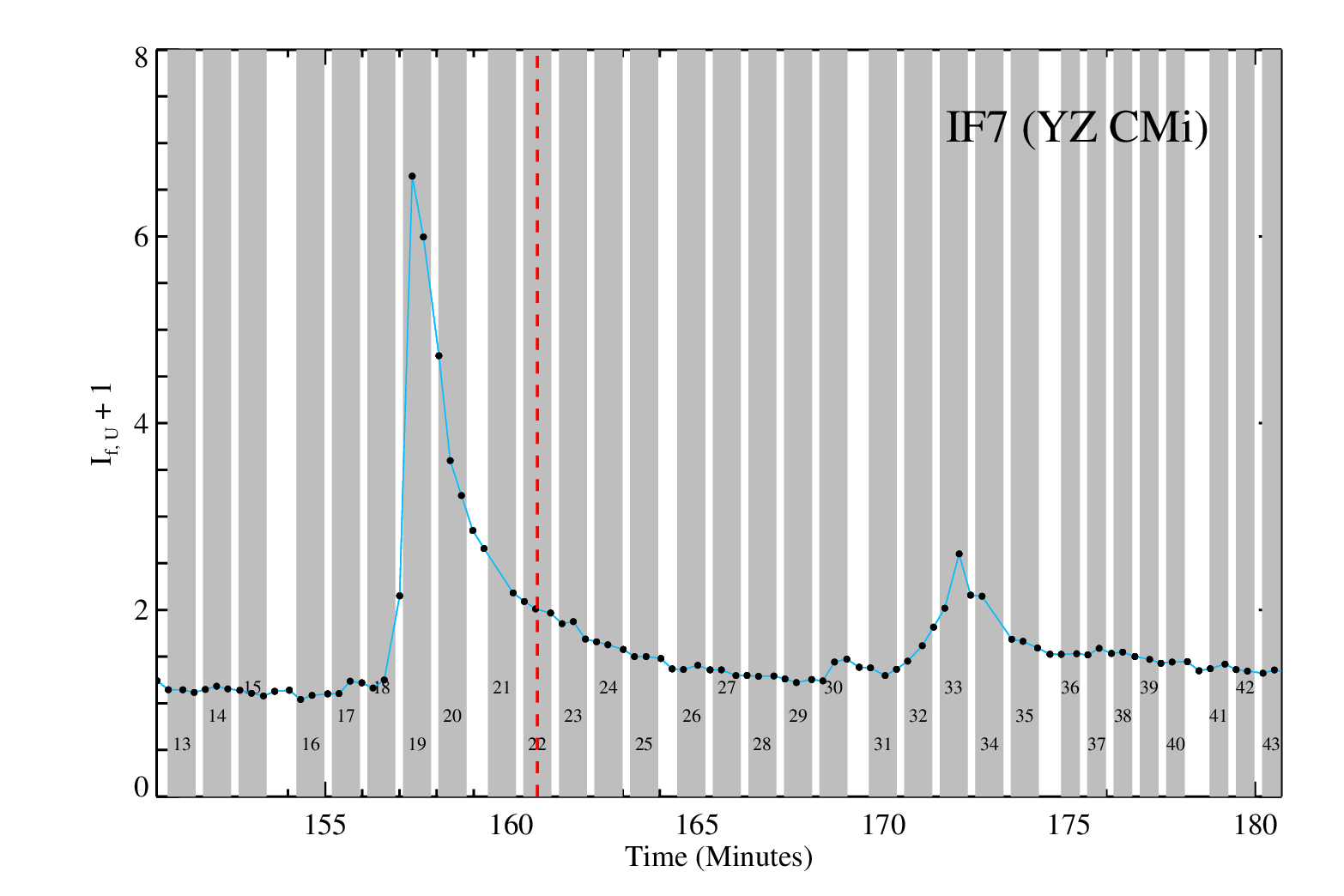}
\caption{Same as Figure \ref{fig:appendix_integ1} but for IF7.   }
\label{fig:appendix_integ8}
\end{figure}

\begin{figure}
\centering
\includegraphics[scale=0.4]{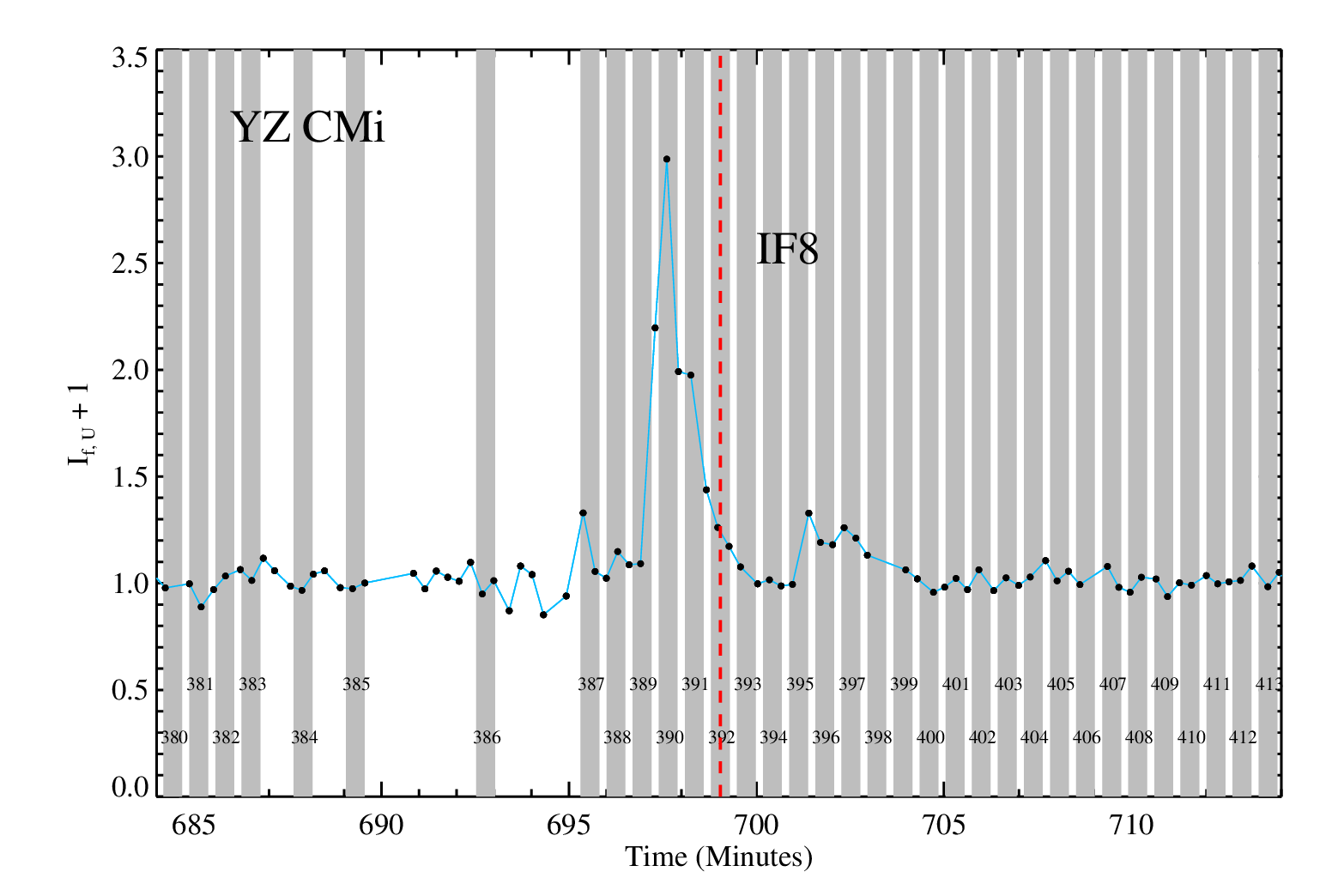}
\caption{Same as Figure \ref{fig:appendix_integ1} but for IF8.   }
\label{fig:appendix_integ15}
\end{figure}

\begin{figure}
\centering
\includegraphics[scale=0.4]{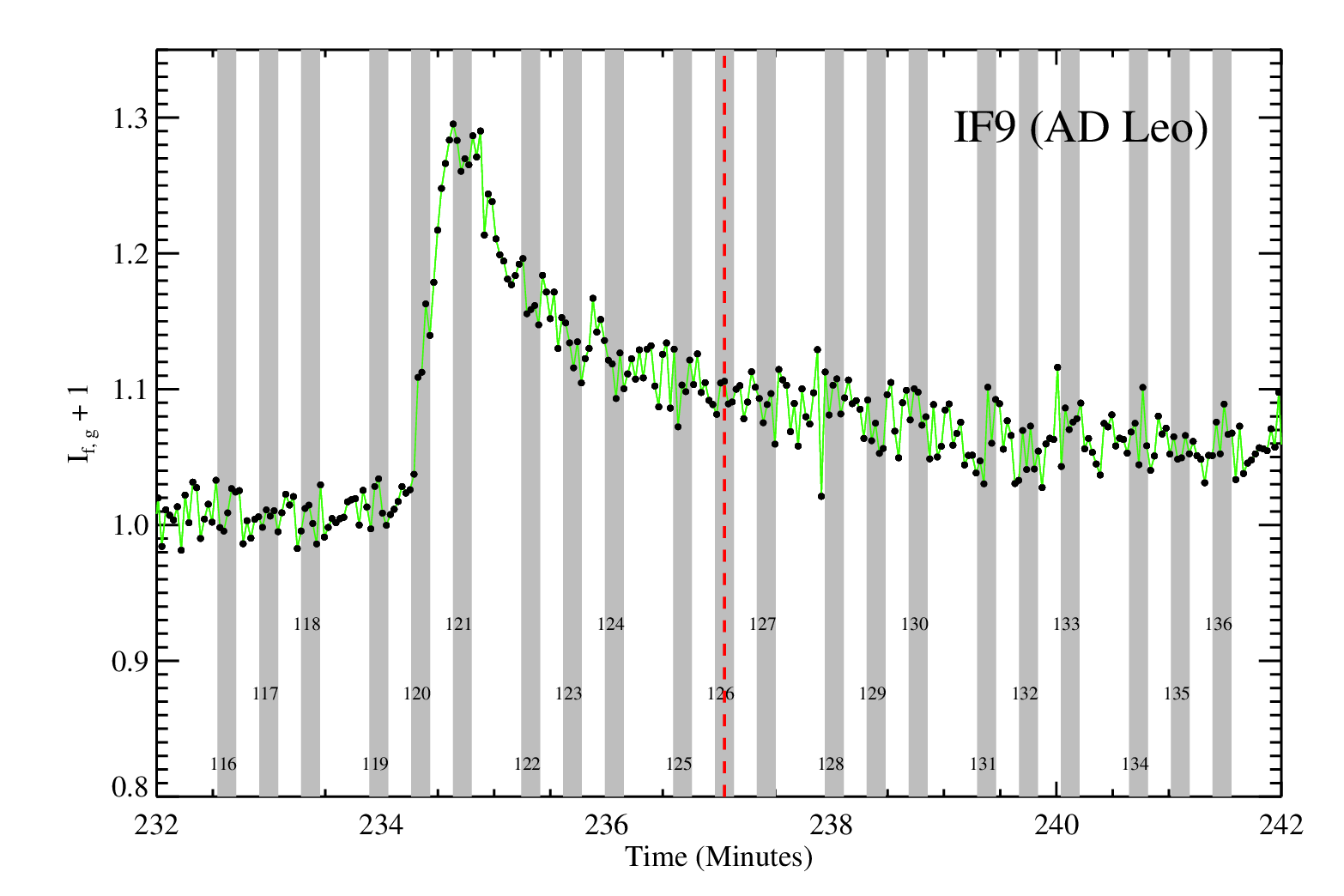}
\caption{Same as Figure \ref{fig:appendix_integ1} but for IF9.   }
\label{fig:appendix_integ10}
\end{figure}

\begin{figure}
\centering
\includegraphics[scale=0.4]{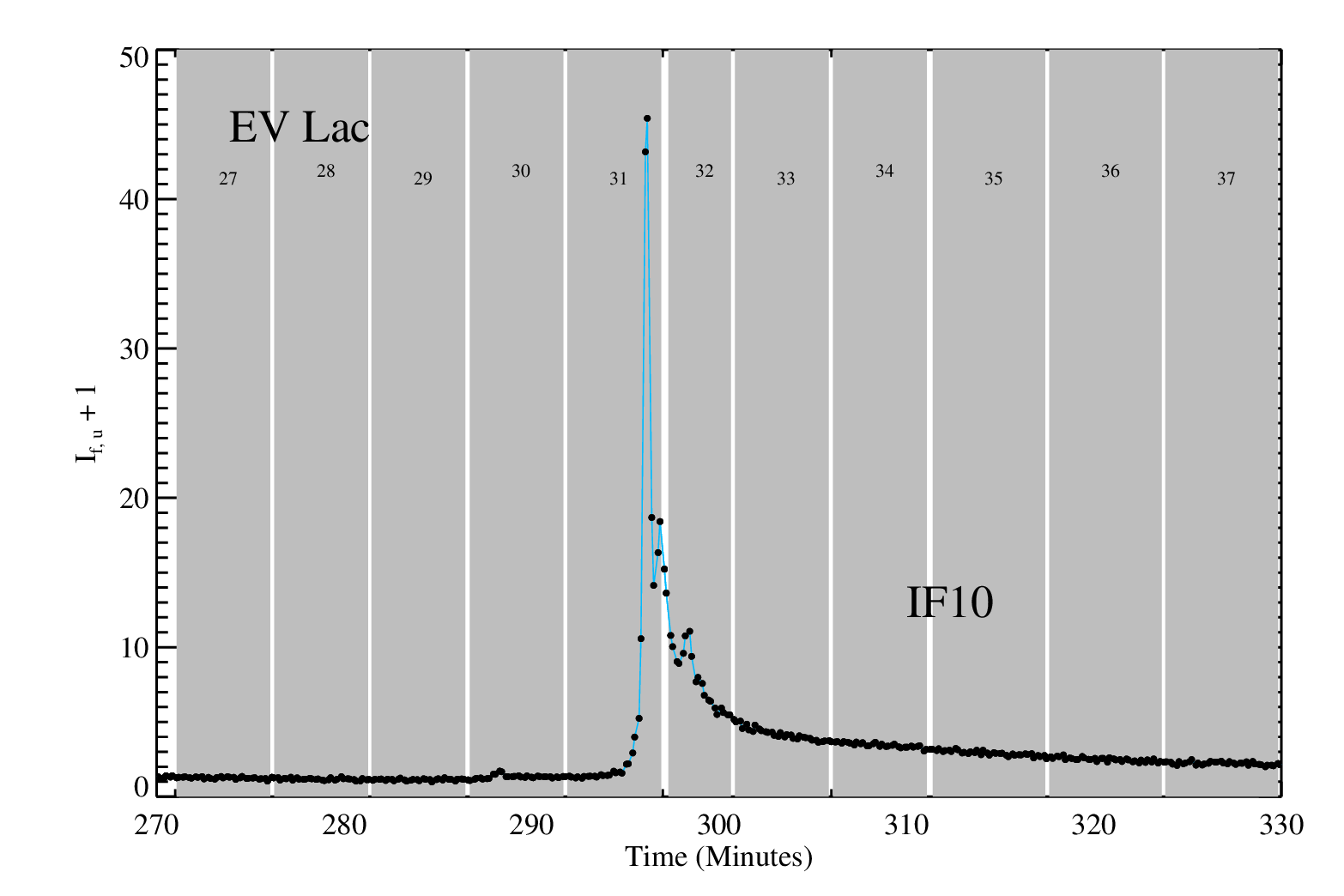}
\caption{Same as Figure \ref{fig:appendix_integ1} but for IF10.   }
\label{fig:appendix_integ21}
\end{figure}

\begin{figure}
\centering
\includegraphics[scale=0.4]{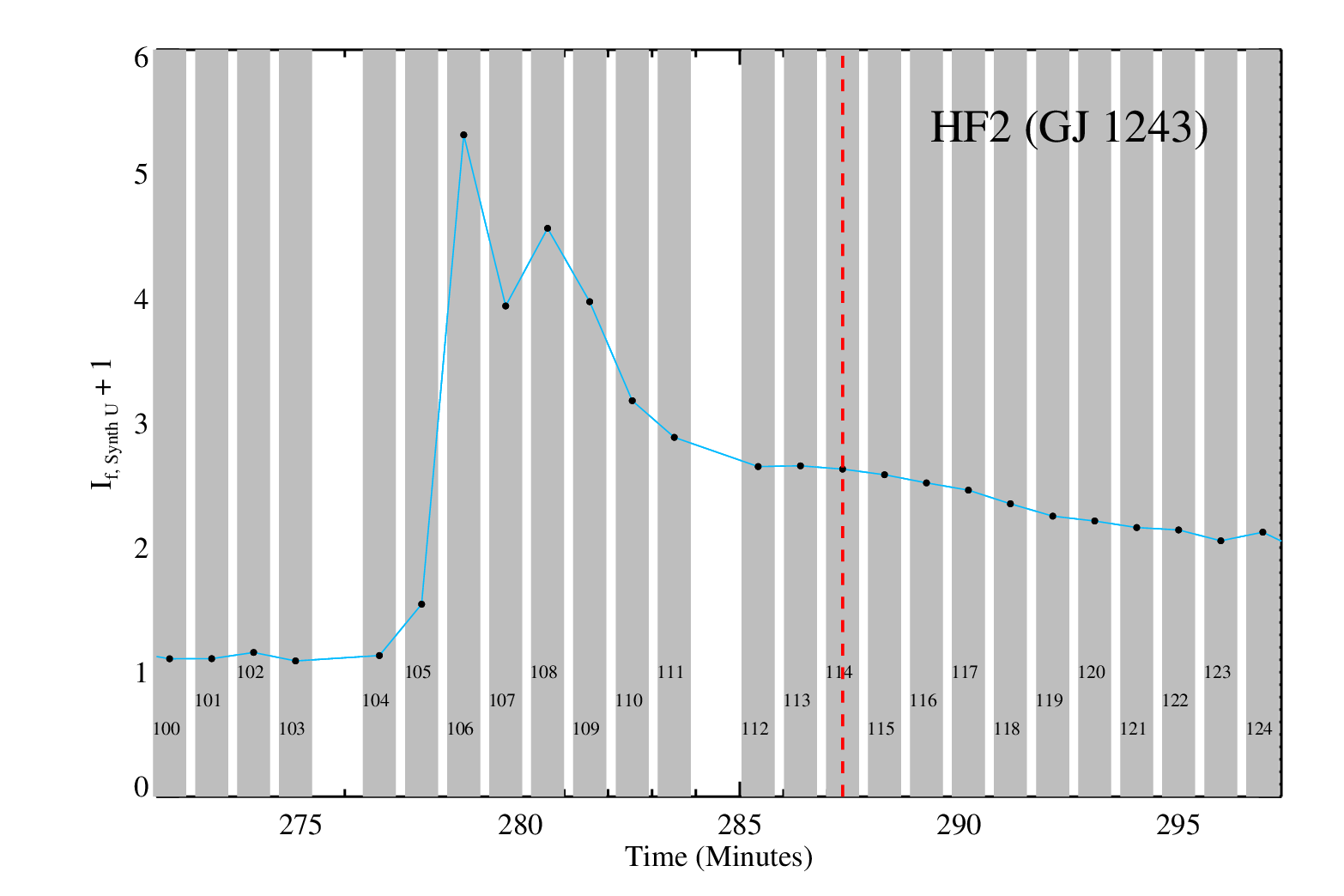}
\caption{Same as Figure \ref{fig:appendix_integ1} but for HF2.   }
\label{fig:appendix_integ13}
\end{figure}

\begin{figure}
\centering
\includegraphics[scale=0.4]{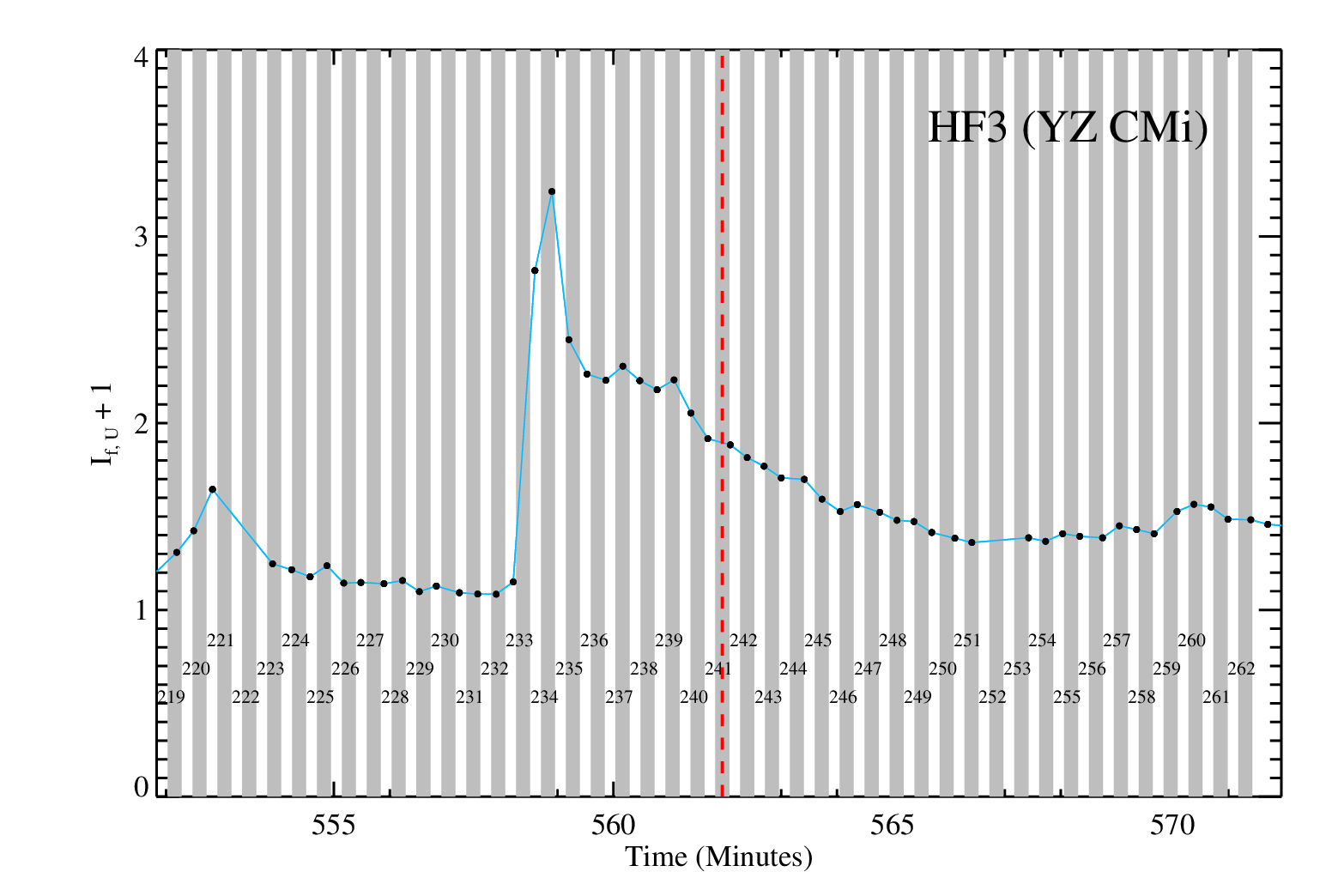}
\caption{Same as Figure \ref{fig:appendix_integ1} but for HF3.   }
\label{fig:appendix_integ14}
\end{figure}

\begin{figure}
\centering
\includegraphics[scale=0.4]{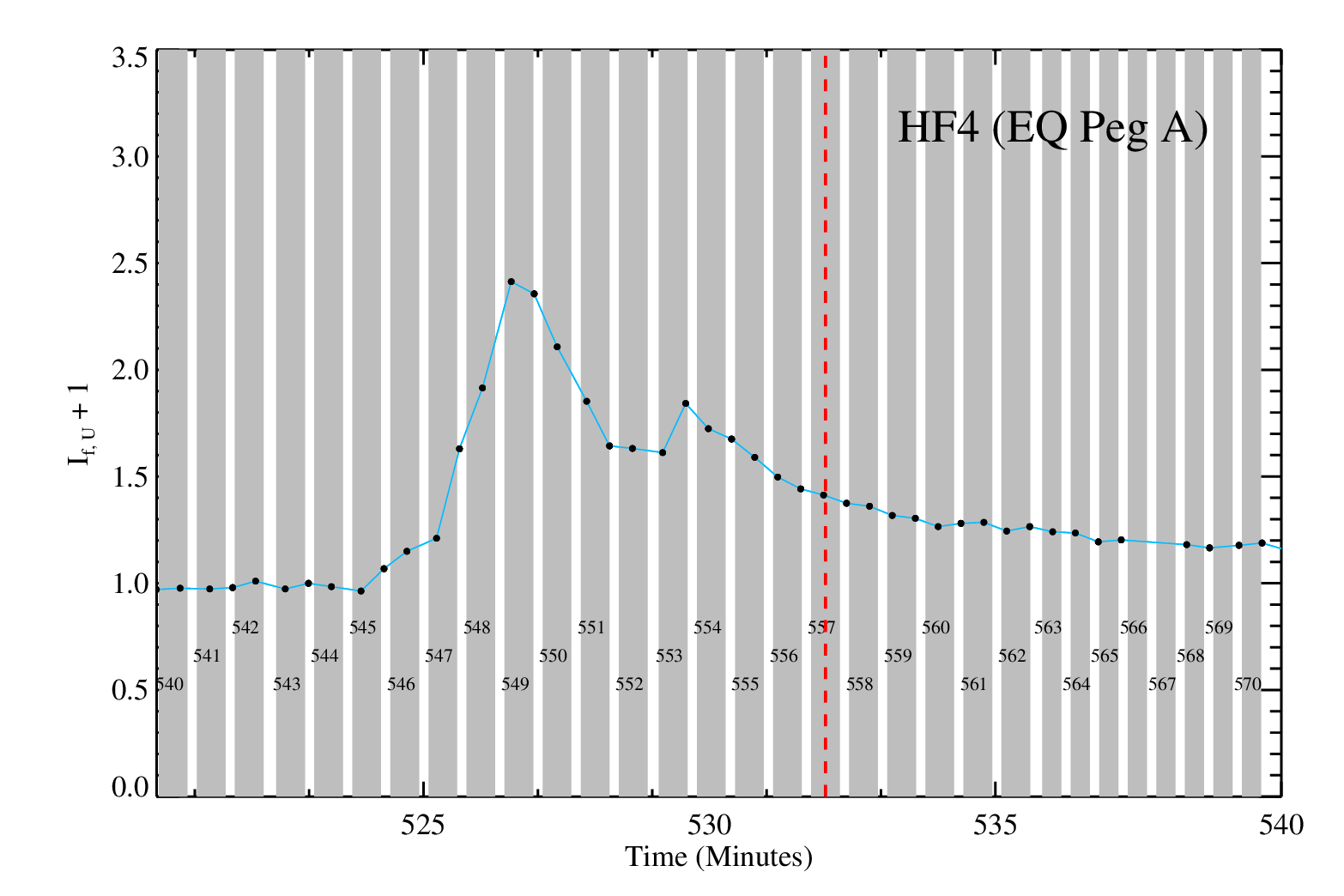}
\caption{Same as Figure \ref{fig:appendix_integ1} but for HF4.   }
\label{fig:appendix_integ11}
\end{figure}

\begin{figure}
\centering
\includegraphics[scale=0.4]{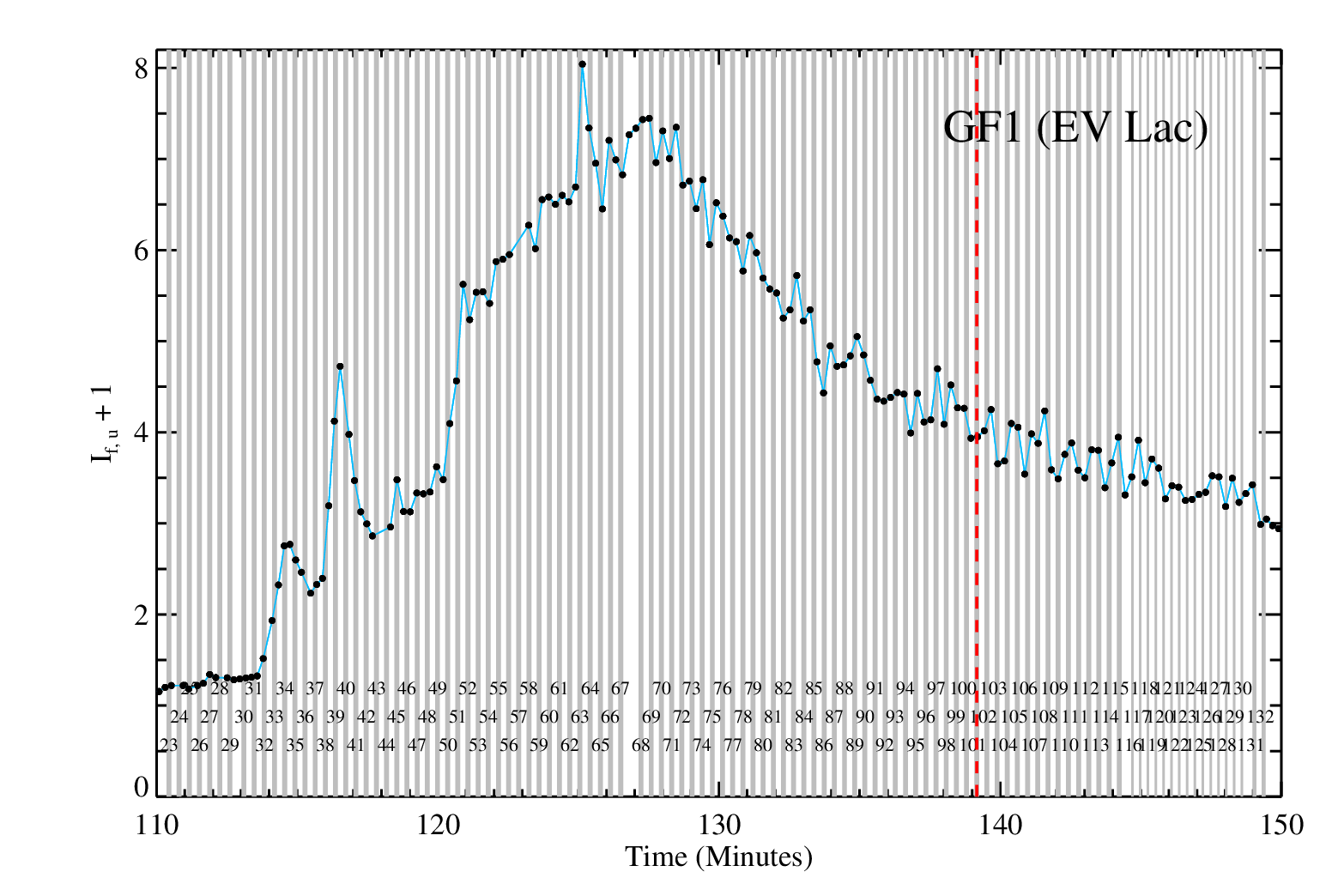}
\caption{Same as Figure \ref{fig:appendix_integ1} but for GF1.   }
\label{fig:appendix_integ2}
\end{figure}

\begin{figure}
\centering
\includegraphics[scale=0.4]{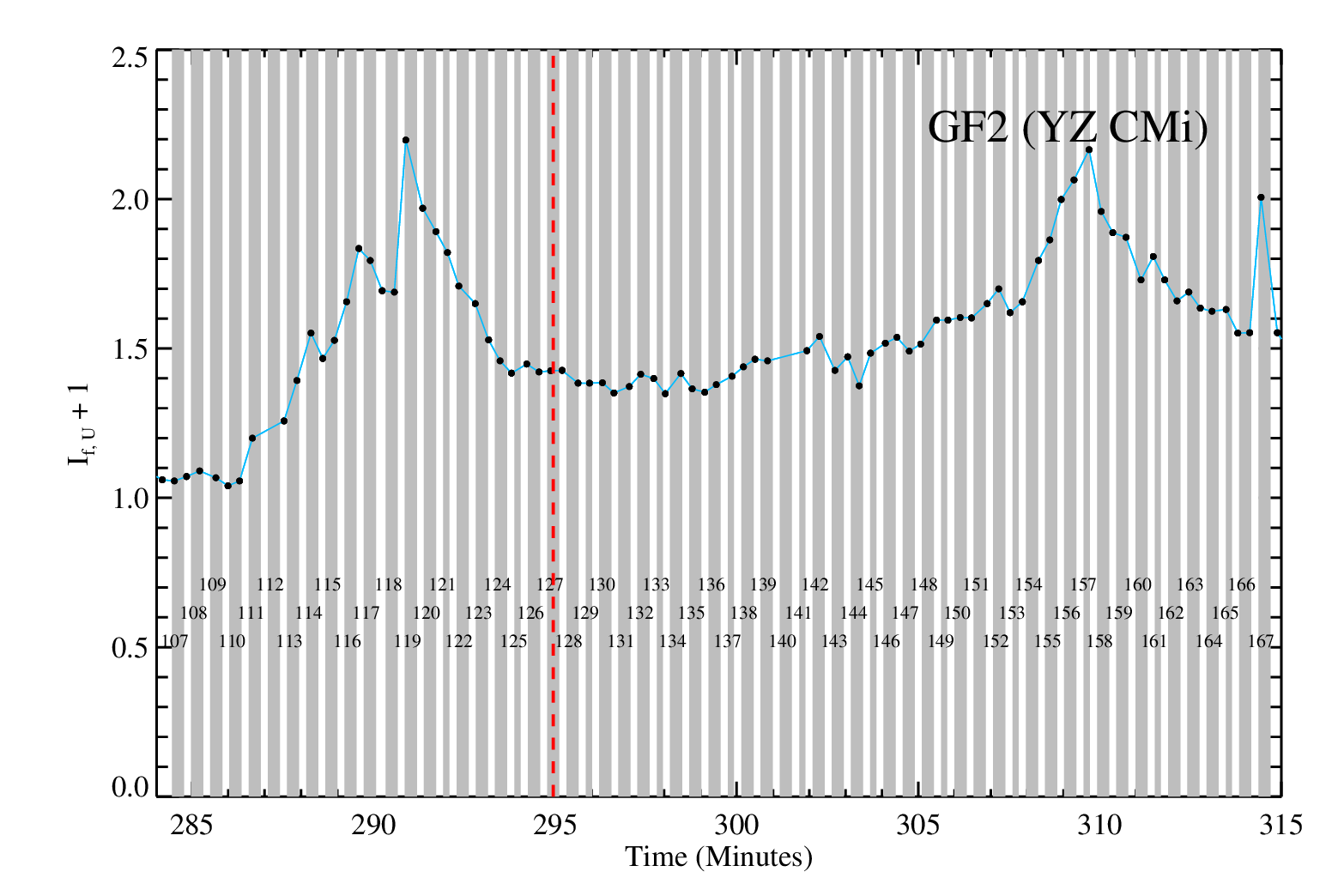}
\caption{Same as Figure \ref{fig:appendix_integ1} but for GF2.   }
\label{fig:appendix_integ19}
\end{figure}

\begin{figure}
\centering
\includegraphics[scale=0.4]{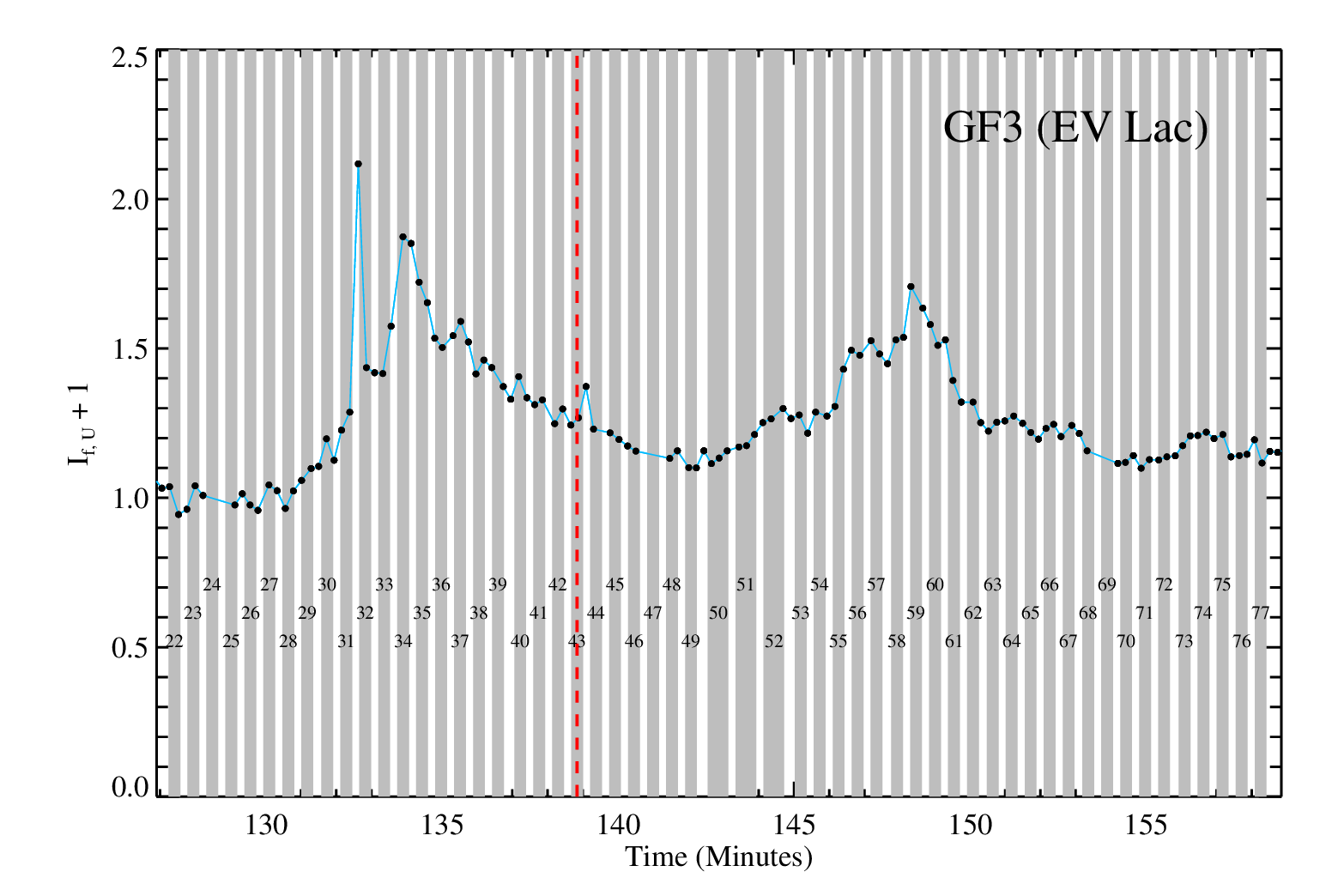}
\caption{Same as Figure \ref{fig:appendix_integ1} but for GF3.   }
\label{fig:appendix_integ7}
\end{figure}

\begin{figure}
\centering
\includegraphics[scale=0.4]{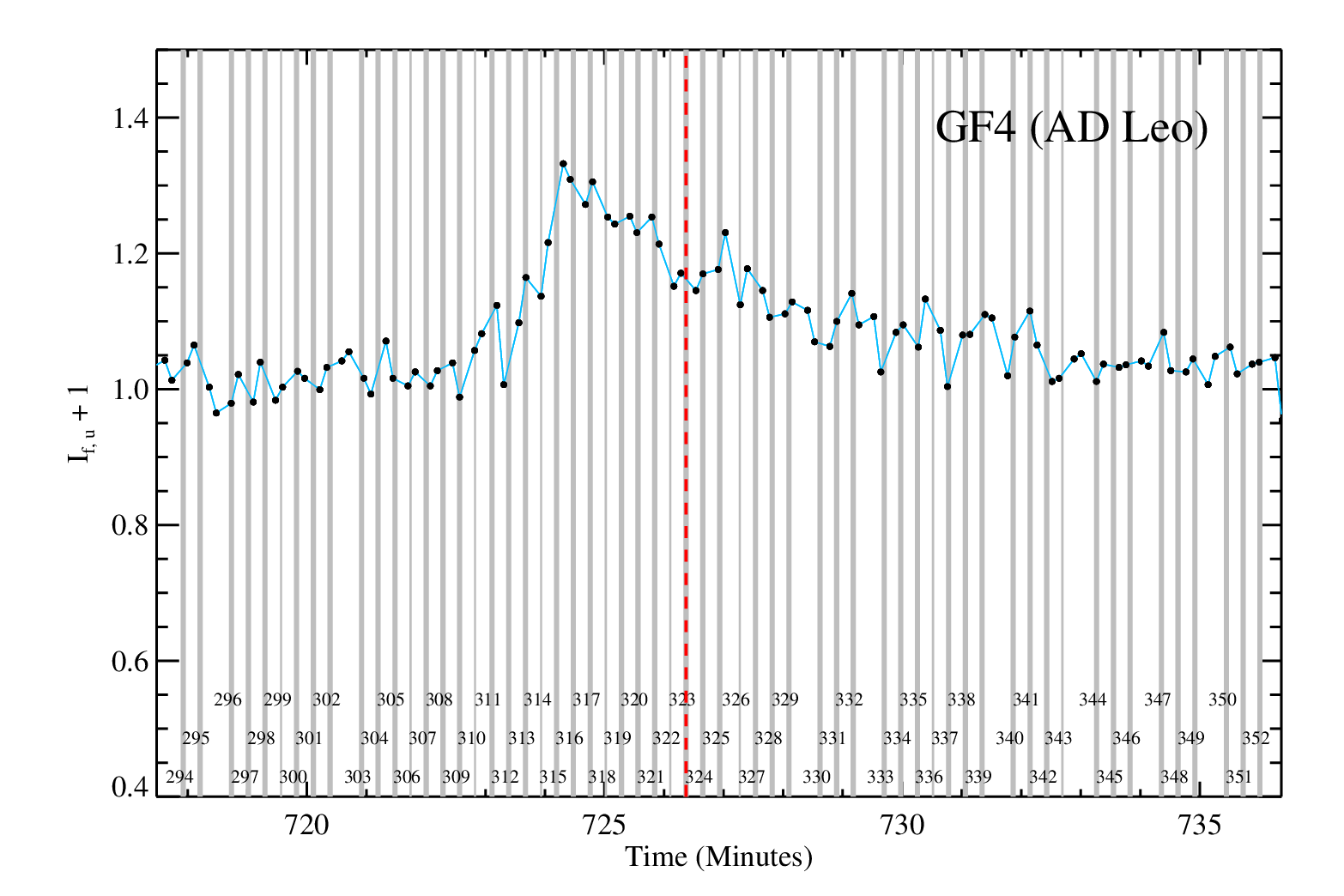}
\caption{Same as Figure \ref{fig:appendix_integ1} but for GF4. }
\label{fig:appendix_integ18}
\end{figure}

\clearpage
\begin{figure}
\centering
\includegraphics[scale=0.4]{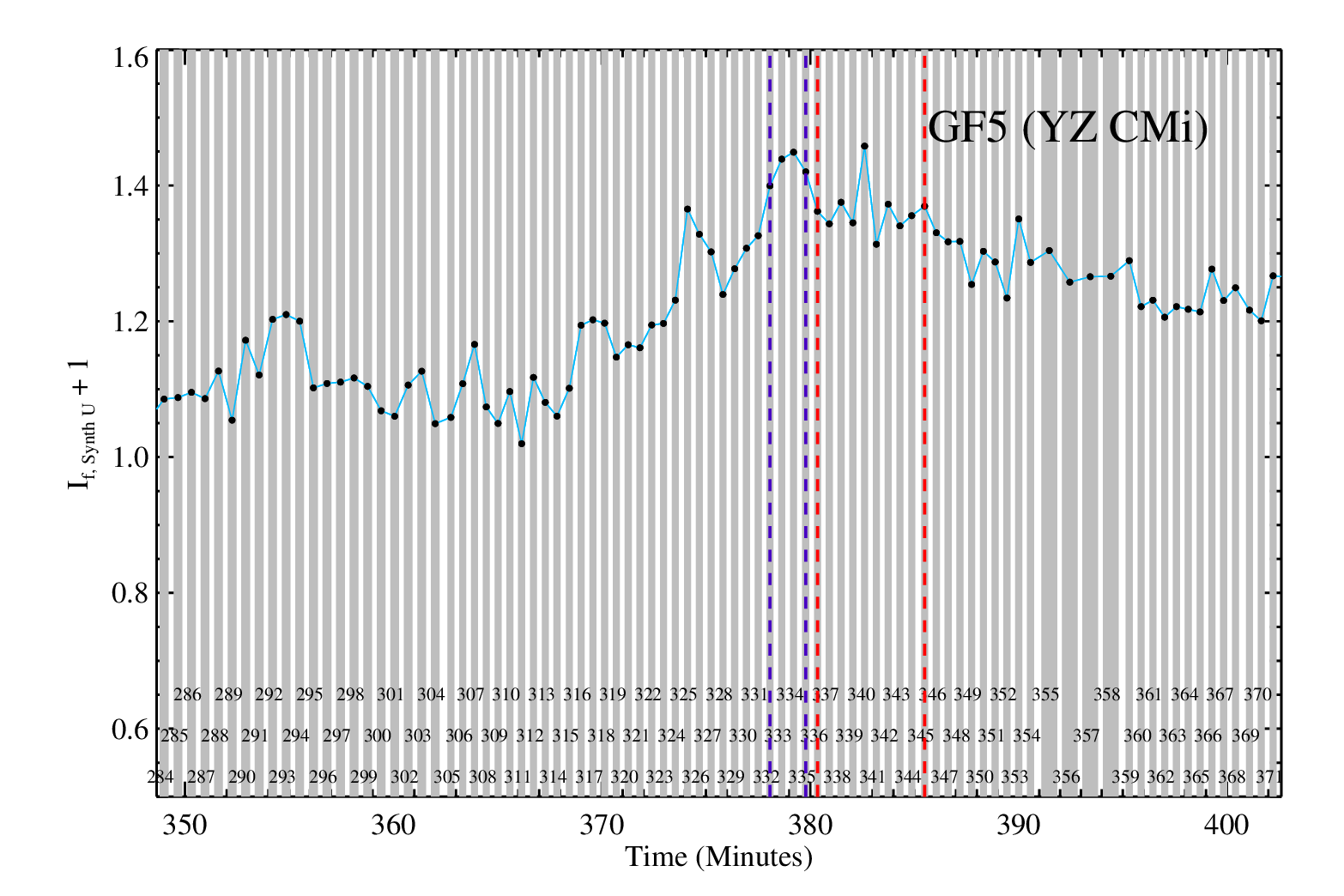}
\caption{Same as Figure \ref{fig:appendix_integ1} but for GF5.   The red lines indicate the boundaries of
  ten spectra used for the gradual decay phase spectrum. The purple vertical lines indicate the four spectra averaged for the peak spectrum. }
\label{fig:appendix_integ17}
\end{figure}

\clearpage

\section{The Spectral Flare Atlas} \label{sec:appendix_atlas}
Figures 62.1\,--\,62.21 show time sequences of the flare-only emission spectra
for each flare in the Flare Atlas (Figure \ref{fig:allspec_IF3} shows a representative
sequence of flare spectra during IF3; 
Figures 62.1\,--\,62.21 are available in the online version of the
Journal).  The top panels show the rise and peak phases, and the
bottom
panels show the fast and gradual decay phases.
 The color of each spectrum indicates the relative timing in the
 sequence (top panel: the order is black to blue to green to yellow to
 red; bottom panel: the order is red to yellow to green to blue to black;
 the number of colors depends on the number of spectra in the sequence), and the total duration of each time-sequence
 is roughly the same as for the 
  light curves shown in Figures
  \ref{fig:lcphot_panel_if}\,--\,\ref{fig:lcphot_panel_gf} (and
  Figures \ref{fig:appendix_integ1}\,--\,\ref{fig:appendix_integ9b}
  for IF1 and MDSF2, respectively).  The
  approximate time duration and range of S\#'s (see Appendix
  \ref{sec:appendix_times}) are indicated in the top
  right of each
  panel. The grey
  shaded area indicates the wavelength range most affected by the DIS
  dichroic.  Note, these spectra correspond to the $n$B subcategory (see
Section \ref{sec:spectra}) unless
otherwise specified in the figure captions.  The exposure times and
dates of the observations are
given in Table \ref{table:obslogAll}.
 The data (original unscaled flux and scaled flare-only flux) are
 available online through the 
VizieR service.

\figsetstart
\figsetnum{62}
\figsettitle{The Spectral Flare Atlas}

\figsetgrpstart
\figsetgrpnum{62.1}
\figsetgrptitle{Flare Spectra of IF3}
\figsetplot{f62_1.eps}
\figsetgrpnote{A time-sequence of flare spectra (quiescent level subtracted)
  during the rise and peak phases (top
  panel, ordered by time from black to red) and fast decay and gradual
  decay phases (bottom panel, ordered by time from red to black) of
  IF3.  For a complete description of the figure, see Appendix \ref{sec:appendix_atlas}.  The emission
  line evolution is shown in Figure \ref{fig:lines_IF3_B}.  A
  time sequence for each flare (Figures 62.1\,--\,62.21) is available in the online version of the Journal. }
\figsetgrpend

\figsetgrpstart
\figsetgrpnum{62.2}
\figsetgrptitle{Flare Spectra of IF1}
\figsetplot{f62_2.eps}
\figsetgrpnote{Same as Figure \ref{fig:allspec_IF3} but for IF1 for the times
  $t=1.27-2.05$ hrs and $t=2.35-2.49$ hrs (all spectra except MDSF2; see Figure \ref{fig:allspec_MDSF2} for the spectral sequence for
  MDSF2 from $t=2.05-2.35$ hrs).  Only one panel is shown
  for this flare because we obtained spectra only during the decay
  phase.  The \Ha\ line ($\lambda = 6538-6584$\AA) is saturated and is
not shown.  }
\figsetgrpend

\figsetgrpstart
\figsetgrpnum{62.3}
\figsetgrptitle{Flare Spectra of MDSF2}
\figsetplot{f62_3.eps}
\figsetgrpnote{Same as Figure \ref{fig:allspec_IF3} but for MDSF2.
  The \Ha\ line ($\lambda = 6538-6584$\AA) is saturated and is not shown. }
\label{fig:allspec_MDSF2}
\figsetgrpend

\figsetend

\begin{figure}
\centering
\caption{(See published article for figure) A time-sequence of flare spectra (quiescent level subtracted)
  during the rise and peak phases (top
  panel, ordered by time from black to red) and fast decay and gradual
  decay phases (bottom panel, ordered by time from red to black) of
  IF3.  For a complete description of the figure, see Appendix \ref{sec:appendix_atlas}.  The emission
  line evolution is shown in Figure \ref{fig:lines_IF3_B}.  A
  time sequence for each flare (Figures 62.1\,--\,62.21) is available in the online version of the Journal. }
\label{fig:allspec_IF3}
\end{figure}

\clearpage

\section{Subtracted Flare Spectra} \label{sec:appendix_astar}
In Section \ref{sec:astar}, we found that newly formed flare emission
(for MDSF2, the flux in spectrum S\#113 with the background gradual
phase emission, S\#101\,--\,103, subtracted) resembles the spectrum of a hot star, like Vega.  It is a
concern, however, whether the background gradual phase emission continues to
evolve (e.g., decrease) during the rise phase of the secondary flare, thereby
resulting in an \emph{oversubtraction} of the background flux.  In this
section, we show how our results change if we estimate the amount by
which the previously decaying flare emission decreases over the course
of the rise phase of MDSF2.

\begin{figure}
\begin{center}
\includegraphics[scale=0.55]{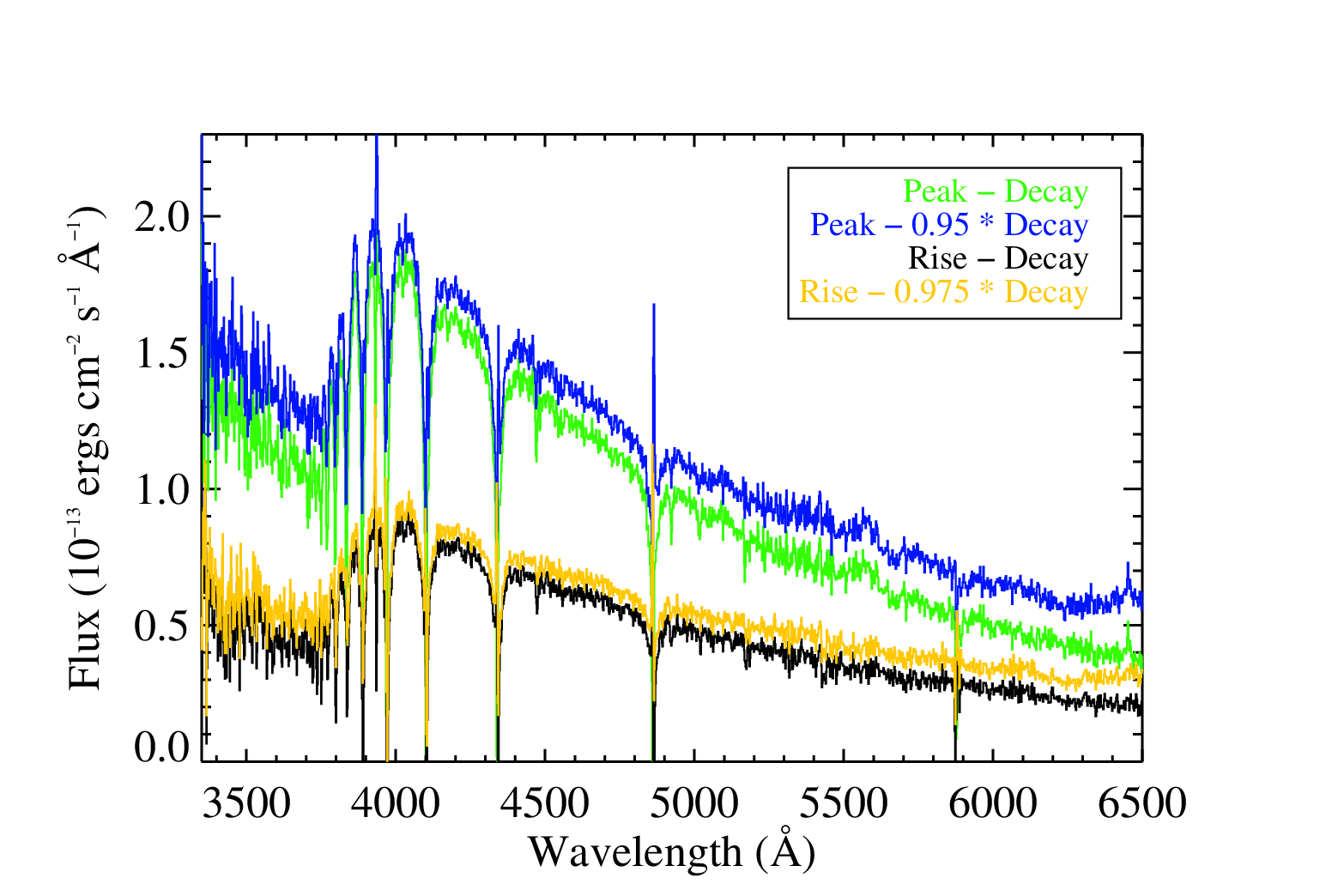} 
\caption{The flare-only spectra during MDSF2 rise (black) and peak
  (green), as shown previously in 
Figure \ref{fig:magnumopus}.  We adjust for an approximate amount by
which the previously decaying
emission evolves to the rise phase time (yellow) and peak time (blue).      }
\label{fig:astar_correct}
\end{center}
\end{figure}

We estimate the
decay phase timescale using a double exponential fit to the overall
decay trend of the $U$ band in IF1 (see Figure \ref{fig:lcphot_panel_if}).  We then scale this fit by 90\% to match the underlying
level of the troughs in the decay phase.  We predict that the $U$ band
of the previously decaying emission would decrease by 5\% from S\#102
to S\#113 (peak) and by 2.5\% from
S\#102 to S\#108 (mid-rise phase) if the secondary flare (MDSF2) were not present.  
Figure \ref{fig:astar_correct} shows the results of subtracting these
adjusted background spectra.  Qualitatively, our conclusions are the
same. However, there are significant temperature differences for flare
emission calculated using 
the unadjusted subtractions and the adjusted subtractions.  For the
unadjusted rise and peak phase spectra, \TBB\ is $14\,900$ K and $17\,700$ K,
respectively.  For the adjusted rise and peak phase spectra, \TBB\ is
$13\,100$ and $15\,000$ K, respectively. Thus, accounting for the
decay of background gradual phase emission from IF1 decreases the derived
temperatures of MDSF2 by \s1800\,--\,2700 K.
Note, the detection of the Balmer continuum and lines in seen, and is
robust result.  The identification of ``hot star'' emission is
therefore confirmed.

\section{Possible Evidence for Wing Absorption at the peak of IF3} \label{sec:IF3_wings}
The flare IF3 has an impulsive phase with the largest continuum
enhancement ($I_{f, u}+1\sim78$) with spectral coverage in the DIS sample.  The inferred
areal coverage of the \TBB\s$12\,000$ K emission 
is large (Section \ref{sec:bbpeak},
Table \ref{table:chitable}), and we search for evidence of wing depressions signifying
the presence of hot star-like Hydrogen Balmer absorption (see Section
\ref{sec:wing_absorption}).  These wing depressions provide new
constraints for RHD models of deep heating in flares and are found in IF3 only 
during the newly formed emission at peak.

\begin{figure}
\begin{center}
\includegraphics[scale=0.4,angle=90]{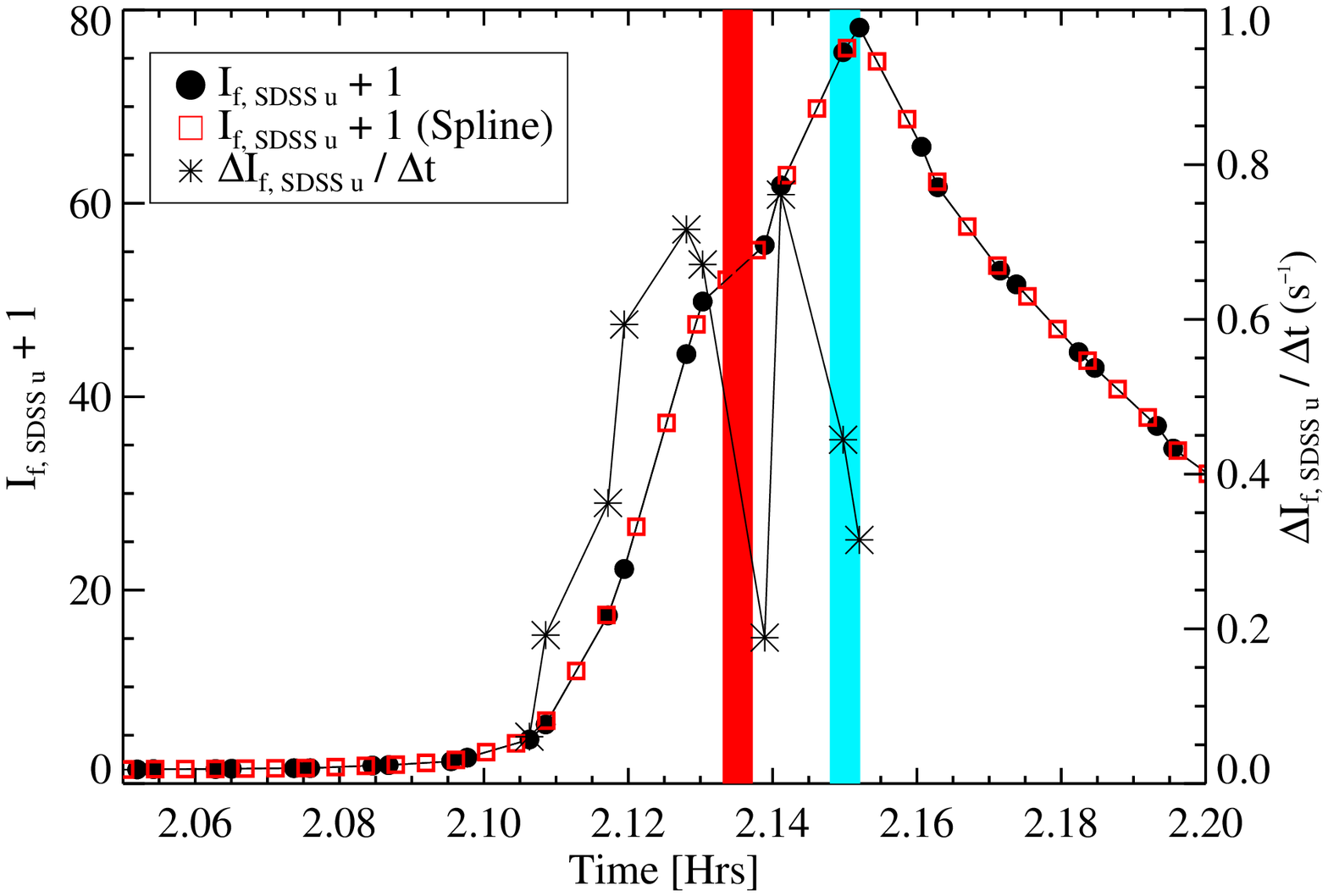}
\caption[Line Profiles for IF3]{The SDSS $u$-band light curve of the rise
  phase of IF3 (left axis, black circles) and the first derivative of the light curve
(right axis, asterisks).  Sequential $u$-band observations have time-intervals of
either 8 or 30 seconds.  The vertical shaded bars correspond to the
times of spectra used to obtain the 
line profiles during the first maximum (red) and peak (turquoise) phases
of the flare shown in Figure \ref{fig:line_width_X}.  The $u$-band light
curve is interpolated using a spline function and binned to 15 seconds
(red squares, left axis) to show the smoothed evolution. }
\label{fig:IF3_burst}
\end{center}
\end{figure}

\begin{figure}
\begin{center}
\includegraphics[scale=0.4,angle=90]{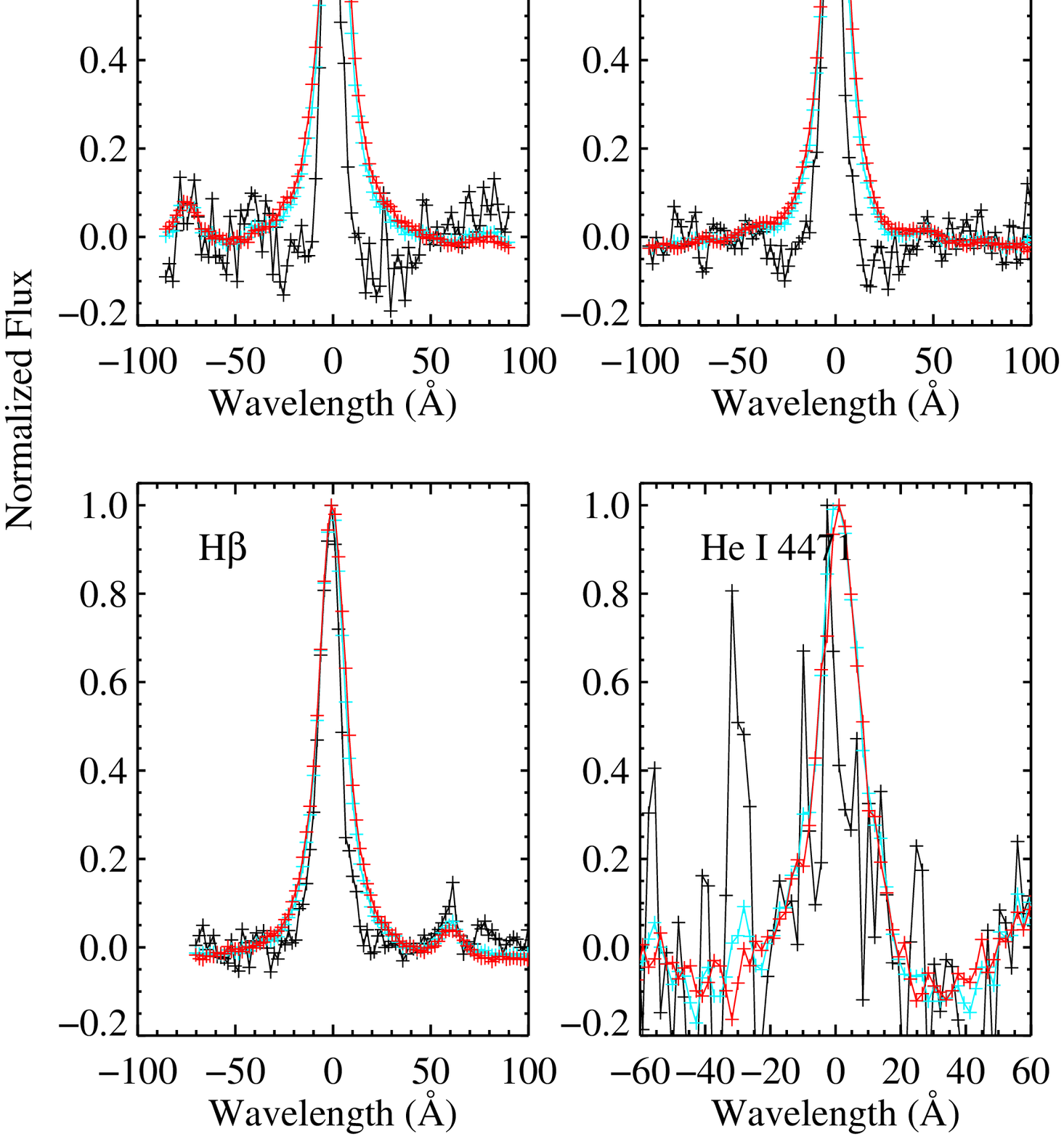}
\caption[Line Profiles for IF3]{The H$\delta$, H$\gamma$,
  H$\beta$, and He I $\lambda$4471 of IF3 at \s80\% of the maximum continuum emission during the rise
  phase (S\#29 normalized, red),
  at maximum continuum emission (S\#31 normalized, turquoise)
  and the newly formed line emission at peak continuum 
  (S\#29 subtracted from S\#31 then normalized; black).  Before the
  normalization, the local continuum was
  subtracted with a linear fit.  The times at which these spectra
  were taken are shown in Figure \ref{fig:IF3_burst}.  The three
  Hydrogen lines all have narrower profiles in the newly formed peak
  flare emission, with \Hd\ showing the largest effect. }
\label{fig:line_width_X}
\end{center}
\end{figure}

In Figure \ref{fig:IF3_burst}, we show the SDSS $u$-band 
photometry during the rise and peak emission.  The black circles are
the photometry data, and the asterisks represent the derivative
of the light curve.  The derivative is useful for diagnosing changes
in this light curve which has an irregular cadence.  At
$t=2.14$ hours (\s80\% of the maximum emission) the derivative has decreased,
indicating the presence of a transient maximum in the photometry; we interpret this as
evidence of the end of a fundamental ``burst''
of emission during the rise phase.  The derivative then increases, signifying a new burst of emission leading to the peak.  The
red squares represent the photometry interpolated (using a spline
function) and rebinned to a constant fifteen second spacing, illustrating the same
effect as the derivative.  In Figure~\ref{fig:line_width_X} we show evidence of wing depressions at
the time of the IF3 peak for the newly formed emission in the peak
emission ``burst''.   We assume that this burst corresponds to
a different flare region on the star (in Section \ref{sec:speeds},
we show that the rise phase corresponds to increasing flare area,
suggesting that this assumption is valid), allowing the rise burst
spectrum (S\#29)
to be subtracted from the peak burst spectrum (S\#31).  The green profiles show the
normalized total flare emission at S\#29 (this corresponds to the rise
phase when the continuum flux is at \s80\% of the
maximum -- before the
second flare burst), the turquoise profiles show the normalized total flare emission at
S\#31 (at maximum continuum emission), and the black profiles show the
difference (using the unnormalized spectra) of S\#31\,--\,S\#29.  The
integration times of these spectra are indicated with green and
turquoise shaded bars in Figure~\ref{fig:IF3_burst}.   The line
emission formed during the second burst has narrower Hydrogen line profiles; the wings are depressed relative to the total flare
emission. The apparent amount of wing depression increases from \Hb\ to
\Hg\ to \Hd, as observed in IF4 and IF1, signifying that the amount of
Hydrogen line absorption increases for higher order Balmer lines as
occurs in the spectra of hot stars.

\section{Corrections to \XBB\ using Hot Star Models} \label{sec:appendix_CK}
Correction factors for \XBB\ are determined from the ATLAS9 grid of 
LTE stellar atmospheres \citep {Kurucz2004} for a range of effective
temperatures, $T_{\mathrm{eff}}=5000-16\,000$ K.
Surface fluxes were calculated from the models with solar metallicity, $v_{\mathrm{turb}} = 2$ km s$^{-1}$, mixing
length parameter l/H$=1.25$, and log $g=5$.  The color temperatures of
all stellar models were measured using the continuum windows in
Table \ref{table:bbwindows}.  These windows were slightly adjusted to effectively match the model
spectral shape from $\lambda=4000-4800$\AA, resulting in a color temperature
grid ranging from 3820
K to $23\,740$ K.  The surface flux of each model spectrum
was then compared to the surface flux of a blackbody (of the same
color temperature) from $\lambda=4170-4210$\AA\ and from
$\lambda=4490-4530$\AA.  The ratio of the surface fluxes gave the correction factor to apply
to the values of \XBB\ calculated from the blackbody fits.  

\section{``Stacking Peaks'': Temperature differences in HF1 vs IF2}  \label{sec:stack}
The flares HF1 and IF2 have respectively a higher ($2.33\pm0.11$) and
lower ($1.74\pm0.04$) value of \chifp.  They also
occurred on the same star (YZ CMi), consecutively on the same night, at similar airmass,
under similar weather conditions; thereby offering robust comparison of two
flares with extremely different flare morphologies (multiple peak
vs. single peak). During the impulsive phase 
of HF1, there were a series of peaks, and we ask the question whether
adding the peaks together would produce the same spectrum as the main
peak in IF2.  If they match, then it is possible that the fast, large peak of IF2 is a result of smaller peaks (like
those seen in HF1) that ``stack'' together very quickly below the time
resolution of our data.

\begin{figure}
\includegraphics[scale=0.35]{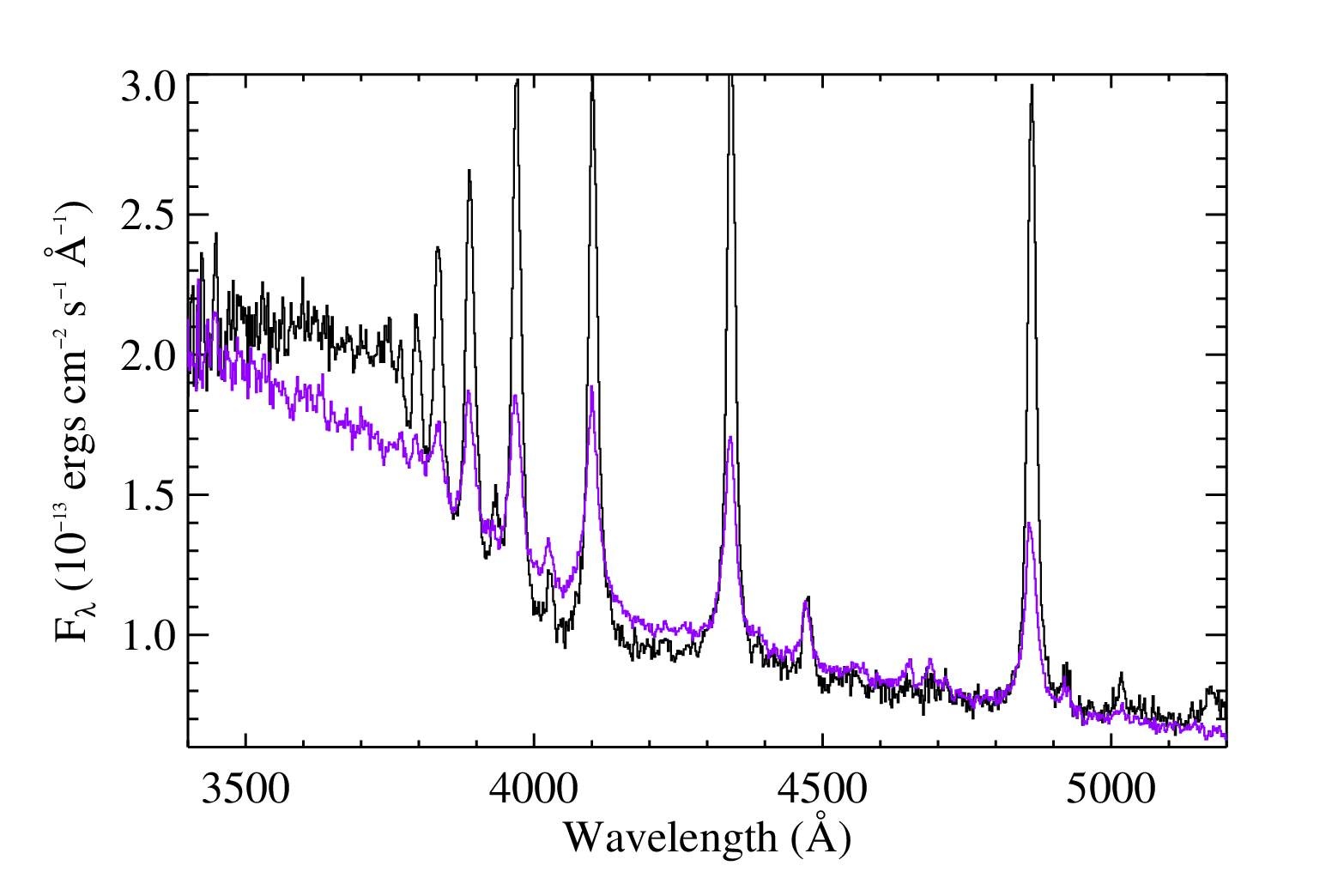}
\includegraphics[scale=0.35]{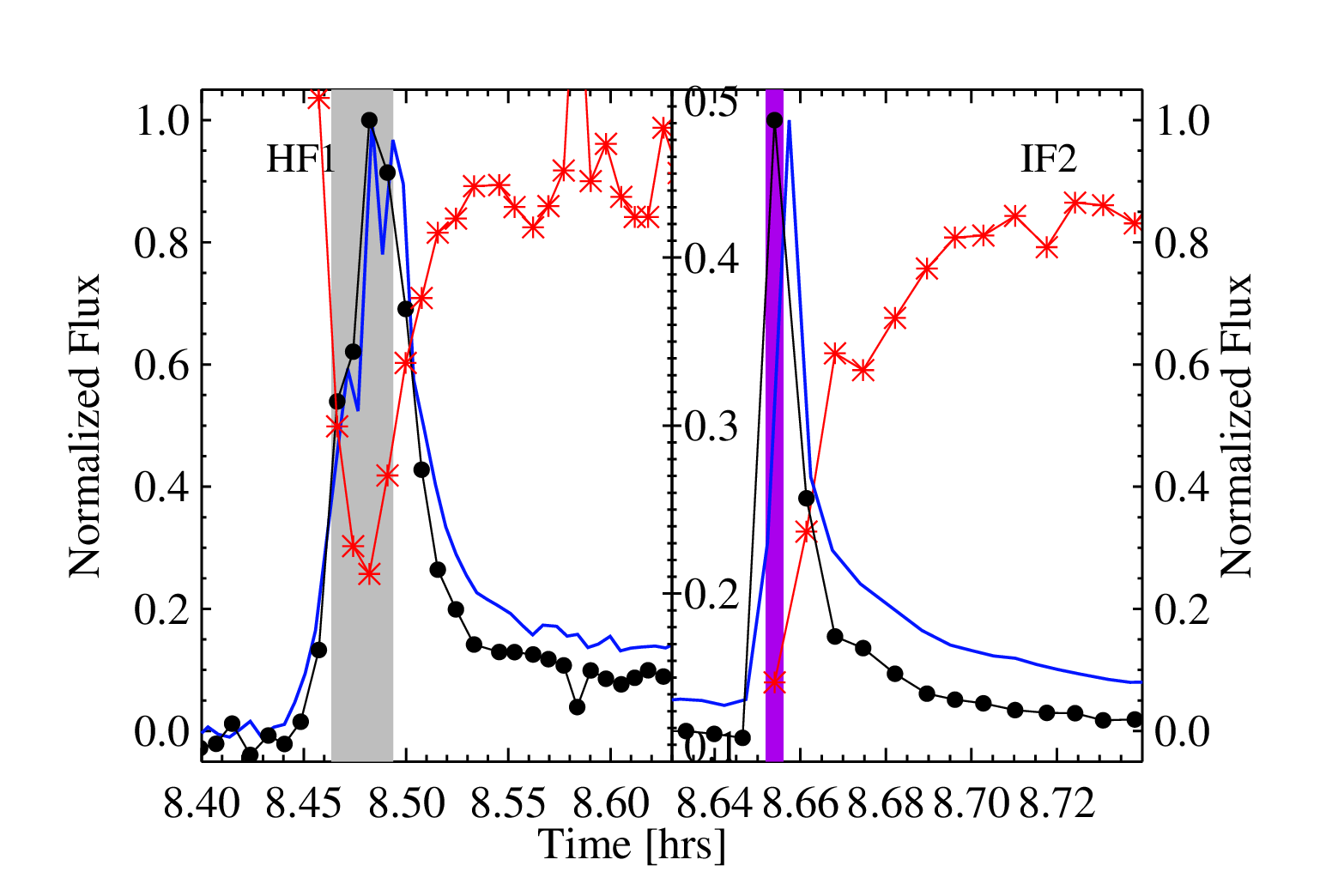}
\caption{(Left) Comparison of stacked spectrum from
  the impulsive phase of HF1 (black) to the peak spectrum of IF2
  (purple).  (Right) The times considered are indicated in the right
  panel (shaded grey area and shaded purple area), which 
  also shows the light curves for C4170 (black circles) and the $U$-band (blue line).  The leftmost and rightmost axes
  are these quantities normalized to their peaks.  The middle axis
  (referring to red asterisks) is the percentage of Hydrogen Balmer
  emission (Hydrogen Balmer flux / total 3420\,--\,5200\AA\ flux; see
  Section \ref{sec:hydrogen}) as a function of time.}
\label{fig:gedanken_stack}
\end{figure}

After adding HF1 S\#518$-$S\#521, the total flare-only emission in C$3615$ is
$2\times10^{-13}$ ergs s$^{-1}$ cm$^{-2}$ \AA$^{-1}$, the same level as in the
peak of IF2.  Figure \ref{fig:gedanken_stack} shows the spectra 
(left panel) and the lightcurves (right panel). 
The spectra show noticeable differences in the blue-optical shape such
that the stacked HF2 spectrum (black) has $T_{\mathrm{BB}} \sim 10\,700$ K
and the peak IF2 spectrum (purple) has a steeper slope with
$T_{\mathrm{BB}} \sim 14\,000$ K.  There is also a 
difference in the relative amounts of \Hg\ to C4170 formed (ratios of
39 for HF1 and 17 for IF2) and the percentage of Hydrogen Balmer (HB)
emission (\s25\% for HF1 and \s15\% for IF2).   The evolution of the percentage of HB emission is shown in red
asterisks in Figure \ref{fig:gedanken_stack} (right panel); in the gradual
phase, a similar percentage is achieved (\s40-45\%), but the gradual
phase value is attained relatively quickly in HF1 and more slowly in IF2.  

This exercise has important implications for flare heating mechanisms.
Larger peaks are not necessarily a straightforward ``areal'' sum of smaller peaks;
instead, larger peaks can result from a larger temperature of the
blackbody component produced from the flare.  Flare heating during very impulsive
events (IF2) may be more concentrated, temporally and spatially, in the
atmosphere whereas several smaller peaks that are spread out in time
(and space) may 
result from diffuse heating of several individual kernels over a larger area thereby producing a
cooler blackbody in each kernel and relatively more HB 
emission.  

We estimate the speed of areal increase (Section \ref{sec:speeds}) to be \s100 \kms\
(rise/peak) and \s15 \kms\ (initial fast decay) for
IF2 and \s40 \kms\ (fast rise and peak phase) and \s5 \kms\ (the
fast decay) of 
of HF1.  The inferred flare temperature (at peak) appears to be 
correlated with the speed for these two flares.  In contrast, we found
that the (rise phase) speed appears inversely correlated with the (rise
and peak) temperature between MDSF2
and IF3.  IF2 has a similar rise phase speed compared to IF3, for
which we concluded that Scenario 2 type heating (Section
\ref{sec:speeds_meaning}) describes the development of the flare area.  We speculate that it may be also
important to consider the role of shock heating as a contribution to the larger
temperature of IF2 (due to the inferred supersonic speed for the stellar
chromosphere).  Although the calculation of speed in
HF1 may be affected by the presence of several distinct (temporally resolved, spatially
unresolved) areas on the surface at a
given time during the impulsive phase, there may be a
difference in the flare mechanism or heating properties between a
flare with slow
speed but lower temperatures (HF1) and a flare with slow speed but higher temperatures (MDSF2).

Individually the area of each peak in HF1 is $< 40-60$\% the area of
the blackbody in IF2 (Table \ref{table:chitable}), but the decay phase value of the percentage of HB
emission is attained slowly in IF2; we speculate that this is
reflecting the larger area of IF2 (hence, longer time to decrease in
size assuming the beam heating turns off gradually).  Ultimately, we
hope to understand both the temperature and areal behavior using RHD models.

\clearpage

\end{document}

%% file: figures.tex
\begin{figure}
\begin{center}
\includegraphics[scale=0.4]{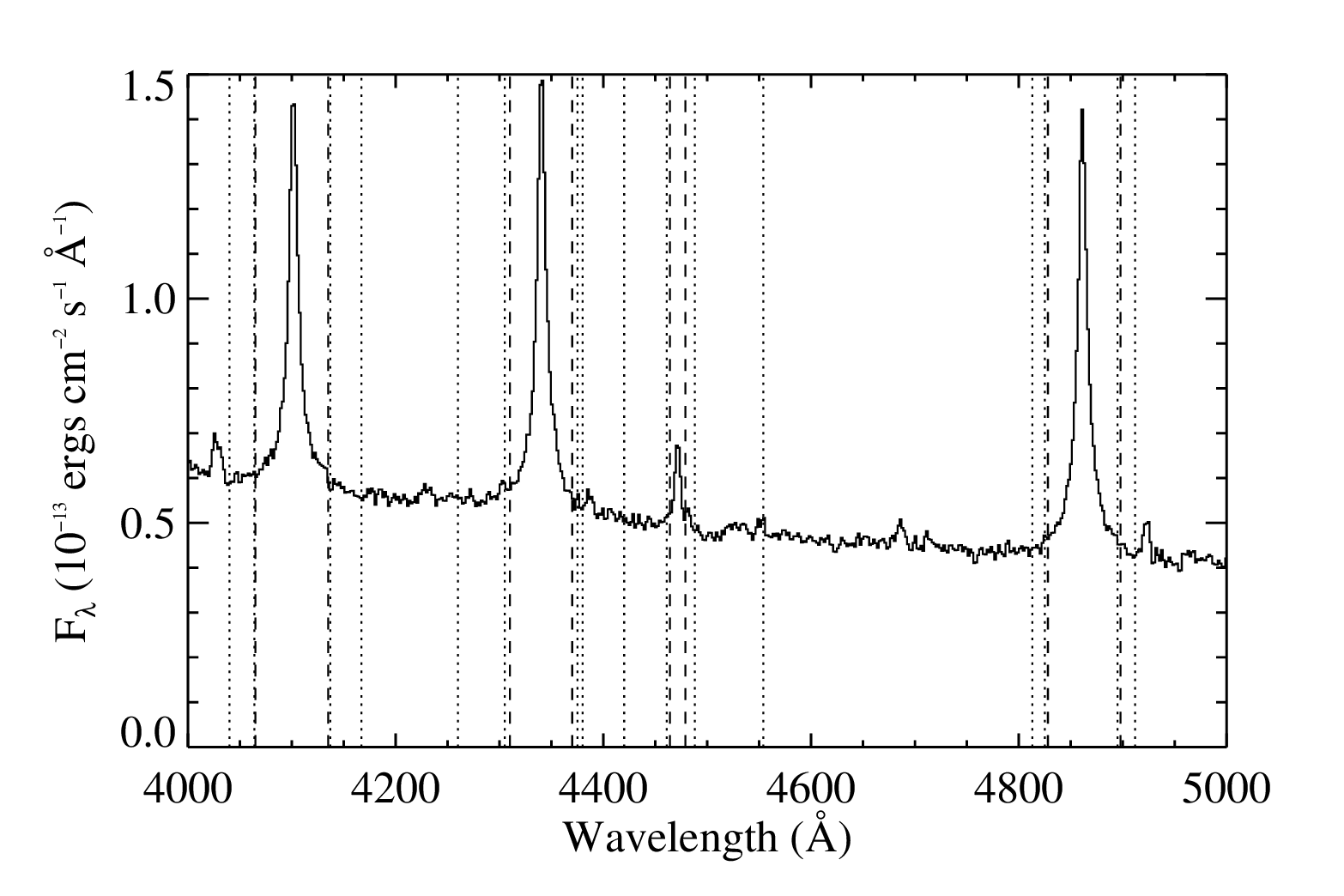}
\caption{The peak (flare-only) spectrum from the flare
  ($t_{\mathrm{peak}} = 2.6245$ hours) on 2010 Feb 14 on YZ CMi
  showing the windows of line flux integration for the \Hd, \Hg, He
  \textsc{i} $\lambda4471$, and \Hb\ emission lines (dashed lines; see
  Table \ref{table:linewindows}).  The underlying continuum level was
  estimated using linear fits to the nearby regions indicated with
  dotted lines.  These data were
  obtained with the 1.5\arcsec\ slit. }
\label{fig:trustme}
\end{center}
\end{figure}

\begin{figure}
\begin{center}
\includegraphics[scale=0.4]{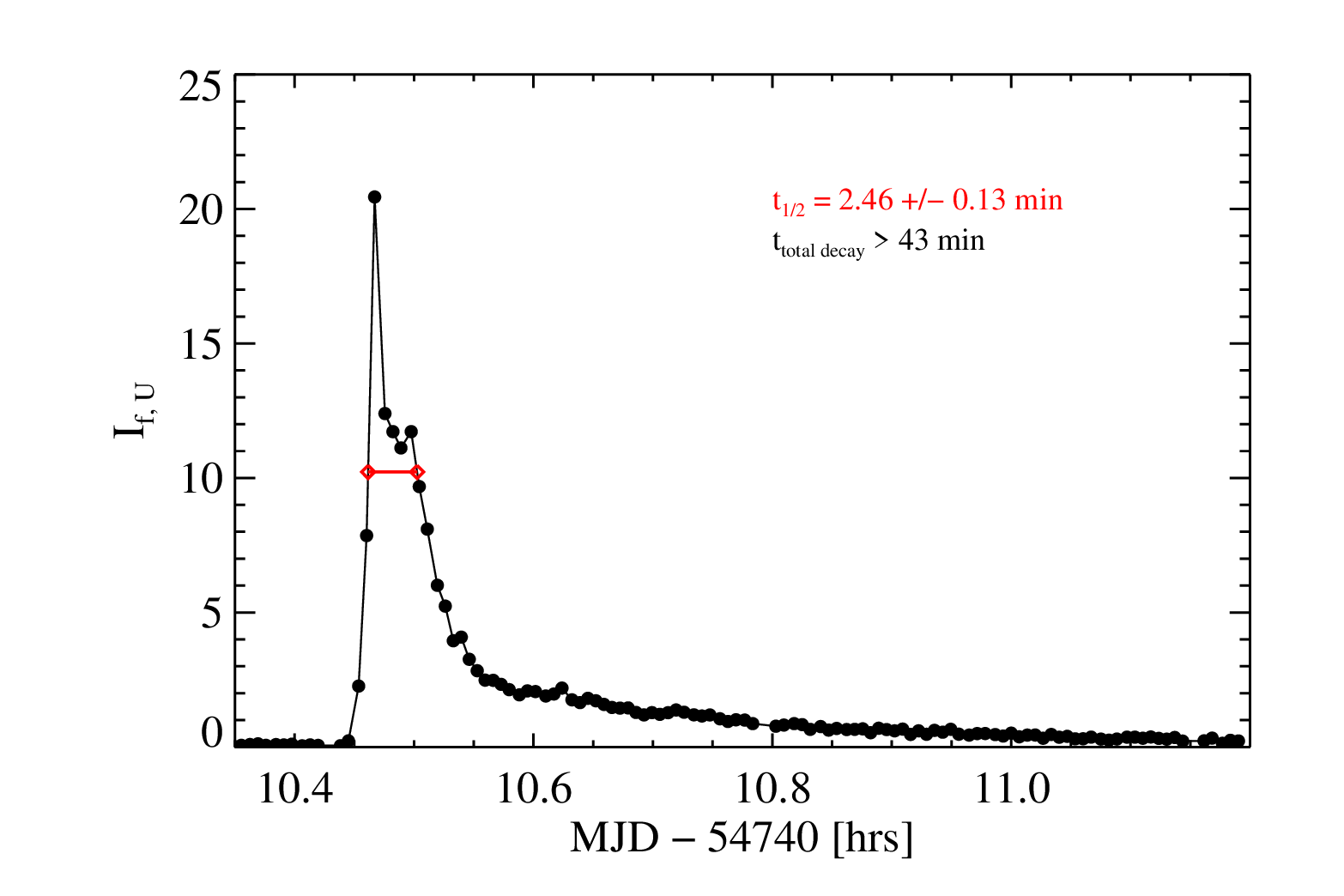}
\caption{Example calculation of $t_{1/2}$ for a $U$-band light
  curve of a flare ($t_{\mathrm{peak}}=$ 10.4686 hours) on EQ Peg A on
  2008 Oct 01.
The time between observations ranges from $24 - 31$ sec.
The $t_{1/2}$ is the FWHM of the light curve, illustrated in red
($t_{1/2}=2.46$ minutes).
The observations ended before the flare finished, so a total decay
time could not be determined.  }
\label{fig:thalf_example}
\end{center}
\end{figure}

\begin{figure}
\begin{center}
\includegraphics[scale=0.40]{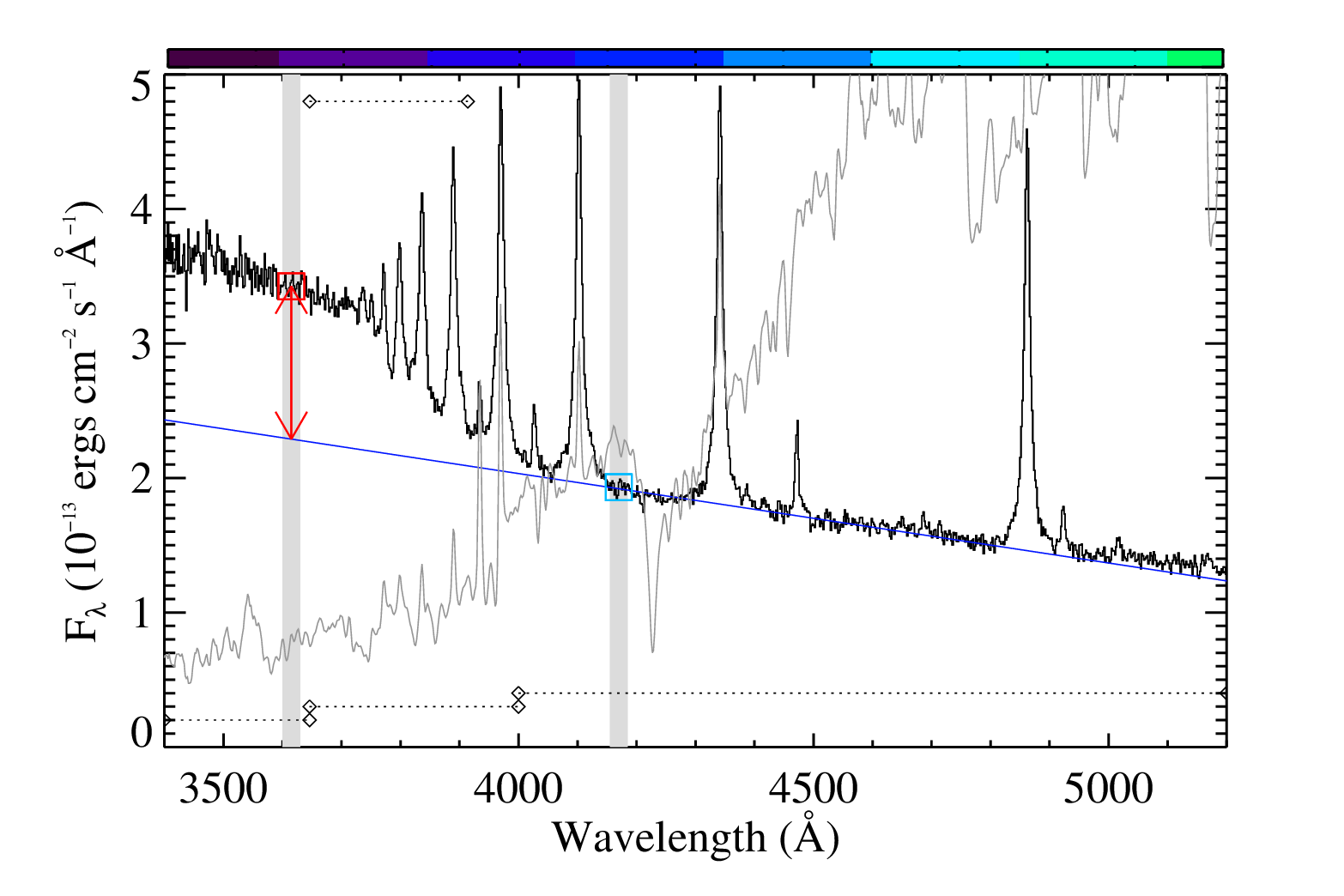}
\caption{ The flare-only emission from the peak of a flare on AD Leo
  from 2010 April 03 ($t_{\mathrm{peak}} = 3.9120$ hours) showing the
  important terms from Section \ref{sec:spectra_param} used in
  this work.  The coverage of the near-UV, intermediate, and blue-optical zones are
  indicated at the bottom with diamonds; the coverage of the PseudoC is indicated
  at the top with diamonds. The quiescent spectrum is
  shown in grey, and the best-fit line using the continuum windows 
  in Table \ref{table:bbwindows} is shown in dark blue.  Vertical
  grey bars indicate the blue and red wavelength regions used to calculate C3615 and
  C4170; squares denote the flare-only flux values in these
 regions.  The \chif\ (Balmer jump ratio) is the flux of the red square divided by the
 flux of the blue square (\chif$=1.8$ in this spectrum).
The red arrows indicate the excess emission, BaC3615, above the
 extrapolation of the linear
 fit to the blue-optical zone.  Although $U$-band photometry was not
 available during this flare, $I_{f, SDSS g} + 1 \sim 1.28$ and
 $I_{f, \mathrm{C3615}} + 1 \sim 5.4$.  For reference the visible colors
 for this wavelength range are indicated with the colorbar.  }
\label{fig:adleo_E10}
\end{center}
\end{figure}

\clearpage
\begin{figure}
\includegraphics[scale=.8]{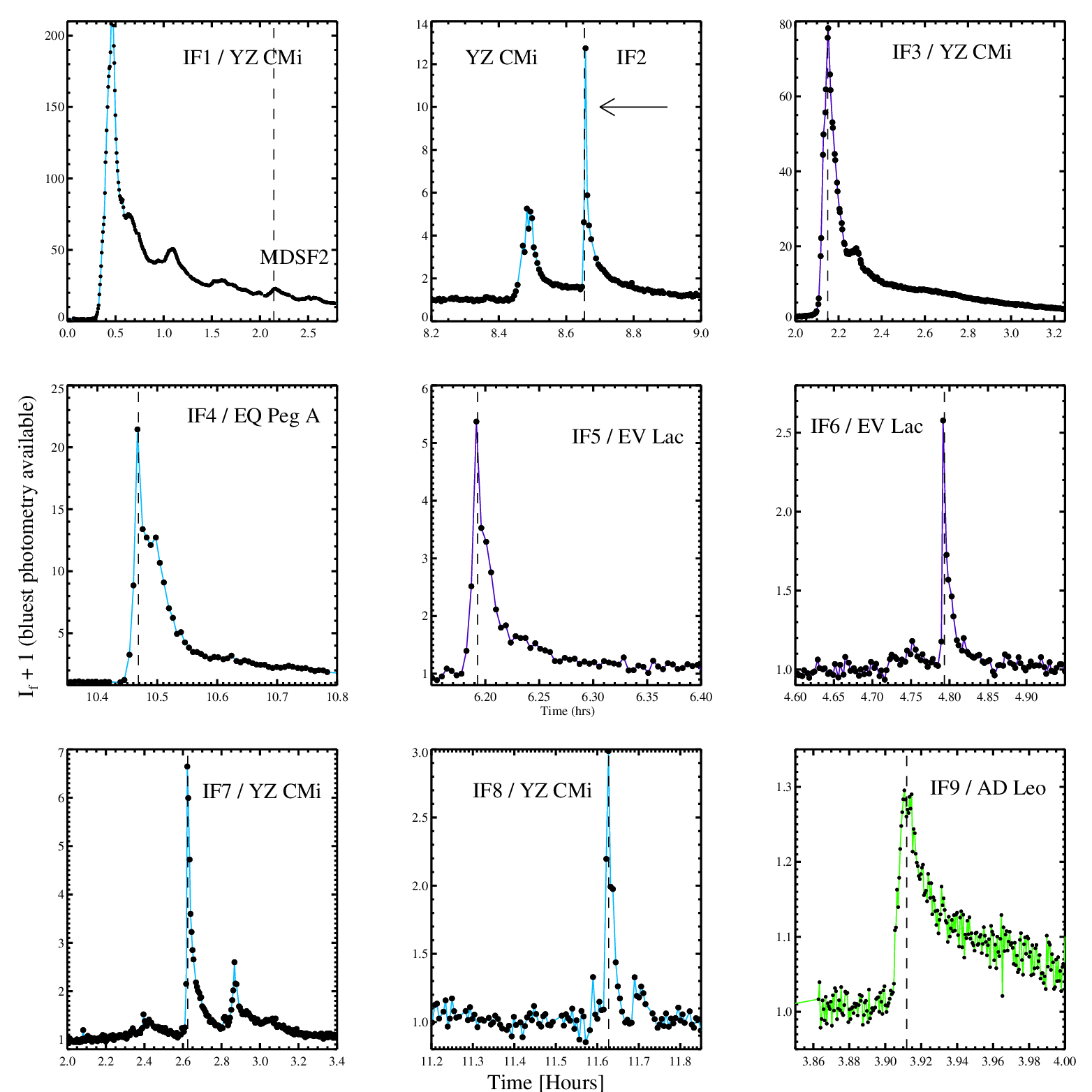}
\caption{The impulsive flare photometry. The purple lines show
  photometry from the NMSU 1m (Johnson $U$), the
  light 
blue shows photometry from ARCSAT/Flarecam (SDSS $u$), and the green
shows photometry from ARCSAT/Flarecam (SDSS $g$). The vertical dashed
black 
lines indicate the time of maximum C3615. An arrow indicates the IF2
event. Time is the \# of hours elapsed on the respective MJD from
Table \ref{table:flare_summary}. }
\label{fig:lcphot_panel_if}
\end{figure}

\clearpage

\begin{figure}
\includegraphics[scale=.8]{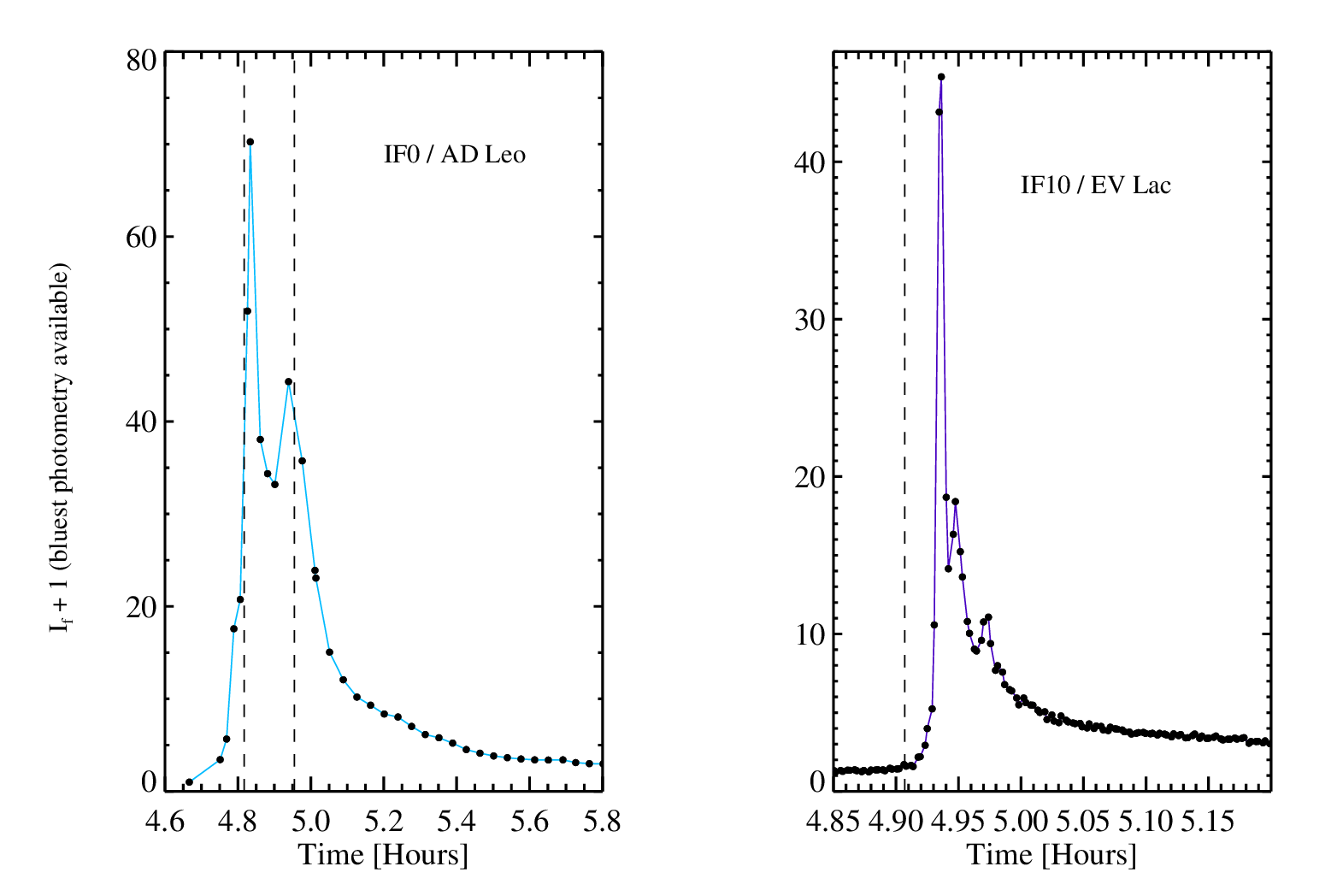}
\caption{Same as in Figure \ref{fig:lcphot_panel_if}, for the
  two impulsive flares with lower time-resolution and limited wavelength coverage
  in the near-UV and blue-optical.  The vertical dashed black 
lines indicate the time of maximum C3615. For IF0, we have indicated
the two times of maximum C3615 from the spectra.  The integration
times for IF10 were long and the midtime of the spectrum with maximum
C3615 is before the rise phase; please refer to Figure
\ref{fig:appendix_integ21} in Appendix \ref{sec:appendix_times}.  }
\label{fig:lcphot_panel_if2}
\end{figure}
\clearpage

\begin{figure}
\includegraphics[scale=.8]{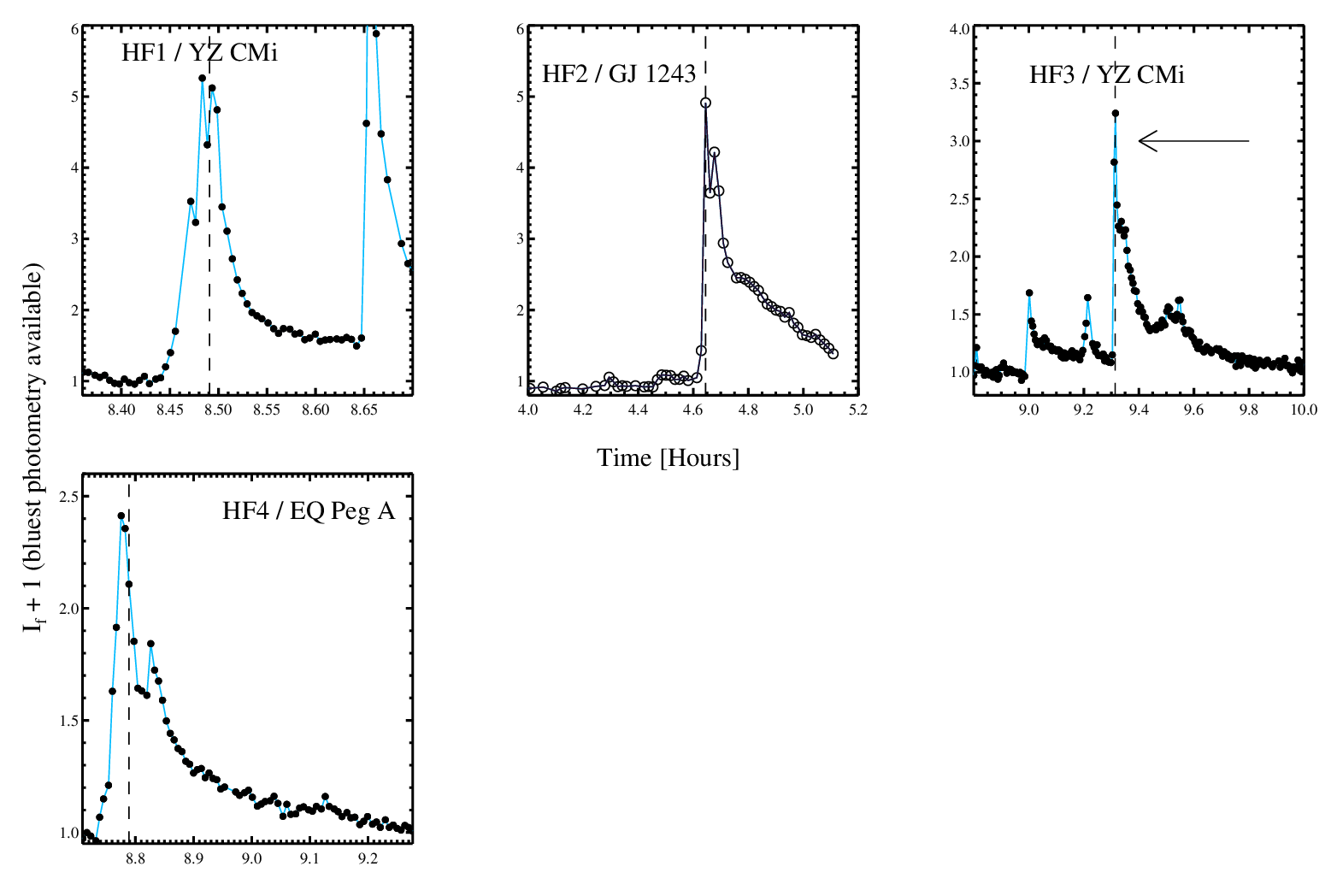}
\caption{ Same as in Figure \ref{fig:lcphot_panel_if}, for the
  hybrid flares.  The vertical dashed black 
lines indicate the time of maximum C3615.  Open circles show
spectrophotometry estimations of the Johnson $U$ band.  An arrow
indicates the HF3 event. }
\label{fig:lcphot_panel_hf}
\end{figure}
\clearpage

\begin{figure}
\includegraphics[scale=.8]{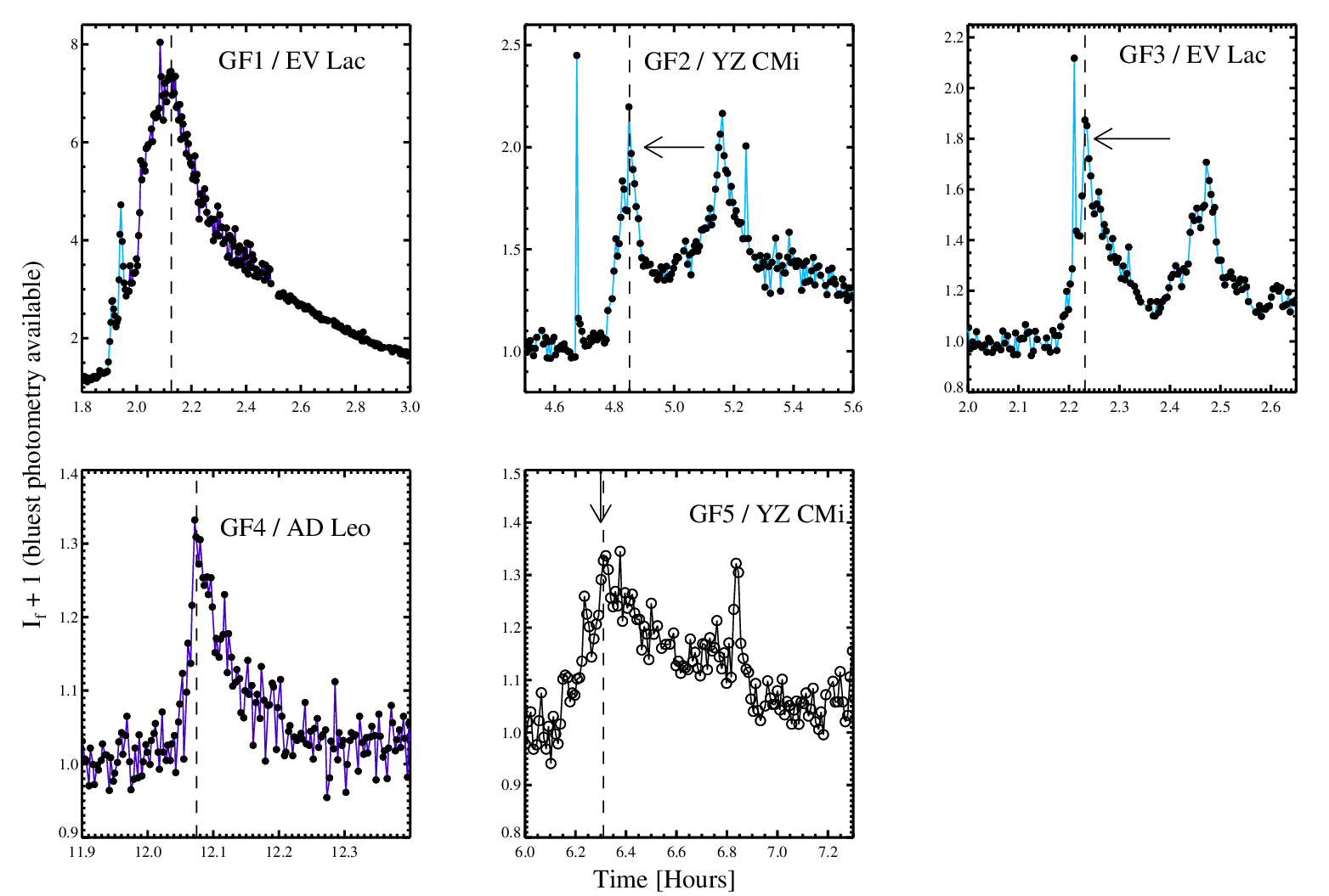}
\caption{Same as in Figure \ref{fig:lcphot_panel_if}, for the
  gradual flares.  The vertical dashed black 
lines indicate the time of maximum C3615.  Open circles show
spectrophotometry estimations of the Johnson $U$ band.  Arrows
indicate the GF2, GF3, and GF5 events. }
\label{fig:lcphot_panel_gf}
\end{figure}
\clearpage

\begin{figure}
\includegraphics[scale=0.8]{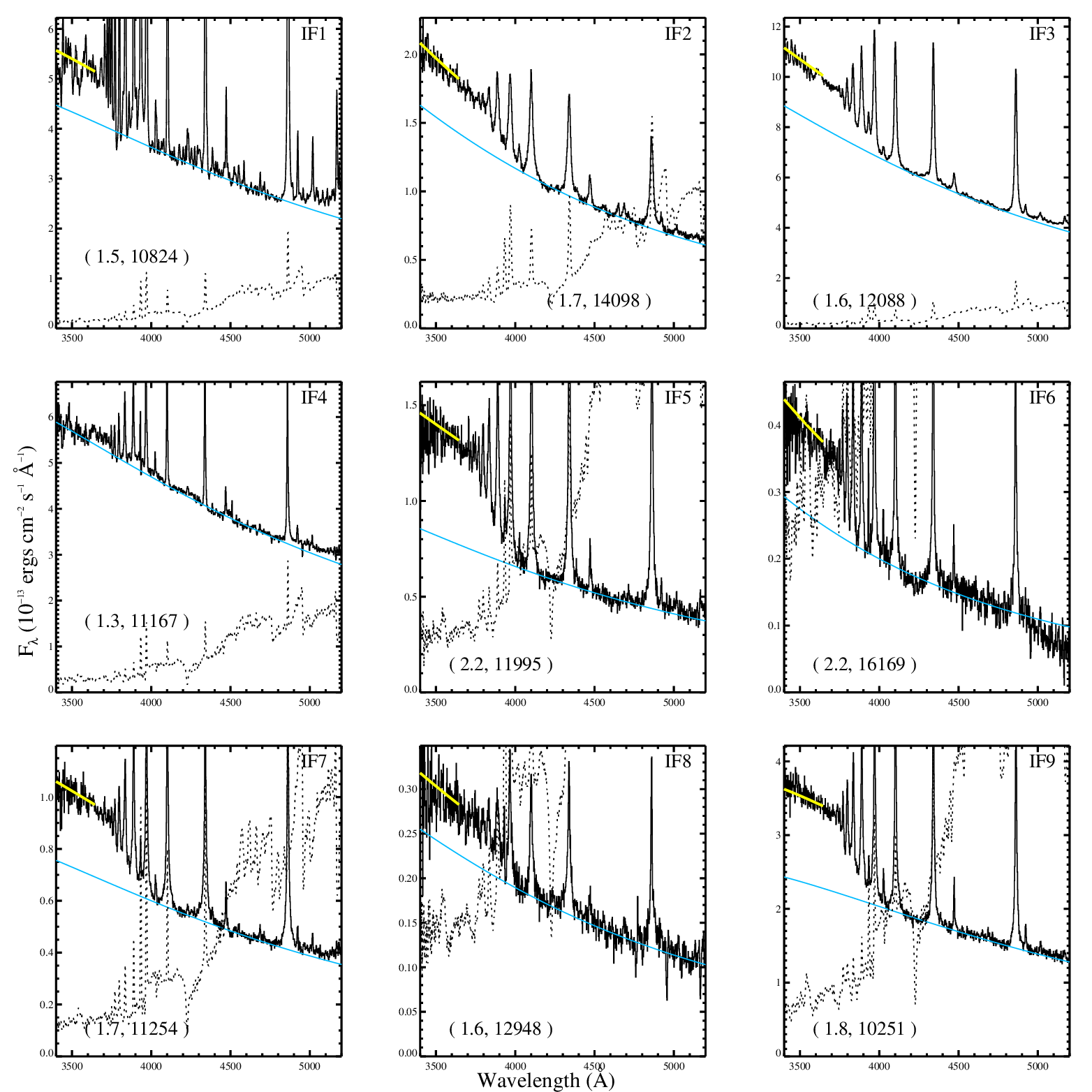}
\caption{ The spectra at maximum continuum emission (C3615).  The black is the
  flare-only emission, the dotted line is the quiescent spectrum. The best-fit Planck
  function to the blue-optical region is shown in light blue.  The
  yellow curve at $\lambda < 3646$\AA\ is the
  best-fit Planck function \emph{scaled} to the C$3615$ flux.
   In parentheses, the \chifp\ and the best-fit color
   temperature are given.  The S\#'s and times of these spectra are given in Table
   \ref{table:times}.  See Section \ref{sec:bluecont} for more details
   about the continuum properties at these times. } 
\label{fig:peak_panels1}
\end{figure}

\clearpage
\begin{figure}
\includegraphics[scale=0.7]{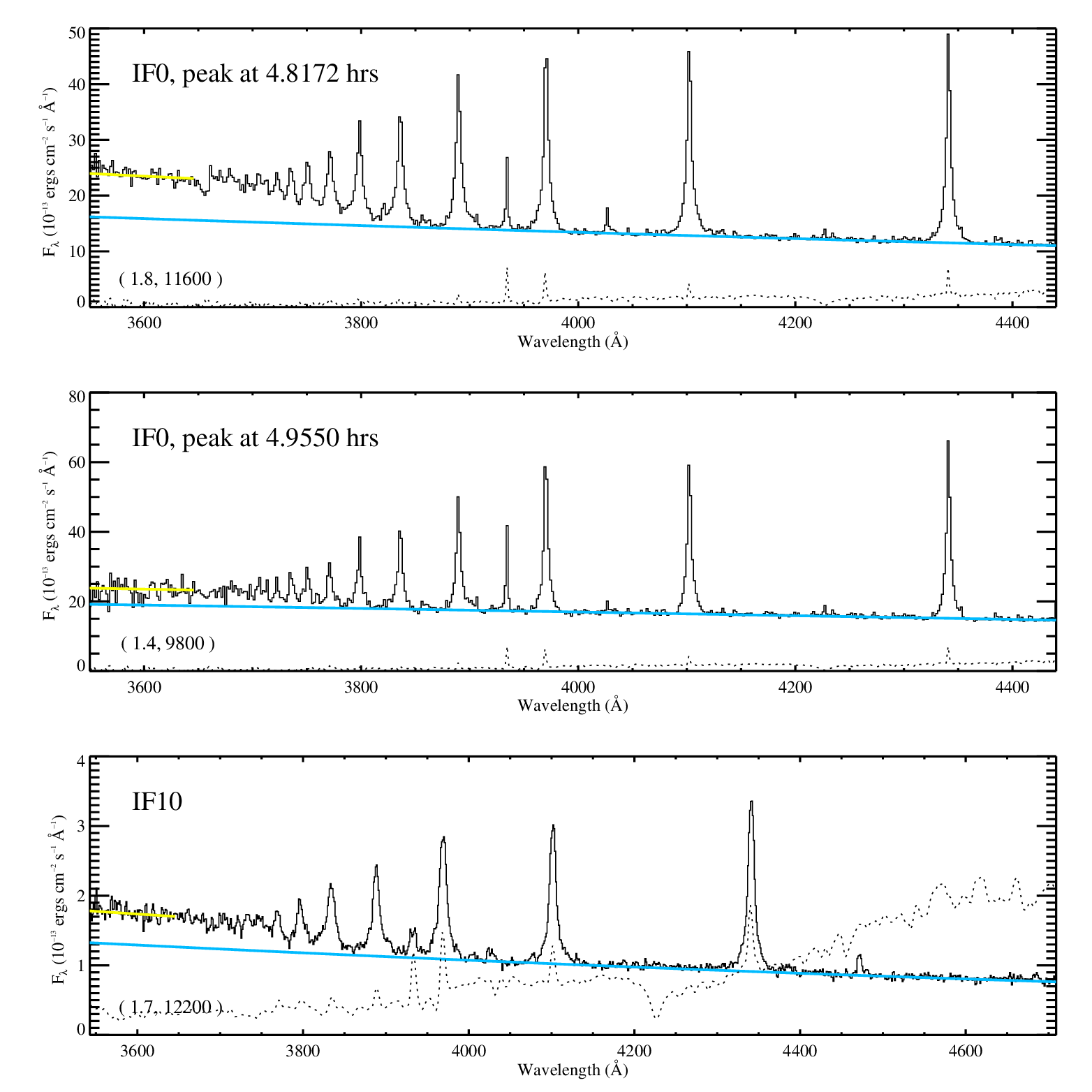}
\caption{Top:  The flare-only spectrum (S\#36) at the first maximum continuum emission (C3615) 
of IF0.  Middle:  The flare-only spectrum (S\#40) at the second maximum continuum emission
(3615) of IF0.  Bottom: The flare-only spectrum (S\#31) at maximum continuum emission
  (C3615) of IF10.  Note the long integration times in Figure
  \ref{fig:appendix_integ21} of Appendix \ref{sec:appendix_times}
  during IF10.  The
  quiescent spectra are shown as dotted lines.  The best-fit Planck
  function to the blue-optical region is shown in light blue.  The
  yellow curve at $\lambda < 3646$\AA\ is the
  best-fit Planck function \emph{scaled} to the C$3615$ flux.
   In parentheses, the \chifp\ and the best-fit color
   temperature are given.   }
\label{fig:peak_panels1b}
\end{figure}

\clearpage
\begin{figure}
\includegraphics[scale=0.8]{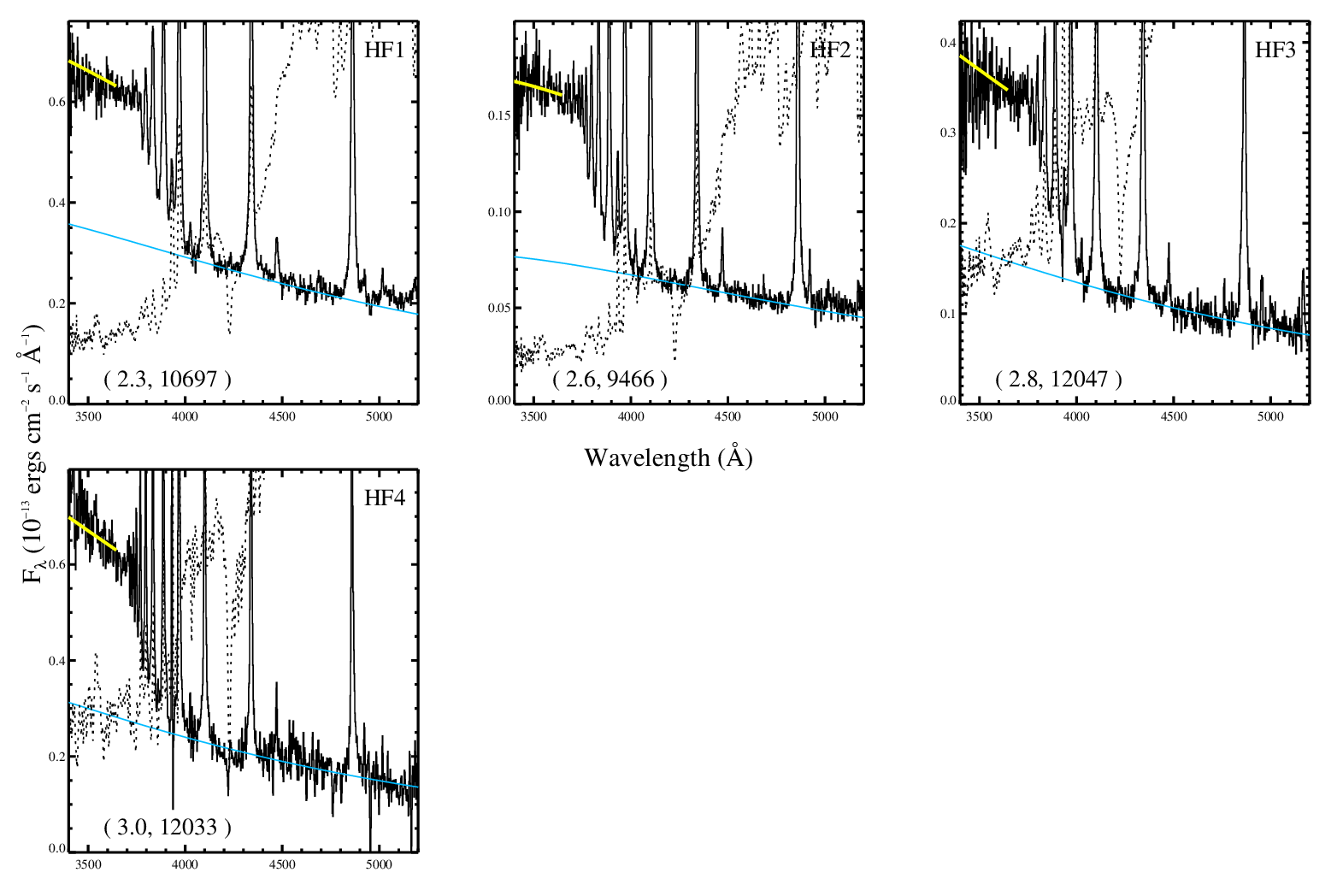}
\caption{ Same as in Figure \ref{fig:peak_panels1}. } 
\label{fig:peak_panels2}
\end{figure}

\clearpage
\begin{figure}
\includegraphics[scale=0.8]{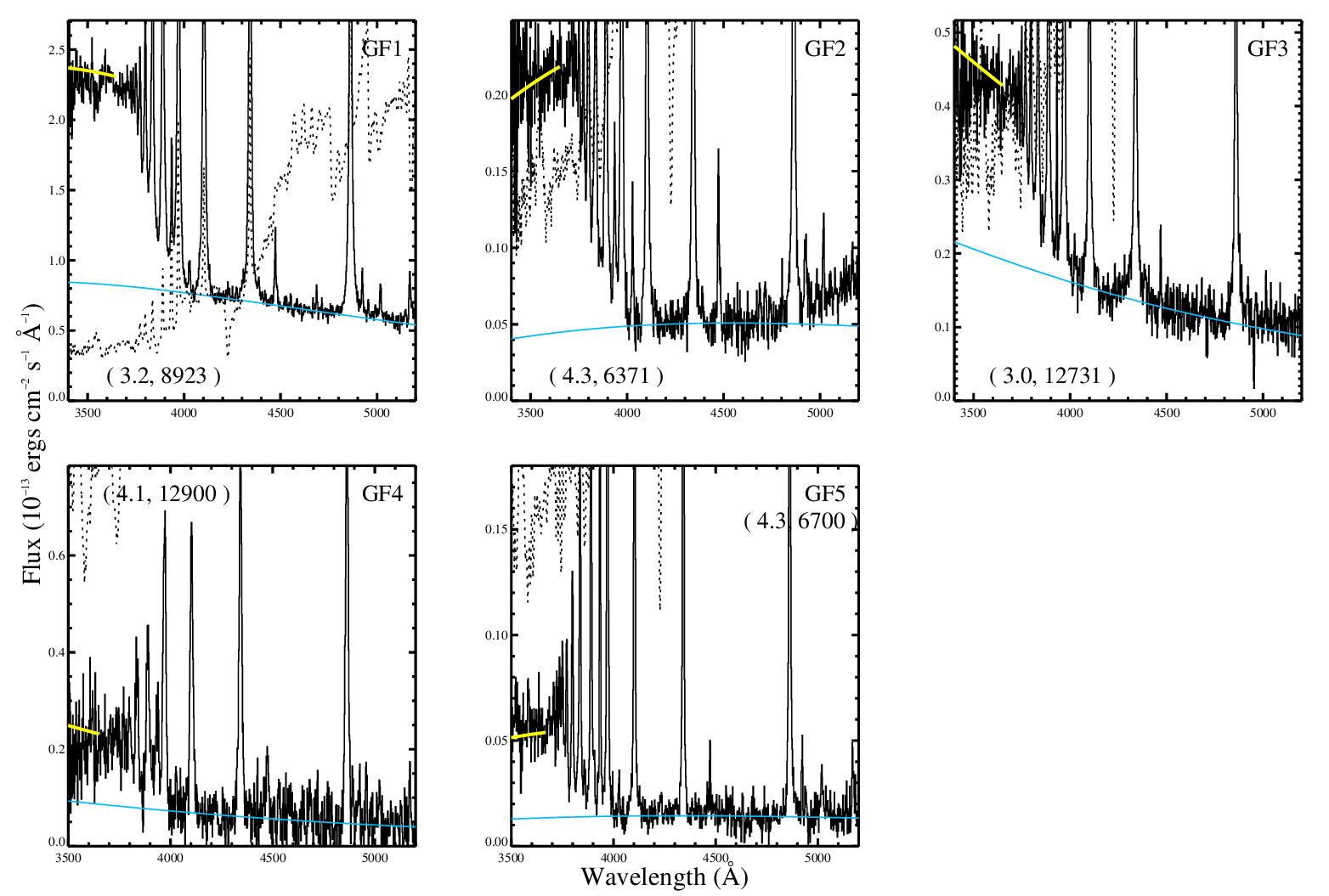}
\caption{ Same as in Figure \ref{fig:peak_panels1}. }
\label{fig:peak_panels3}
\end{figure}

\clearpage

\begin{figure}
\begin{center}
\includegraphics[scale=0.40]{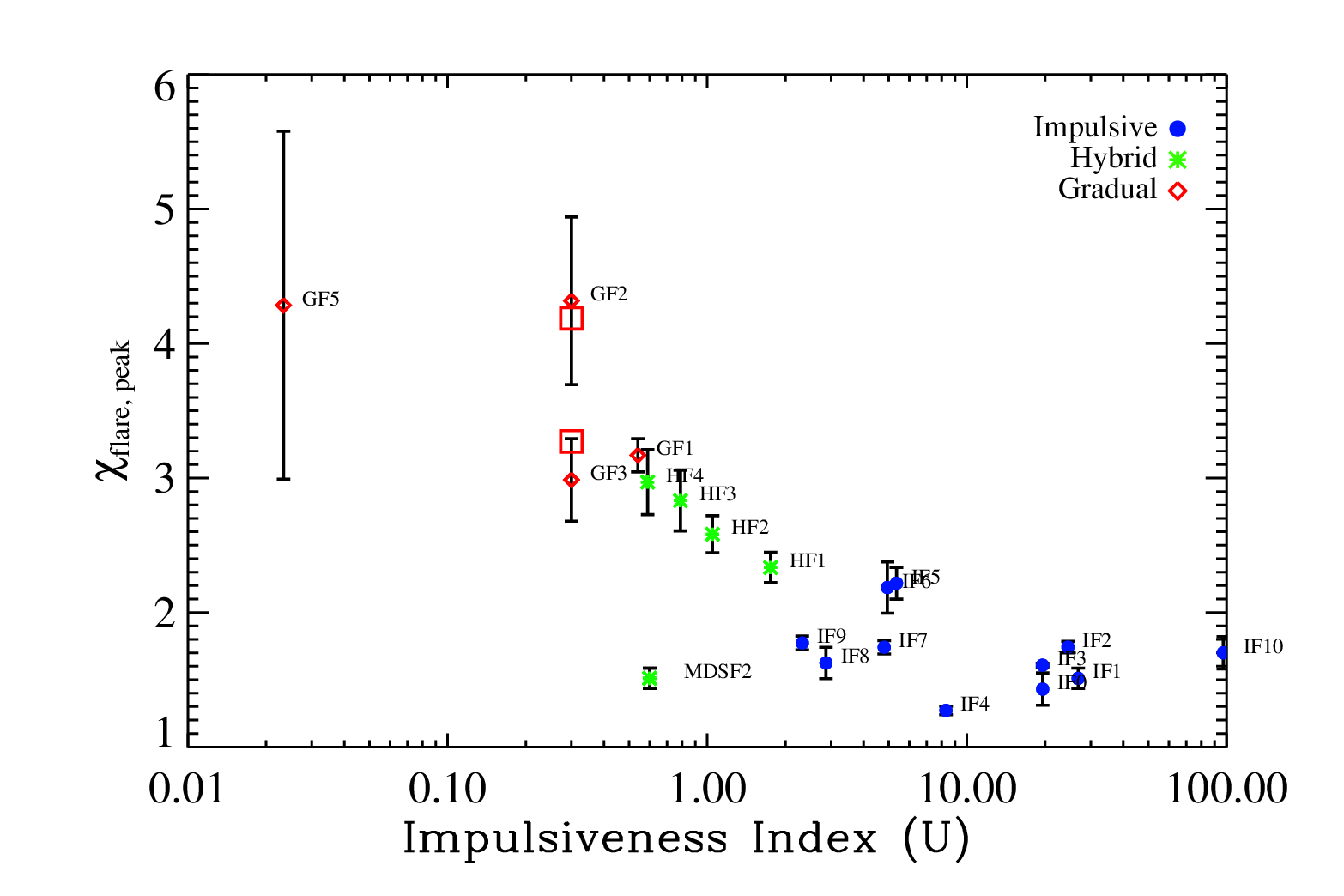} 
\caption{
 The peak properties of the flare sample:
  $\chi_{\mathrm{flare, peak}}$ (or Balmer jump ratio) shows a global trend with the
  impulsiveness index, $\mathcal{I}$ of the flare.  The flares are
  colored according to their IF/HF/GF designation.  We show a point
  for $\mathcal{I}\sim0.6$ for MDSF2 and $\mathcal{I}\sim27$ for the entire IF1 event given the same
  \chifp$\sim1.5$.  Note that the \chifp\ and $\mathcal{I}$ for IF10
  were obtained at a largely different cadence.   For GF3 and GF2,
  we show \chifp\ for the secondary peaks in the events (\chifp$=$3.3 and 4.2, respectively) as red square
  symbols.  The \chifp\ for GF4 is exluded from Figures
  \ref{fig:if_vs_chi}, \ref{fig:FEU_HG}, and \ref{fig:FEU_HG_inset}
  because of its large uncertainty (see text).}
\label{fig:if_vs_chi}
\end{center}
\end{figure}

\begin{figure}
\begin{center}
\includegraphics[scale=0.4]{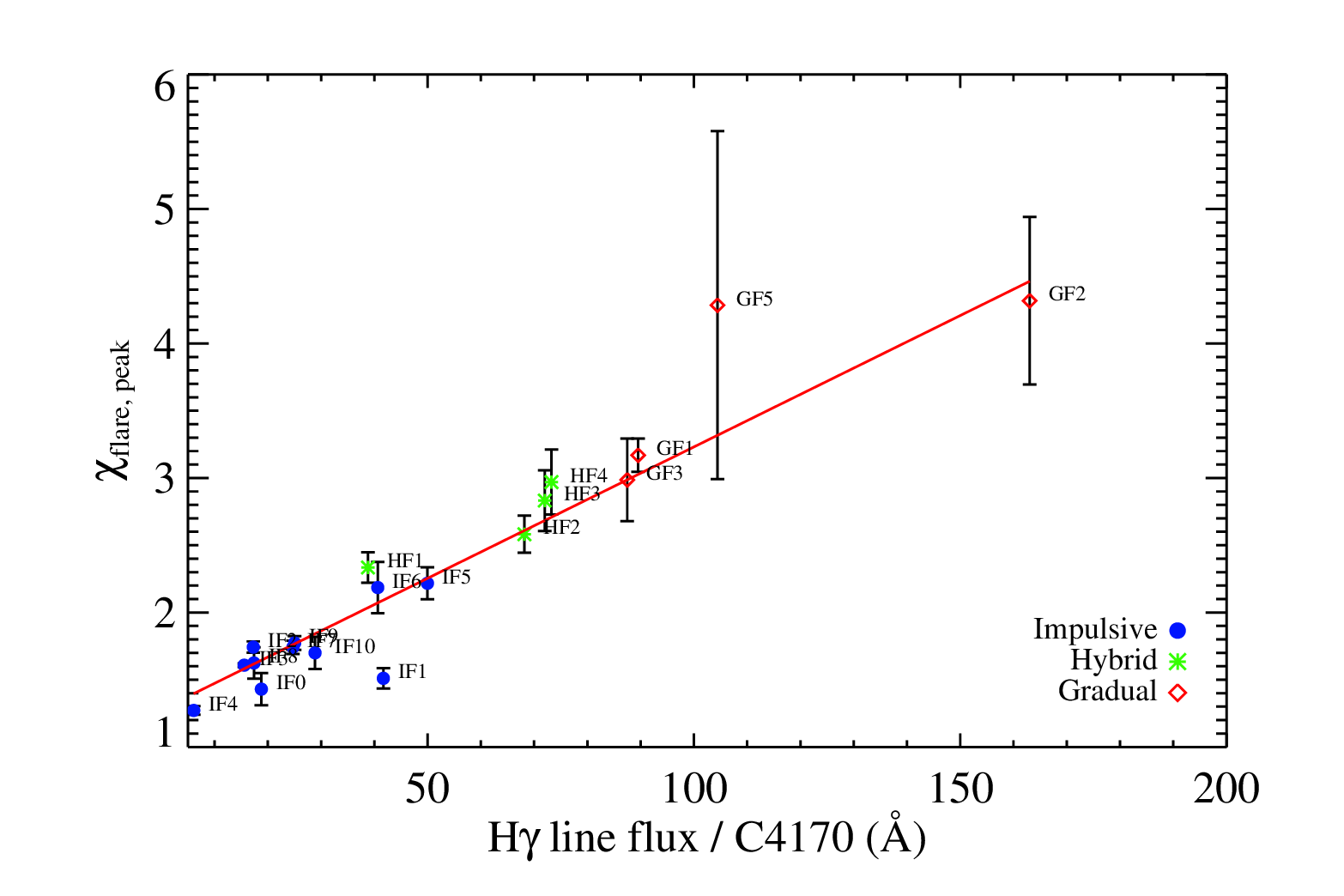}
\caption{ The \chifp\ vs the ratio of \Hg\ to C4170 (at peak C4170). 
 There is nearly a linear relationship, which is shown as a red line
 (see text).  The point for IF1 corresponds to the peak C4170 of
 MDSF2. 
  }
\label{fig:FEU_HG}
\end{center}
\end{figure}

\begin{figure}
\begin{center}
\includegraphics[scale=0.4]{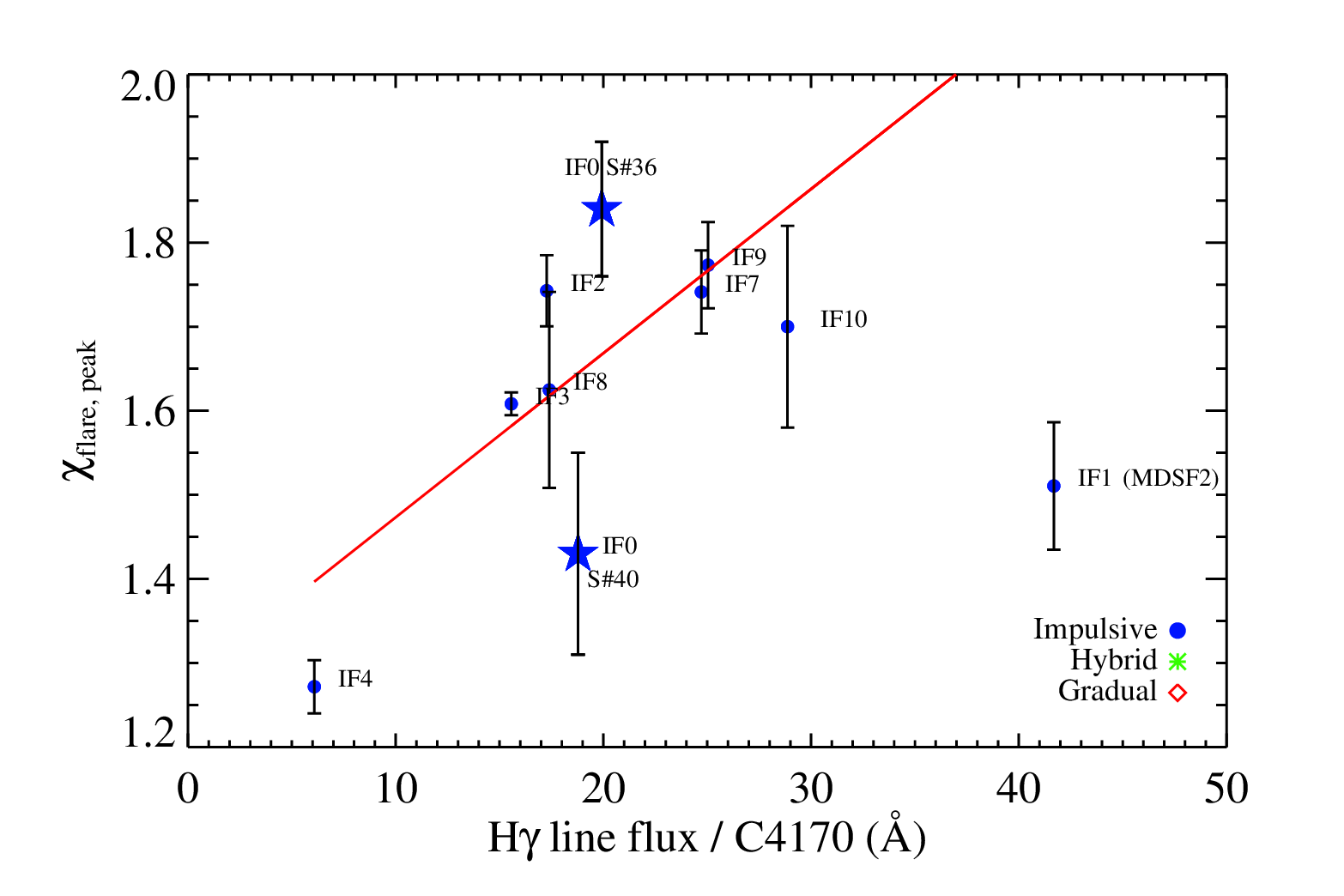}
\caption{ The \chifp\ vs the ratio of \Hg\ to C4170 (at peak C4170) for the low values of \chifp\ and \Hg\ to
  C4170 ratio in Figure \ref{fig:FEU_HG}.  The linear relationship fit
  to the entire sample is shown as a red line (see text).   The point for IF1
  corresponds to the peak C4170 during MDSF2.
   This figure shows the two main peaks (S\#36 and S\#40) of IF0 as star symbols.  
 }
\label{fig:FEU_HG_inset}
\end{center}
\end{figure}

\begin{figure}
\begin{center}
\includegraphics[scale=0.4,angle=90]{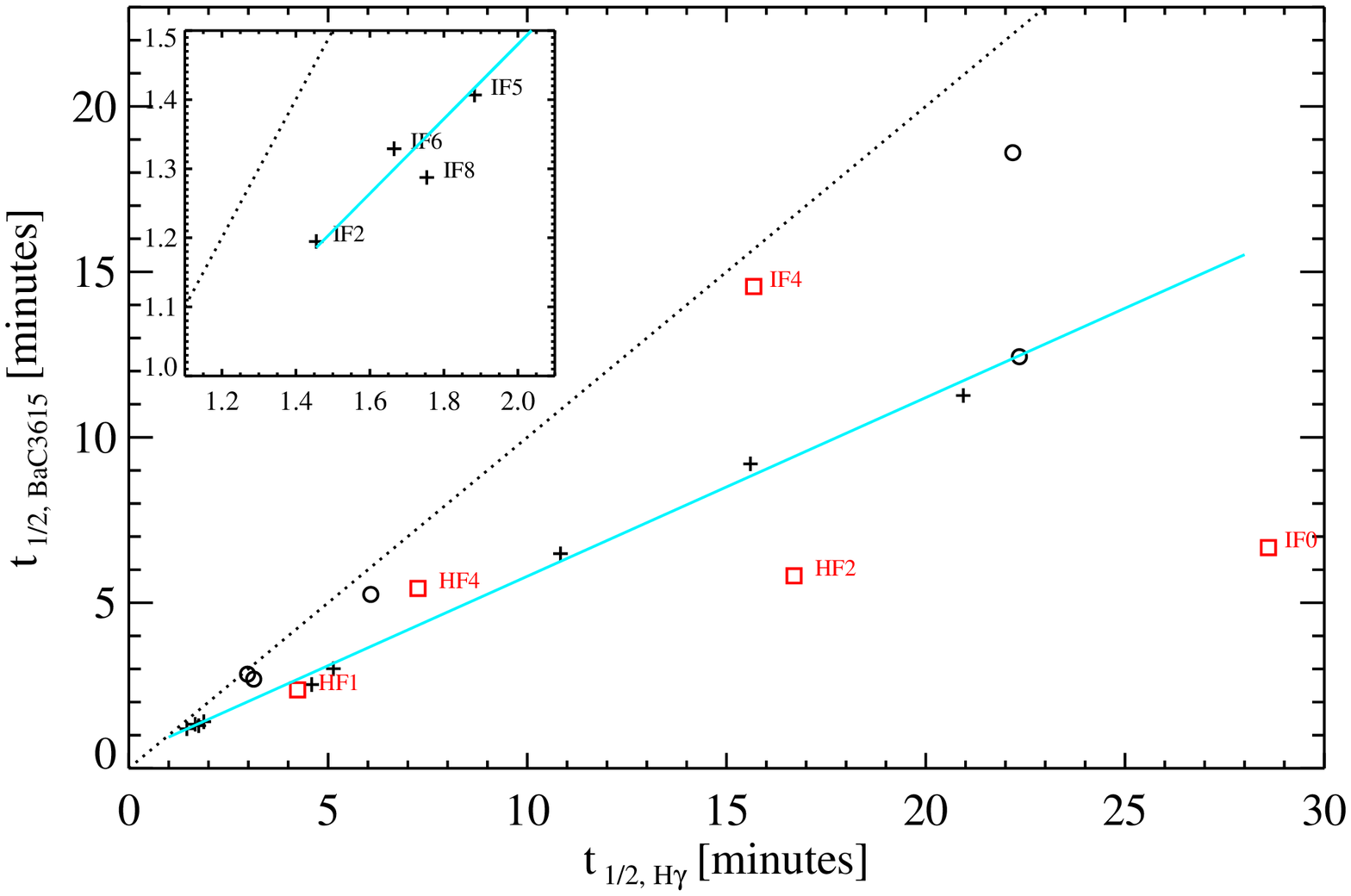}
\caption{The $t_{1/2}$ values of   
  BaC3615 versus the values for the H$\gamma$ line.  The inset has the same axes, showing the flares with
  the smallest values of  $t_{1/2}$. The red squares with labels
  represent impulsive or hybrid flares with 2\,--\,3 peaks; the crosses are
  impulsive or hybrid flares with single peaks, and the open circles are gradual flares.
  The
  dotted line is the 1:1 line and the solid light blue line is the linear fit
to the IF and HF events with single peaks (see text).  Note that only
an upper limit of $t_{1/2,\mathrm{BaC3615}} < $400 sec was determined for IF0. }
\label{fig:thalf_UHg}
\end{center}
\end{figure}

\begin{figure}
\includegraphics[scale=0.35]{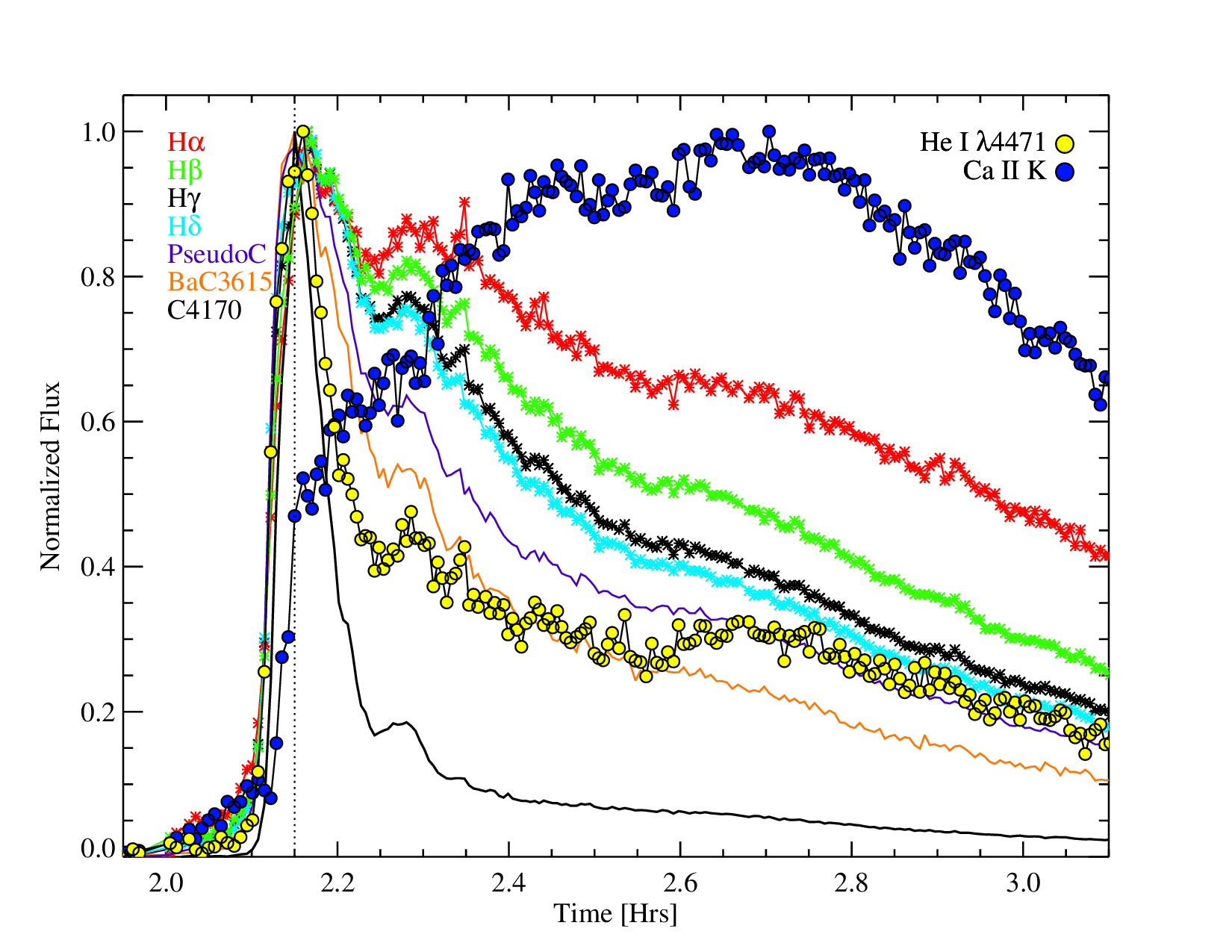}
\includegraphics[scale=0.35]{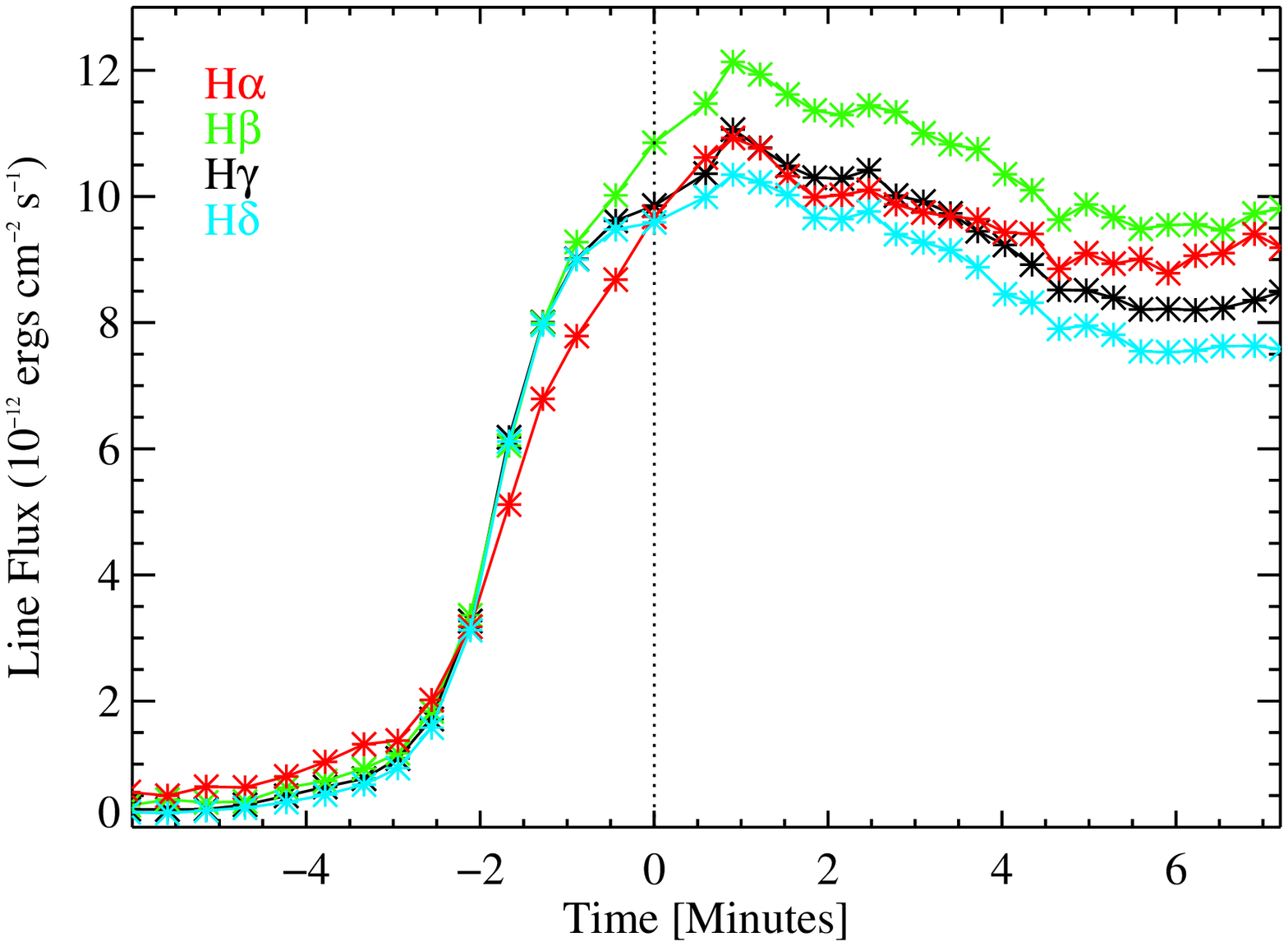}
\caption{(Left) Line and continuum evolution for IF3.  C4170,
  BaC3615, PseudoC, \Ha, \Hb, \Hg, \Hd, Ca II K, and He I$\lambda$4471
  light curves are plotted normalized to their peak fluxes.  The
  vertical dotted line corresponds to the time of peak continuum (S\#31).
(Right) Expanded view of the rise phase, peak, and initial decay of IF3 for
the \Ha, \Hb, \Hg, and \Hd\ line fluxes.  Note the ``S''-shape in the rise
phase morphology of \Hb, \Hg\ and \Hd.  All four lines reach maximum
in the same spectrum.  Note, that the \Hb, \Hg\ and \Hd\ lines diverge
from a common flare flux at S\#29, just before the peak.}
\label{fig:lines_IF3_B}
\end{figure}

\begin{figure}
\begin{center}
\includegraphics[scale=0.35]{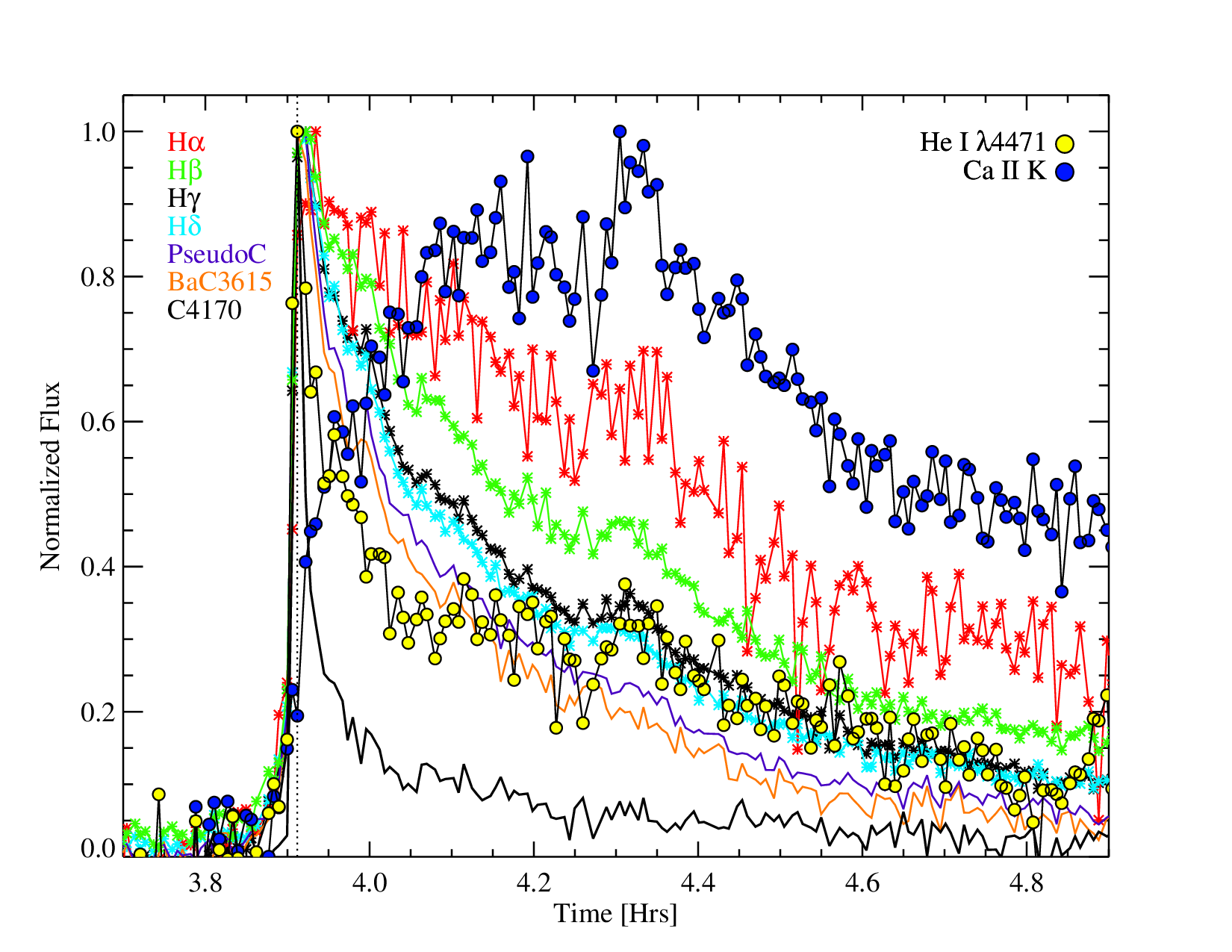}
 \caption{ Line and continuum evolution for IF9. Symbols
   same as in Figure \ref{fig:lines_IF3_B}.  }
\label{fig:lines_IF9_B}
\end{center}
\end{figure}

\begin{figure}
\begin{center}
\includegraphics[scale=0.35]{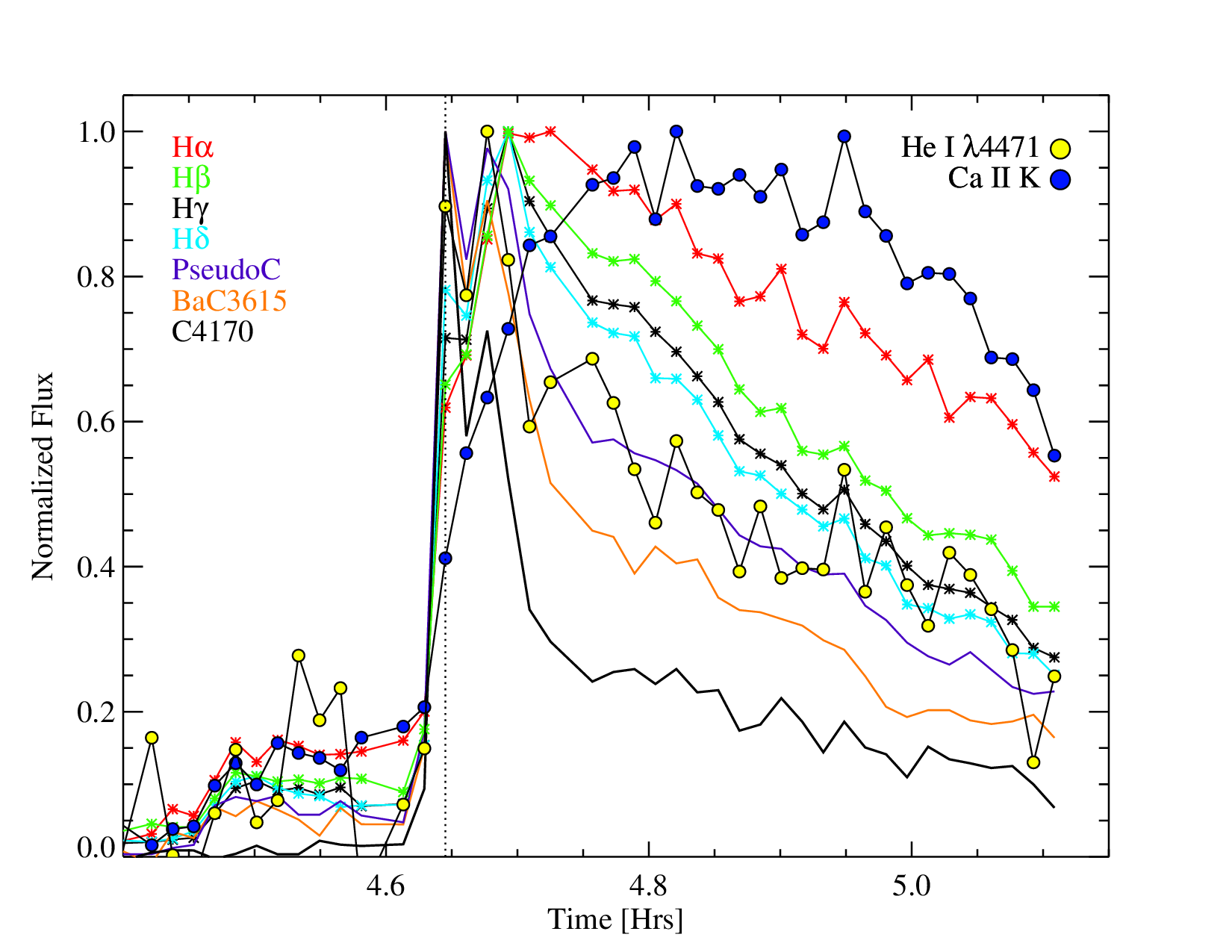}
 \caption{ Line and continuum evolution for HF2.  Symbols
   same as in Figure \ref{fig:lines_IF3_B}.}
 \label{fig:lines_hf2_B}
\end{center}
\end{figure}

\begin{figure}
\begin{center}
\includegraphics[scale=0.35]{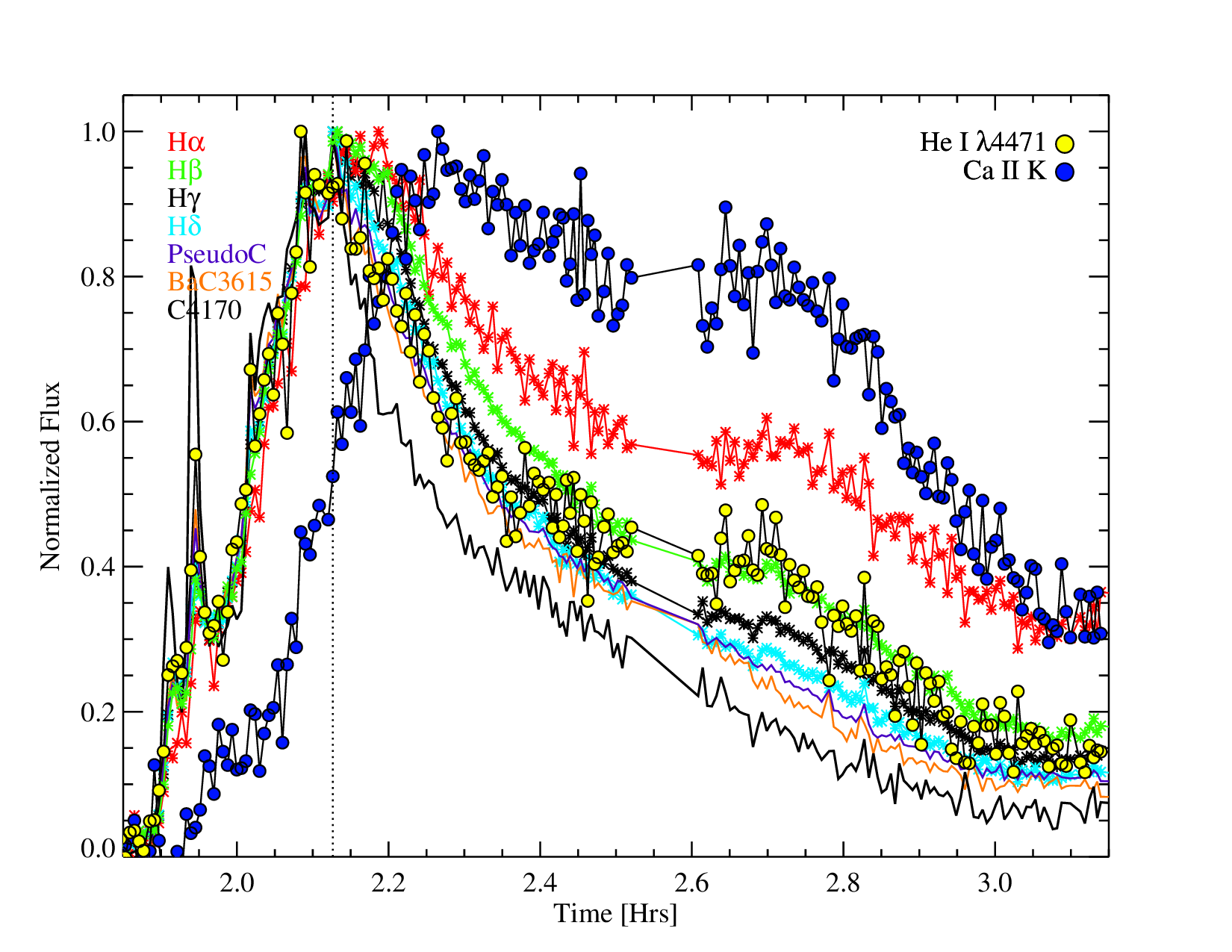}
\caption{Line and continuum evolution for GF1.  Symbols
   same as in Figure \ref{fig:lines_IF3_B}. }
\label{fig:lines_GF1_B}
\end{center}
\end{figure}

\begin{figure}
\begin{center}
\includegraphics[scale=0.4]{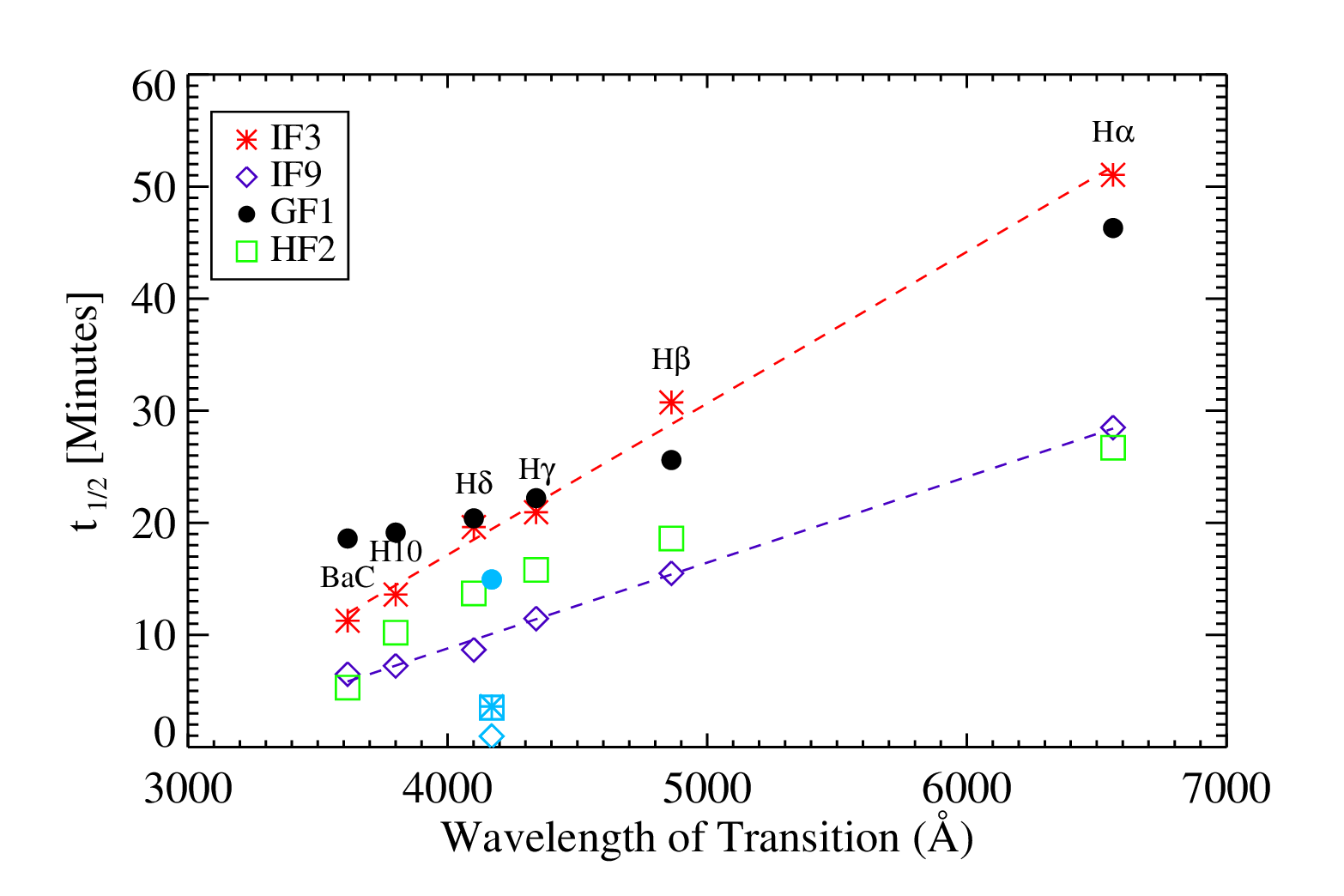}
\caption{The $t_{1/2}$ vs. wavelength (time-decrement) of the Balmer emission
  features in flares IF3, IF9, and GF1 and HF2.   The $t_{1/2}$ value for
  H10 is shown as a representative member
  of the PseudoC.  The higher order (shorter wavelength) lines evolve
  faster, and the evolution timescale is 
  inversely proportional to the energy of the transition. Linear fits 
  to the Hydrogen Balmer features are show as a red dashed
  line for IF3 and as a purple dashed line for IF9. The $t_{1/2, \mathrm{C4170}}$ values are shown as
  light blue symbols.}
\label{fig:nature_baby}
\end{center}
\end{figure}

\begin{figure}
\begin{center}
\includegraphics[scale=0.35]{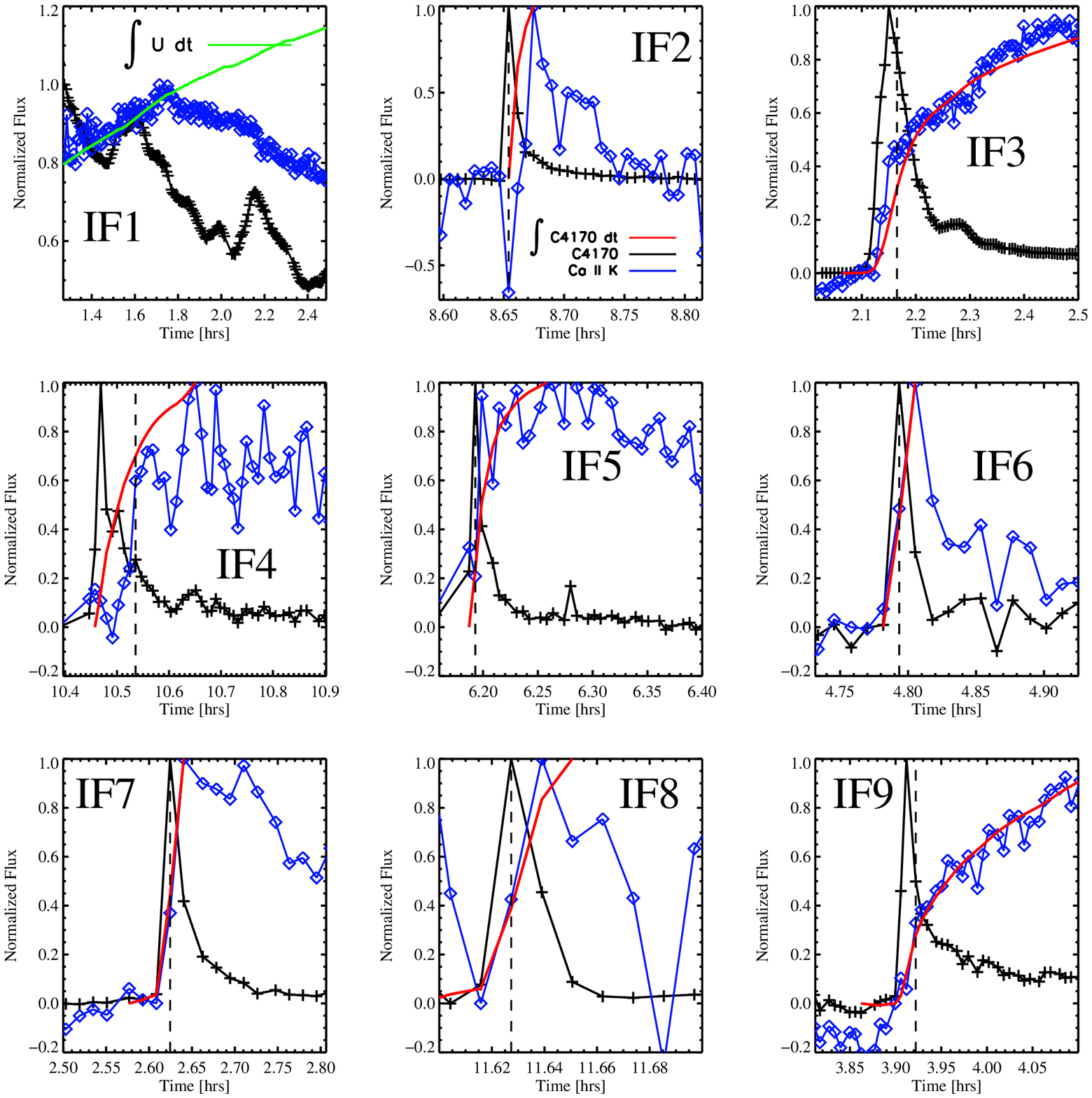}
 \caption{Ca \textsc{ii} K fluxes compared to the flux of C4170 
   and the cumulative integral of C4170 (solid red line) for the IF
   events with DIS data.  Because IF1 did not have complete spectral coverage, we
   plot the Ca \textsc{ii} K flux against the $U$-band flux and the
   cumulative integtral of the $U$-band (solid green line).
   The peak fluxes are normalized to 1.  The times of
   maximum \Hg\ line emission are indicated by vertical dashed lines. 
  If elevated, the preflare value in Ca \textsc{ii} K was subtracted from the light
  curves.  Note that the Ca \textsc{ii} K
  variations in the light curve of IF8 are not significant.  The Ca \textsc{ii} K flux follows the
   cumulative
integral of C4170 particularly well (for times before the maximum Ca
\textsc{ii} K value) for most IF events, except for IF2 and IF4.}
 \label{fig:CaIIK1}
\end{center}
\end{figure}

\begin{figure}
  \begin{center}
\includegraphics[scale=0.35]{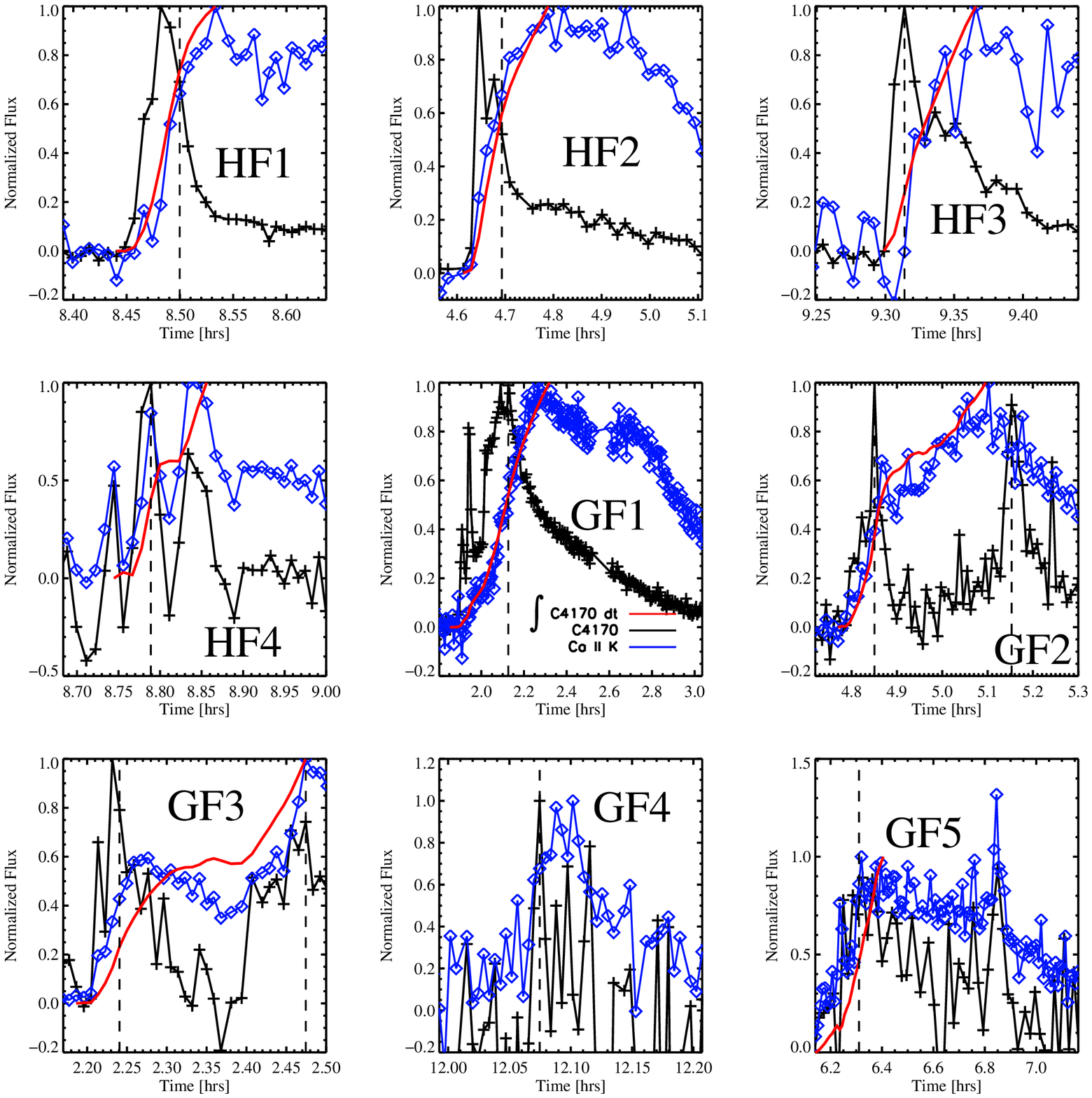}
 \caption{ Same as for Figure \ref{fig:CaIIK1} for the HF and GF
   events.   
The integral of C4170 is not shown for GF4, which has noisy continuum
data. 
The Ca \textsc{ii} K flux follows the
   cumulative
integral of C4170 particularly well (for times before the maximum of Ca \textsc{ii} K) for HF1, HF2, HF3, GF1, and GF2.}
 \label{fig:CaIIK2}
\end{center}
\end{figure}

\begin{figure}
\plottwo{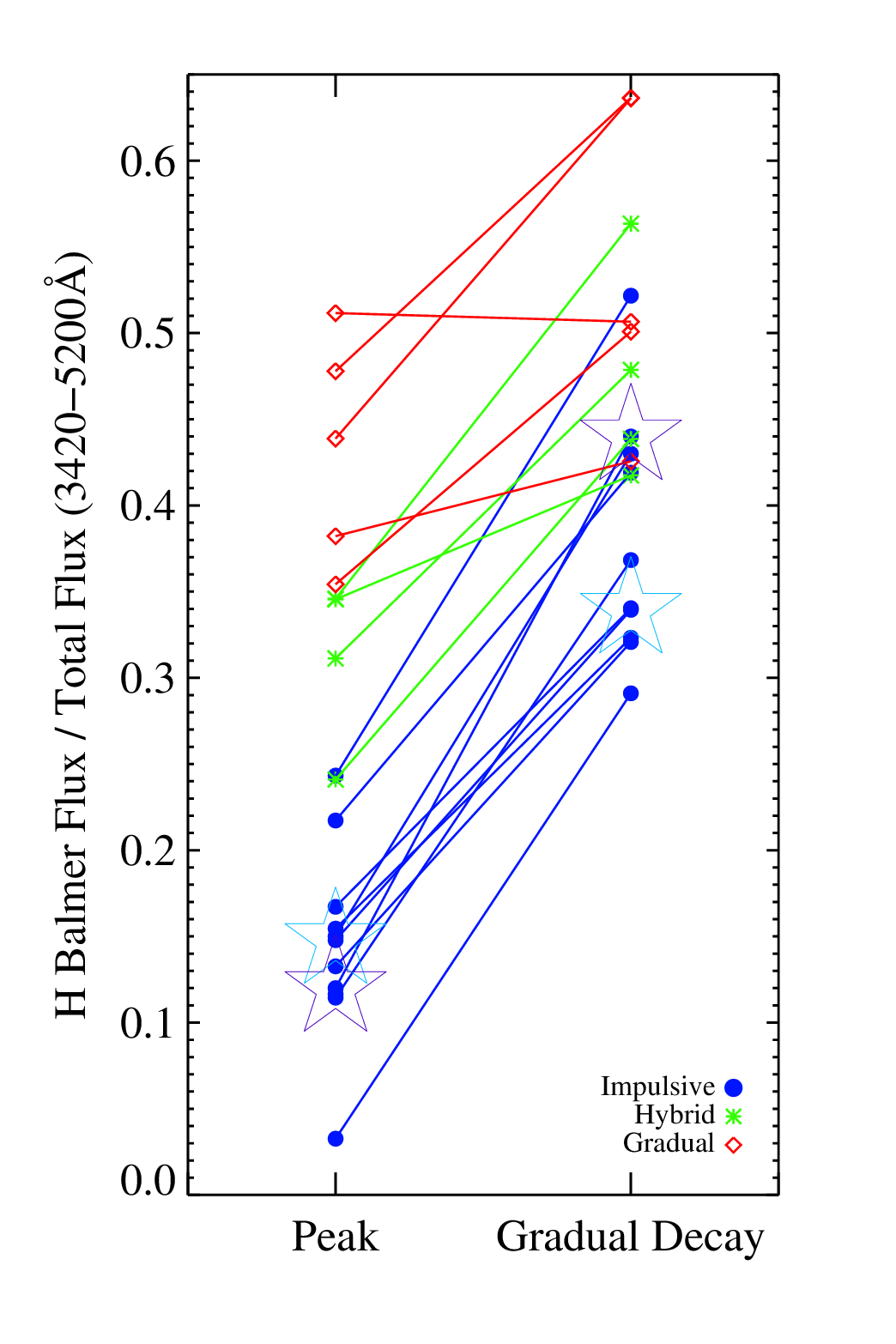}{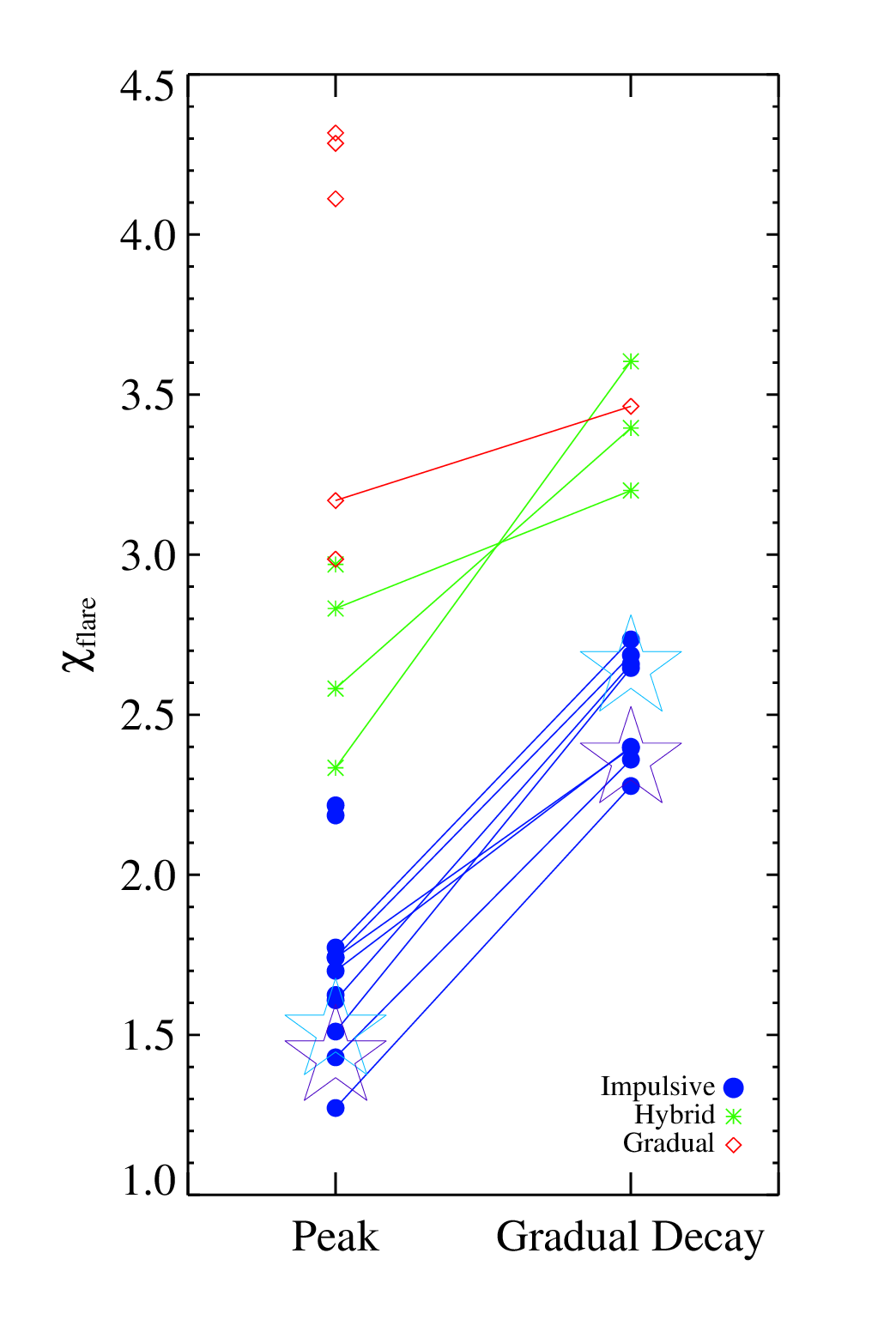}
 \caption{ (Left panel) The
   relative contribution from
   the Hydrogen Balmer component during the peak and gradual phases.  Light blue stars show the decay
 phase (S\#23\,--\,25) and secondary flare (S\#103) values for IF1; purple stars are the peak
 of the second continuum maximum ($t=1038$s, S\#40) and the beginning gradual
 phase for the Great Flare (IF0). (Right panel) The evolution of \chif\ from peak
to gradual phases, calculated at the same
times as in the panel to the left.  The \chifd\ values are not
included if they have $>$20\% errors (due to very low levels of
emission in the decay phase).  This cut excludes IF5, IF6, GF3,
IF8, GF5, GF4, and GF2. }
 \label{fig:hb_phases}
\end{figure}

\begin{figure}
  \begin{center}
\includegraphics[scale=0.4]{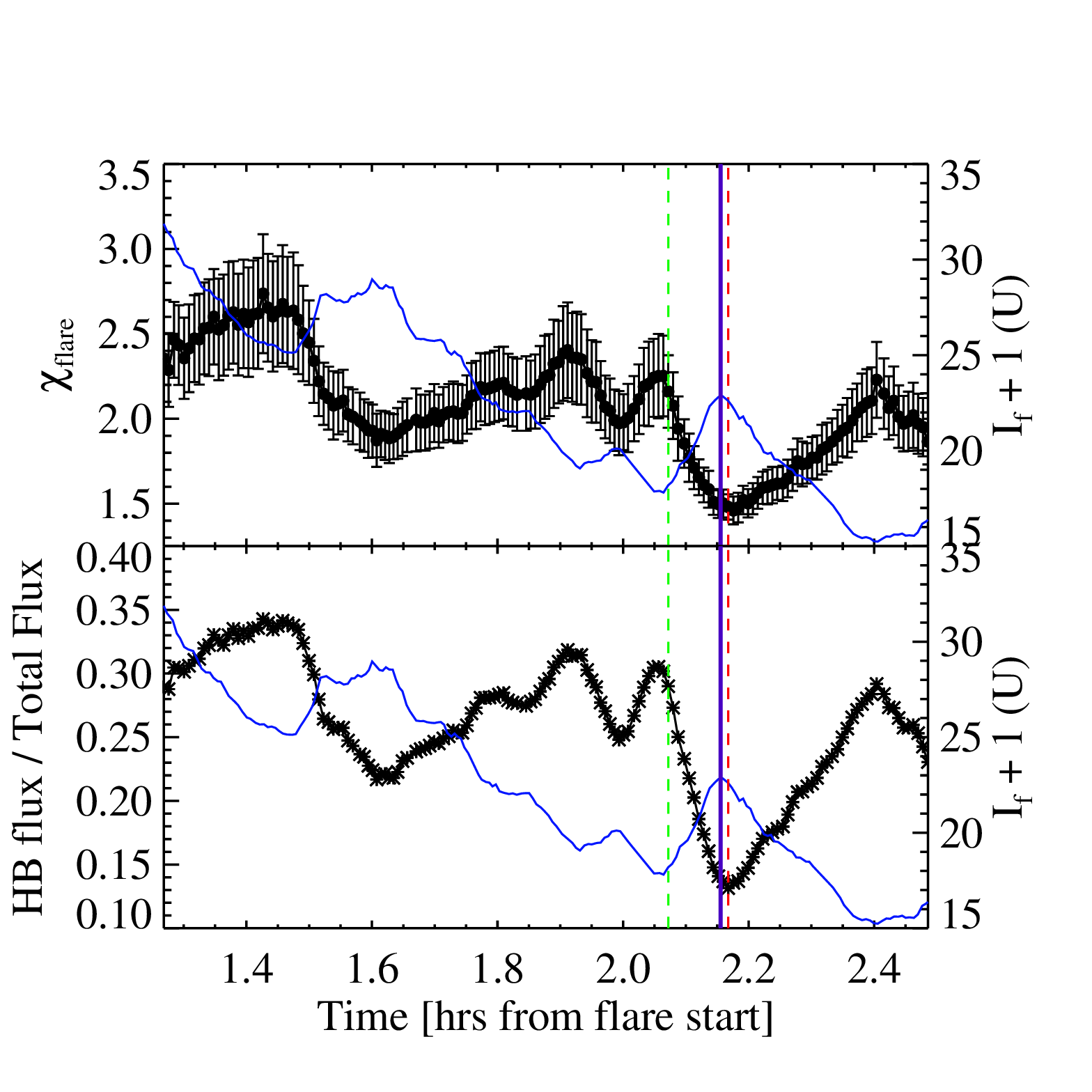}
 \caption{ (Top Panel) The \chif\ as a
   function time (black circles) and the $U$-band evolution (blue) are shown.  (Bottom Panel) The time-evolution of the ratio of HB flux to total
   flux (dark asterisks).  The ratio varies significantly and is anti-correlated with the
$U$-band, strikingly similar to the evolution of \chif. The secondary
flare, MDSF2, begins at $t\sim2.072$ hours (vertical green line), 
peaks in the $U$ band between $t=2.1523$ and $t=$2.1587 hours (vertical purple line), and
the minimum percentage of HB emission does not occur until slightly after the peak at
$t=$2.1674 hours 
(vertical red line). 
Error bars on the bottom
panel are the estimated statistical errors (not visible compared to
the symbols).  They assume that the
uncertainties in the levels of the BaC flux and PseudoC flux are given by
the difference in blackbody fits and straight line fits.  }
 \label{fig:megaflare_All1}
\end{center}
\end{figure}
\clearpage

\begin{figure}
  \begin{center}
\includegraphics[scale=0.4]{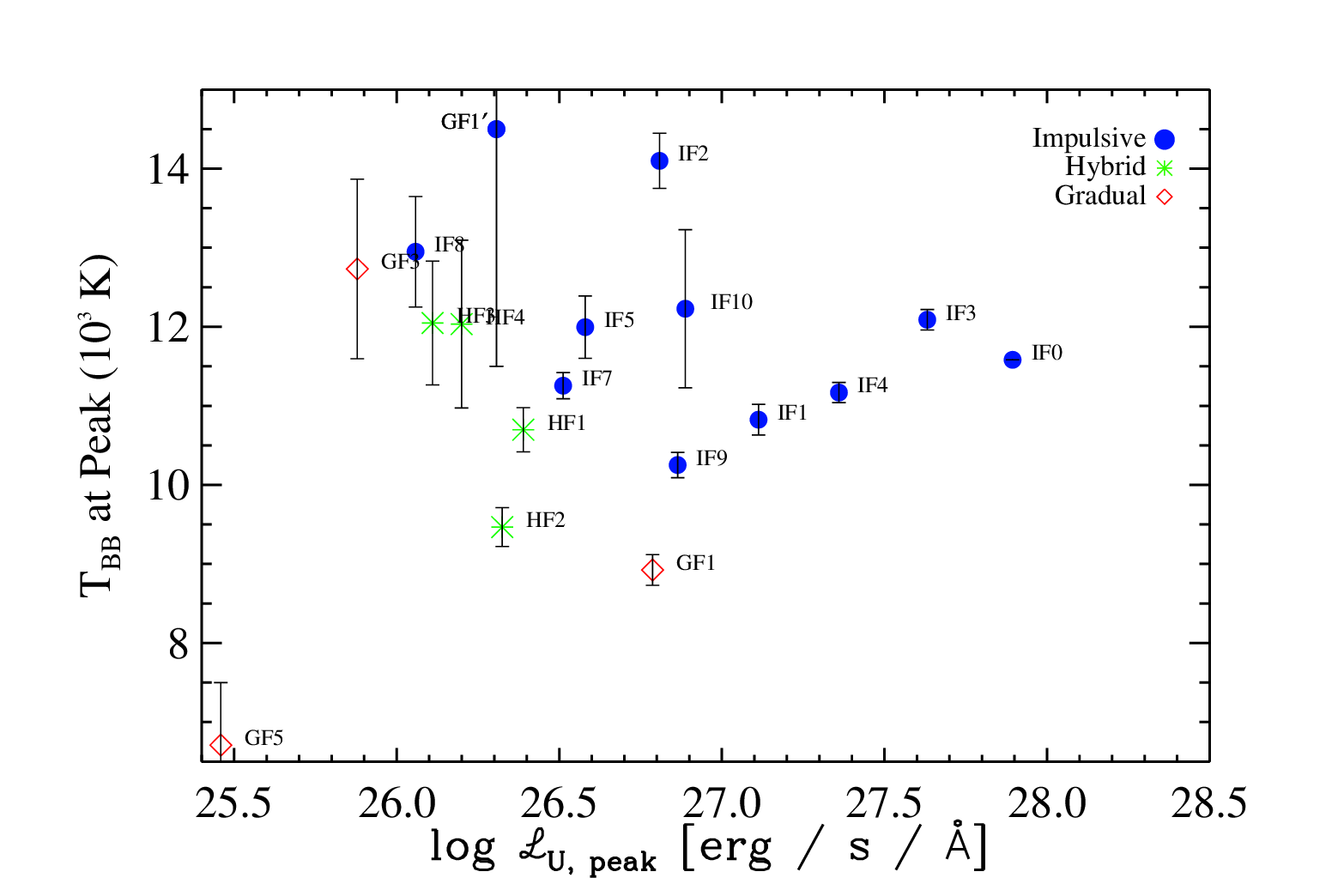}
\caption{The distribution of \TBB\ at the time of maximum continuum (C3615)
  emission vs. the peak specific $U$-band luminosity for the seventeen flares with
  well-determined color temperatures.  }
\label{fig:TXdist}
\end{center}
\end{figure}

\begin{figure}
  \begin{center}
  \includegraphics[scale=0.4]{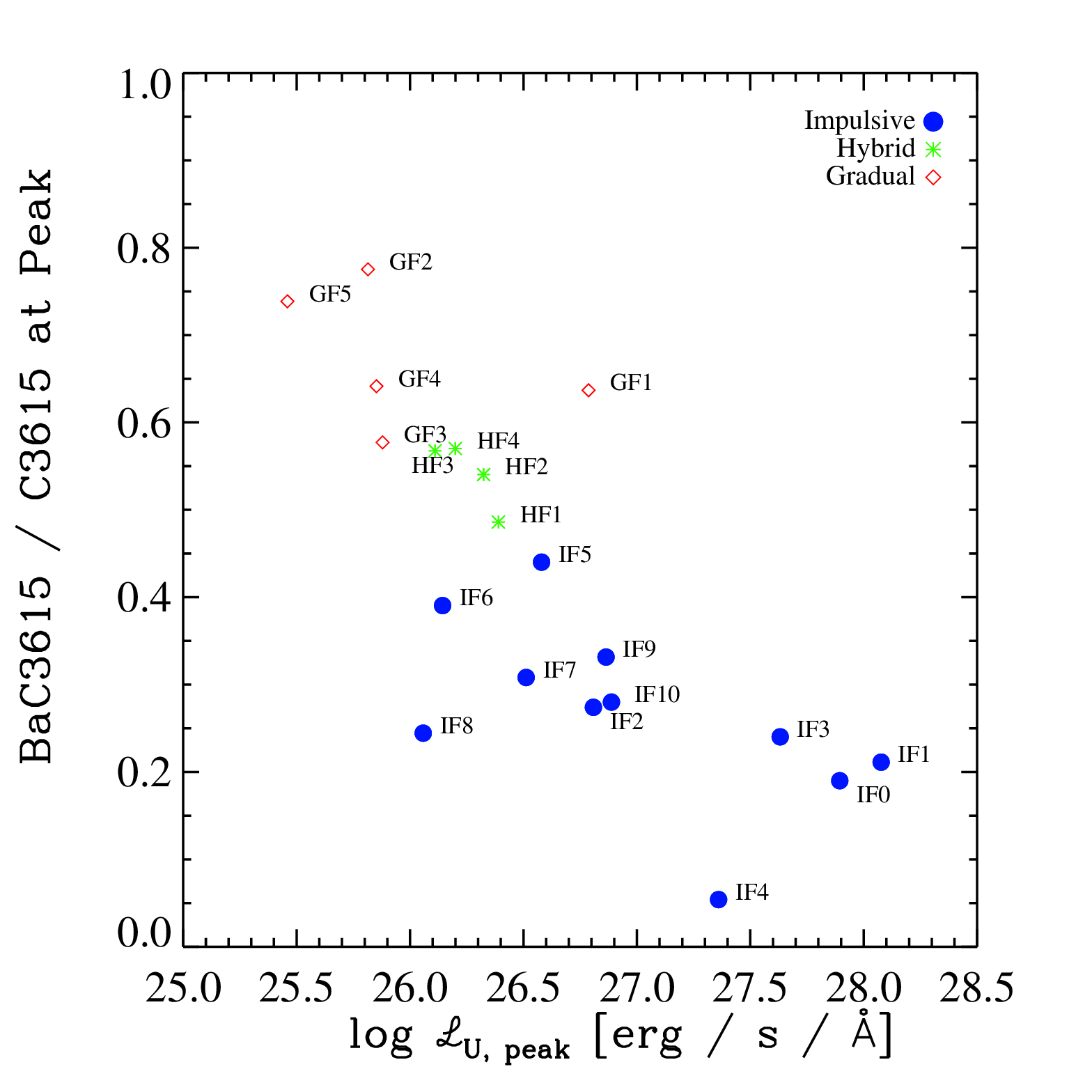}   
  \caption{The fraction of the C3615 flux that is contained in the 
    BaC3615 component.  Note the decreased contribution from
    BaC3615 in larger amplitude flares and also in more impulsive flares.   } 
  \label{fig:master0}
\end{center}
\end{figure}

\begin{figure}
\begin{center}
\includegraphics[scale=0.4]{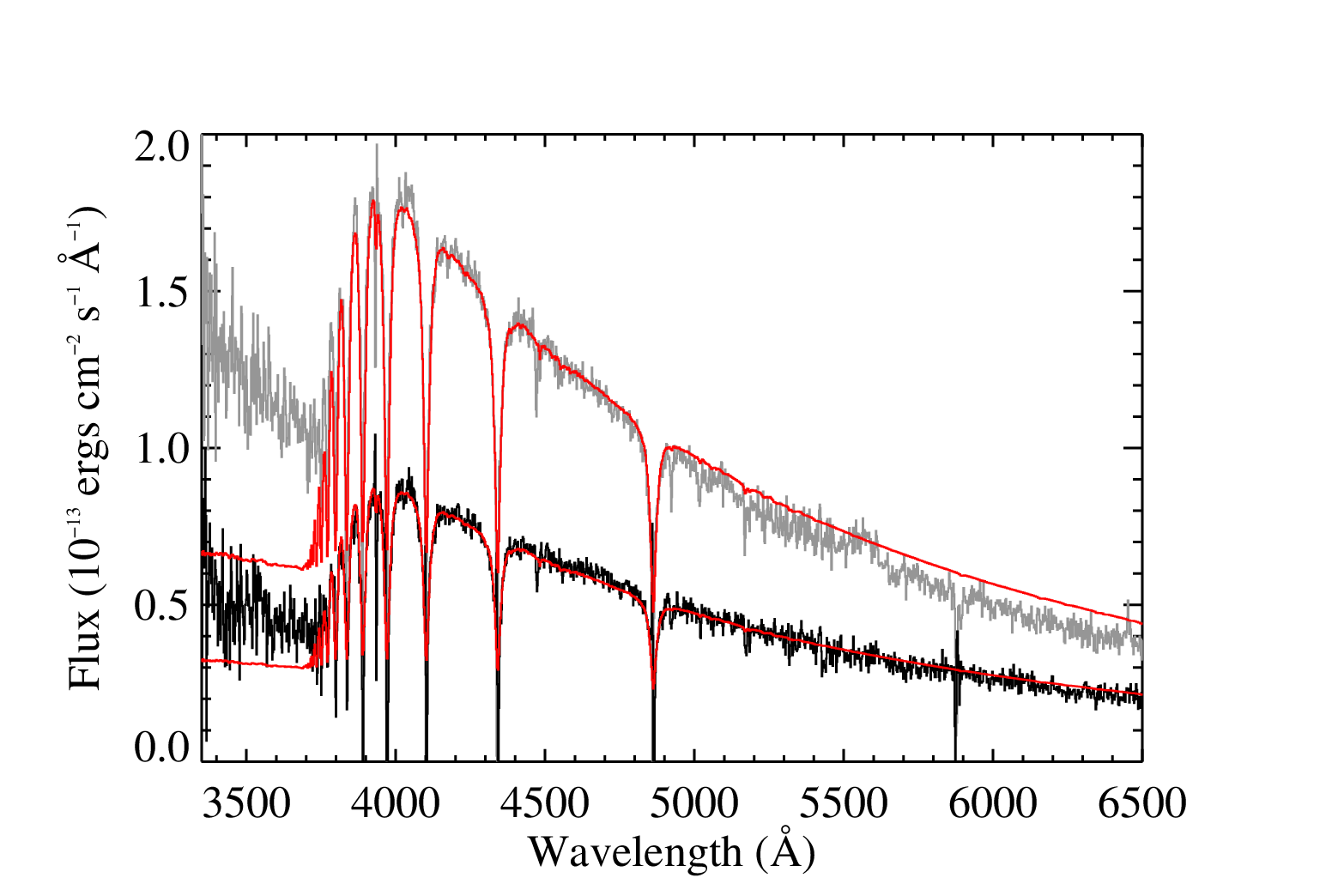}
\caption{The
  black is the newly-formed flare emission approximately half way up the rise of the
  secondary flare MDSF2 (an average spectrum of S\#108 and S\#109) and the grey is the newly-formed flare
  emission just prior to the sub-peak (S\#113).  The red is the spectrum
  of Vega with a scaling applied to match the flare spectra at C4170.   }
\label{fig:magnumopus}
\end{center}
\end{figure}

\begin{figure}
\begin{center}
\includegraphics[scale=0.4,angle=90]{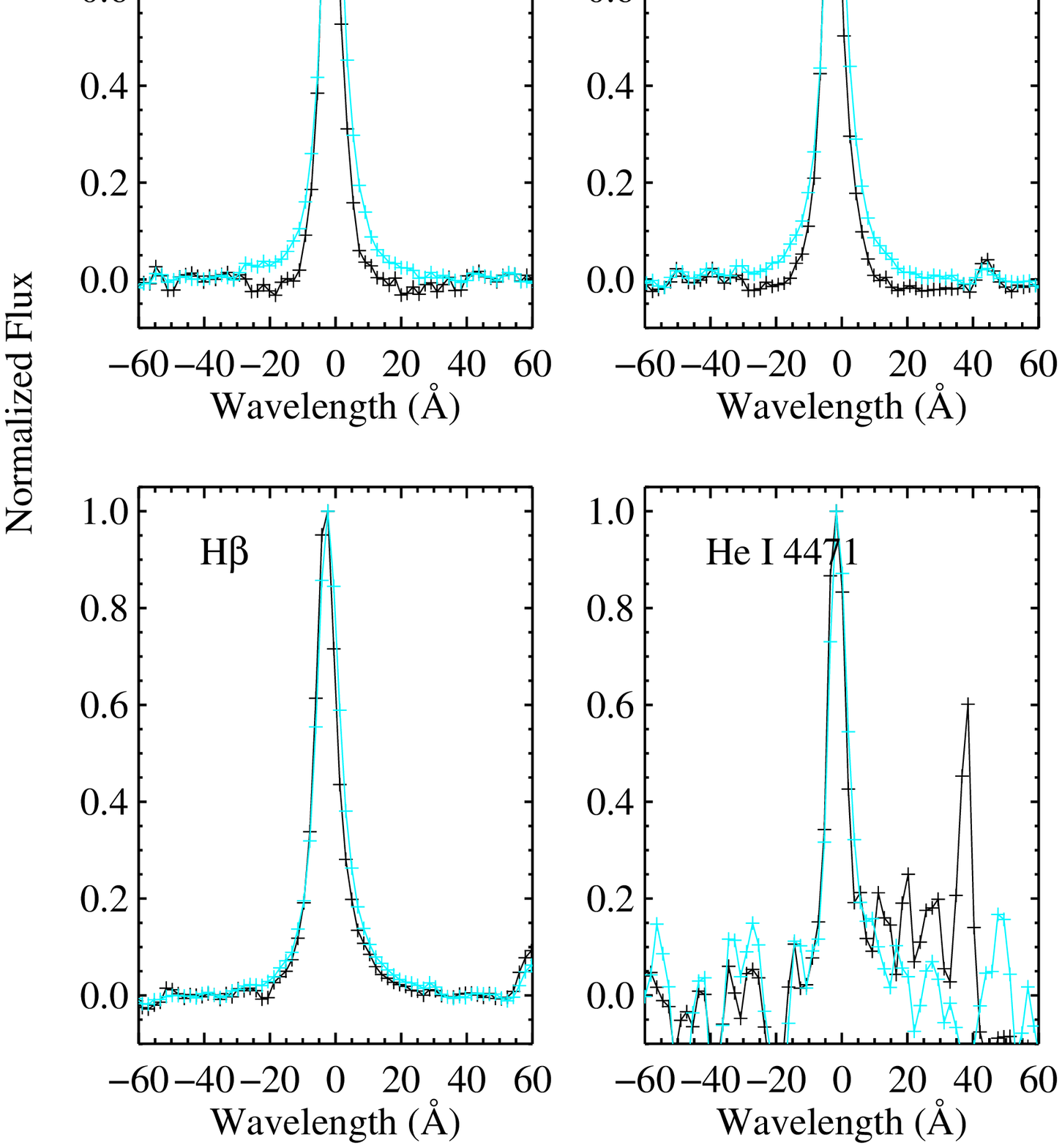}
\caption{The H$\delta$, H$\gamma$, H$\beta$ and He I $\lambda$4471 profiles of IF4 at peak
  continuum emission (S\#665, black) and peak H$\beta$ emission
  (S\#672, turquoise),
  normalized to the maxima of the line profiles.  A fit to the local
  continuum was removed before normalization.  For \Hd, \Hg, and
  \Hb\ respectively, the widths are (15.5\AA, 16.7\AA,
  21.4\AA) at maximum continuum and (21.8\AA, 22.6\AA, 23.7\AA) at maximum
  line emission.  The maximum line emission occurs at the same time
  for the three lines, \s 4.5 minutes after the maximum continuum
  emission. The feature at $+40$\AA\ in the He \textsc{i} panel is
  likely a cosmic ray.  We attribute the deficit in flux at
  $\pm$20\AA\ in the black spectrum, which is especially apparent around \Hd, to the formation of
  A-type star absorption wings during the flare.
}
\label{fig:line_width_5}
\end{center}
\end{figure}

\begin{figure}
\begin{center}
\includegraphics[scale=0.4,angle=90]{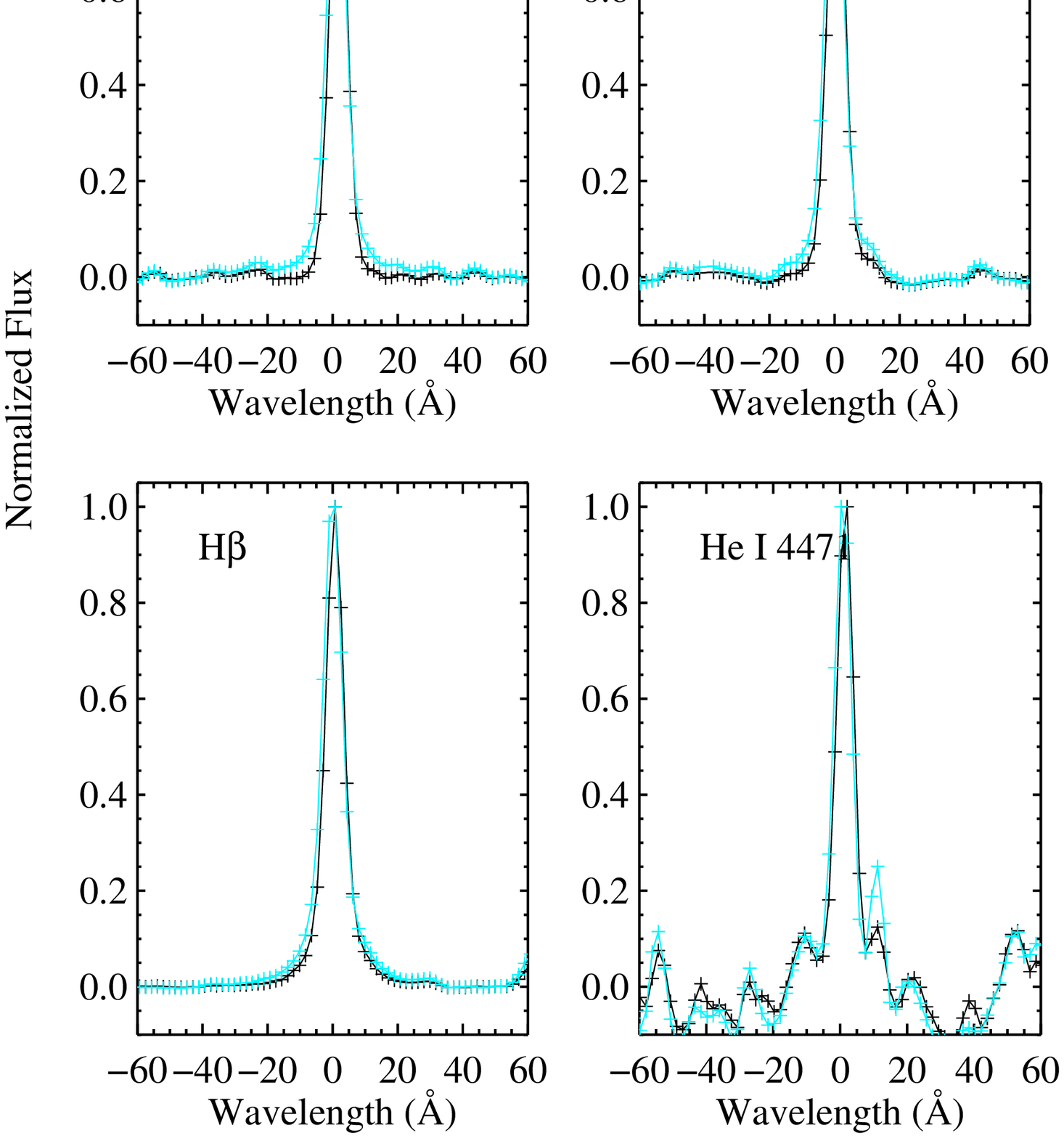}
\caption{The H$\delta$, H$\gamma$,
  H$\beta$ and He I $\lambda$4471 profiles of IF1 at peak continuum of
  MDSF2 (S\#113, black) and peak \Hb\ emission (S\#24, turquoise), normalized to the
  maxima of the line profiles.  A fit to the local
  continuum was removed before normalization.  In the peak continuum
  spectrum, the line wings are
  depressed relative to the nearby continuum level, the 
  depressions are more
  apparent for the higher order lines, and the line widths are
  smaller.  These effects are similarly seen in IF4 (Figure \ref{fig:line_width_5}). The widths of \Hd, \Hg, and
  \Hb\ are 12.2, 12.6, and 15.2\AA\ at maximum continuum emission and
  14.8, 14.9, and 18.2\AA\ at maximum line emission.
  The feature at $+10$\AA\ from Helium is likely Mg \textsc{ii} $\lambda$4481. }
\label{fig:line_width_6}
\end{center}
\end{figure}

\clearpage
\begin{figure}
\begin{center}
\includegraphics[scale=0.4]{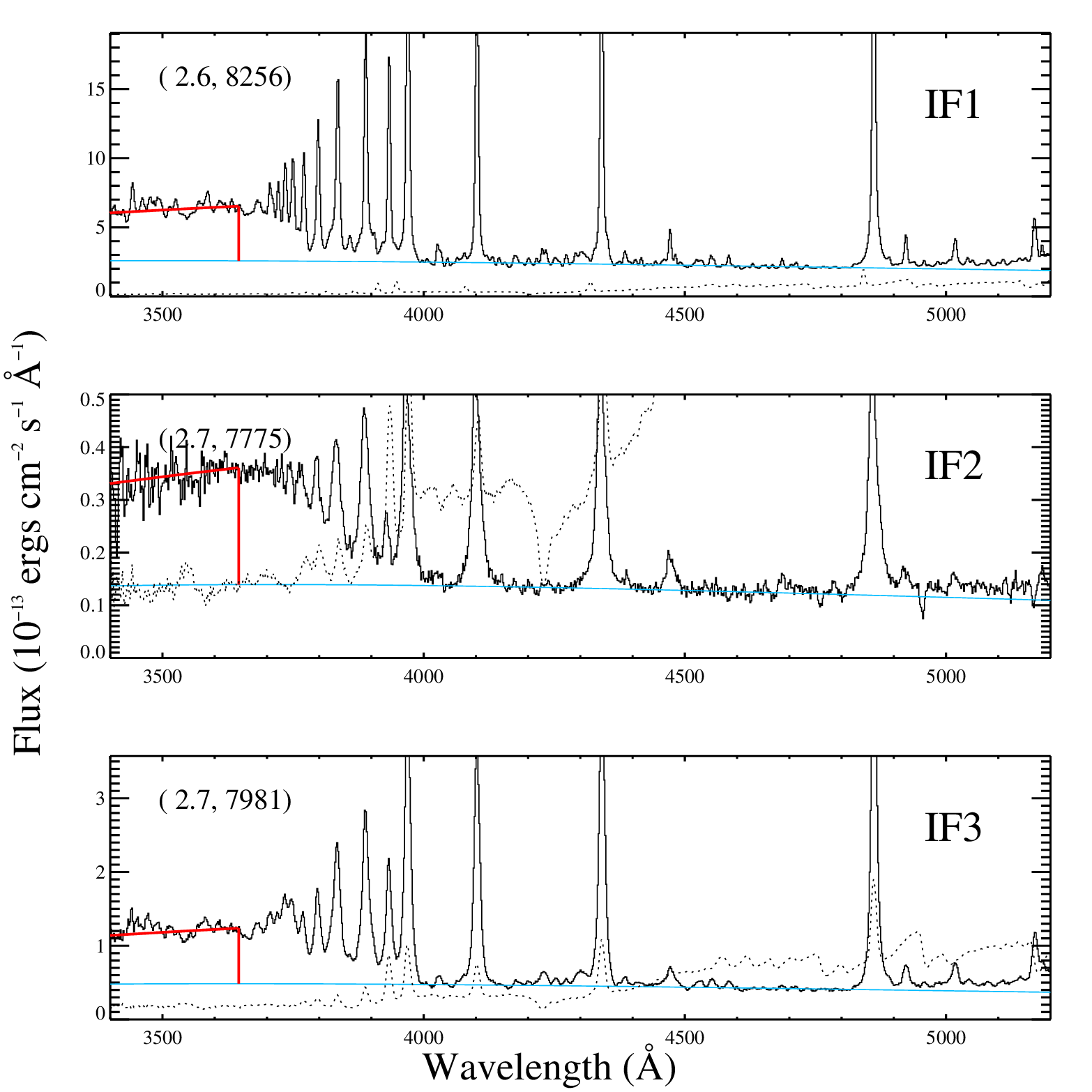}
\caption{The flare-only gradual phase emission
  (black) compared to the quiescent emission (dotted line). 
  Two-component model continuum fits are shown:  the BaCF11 model in red, the best-fit
blackbody in light blue.  In parentheses, the \chifd\ and \TBB\ are listed.  }
\label{fig:grad1}
\end{center}
\end{figure}

\begin{figure}
\begin{center}
\includegraphics[scale=0.4]{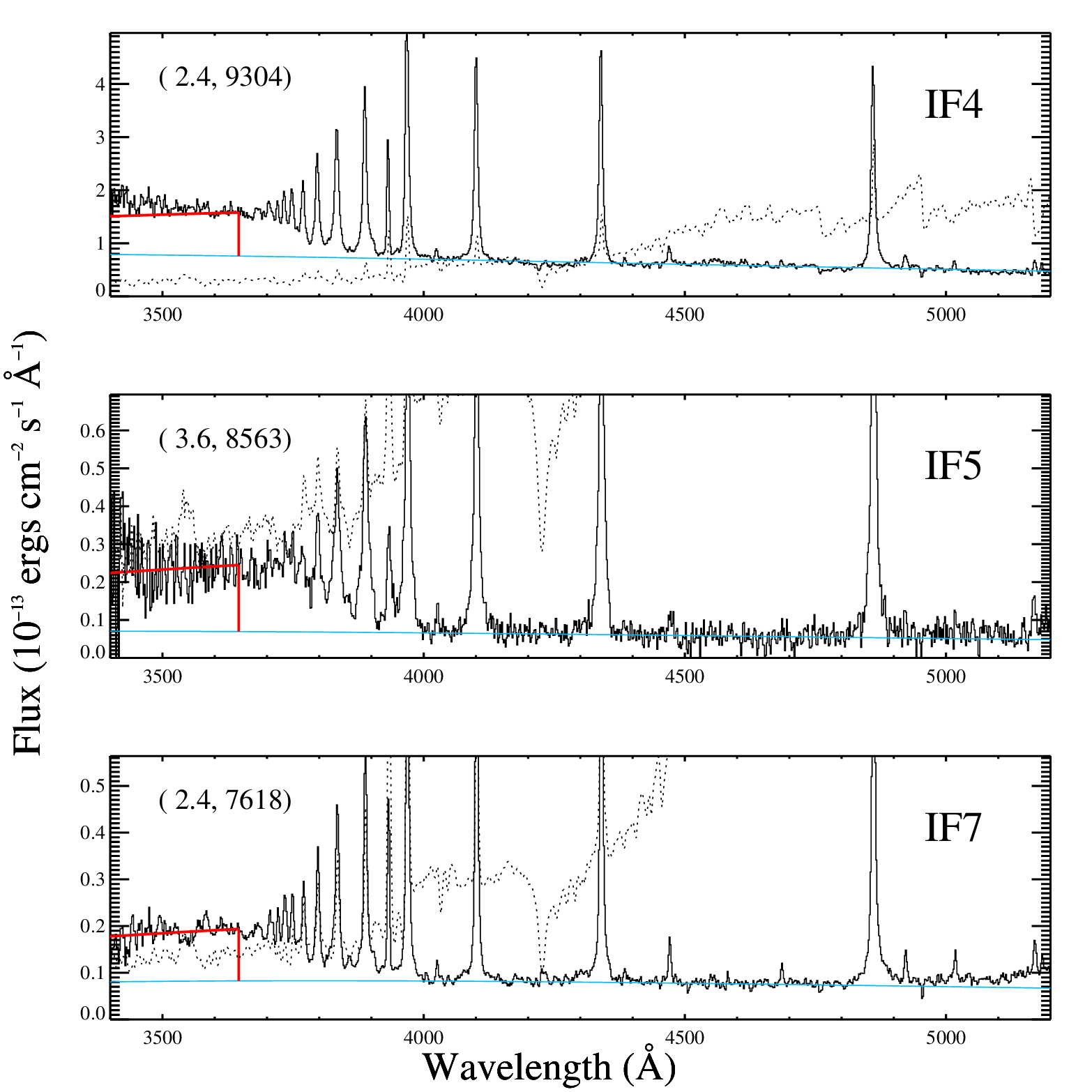}
\caption{Same as for Figure \ref{fig:grad1}.  }
\label{fig:grad2}
\end{center}
\end{figure}

\begin{figure}
\begin{center}
\includegraphics[scale=0.4]{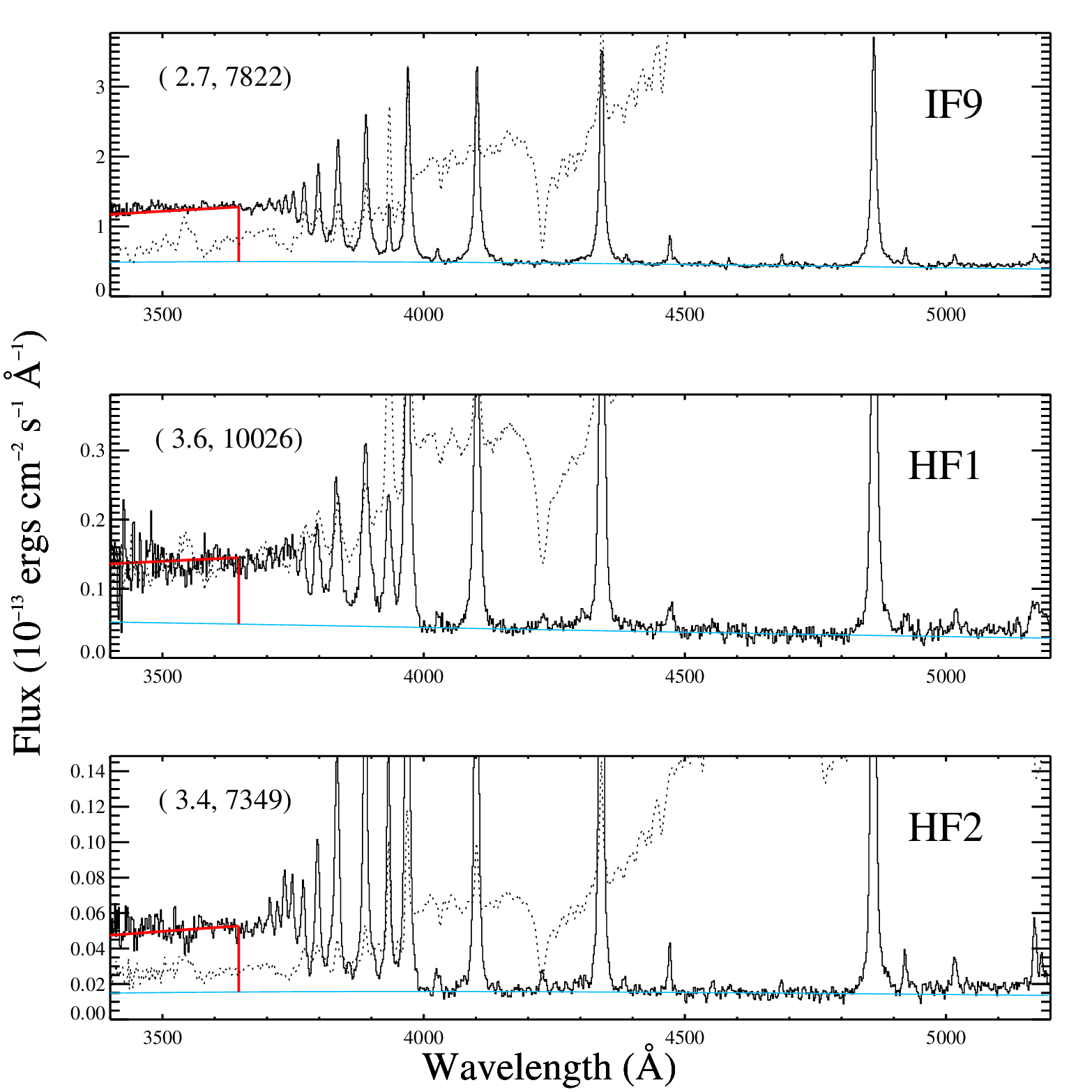}
\caption{Same as for Figure \ref{fig:grad1}.  }
\label{fig:grad3}
\end{center}
\end{figure}

\begin{figure}
\begin{center}
\includegraphics[scale=0.4]{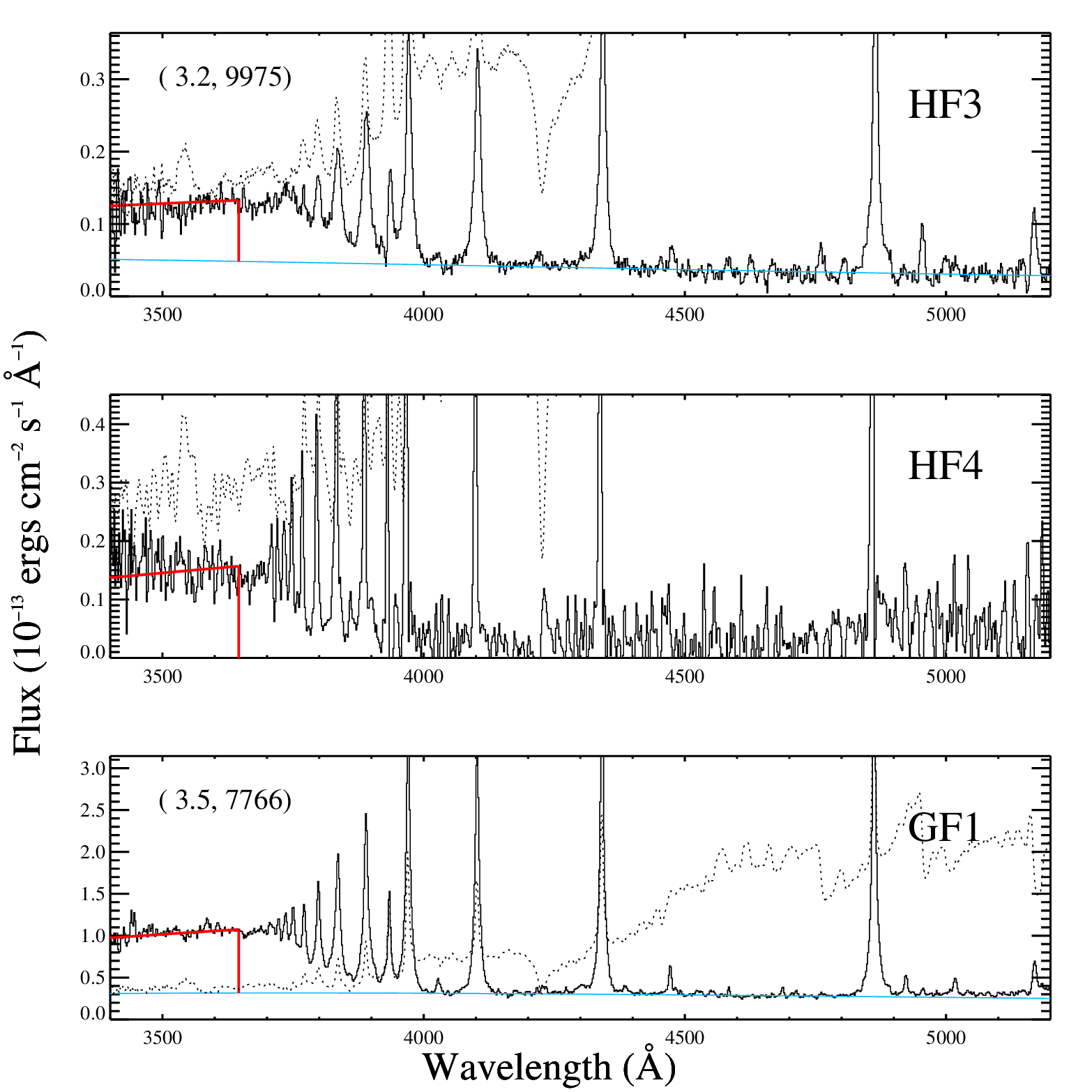}
\caption{Same as for Figure \ref{fig:grad1}.
  HF4 does not have well-determined values of \chifd\ and \TBB; only the BaCF11 model is plotted for
this flare. }
\label{fig:grad4}
\end{center}
\end{figure}

\begin{figure}
\begin{center}
\includegraphics[scale=0.4]{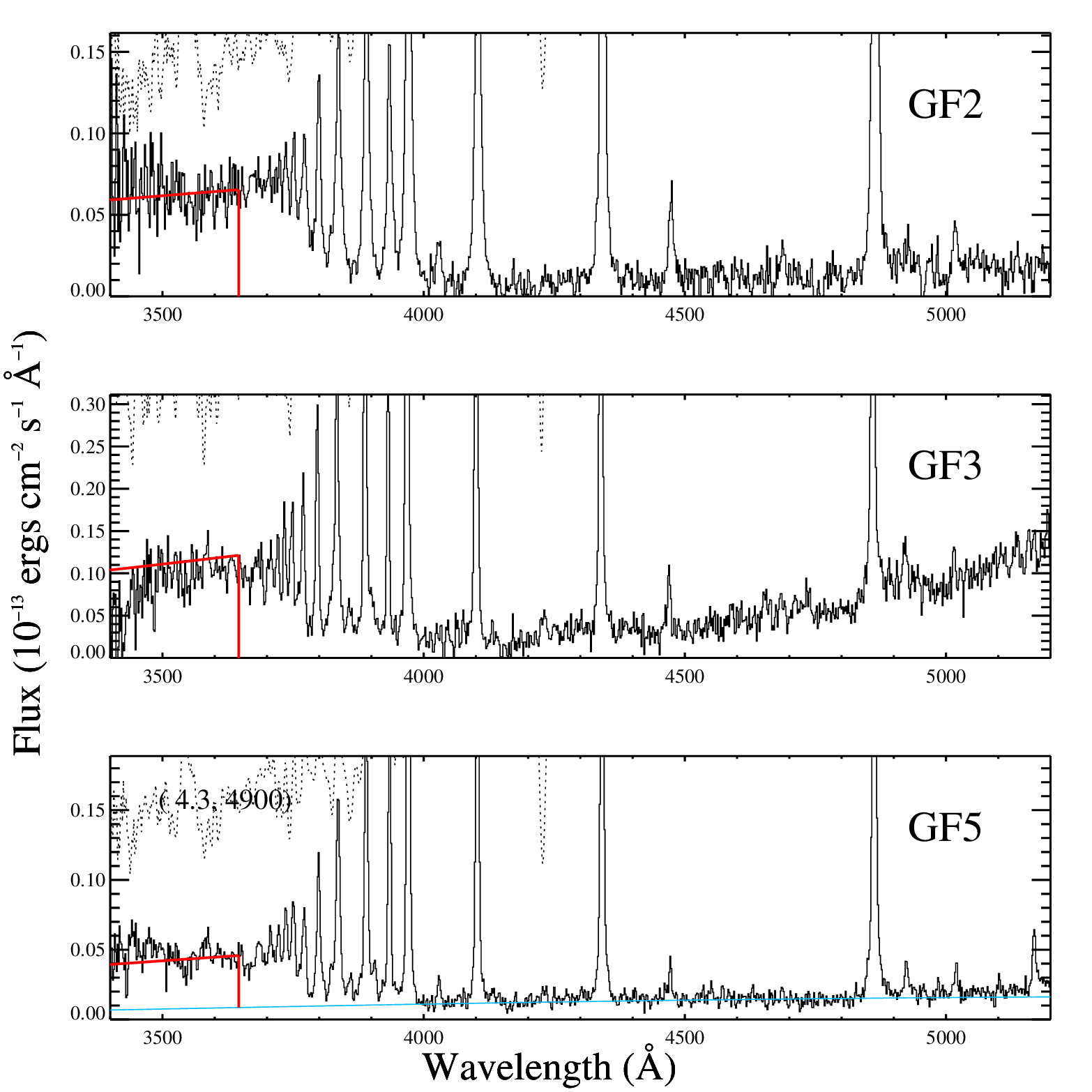}
\caption{Same as for Figure
  \ref{fig:grad1}. GF2 and GF3 do not have well-determined values of 
\chifd\ and \TBB; only the BaCF11 model is plotted for
these flares.  }
\label{fig:grad5}
\end{center}
\end{figure}

\begin{figure}
\begin{center}
\includegraphics[scale=0.4]{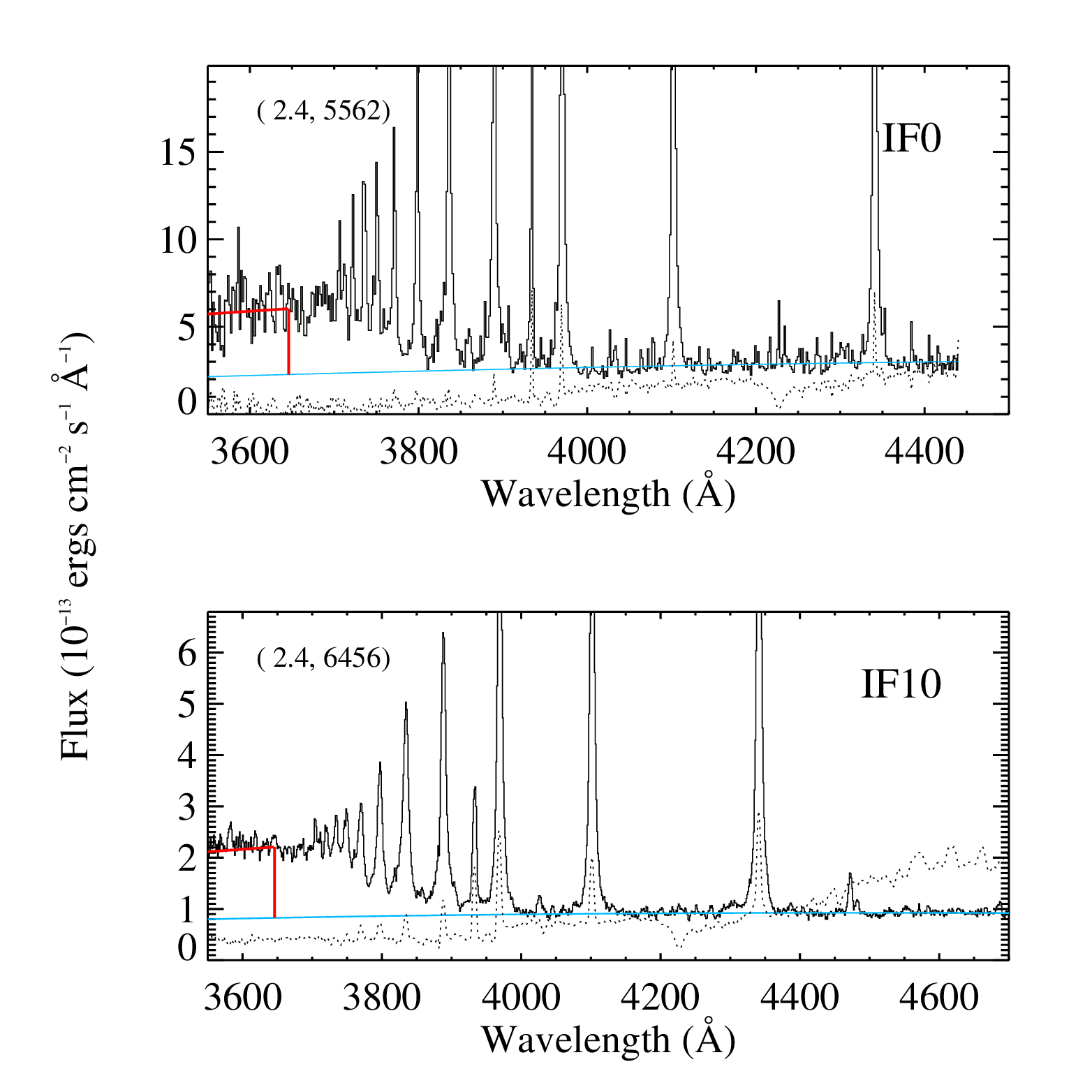}
\caption{Same as for Figure
  \ref{fig:grad1}.  Note the different wavelength ranges. For IF10, we show
  spectrum S\#32 (Table \ref{table:times}) which encompasses a secondary flare and the beginning
  of the gradual decay phase. }
\label{fig:grad1b}
\end{center}
\end{figure}

\begin{figure}
\begin{center}
\includegraphics[scale=0.4]{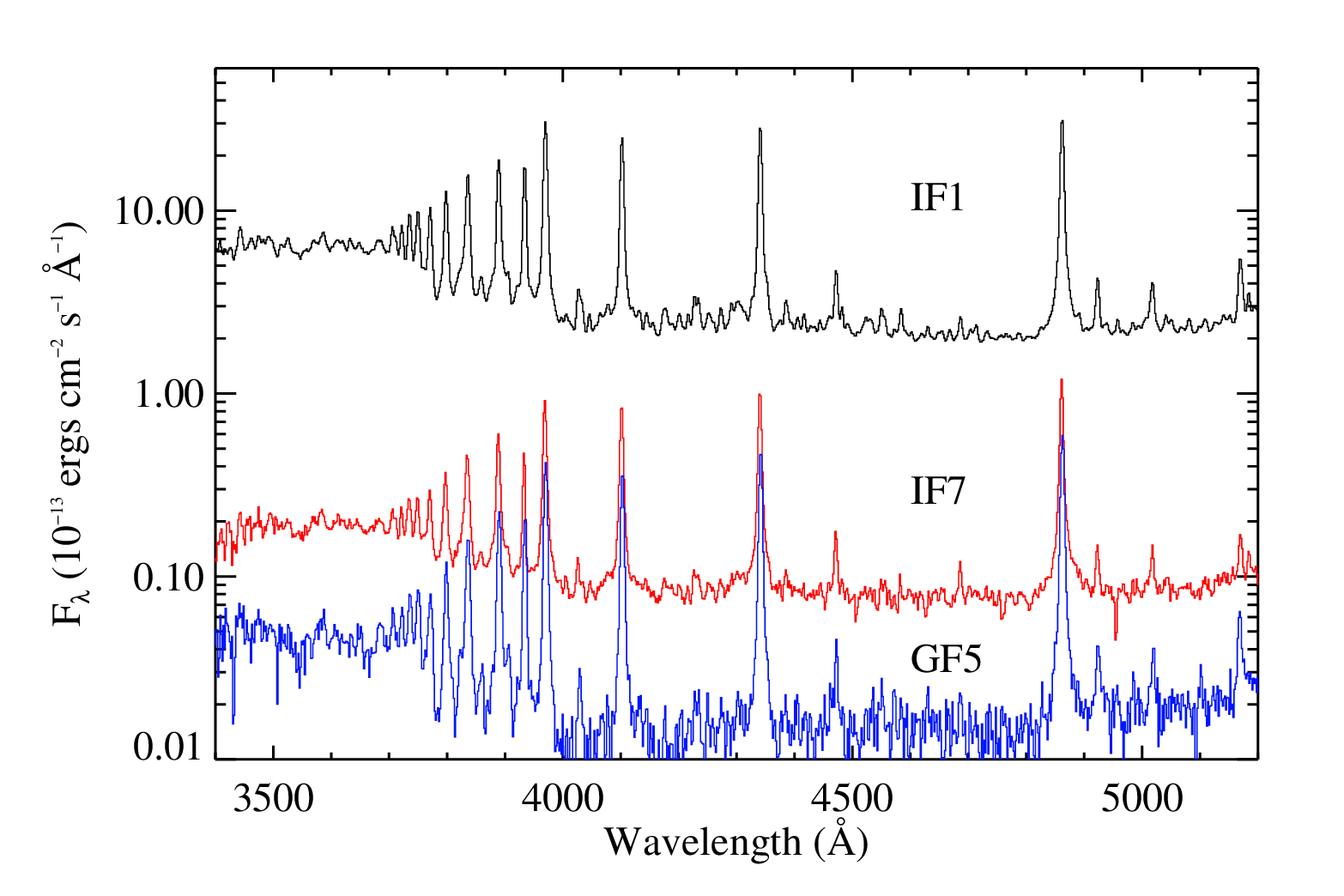}
\caption{Flare-only gradual decay phase spectra from
  IF1 (black), IF7 (red), and GF5 (blue) on YZ CMi.  These data have the same
  spectral resolution, revealing many features in common between the
  smallest and largest flares in the sample.  }
\label{fig:magnumopus2}
\end{center}
\end{figure}

\begin{figure}
\begin{center}
\includegraphics[scale=0.45]{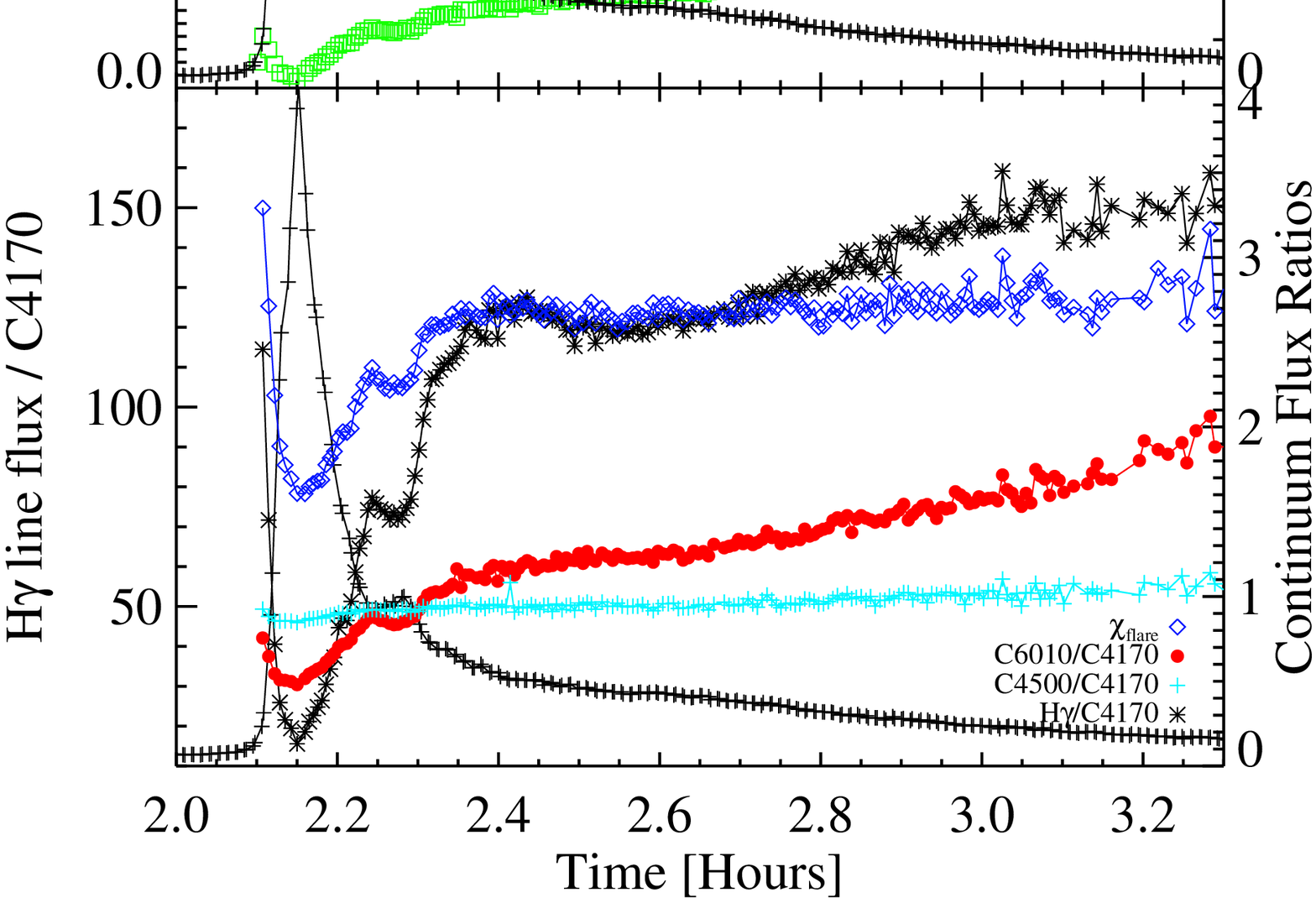} 
\caption{(Top) Time evolution
  of the Hydrogen Balmer (HB) flux ratio in the flare spectrum from
  $\lambda=3420-5200$\AA\ compared to the time evolution of Conundruum
flux ratio (green) for the IF3 event.  The SDSS $u$-band light curve is
shown in crosses (right axis).  The Conundruum flux contribution increases as the gradual decay
phase evolves. (Bottom) The evolution of the continuum flux ratios, 
  C6010/C4170 (red), C4500/C4170 (turqouise), and C3615/C4170 ($=$\chif,
  blue) for IF3 shown on right axis. The smallest value of C4500/C4170 is
  0.85 at peak (S\#31).  The line flux of \Hg\ (relative
  to C4170, asterisks, left axis) and the
  SDSS $u$-band light curve (black crosses) show much stronger reaction
  to the impulsive phase near flare peak.}
\label{fig:Conundruum}
\end{center}
\end{figure}

\begin{figure}
\includegraphics[scale=0.2]{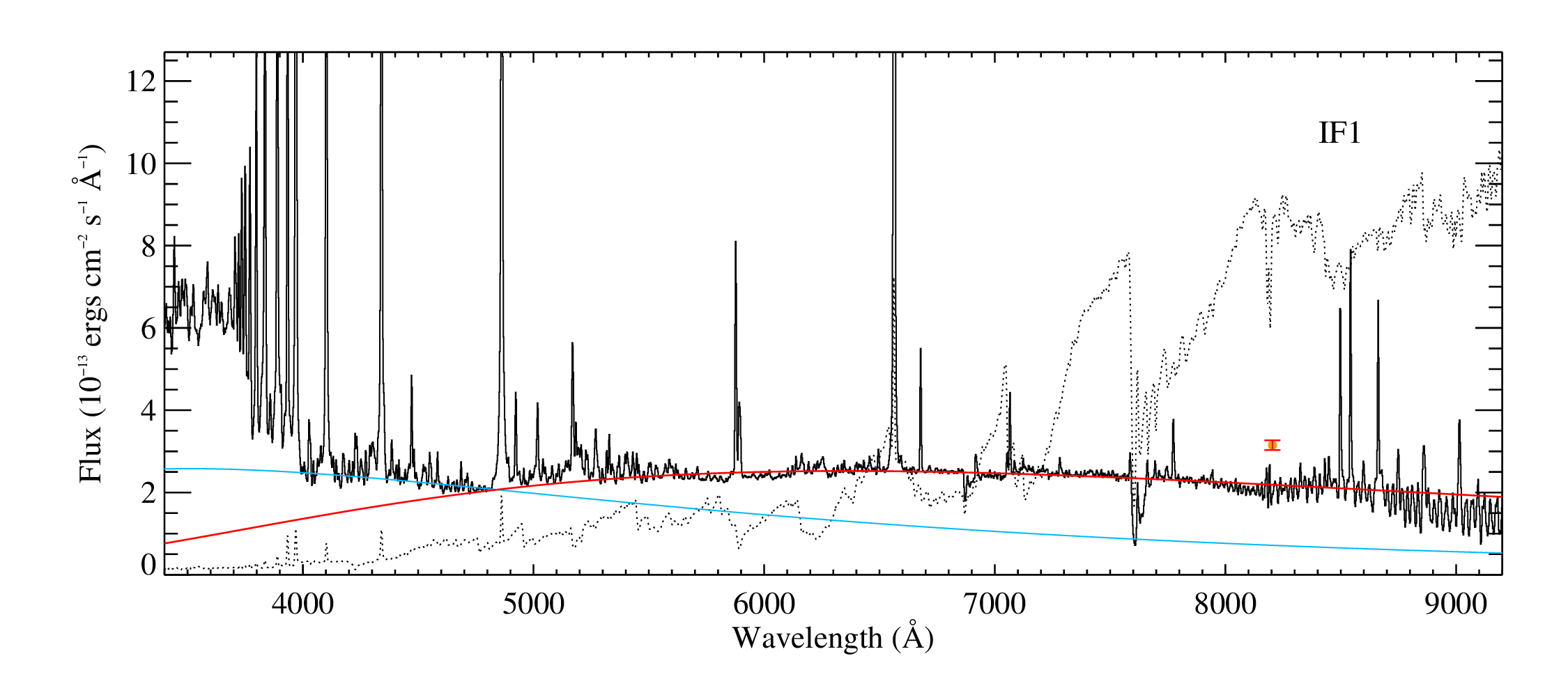}
\includegraphics[scale=0.2]{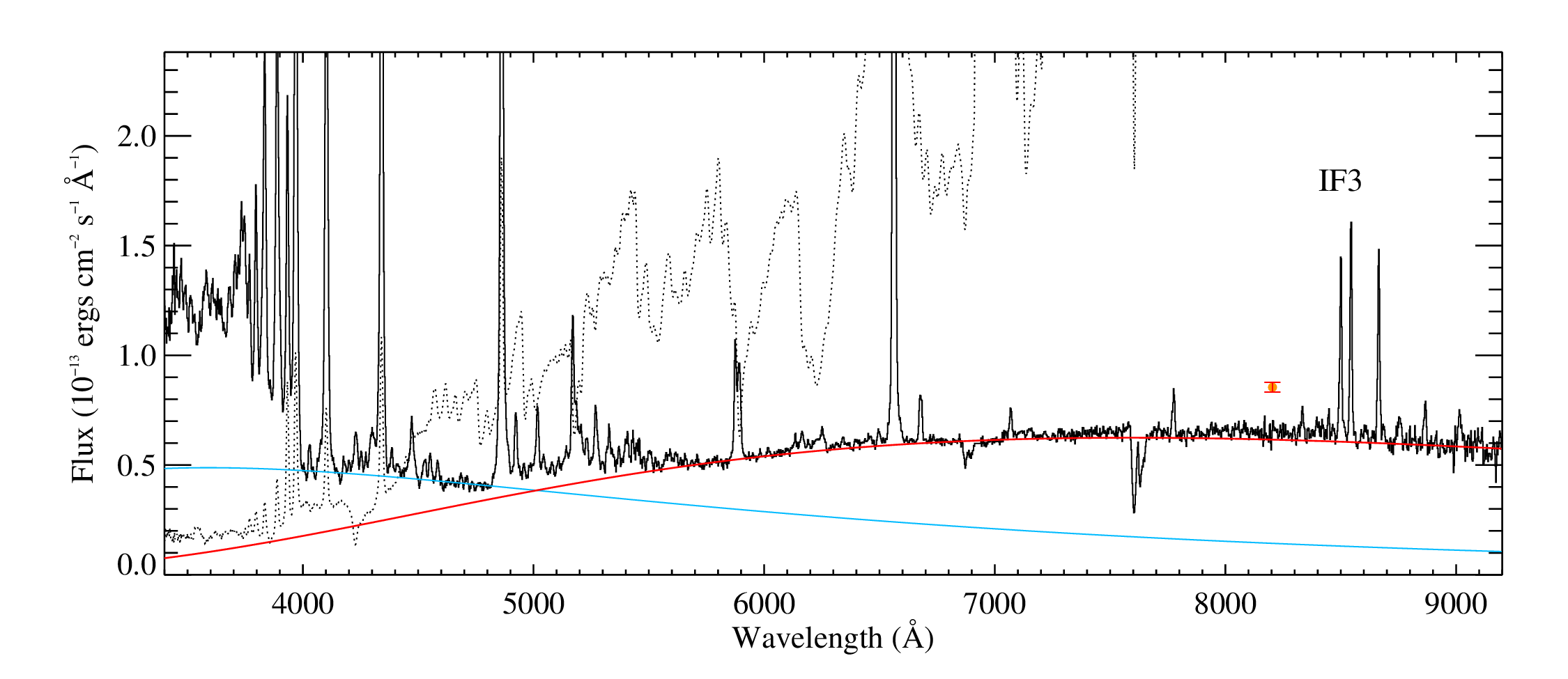}
\caption{(Left) The $\lambda=3400-9200$\AA\ flare-only spectrum
  during the gradual decay phase of IF1.  The quiescent
  spectrum is shown as a dotted line.  The blackbody fit to the blue
  optical is shown as the light blue line, the blackbody fit to the red shown
  as the red line.  The red error bars is the height of the
  Paschen jump, $F_{\lambda=8200} - F_{\lambda=8300}$ offset
  vertically from the spectrum, predicted from the RHDF11 spectrum.  (Right) The $\lambda=3400-9200$\AA\ flare-only
  spectrum during the gradual decay phase of IF3 (with symbols shown
  as in top panel).  In both panels, the Conundruum becomes apparent
  as the break in the spectrum where the blue and red lines cross at
  $\lambda \sim 5000$\AA. }
\label{fig:megaflare_fullSED}
\end{figure}

\begin{figure}
\begin{center}
\includegraphics[scale=0.45]{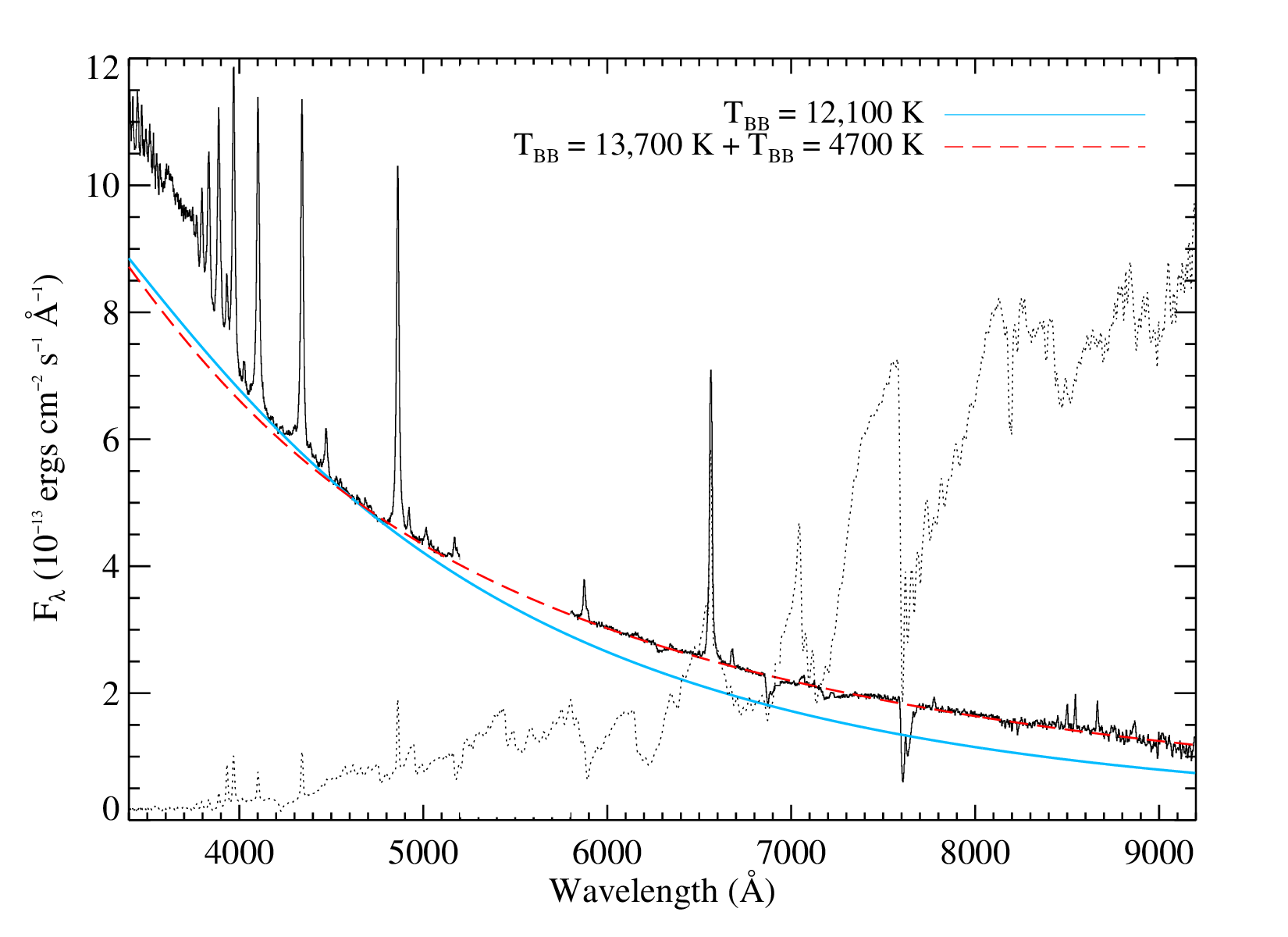}
\caption{The peak flare-only spectrum of IF3 (S\#31, black) from $\lambda=3400-9200$\AA.  The blackbody
  fit to the blue-optical zone (BW1\,--\,BW6; Table \ref{table:bbwindows}) is shown in light blue.
  A fit using the sum of two blackbody curves (BW3\,--\,BW6,
  RW1\,--\,RW3) is shown as a red dashed
  line.  The quiescent spectrum is shown as a dotted line. }
\label{fig:XXX}
\end{center}
\end{figure}

\begin{figure}
\begin{center}
\includegraphics[scale=0.4]{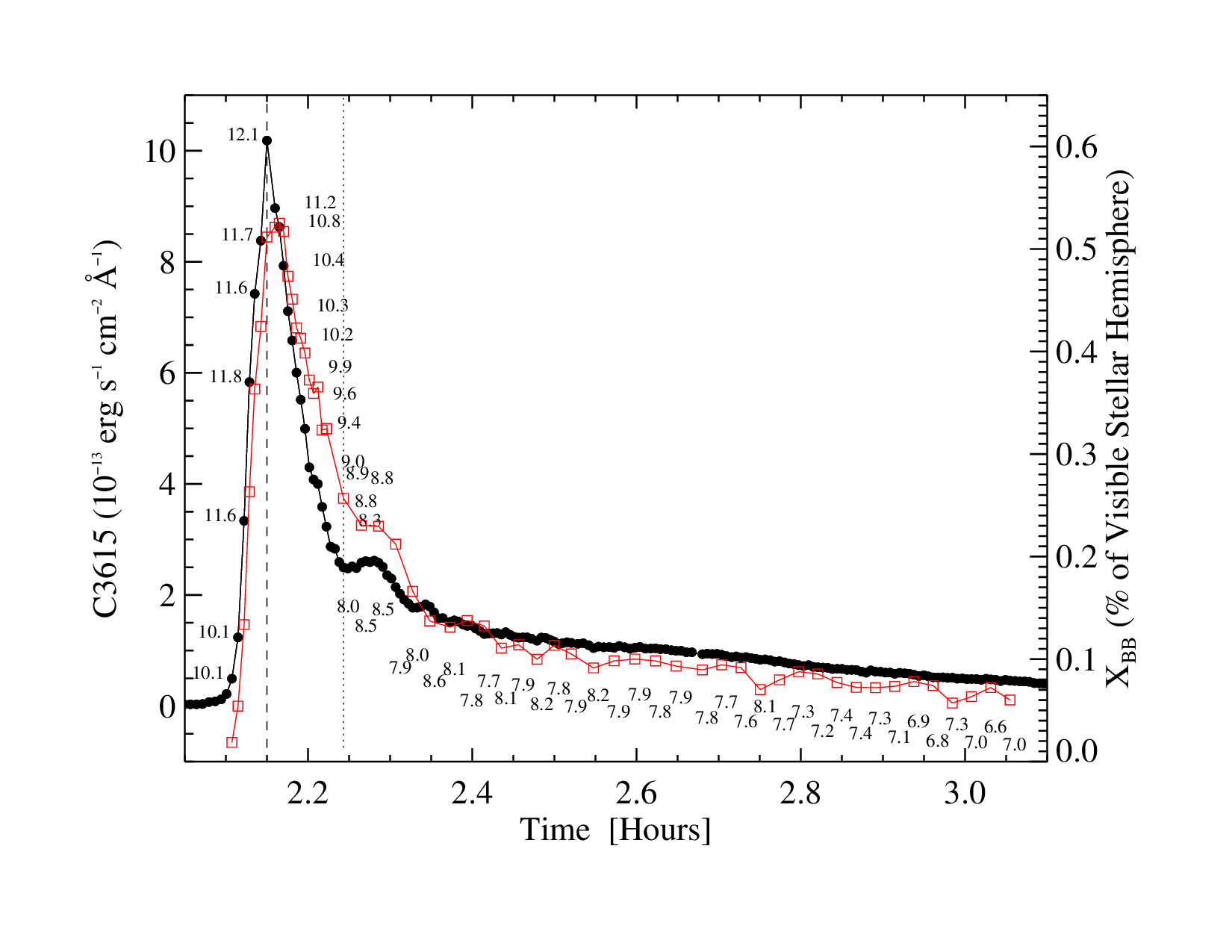}
\caption{
  The evolution of derived and measured quantities for IF3:  C3615
  (black circles), \TBB\ ($\times10^3$ K, numbered labels), and \XBB\
  (right axis, red squares).  The values of \TBB\ and \XBB\ are shown
  for every spectrum from S\#25\,--\,44 and every four spectra after
  the end of 
  the fast decay phase (indicated by a dotted line).  For clarity, the values of \TBB\ are offset to
  the left of the respective C3615 points during the rise and peak, to the
  right of the respective C3615 points during the fast decay, and centered
  on the times of the respective C3615 points during the secondary
  flare ($t\sim2.25$ hours) and the gradual decay phase.   The vertical dashed line
  indicates the time of maximum C3615, and the vertical dotted line
  indicates the end of the fast decay phase. }
\label{fig:IF3_evol}
\end{center}
\end{figure}

\begin{figure}
\begin{center}
\includegraphics[scale=0.4]{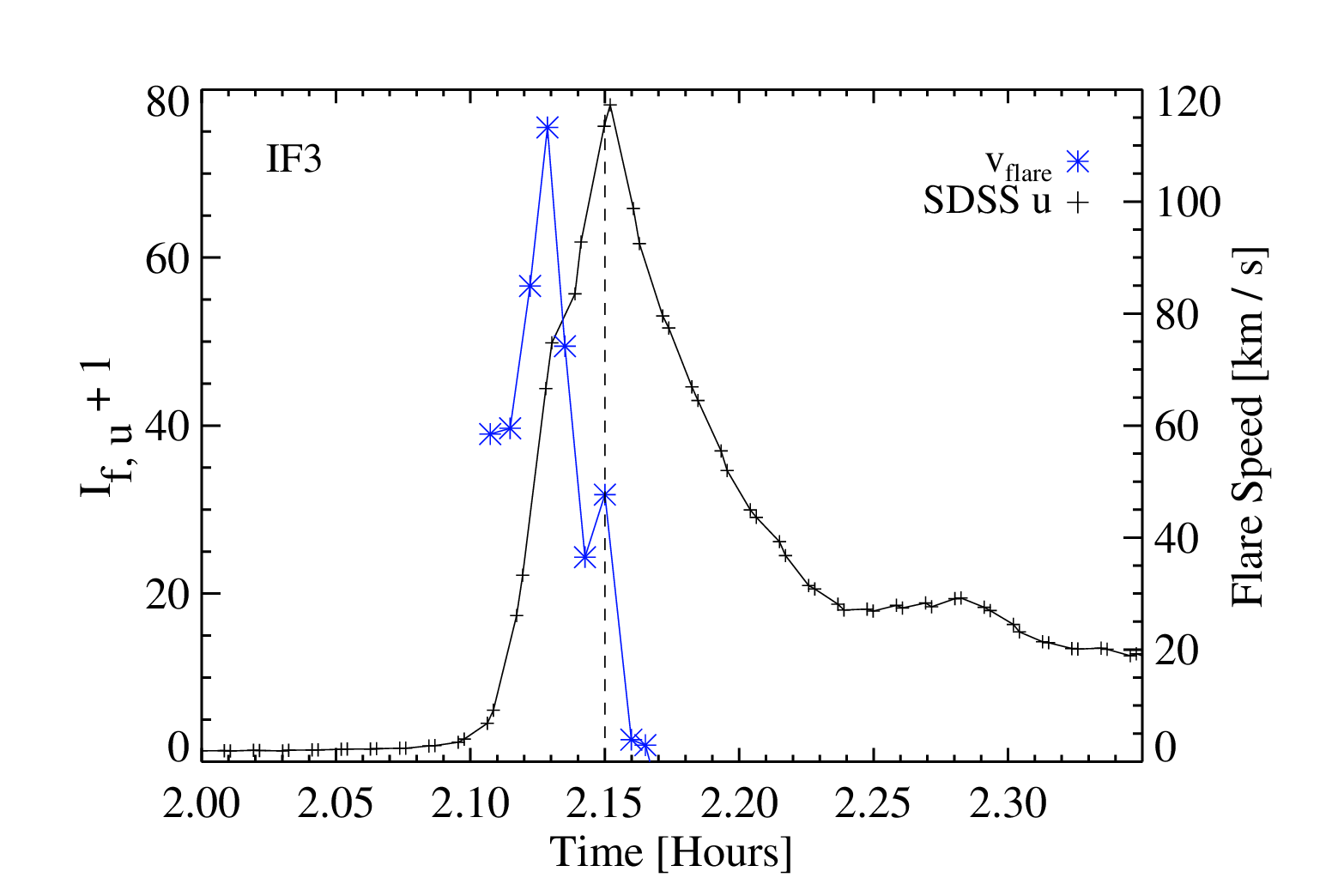}
\caption{
  SDSS $u$ band enhancements (crosses, solid black line) and derived speed
  (asterisks, solid blue line) of
  the expanding flare perimeter during the rise phase of IF3. The
  dashed vertical line indicates maximum C3615 emission.  See text
  for details. }
\label{fig:speed}
\end{center}
\end{figure}

%% file: tables_final.tex
\clearpage

\begin{deluxetable}{lllllllll}
\tabletypesize{\scriptsize}
\tablewidth{0pt}
\tablecaption{Target Star Basic Parameters}
\tablehead{
\colhead{Star} &
\colhead{Spectral Type} &
\colhead{U} &
\colhead{B} &
\colhead{V} &
\colhead{dist [pc]}  &
\colhead{R [R$_{\mathrm{Sun}}$]} &
\colhead{log $L_U$ [erg s$^{-1}$]} &
\colhead{$F_{U, q}$$^{\dagger\dagger}$}
}
\startdata
YZ CMi (Gl 285) & dM4.5e & 13.77 & 12.80 & 11.19 & 5.97 & 0.30 & 28.6
& 1.35 \\
EV Lac (Gl 873) & dM3.5e & 12.96 & 11.86 & 10.28  & 5.05 & 0.36 & 28.8
& 2.85\\
AD Leo (Gl 388) & dM3e  & 11.91 & 10.85 & 9.32 & 4.89 & 0.43 & 29.2 & 7.49 \\ 
EQ Peg A (Gl 896 A) & dM3.5e & 13.18 & 12.12 & 10.41 & 6.34 & 0.36 &
28.9 & 2.33\\
GJ 1243 & dM4e & $\sim$15.5$^\dagger$ & 14.47 & 12.83 & 12  & 0.36 & $\sim$28.5$^\dagger$
& 0.27 
\enddata
\tablecomments{Magnitudes are obtained from \cite{Reid2005};
  magnitudes for GJ 1243 are obtained from \cite{Reid2004}.  $^\dagger$The
  $U$-band magnitude and luminosity
for GJ 1243 assumes that $U-B= 1$. $L_U$ is the quiescent $U$-band
luminosity and F$_{U, q}$ is the quiescent $U$-band flux
density. $^{\dagger\dagger}$Units of quiescent $U$-band flux density
are 10$^{-14}$ erg s$^{-1}$ cm$^{-2}$ \AA$^{-1}$.}
\label{table:starssummary}
\end{deluxetable}

\clearpage
\begin{deluxetable}{clllll}
\tabletypesize{\scriptsize}
\tablewidth{0pt}
\rotate
\tablecaption{Observing Log}
\tablehead{
\colhead{UT Date [MJD]} &
\colhead{Time [Hrs]} &
\colhead{Exp time [s] (\# exp\tablenotemark{*})} &
\colhead{Slit [\arcsec] (Resolution\tablenotemark{**} [\AA])} &
\colhead{Photometry} & 
\colhead{Flares}
}
\startdata
YZ CMi & & & & &\\
\hline
\hline

 2009 Jan 16 (54847) & 1.3  & 10 (157B, 163T\tablenotemark{***}, 0R) & 1.5 (6, 13, 6.4)
 & 1m ($U$) & IF1\\
 2009 Jan 26 (54857) & 6.183 & 20\,--\,45, 60, 300 (485B, 524T, 0R) & 1.5
 (5.7, 11, 6) & n/a & GF5 \\
 2010 Feb 14 (55241) & 4.856 & 5, 10, 30, 45, 150 (332B, 424T, 83R) &
 1.5 (5.9, 12, 6.1) & 1m ($U$) & IF7\\
 2010 Dec 11 (55541) & 5.028 & 18,30 (438B, 468T, 0R) & 5 (17, 30,
 12.5) & 1m ($U$) & IF8 \\
 2010 Dec 13 (55543) & 5.191 & 15, 18, 90 (574B, 606T, 0R) & 5 (9.6,
 30, 11.4)& 1m ($U$) & HF3 \\
 2011 Feb 08 (55600) & 5.580 & 8, 12, 15, 20 (610B, 703T, 0R) & 5
 (18, 31, 13) & 1m ($U$), 0.5m ($g$) & IF2, HF1 \\
 2011 Feb 24 (55616) & 3.988 & 7, 10, 12, 15 (437B, 462T, 0R) & 5 (19,
 32, 13.5) & 0.5m ($ugri$) & IF3\\
  2011 Mar 02 (55622) & 3.921 & 20 (235B, 403T, 32R) & 1.5 (7.3, 19,
  7.4) & 1m ($U$) & GF2\\ 
\hline
\hline
EV Lac & & & & &\\
\hline
\hline
2009 Oct 10 (55114) & 8.202  & 6, 15, 20, 30 (572B, 626T, 0R)  & 1.5
(5.5, 12, 5.5) & 1m ($U$) 0.5m ($u$) & IF6, GF3 \\
2010 Oct 11 (55480)  & 4.868  & 3, 5, 9, 10 (621B, 686T, 108R) &1.5
(7.3, 18, 6.5-7) & 1m ($U$) 0.5m ($ugr$) & IF5, GF1\\
\hline
\hline
AD Leo & & & & &\\
\hline
\hline
2010 Apr 03 (55289) & 3.973 & 1, 2, 10 (431B, 581T, 145R) & 1.5 (6.5,
15, 6-6.5) & 0.5m ($g$) & IF9\\
2011 Feb 08 (55600) & 2.856 & 2, 5 (409B, 446T, 109R) & 5  (18, 31,
13)& 0.5m ($u$) & GF4\\
\hline
\hline
EQ Peg A & & & & &\\
\hline
\hline
2008 Oct 01 (54740) &  9.215 & 20, 30, 40 (722B, 786T, 0R) &1.5 (6,
12, 6.1)  & 1m ($U$) & HF4, IF4 \\
\hline
\hline
GJ 1243 & & & & &\\
\hline
\hline
2011 Oct 20 (55854) & 2.830 & 45, 60 (135B, 159T, 0R) & 5 (20, 31, 14)
& n/a & HF2
\enddata
\tablecomments{Time is the total monitoring time for the night.}
\tablenotetext{*}{
  $n$B indicate the number of exposures of sufficient quality for blue
  analysis; $n$T indicate total number of exposures
  obtained; $n$R indicate the number of short exposures obtained at a
  lower cadence.  See Section \ref{sec:spectra} for more information.
  The data from 2010 Feb 14 exclude the short $n$R exposure data from the
   $n$B total.  The data from 2010 Apr 03 exclude the short $n$R
  exposure data from the $n$B total.}
\tablenotetext{**}{
   In parentheses next to the slit width, we have indicated the
   following:  (FWHM of arc line He
   \textsc{i} $\lambda$4471, full width at 0.1 max of 
   arc line He \textsc{i} $\lambda$4471, and the FWHM of \Hg\ for a sample target
   spectrum) and units are in \AA.}
\tablenotetext{***}{Does not include two spectra that were found to 
   to have spurious flux values and does not include the four spectra at the beginning
   of the night that had 30 sec exposure times and saturated red flux
   values.}
\label{table:obslogAll}
\end{deluxetable}
\clearpage

\clearpage
\begin{deluxetable}{lcc}
\tabletypesize{\scriptsize}
\tablewidth{0pt}
\tablecaption{Line and Continuum Windows}
\tablehead{
\colhead{Line} &
\colhead{Line Integration Window [\AA]} &
\colhead{Continuum Region [\AA]}
}
\startdata
H$\alpha$ & $6520-6610$ & $6465-6484, 6610-6665$ \\
H$\beta$ & $4828-4898$ & $4813-4825$, $4895-4912$ \\
H$\gamma$ & $4310-4370$ & $4260-4305$, $4375-4420$\\
H$\gamma$ (DAO) & $4321-4361$ & $4260-4305$, $4375-4420$\\
H$\delta$ & $4065-4135$ & $4040-4064$, $4137-4167$ \\
H$\delta$ (DAO) & $4084-4122$ & $4040-4064$, $4137-4167$ \\
H$\epsilon + $ Ca \textsc{ii} H & $3944-3997$ & $3914-3923$, $4000-4017$ \\
Ca \textsc{ii} K & $3923-3945$ & $3914-3923$, $3946-3955$ \\
He \textsc{i} $\lambda$4471 (1.5'' slit) & $4464.6-4481$ & $4380-4461$, $4488-4554$\\
He \textsc{i} $\lambda$4471 (5'' slit) & $4460-4495$ & $4380-4461$,
$4488-4554$ \\
\hline
Continuum Measure & Spectral Region [\AA] &  \\
\hline
C$3615$ & $3600-3630$ &\\
C$4170 $ & $4155-4185$ &\\
C$4500$ & $4490-4520$ &\\
C$6010$ & $5990-6030$ &
\enddata
\label{table:linewindows}
\end{deluxetable}

\clearpage

\begin{deluxetable}{lcl}
\tabletypesize{\scriptsize}
\tablewidth{0pt}
\tablecaption{Continuum Windows for Blackbody and Line Fitting}
\tablehead{
\colhead{Name} &
\colhead{Window [\AA]} &
\colhead{Comments} }
\startdata
BW1 & $4000-4015$ & $n\lambda=8$ \\
BW2 & $4040-4057$ &  $n\lambda=9$ \\
BW3 & $4152-4210$ &   $n\lambda=31$, weighted by 0.97 \\
BW4 & $4495-4520$ & $n\lambda=13$ \\
BW5 & $4562-4630$ & $n\lambda=37$,  line at 4584\AA\ in IF1 decay \\
BW6 & $4731-4800$ & $n\lambda=37$, not used for fitting spectra from
DAO \\
\hline
RW1 & $5920-6120$ & $n\lambda=71$ \\
RW2 & $6360-6470$ & $n\lambda=39$ \\
RW3 & $6705-6845$ & $n\lambda=50$ 
\enddata
\label{table:bbwindows}
\tablecomments{For the Great Flare (IF0), the
  higher spectral resolution data (and limited wavelength range) allowed
  fitting in BW1, BW2, BW3 and also in the windows,
  $3915-3922$\AA\, $4242-4287$\AA, and $4400-4420$\AA.  $n\lambda$
  is the number of wavelength points within each spectral window.}
\end{deluxetable}
\clearpage

\begin{deluxetable}{lcc}
\tabletypesize{\scriptsize}
\tablewidth{0pt}
\tablecaption{Peak and Gradual Decay Phase Times}
\tablehead{
\colhead{Flare ID} &
\colhead{ Peak Time [Hr] (S\#)} &
\colhead{ Decay Time [Hr] (S\#)} 
}
\startdata
IF0  & 4.8172, 4.9550 (36, 40) & 5.0825 (45) \\
IF1  & 2.1441 (113) & 1.4576 (24) \\
IF2  & 8.6540 (542) & 8.6747 (545) \\
IF3  & 2.1500 (31) &   2.4512 (87) \\
IF4  & 10.4686 (665) & 10.5690 (674) \\
IF5  & 6.1929 (516) & 6.2205 (520) \\
IF6  & 4.7934 (261) & 4.8300 (264) \\
IF7  & 2.6245 (19) & 2.6785 (22) \\
IF8  & 11.6274 (390)  &  11.6506 (392) \\
IF9  & 3.9120 (121) & 3.9508 (126) \\
IF10  & 4.9070 (31)  & 4.9828 (32), 5.1431 (34) \\
\hline
HF1  & 8.4909 (521) & 8.5455 (527) \\
HF2  & 4.6452 (106) & 4.7891 (114) \\
HF3  & 9.3138 (234) & 9.3658 (241) \\
HF4  & 8.7889 (550) &  8.8672 (557) \\
\hline
GF1 & 2.1265 (69) & 2.3194 (101) \\
GF2 & 4.8504 (119) & 4.9158 (127) \\
GF3 & 2.2317 (34) & 2.3140 (43) \\
GF4 & 12.0746 (316) & 12.1062 (323) \\
GF5 & 6.3106 (332\,--\,335) & 6.3769 (336\,--\,345) 
\enddata
\tablecomments{The time and spectrum number for the peak and
  gradual decay phases.  The times are given as hours on the MJD
  of the event.  Note that
  three spectra around the gradual decay time listed above are
    averaged for the gradual decay phase values in the figures and
    text (the spectra numbers and times before and after are not
    given in the table).  The times for IF1
  are given in ``elapsed hours from flare start'' as presented in
  \cite{Kowalski2010}; see note on times in Section \ref{sec:spectra_param}. }
\label{table:times}
\end{deluxetable}

\clearpage
\begin{deluxetable}{llllllllll}
\tabletypesize{\scriptsize}
\tablewidth{0pt}
\rotate
\tablecaption{Flare Summary Table (1):  Key Photometry Properties}
\tablehead{
\colhead{ID} &
\colhead{Star} &
\colhead{Date} &
\colhead{$t_{peak}$***} &
\colhead{$I_{f, U, \mathrm{peak}}+1$} &
\colhead{ED [s]} &
\colhead{$E_U$ [$10^{32}$ ergs]} &
\colhead{$L_{U, peak}$ [$10^{29}$ ergs s$^{-1}$]} &
\colhead{$t_{1/2,U}$ [min]} &
\colhead{$\mathcal{I}$}
}
\startdata
IF1\tablenotemark{\dag} & YZ CMi & 2009 Jan 16 & 2.1441 & 208 & $93\,690$ & 38& 84.63 & 7.73 & 27.0 \\
IF2 & YZ CMi & 2011 Feb 08 &  8.6540 & 12.2 &690&6.2& 4.50 & 0.48 & 23.3 \\
IF3 & YZ CMi & 2011 Feb 24 &  2.1500 & 78.2 & $45\,810$ &18.5& 29.98 & 3.82 & 20.2 \\
IF4 & EQ Peg A & 2008 Oct 01 &  10.4686 & 21.4&4710&3.7& 16.04 & 2.46 & 8.3 \\
IF5 & EV Lac & 2010 Oct 11 & 6.1929 & 5.4 & 460 &0.28&2.7 & 0.84 & 5.2 \\
IF6 & EV Lac & 2009 Oct 10 & 4.7934 & 2.6 & 70 &0.04& 0.97 &0.32 & 5 \\
IF7 & YZ CMi & 2010 Feb 14 & 2.6245 & 6.4 & 810 &0.33&2.19 & 1.18 & 4.6 \\
IF8 & YZ CMi & 2010 Dec 11 & 11.6274 & 2.9 & 130 &0.05& 0.78 &0.69 & 2.8 \\
IF9\tablenotemark{*} & AD Leo & 2010 Apr 03 & 3.9120 & 4.4 & 1300 &2-4 & 5.11  & 1.47  & 2.3 \\
\hline
HF1 & YZ CMi & 2011 Feb 08 & 8.4909 & 5.3 & 920 &0.37& 1.72 & 2.43 & 1.8 \\
HF2\tablenotemark{*} & GJ 1243  & 2011 Oct 20 &  4.6452 & 5.4 & 500 &
0.86 & 1.29 & 4.19 & 0.93 \\
HF3 & YZ CMi & 2010 Dec 13 & 9.3138 & 3.2 & 440 &0.18& 0.88 & 2.84 & 0.77 \\
HF4 & EQ Peg A & 2008 Oct 01 & 8.7889 & 2.4 &530& 0.41& 1.11 & 2.40 &0.58 \\
\hline
GF1 & EV Lac & 2010 Oct 11 & 2.1265 & 8.04 &$10\,180$ &6.20& 4.29 & 13.02 & 0.54 \\
GF2 & YZ CMi & 2011 Mar 02 & 4.8504 & 2.1 & $2020$ &0.81& 0.43 & 3.68& 0.3 \\
GF3 & EV Lac & 2009 Oct 10 & 2.4744 & 1.87 & 570 & 0.35 &0.38 & 2.88 & 0.3 \\
GF4 & AD Leo & 2011 Feb 08 & 12.0746 & 1.32 & 80 &0.12& 0.48 & 2.13 &0.15 \\
GF5\tablenotemark{*} & YZ CMi & 2009 Jan 26 & 6.3106 & 1.3-1.5 & 170 & 0.36& 0.13 & 21 & 0.02 \\
\hline
IF0 & AD Leo & 1985 April 12 & 4.8172, 4.9550 &  70.2
(43\tablenotemark{**}) & $50\,000$ & 79 & 88.06 & 3.53 & 19.6\\
IF10 & EV Lac & 2009 Oct 27 & 4.9069 & 45.4 (9\tablenotemark{**}) &
8720 &5.3 & 27.0 & 0.46& 96 \\
\enddata
\label{table:flare_summary}
\tablenotetext{*}{$U$-band properties were estimated from spectra
  because no $U$-band data were obtained. The $t_{1/2}$ for GF5 estimated by smoothing the lightcurve over
  three spectra.  The energies for HF2 are lower limits because the
  observations ended before the end of the gradual phase. }
\tablenotetext{**}{Estimated from spectra with longer integration
  time than the photometry.}
\tablenotetext{***}{Time in hours from the beginning of the MJD obtained from spectra using the mid exposure of maximum
  C3615, except for IF1 (see note on times in Section \ref{sec:spectra_param}).}
\tablenotetext{\dag}{All properties pertain to spectral observation window except peak
  amplitude of $U$, $t_{1/2,U}$, and $\mathcal{I}$.}
\end{deluxetable}
\clearpage

\begin{deluxetable}{l|ll|ll|ll|l|cccc}
\tabletypesize{\scriptsize}
\tablewidth{0pt}
\rotate
\tablecaption{Flare Summary Table (2):  Key Continuum Observables and
  Derived Properties from
Spectra}
\tablehead{
\colhead{ID} &
\multicolumn{2}{c}{C4170} &
\colhead{$\chi_{\mathrm{flare,peak}}$} &
\colhead{$\chi_{\mathrm{flare,decay}}$} &
\multicolumn{2}{c}{peak} & 
\multicolumn{1}{c}{decay} &
 \multicolumn{4}{c}{NUV Slope ($m_{\mathrm{NUV}}$)\dag}
}
\startdata 
   & $I_{f,\mathrm{peak}}+1$ & $t_{1/2}$ [min] & & & \TBB [K] & \XBB [\%] & \TBB [K] & Peak & Decay & Peak \,--\, BB & Decay \,--\, BB \\
IF1\tablenotemark{*} & 11.5 (6) & 8.8 & 1.51 (0.08) & 2.65 (0.20) & $10\,800$ & 0.207 & 8300 &   0.1 (3.5) &   1.5 (4.6) &  14.9 (18.4) &    4.1 (10.6) \\
IF2 & 4.1 & 0.6 & 1.74 (0.04) & 2.67 (0.18) & $14\,100$ & 0.029&  7800 &   -5.9 (0.0) &   4.8 (1.2) &  -6.4 (0.3) &    8.3 (2.1) \\
IF3 & 20.1 & 3.6 & 1.61 (0.01) & 2.66 (0.16) & $12\,100$ & 0.269 &8000 &-4.1 (0.1) &   3.0 (8.1) &  -4.2 (0.5) &    4.7 (13.4) \\
IF4  & 7.3  & 1.1 & 1.27 (0.03) & 2.28 (0.09) & $11\,200$ & 0.188 &9300  &   -2.5 (0.1) &  -6.9 (1.0) &  34.5 (2.7) &  -10.8 (1.3)\\
IF5  & 1.8  & 0.5 & 2.22 (0.11) & \nodata & $12\,000$ & 0.013 & 8600  &   -0.5 (0.3) &  -0.3 (7.7) &   5.0 (0.8) &    6.8 (8.7)\\
IF6 & 1.2  & 0.9 & 2.19 (0.20) & \nodata & \nodata & \nodata & \nodata &   -4.1 (1.4) &  13.4 (9.1) &   1.3 (4.5) &   20.2 (15.2)\\
IF7  & 2.7  & 1.3 &  1.74 (0.05) & 2.40 (0.19) & $11\,300$ & 0.030 & 7600 &   -1.9 (0.3) &   6.3 (2.0) &   2.8 (1.1) &    9.1 (3.3)\\
IF8  & 2.9  & 1.0 & 1.62 (0.11) & \nodata & $13\,000$ & 0.006 & \nodata   &   -3.3 (0.8) &  40.7 (4.4) &   3.9 (4.3) &   44.3 (11.4) \\
IF9   & 1.8  & 1.0 & 1.77 (0.05) & 2.74 (0.12) & $10\,300$ & 0.046 & 7800 &   -4.1 (0.0) &   3.0 (1.4) &  -7.0 (0.0) & 4.9 (2.4)\\
\hline
HF1 & 1.8  & 2.4 & 2.33 (0.11) & 3.60 (0.51) & $10\,700$ & 0.017 & $10\,000$&   -0.8 (0.1) &   1.2 (8.9) &   1.9 (0.2) &   -4.7 (18.0) \\
HF2 & 1.9  & 3.5 &  2.58 (0.13) & 3.40 (0.38) & $9500$ & 0.018 & 7300 &   -1.5 (1.7) &   4.1 (3.7) &  -1.2 (3.1) &    7.3 (3.9)\\
HF3 & 1.4 & 1.4 & 2.83 (0.23) & 3.20 (0.46) & $12\,000$ & 0.005 & $10\,000$&   -0.3 (0.2) &   9.3 (3.7) &   3.1 (0.5) &   15.8 (3.5)\\
HF4 & 1.3 & 1.5 & 2.97 (0.24)  & \nodata & $12\,000$ & 0.008 &\nodata    &   -8.1 (0.3) &  -8.3 (1.8) & -11.2 (0.6) &   -9.2 (4.0)\\
\hline
GF1  & 1.9  & 15.0 & 3.17 (0.12) &3.46 (0.23) & $8900$ & 0.045 & 7800 &    2.1 (1.9) &   4.6 (0.9) &   3.9 (3.0) &    5.9 (1.3) \\
GF2 & 1.15  & 0.7 &  4.32 (0.62) & \nodata & \nodata & \nodata &\nodata &    1.9 (1.7) &  13.8 (1.5) &   1.4 (2.1) &    8.1 (13.5)\\
GF3 & 1.87 & 2.2 & 2.99 (0.31) & \nodata & $12\,700$ & 0.003 & \nodata  &    1.1 (2.6) &  18.1 (12.4) &   5.9 (4.8) &   22.6 (9.5)\\
GF4 & 1.03  & 0.5 & \nodata & \nodata & \nodata & \nodata & \nodata &    5.1 (6.3) &  19.0 (9.0) &  11.1 (10.1) &   17.2 (5.7)\\
GF5  & 1.04 &0.9 & 4.28 (1.29) & 4.26 (1.63) & 6700 & 0.006 & 4900 &   11.0 (8.8) &  -6.1 (4.0) &  12.7 (10.3) &  -29.9 (10.6) \\
\hline
IF0\tablenotemark{**} & 8.7 & 11.7 & 1.8 (0.08)& 2.4 (0.8) & $11\,600$&  0.20& 5600 & -9.2 (3.2)&  15.8 (9.8) & -19.8 (9.7)&  9.7 (31.3)\\
                                   &      &           & 1.4 (0.12)  & &9800&0.44& &8.1 (1.4) &&52 (7) & \\
IF10\tablenotemark{***} & 1.3 & \nodata & 1.7 (0.12) & 2.4 (0.2) &
$12\,200$ & 0.020 & 6500  &   -11.8 (0.5) &  0.0 (6.7) & -32.7 (2.3) &   -5.0 (11.5)
\enddata
\tablecomments{
 Errors on \chif\
 values are given in parentheses.  No values are given for those with
 large errors (\s100\%) on \chif. The units of \XBB\ are \% of the
 visible stellar hemisphere; these values are derived from a Planck
 function and are not corrected using Appendix \ref{sec:appendix_CK}.  }
\tablenotetext{*}{The value of $I_{f, \mathrm{C4170}, \mathrm{peak}}+1$ for IF1
  is obtained
  at S\#113 (at the peak of MDSF2), but this value includes gradual
  decay phase emission from IF1; the value for
  MDSF2 obtained by subtracting the gradual decay emission at S\#102 is given in
  parentheses. The $t_{1/2, \mathrm{C4170}}$ of IF1 is given for
  MDSF2 after subtracting the emission from S\#102.  The values for \chif\
  are given for the total emission at S\#113 and S\#124 for the peak
  and decay, respectively.}
\tablenotetext{**}{Two values for IF0 were calculated at
the first (S\#36; top row) and
second (S\#40; bottom row) peaks of C3615, respectively, during the event. }
\tablenotetext{***}{Due to the low cadence of the spectra during IF10, a value of 
$t_{1/2,C4170}$ could not be determined. }
\tablenotetext{\dag}{NUV slopes from $\lambda=3420-3630$\AA; errors are
  given in parentheses.  The 
spectra are normalized by C3615 before calculating slopes.  The
fractional change in flux at a given wavelength, $\lambda$, in the Balmer continuum
relative to the flux in C3615 is given by
$\frac{F_{\lambda}-F_{\lambda=3615}}{F_{\lambda=3615}} = (\lambda-3615)
\times m \times 10^{-4}$.
Positive values of $m_{\mathrm{NUV}}$ indicate red slopes, negative values of $m_{\mathrm{NUV}}$
indicate blue slopes.  $m_{\mathrm{NUV}}$ is essentially the
``fractional change in flux per \AA\ relative to the flux C3615''.}
\label{table:chitable}
\end{deluxetable}

\clearpage

\begin{deluxetable}{l|cc|cccc|cc|cccc}
\tabletypesize{\scriptsize}
\tablewidth{0pt}
\rotate
\tablecaption{Hydrogen Balmer (HB) Properties}

\tablehead{
\colhead{Flare ID} &
\multicolumn{2}{c}{HB flux / Total flux} &
\multicolumn{1}{c}{\Hd\ / Total flux} &
\multicolumn{1}{c}{\Hg\ / Total flux} &
\multicolumn{1}{c}{\Hb\ / Total flux} &
\multicolumn{1}{c}{\Ha\ / Total flux} &
\multicolumn{2}{c}{BaC3615 / C3615} &
}
\startdata
& Peak & Decay & Peak & Peak & Peak & Peak & Peak & Decay \\
 
IF0\tablenotemark{***}  & 0.19, 0.12 & 0.44 & 0.017, 0.016 & 0.016, 0.018 &\nodata & \nodata & 0.34, 0.19& 0.66 \\
 IF1  & 0.15 & 0.34 & 0.015 & 0.020 & 0.030 & \nodata & 0.21 & 0.59 \\
 IF2  & 0.13 & 0.32 & 0.009 & 0.009 & 0.007 & 0.005 & 0.27 & 0.58 \\
 IF3  & 0.11 & 0.37 & 0.008 & 0.008 & 0.009 & 0.008 &  0.24  & 0.59 \\
 IF4  & 0.03 & 0.29 & 0.003 & 0.003 & 0.005  & \nodata & 0.05 &  0.51 \\
 IF5  & 0.24 & 0.52 & 0.018 & 0.020 & 0.020 & 0.017 &  0.44 &  0.71\\
 IF6  & 0.22 & 0.42 & 0.017 & 0.018 & 0.018 & \nodata & 0.39 & 0.83\\
 IF7  & 0.15 & 0.32 & 0.011 & 0.012 & 0.013 & 0.010 &  0.31  & 0.57  \\
 IF8  & 0.12 & \nodata\tablenotemark{**} & 0.008 & 0.009 & 0.008 &
 0.0005 &  0.24 &  \nodata \\
 IF9  & 0.17 & 0.34 & 0.012 & 0.012 & 0.012 & 0.009 &  0.33 & 0.59\\
 IF10\tablenotemark{***}  & 0.15 & 0.43 & 0.017 & 0.020 & \nodata &
 \nodata & 0.28 &  0.61\tablenotemark{\ddag} \\
 \hline
 HF1  & 0.24\tablenotemark{*} & 0.44 & 0.021 & 0.022 & 0.022 & 0.016 & 0.49 & 0.67 \\
 HF2  & 0.31 & 0.48 & 0.023 & 0.025 & 0.029 & 0.027 & 0.54 & 0.66  \\
 HF3  & 0.35 & 0.42 & 0.026 & 0.026 & 0.026 & 0.022 & 0.57  & 0.62  \\
 HF4  & 0.35 & 0.56 & 0.024 & 0.026 & 0.023  & \nodata & 0.57 & 0.89  \\
 \hline
 GF1 & 0.38 & 0.43 & 0.030 & 0.031 & 0.032 & 0.027 & 0.64 & 0.69   \\
 GF2 & 0.48 & 0.64 & 0.035 & 0.039 & 0.046 & 0.054 & 0.78 & 1.0  \\
 GF3 & 0.35 & 0.50 & 0.030 & 0.030 & 0.030 &  \nodata & 0.58 & 1.0\\
 GF4 & 0.51 & 0.64 & 0.044 & 0.057 & 0.065 & 0.068 & 0.64 & 0.90  \\
 GF5 & 0.51 & 0.51 & 0.057 & 0.066 & 0.088 & 0.100 & 0.74 & 0.80 
\enddata
\tablecomments{All values in the table are given as fractions,
  although they are often discussed as percentages in the text. }
\tablenotetext{*}{Peak in C4170 one spectrum earlier, so averaged
  S\#520\,--\,521. }
  \tablenotetext{**}{There is no gradual decay phase for IF8 (or it is at a very
  low level of flux). } 
 \tablenotetext{***}{The flares IF0 and IF10 have a narrower wavelength coverage
  because the data were obtained with different instruments than DIS.  The IF0 data
  do not include continuum flux at $\lambda > 4440$\AA\ and \Hb, and
  the IF10 data do not include continuum flux at $\lambda > 4700$\AA\
  and \Hb.  Peak values for IF0 are given for first (S\#36) and second
  (S\#40) maxima
  in the flare, respectively. }
 \tablenotetext{\ddag}{Decay value for IF10 given for spectrum immediately following the
  peak spectrum (S\#32); the value for S\#34 is 0.69.}
\label{table:hb_phasez}
\end{deluxetable}
\clearpage


\begin{deluxetable}{lccc}
\tabletypesize{\scriptsize}
\tablewidth{0pt}
\tablecaption{Color temperature diagnostics of MDSF2}

\tablehead{
\colhead{Spectrum \#} &
\colhead{Method 1} &
\colhead{Method 2} &
\colhead{Method 3} 
}
\startdata
$t=0$ min; (S\#101$+$S\#102$+$S\#103) / 3 & $8\,400$ & \nodata & \nodata \\
$t=2.7$ min; (S\# 108$+$S\#109) / 2 & $9\,500$  &  $14\,900$ & $13\,100$  \\
$t=4.8$ min; S\#113 & $10\,800$  & $17\,700$ & $15\,000$ \\ 
\enddata
\label{table:color_astar}
\tablecomments{See Section \ref{sec:astar} for a description of the methods.}
\end{deluxetable}

\begin{deluxetable}{l|cc|cc}
\tabletypesize{\scriptsize}
\tablewidth{0pt}
\tablecaption{RHD Model Predictions}

\tablehead{
\colhead{Model} &
\colhead{\TBB\ [K]} & 
\colhead{\chifp} &
\multicolumn{2}{c}{NUV Slope ($m_{\mathrm{NUV}}$)}
}
\startdata
 & & & Peak & Peak \,--\, BB \\
F10 ($t=$ 230 s) & 5000 & 4.8 & 7.3 & 5.5 \\
F11 ($t=$ 15.9 s) & 5300 & 8.2 & 5.4 & 5.1 \\
\enddata
\label{table:model_table}
\end{deluxetable}

\clearpage

\begin{deluxetable}{|p{2.35in}|p{2.35in}|p{2.35in}|}
\tabletypesize{\scriptsize}
\rotate
\tablewidth{0pt}
\tablecaption{Summary Table}
\tablehead{
\colhead{Observed Property}  &
\colhead{Physical Meaning} &
\colhead{Modeling Goal} 
}
\startdata
IF events have lower \chifp\ ($\lesssim 2.2$) than HF or
GF events (\chifp $\gtrsim 2.3$; Section \ref{sec:general}); IF events
have smaller BaC3615/C3615 values but over a relatively large range (\s0.2\,--\,0.35; Figure
\ref{fig:master0}). & There is less Balmer
continuum radiation relative to the faster evolving $T \sim 10^4$ K blackbody radiation 
at peak in the flares with faster impulsive phases (IF events); slower impulsive
phases (HF and GF events) have relatively larger amounts of Balmer
emission which evolves slower than the hot, blackbody component. & \chifp\ of
1.6\,--\,1.8;  BaC3615/C3615 \s 0.25 at peak. \\
\hline
There is a large spread of \Hg/C4170 at peak in the flares:
\s6\,--\,165, which is correlated with the type of flare (IF, HF, or
GF) and with \chifp\ (Section \ref{sec:general}).  The fraction of
total flare flux at peak in an individual Balmer line is also
correlated with the type of flare (Section \ref{sec:hydrogen}). & \chif\ is a diagnostic of the relative amount
of Balmer line and
continuum radiation compared to $T \sim 10^4$ K blackbody radiation.
The Balmer lines follow the strength of  the BaC radiation. &
\Hg\ line flux/C4170 \s 15\,--\,25 at peak; \Hg\ line flux/total flare
flux is $<1-2$\% at peak. \\ 
\hline
The percentage of Hydrogen Balmer emission is anti-correlated with the
$U$-band time-evolution (Section \ref{sec:hydrogen}). & The blackbody
continuum is relatively strongest in the
impulsive phase but decays quickly.  The Balmer emission, although
bright at peak, decays less quickly than the blackbody.  This explains the small Balmer
jumps at peak, and larger Balmer jumps in the gradual decay phase. & At peak, the percentage of flare flux emitted in the Hydrogen Balmer
emission from $\lambda=3420-5200$\AA\ is \s15\%.  In the gradual
decay phase, it increases to \s35\%. \\
\hline
The timescale of the Balmer continuum is about twice as fast as the
timescale of \Hg\ for single-peaked events ($t_{1/2, \mathrm{H}\gamma} = 1.5-20$
minutes; Section \ref{sec:Hgamma}). & The Balmer continuum
emission originates from different atmospheric conditions with shorter cooling
timescales than the \Hg\ line.  & Reproduce BaC and Balmer line timescales. \\

\enddata
\label{table:final_table}
\end{deluxetable}

\begin{deluxetable}{|p{2.35in}|p{2.35in}|p{2.35in}|}
\tablenum{11}
\tabletypesize{\scriptsize}
\rotate
\tablewidth{0pt}
\tablecaption{Summary Table, cont...}
\tablehead{
\colhead{Observed Property}  &
\colhead{Physical Meaning} &
\colhead{Modeling Goal} 
}
\startdata
For the simple high-energy, impulsive flares (e.g., IF3 and IF9), the Balmer line emission
and blackbody flux have peaks that occur within 1 minute of each another; 
flares with complex morphology (e.g., double-peaked impulsive phases)
show larger separations in \Hg\ and blackbody peaks ($\Delta t
\gtrsim3$ minutes; Section \ref{sec:Hgamma}).  & Balmer
line emission and the blackbody continuum are produced from a similar heating source
during the rise phase; the late peaks of the Balmer lines may result
from the superposition of two separate flares. & Tens of seconds ($<$ 1
minute) lag
between blackbody peak (first) and \Hg\ peak (second) for simple, high energy events. \\
\hline

For the classical, IF events of relatively large energy (IF3, IF9), there is a linear \emph{time-decrement}
relation between the timescale and wavelength of the Balmer line
transition (Section \ref{sec:timing}): higher order transitions (e.g.,
those in the PseudoC and the BaC) evolve faster. & Balmer line evolution is scaled by
a multiplicative factor during simple, large flares. & Reproduce
time-decrement Balmer line behavior. \\
\hline
The blackbody flux is the fastest evolving spectral component (Section
\ref{sec:timing}).  For the impulsive and hybrid events,
$t_{1/2,\mathrm{C}4170} \sim $0.5\,--\,3.5
 minutes (but can be much larger, \s15 minutes, when it is
similar to the evolution of the Hydrogen Balmer emission). & 
The heating threshold required to produce and sustain the blackbody
continuum flux must be higher than the threshold for the Balmer line
and continuum emission. The blackbody flux likely 
originates from higher densities with shorter cooling timescales than
the Balmer line and continuum emission. & $t_{1/2, \mathrm{C}4170}$
0.5\,--\,3.5 minutes. \\
\hline

The cumulative integral of the C4170 (blackbody continuum flux) is
similar to the Ca \textsc{ii} K line flux until the time of maximum Ca
\textsc{ii} K line flux (Section \ref{sec:caiik}); delays and decreases in the Ca \textsc{ii} K flux 
are observed during the early impulsive phase of some flares.  & The cumulative blackbody emission
is tied to the evaporation of the lower atmosphere to coronal
temperatures, producing enhanced XEUV radiation that heats a large
area of neighboring chromospheric gas to the temperature of Ca \textsc{ii} K
production. & Reproduce relation which holds well for IF6, IF7, IF9, HF1, HF2, GF1, and
GF2. \\

\enddata
\end{deluxetable}

\begin{deluxetable}{|p{2.35in}|p{2.35in}|p{2.35in}|}
\tablenum{11}
\tabletypesize{\scriptsize}
\rotate
\tablewidth{0pt}
\tablecaption{Summary Table, cont...}
\tablehead{
\colhead{Observed Property}  &
\colhead{Physical Meaning} &
\colhead{Modeling Goal} 
}
\startdata
The peak emission in the blue-optical continuum
($\lambda=$4000\,--\,4800\AA) is well-represented by
 a blackbody with temperatures between 
9000 K and 14\,000 K; applies over a range of peak amplitudes and 
morphological type (Section \ref{sec:bbpeak}). & A large amount of heating
occurs at high densities during dMe flares. This is a major
shortcoming of solar-type non-thermal
electron beam heating model predictions. & \TBB$=10\,000-12\,000$ K (Section
\ref{sec:bbpeak}) and
$F_{\mathrm{BB, Bol}} \sim 2-4\times10^{11}$\ergscm\ (Section
  \ref{sec:fillingfactors}). \\
\hline
Interflare variations of \TBB\ are significant (Appendix
\ref{sec:stack}, Section \ref{sec:astar}). & Temperature variations 
between peaks of different flares provide important constraints on the heating
mechanism parameters & Full range of color temperatures at peak is
6700 K (GF5) to $15\,000$ K (MDSF2). \\
\hline
Other continua are needed to match the jump in flux from
$\lambda=4000$\AA\ to $\lambda<3600$\AA:  the pseudo-continuum of
blended Hydrogen lines (PseudoC) and the Balmer
continuum emission (BaC).  At peak, the spectral shape of the BaC is similar to
the hot blackbody shape (Section \ref{sec:peak_bac}). & Some continuum
emission in the NUV is due to Hydrogen recombination from the
mid-to-upper flare
chromosphere, but the total
emission in the NUV is dominated by the shape of the hot blackbody. &
At peak, the shape of the NUV flux ($\lambda < 3630$\AA) is similar to
that of a blackbody with \TBB \s10,000 K. \\
\hline
Some flares show direct evidence of Hydrogen Balmer line and continuum \emph{absorption} & 
The white-light continuum is formed from heating at high densities, within a non-uniform temperature
stratification where Hydrogen Balmer and Paschen continuum opacities
are large, as in a hot star like Vega. & C3615 $<$ C4170 for \TBB
\s$13\,000$\,--\,$15\,000$ K during slow, secondary flares with a $t_{1/2,\mathrm{C}4170}$ of nearly 9 minutes. \\
\enddata
\end{deluxetable}

\begin{deluxetable}{|p{2.35in}|p{2.35in}|p{2.35in}|}
\tablenum{11}
\tabletypesize{\scriptsize}
\rotate
\tablewidth{0pt}
\tablecaption{Summary Table, cont...}
\tablehead{
\colhead{Observed Property}  &
\colhead{Physical Meaning} &
\colhead{Modeling Goal} 
}
\startdata

Flux depressions in the Hydrogen Balmer wings are detected at $\sim
\pm$10\,--\,30\AA\ from line center at the peaks of some flares. & Extreme
heating at peak produces Balmer absorption, which adds to
the Balmer emission originating elsewhere in the atmosphere -- at different heights
or in distant regions on the surface. & Narrower line widths at maximum continuum compared
to maximum line flux, with the wing flux depressions increasing from
\Hb\ to \Hd.  Flares with very small Balmer jump
ratios of \chifp\s 1.5 typically show the most conspicuous wing
depressions; these flares are either large amplitude or have large
values of \TBB.\\
\hline

The gradual decay phase begins at the point when the flux in Hydrogen
Balmer components (as a percentage of total flare flux) has increased by \s20\,--\,30\% compared to peak.  Also, the Balmer jump ratio (\chif)
becomes larger and the shape of the RHD model
Balmer continuum spectrum is consistent with the observed BaC spectral slope (Section \ref{sec:gradspec}, \ref{sec:modelcomparison}). & Hydrogen recombination from the
mid-to-upper flare chromosphere becomes more important as a radiative
cooling process in the gradual
decay phase; heating from a solar-type nonthermal
electron beam is likely present in the gradual phase. & \TBB\ is 8000 K at the beginning of the
gradual decay phase. \\
\hline
In the gradual decay phase, the specific luminosities of C4170 and BaC3615 increase in equal
percentages between flares of different sizes (Section \ref{sec:gradspec}). & The production of Balmer and
blackbody continua may be driven by a similar heating mechanism
(solar-type nonthermal electrons) during
the gradual decay phase; however, RHD models of solar-type nonthermal
electron beams have not produced the
\TBB \s 8000 K component. & $\mathrm{log_{10}} \mathcal{L}_{\mathrm{BaC3615}} = -0.63(\pm0.86) + 1.03(\pm0.03) \times
\mathrm{log_{10}} \mathcal{L}_{\mathrm{C4170}}$ at the beginning of
the gradual decay phase when \TBB \s 8000 K.\\

\enddata
\end{deluxetable}

\begin{deluxetable}{|p{2.35in}|p{2.35in}|p{2.35in}|}
\tablenum{11}
\tabletypesize{\scriptsize}
\rotate
\tablewidth{0pt}
\tablecaption{Summary Table, cont...}
\tablehead{
\colhead{Observed Property}  &
\colhead{Physical Meaning} &
\colhead{Modeling Goal} 
}
\startdata
The best fit blackbody to the blue-optical zone does not account for
all the continuum flux at redder wavelengths ($\lambda > 4900$\AA)
during some flares.  This
``Conundruum'' exhibits a color temperature in the red-optical zone of 3500\,--\,5500 K
and is more important in the gradual decay phase than in the peak
phase (Section \ref{sec:conundruum}). & RHD model spectra are
consistent with this red continuum; higher order Hydrogen recombination
(e.g., Paschen, Brackett) and
possibly increased photospheric (H$^-$) emission from UV backwarming 
are likely important contributions to the red flare flux in the
gradual decay phase. & $T_{\mathrm{red}}<$\TBB K in peak and decay
phases,
$T_{\mathrm{red}}\lesssim5500$ K in gradual decay phase.\\ 
\hline

The rise phase of flares (e.g., MDSF2, IF3) features an increase in
\TBB \s 2000 K. & Areal increase is the dominant effect during the rise
phase; a narrow
  range of interflare and intraflare temperature variation over a
  large range of $U$-band luminosities could be
  due to a temperature threshold that is reached during flare
  heating (see Section \ref{sec:temp_meaning}); heating threshold is reached
relatively early in the flare.  & $\Delta$\TBB \s 2000 K during rise;
\TBB\ independent of peak amplitude over \s 2 orders of magnitude. \\
\hline
The speed of the leading edge of a circular magnetic ``footpoint'' of
\TBB \s $10^4$ K blackbody
emitting area is calculated; during classical high energy IF events
(IF3, IF9), large speeds of \s100 \kms\ occur
in the rise phase before decreasing to $\lesssim 10$ \kms in the initial fast decay (Section
\ref{sec:speeds}). & Fast speeds are
related to the rate at which individual, neighboring footpoints are
heated (Scenario 2), whereas slow
speeds indicate prolonged heating of a given footpoint or group of
footpoints (Scenario 1); a similar two-ribbon development process
(with parallel and perpendicular footpoint motions) on the Sun
and on dMe stars (see Section \ref{sec:solarstellar}). & Flare
models with spatial evolution. \\
\hline
A slow speed ($\lesssim 10$\kms) is observed in both the rise phase
(MDSF2) and initial fast decay phase (IF2, IF3, IF9, HF1) and
generates a range of color temperatures (Section
\ref{sec:speeds}). & There is a difference in
heating mechanism between the slowly increasing area and temperature
and slowly increasing area and quickly decreasing temperature. &
Flare models with spatial evolution. \\

\enddata
\end{deluxetable}

\clearpage